%% file: letter.tex
\documentclass[twocolumn,twocolappendix]{aastex63}
\usepackage{amsmath}
\usepackage{afterpage}
\received{}
\revised{}
\accepted{}

\usepackage{xr}

\usepackage[maxfloats=1000]{morefloats}
\submitjournal{}

\makeatletter
\newcommand*{\addFileDependency}[1]{
  \typeout{(#1)}
  \@addtofilelist{#1}
  \IfFileExists{#1}{}{\typeout{No file #1.}}
}
\makeatother

\newcommand*{\myexternaldocument}[1]{%
    \externaldocument{#1}%
    \addFileDependency{#1.tex}%
    \addFileDependency{#1.aux}%
}

\myexternaldocument{supplement}

\begin{document}

\title{An isolated mass gap black hole or neutron star detected with astrometric microlensing}

\author[0000-0002-6406-1924]{Casey Y. Lam}
\correspondingauthor{Casey Y. Lam} 
\email{casey$\_$lam@berkeley.edu}
\affiliation{University of California, Berkeley, Department of Astronomy, Berkeley, CA 94720}

\author[0000-0001-9611-0009]{Jessica R. Lu}
\affiliation{University of California, Berkeley, Department of Astronomy, Berkeley, CA 94720}

\author[0000-0001-5207-5619]{Andrzej Udalski}
\altaffiliation{OGLE collaboration}
\affiliation{Astronomical Observatory, University of Warsaw, Al.~Ujazdowskie~4,00-478~Warszawa, Poland}

\author{Ian Bond}
\altaffiliation{MOA collaboration}
\affiliation{School of Mathematical and Computational Sciences, Massey University, Private Bag 102-904 North Shore Mail Centre, Auckland 0745, New Zealand}

\author[0000-0001-8043-8413]{David P. Bennett}
\altaffiliation{MOA collaboration}
\affiliation{Code 667, NASA Goddard Space Flight Center, Greenbelt, MD 20771, USA}
\affiliation{Department of Astronomy, University of Maryland, College Park, MD 20742, USA}

\author[0000-0002-2335-1730]{Jan Skowron}
\altaffiliation{OGLE collaboration}
\affiliation{Astronomical Observatory, University of Warsaw, Al.~Ujazdowskie~4,00-478~Warszawa, Poland}

\author[0000-0001-7016-1692]{Przemek Mr{\'o}z}
\altaffiliation{OGLE collaboration}
\affiliation{Astronomical Observatory, University of Warsaw, Al.~Ujazdowskie~4,00-478~Warszawa, Poland}

\author[0000-0002-9245-6368]{Radek Poleski}
\altaffiliation{OGLE collaboration}
\affiliation{Astronomical Observatory, University of Warsaw, Al.~Ujazdowskie~4,00-478~Warszawa, Poland}

\author{Takahiro Sumi}
\altaffiliation{MOA collaboration}
\affiliation{Department of Earth and Space Science,
Graduate School of Science,
Osaka University, Toyonaka,
Osaka 560-0043, Japan}

\author[0000-0002-0548-8995]{Micha\l \,K. Szyma{\'n}ski}
\altaffiliation{OGLE collaboration}
\affiliation{Astronomical Observatory, University of Warsaw, Al.~Ujazdowskie~4,00-478~Warszawa, Poland}

\author[0000-0003-4084-880X]{Szymon Koz{\l}owski}
\altaffiliation{OGLE collaboration}
\affiliation{Astronomical Observatory, University of Warsaw, Al.~Ujazdowskie~4,00-478~Warszawa, Poland}

\author[0000-0002-2339-5899]{Pawe{\l} Pietrukowicz}
\altaffiliation{OGLE collaboration}
\affiliation{Astronomical Observatory, University of Warsaw, Al.~Ujazdowskie~4,00-478~Warszawa, Poland}

\author[0000-0002-7777-0842]{Igor Soszy{\'n}ski}
\altaffiliation{OGLE collaboration}
\affiliation{Astronomical Observatory, University of Warsaw, Al.~Ujazdowskie~4,00-478~Warszawa, Poland}

\author[0000-0001-6364-408X]{Krzysztof Ulaczyk}
\altaffiliation{OGLE collaboration}
\affiliation{Astronomical Observatory, University of Warsaw, Al.~Ujazdowskie~4,00-478~Warszawa, Poland}
\affiliation{Department of Physics, University of Warwick, Gibbet Hill Road,Coventry, CV4~7AL,~UK}

\author[0000-0002-9658-6151]{{\L}ukasz Wyrzykowski}
\altaffiliation{OGLE collaboration}
\affiliation{Astronomical Observatory, University of Warsaw, Al.~Ujazdowskie~4,00-478~Warszawa, Poland}

\author{Shota Miyazaki}
\altaffiliation{MOA collaboration}
\affiliation{Department of Earth and Space Science,
Graduate School of Science,
Osaka University, Toyonaka,
Osaka 560-0043, Japan}

\author{Daisuke Suzuki}
\altaffiliation{MOA collaboration}
\affiliation{Department of Earth and Space Science,
Graduate School of Science,
Osaka University, Toyonaka,
Osaka 560-0043, Japan}

\author{Naoki Koshimoto}
\altaffiliation{MOA collaboration}
\affiliation{Code 667, NASA Goddard Space Flight Center, Greenbelt, MD 20771, USA}
\affiliation{Department of Astronomy, University of Maryland, College Park, MD 20742, USA}
\affiliation{Department of Astronomy, Graduate School of Science, The University of Tokyo, 7-3-1 Hongo, Bunkyo-ku, Tokyo 113-0033, Japan}

\author[0000-0001-5069-319X]{Nicholas J. Rattenbury}
\altaffiliation{MOA collaboration}
\affiliation{Department of Physics, University of Auckland, Private Bag 92019, Auckland, New Zealand}

\author[0000-0003-2874-1196]{Matthew W. Hosek, Jr.}
\affiliation{University of California, Los Angeles, Department of Astronomy, Los Angeles, CA 90095, USA}

\author{Fumio Abe}
\altaffiliation{MOA collaboration}
\affiliation{Institute for Space-Earth Environmental Research, Nagoya University, Nagoya 464-8601, Japan}

\author{Richard Barry}
\altaffiliation{MOA collaboration}
\affiliation{Code 667, NASA Goddard Space Flight Center, Greenbelt, MD 20771, USA}

\author{Aparna Bhattacharya}
\altaffiliation{MOA collaboration}
\affiliation{Code 667, NASA Goddard Space Flight Center, Greenbelt, MD 20771, USA}
\affiliation{Department of Astronomy, University of Maryland, College Park, MD 20742, USA}

\author{Akihiko Fukui}
\altaffiliation{MOA collaboration}
\affiliation{Department of Earth and Planetary Science, Graduate School of Science, The University of Tokyo, 7-3-1 Hongo, Bunkyo-ku, Tokyo 113-0033, Japan}
\affiliation{Instituto de Astrof\'isica de Canarias, V\'ia L\'actea s/n, E-38205 La Laguna, Tenerife, Spain}

\author{Hirosane Fujii}
\altaffiliation{MOA collaboration}
\affiliation{Institute for Space-Earth Environmental Research, Nagoya University, Nagoya 464-8601, Japan}

\author{Yuki Hirao}
\altaffiliation{MOA collaboration}
\affiliation{Department of Earth and Space Science, Graduate School of Science, Osaka University, Toyonaka, Osaka 560-0043, Japan}

\author{Yoshitaka Itow}
\altaffiliation{MOA collaboration}
\affiliation{Institute for Space-Earth Environmental Research, Nagoya University, Nagoya 464-8601, Japan}

\author{Rintaro Kirikawa}
\altaffiliation{MOA collaboration}
\affiliation{Department of Earth and Space Science, Graduate School of Science, Osaka University, Toyonaka, Osaka 560-0043, Japan}

\author{Iona Kondo}
\altaffiliation{MOA collaboration}
\affiliation{Department of Earth and Space Science, Graduate School of Science, Osaka University, Toyonaka, Osaka 560-0043, Japan}

\author{Yutaka Matsubara}
\altaffiliation{MOA collaboration}
\affiliation{Institute for Space-Earth Environmental Research, Nagoya University, Nagoya 464-8601, Japan}

\author{Sho Matsumoto}
\altaffiliation{MOA collaboration}
\affiliation{Department of Earth and Space Science, Graduate School of Science, Osaka University, Toyonaka, Osaka 560-0043, Japan}

\author{Yasushi Muraki}
\altaffiliation{MOA collaboration}
\affiliation{Institute for Space-Earth Environmental Research, Nagoya University, Nagoya 464-8601, Japan}

\author{Greg Olmschenk}
\altaffiliation{MOA collaboration}
\affiliation{Code 667, NASA Goddard Space Flight Center, Greenbelt, MD 20771, USA}

\author[0000-0003-2388-4534]{Cl\'ement Ranc}
\altaffiliation{MOA collaboration}
\affiliation{Zentrum f{\"u}r Astronomie der Universit{\"a}t Heidelberg, Astronomiches Rechen-Institut, M{\"o}nchhofstr.\ 12-14, 69120 Heidelberg, Germany}

\author{Arisa Okamura}
\affiliation{Department of Earth and Space Science, Graduate School of Science, Osaka University, Toyonaka, Osaka 560-0043, Japan}

\author{Yuki Satoh}
\affiliation{Department of Earth and Space Science, Graduate School of Science, Osaka University, Toyonaka, Osaka 560-0043, Japan}

\author{Stela Ishitani Silva}
\altaffiliation{MOA collaboration}
\affiliation{Department of Physics, The Catholic University of America, Washington, DC 20064, USA}
\affiliation{Code 667, NASA Goddard Space Flight Center, Greenbelt, MD 20771, USA}

\author{Taiga Toda}
\altaffiliation{MOA collaboration}
\affiliation{Department of Earth and Space Science, Graduate School of Science, Osaka University, Toyonaka, Osaka 560-0043, Japan}

\author{Paul J. Tristram}
\altaffiliation{MOA collaboration}
\affiliation{University of Canterbury Mt.\ John Observatory, P.O. Box 56, Lake Tekapo 8770, New Zealand}

\author{Aikaterini Vandorou}
\altaffiliation{MOA collaboration}
\affiliation{Code 667, NASA Goddard Space Flight Center, Greenbelt, MD 20771, USA}
\affiliation{Department of Astronomy, University of Maryland, College Park, MD 20742, USA}

\author{Hibiki Yama}
\altaffiliation{MOA collaboration}
\affiliation{Department of Earth and Space Science, Graduate School of Science, Osaka University, Toyonaka, Osaka 560-0043, Japan}

\author[0000-0002-0287-3783]{Natasha S. Abrams}
\affiliation{University of California, Berkeley, Department of Astronomy, Berkeley, CA 94720}

\author[0000-0002-2350-4610]{Shrihan Agarwal}
\affiliation{University of California, Berkeley, Department of Astronomy, Berkeley, CA 94720}

\author[0000-0003-4725-4481]{Sam Rose}
\affiliation{University of California, Berkeley, Department of Astronomy, Berkeley, CA 94720}

\author[0000-0002-5029-3257]{Sean K. Terry}
\affiliation{University of California, Berkeley, Department of Astronomy, Berkeley, CA 94720}

\begin{abstract}

We present the analysis of five black hole candidates identified from gravitational microlensing surveys.
Hubble Space Telescope astrometric data and densely sampled lightcurves from ground-based microlensing surveys are fit with a single-source, single-lens microlensing model in order to measure the mass and luminosity of each lens and determine if it is a black hole.
One of the five targets (OGLE-2011-BLG-0462/MOA-2011-BLG-191 or OB110462 for short) shows a significant $>1$~mas coherent astrometric shift, little to no lens flux, and has an inferred lens mass of 1.6 - 4.4 $M_\odot$.
This makes OB110462 the first definitive discovery of a compact object through astrometric microlensing and it is most likely either a neutron star or a low-mass black hole.
This compact object lens is relatively nearby (0.70-1.92~kpc) and has a slow transverse motion of $<$30~km/s.
OB110462 shows significant tension between models well-fit to photometry vs.~astrometry, making it currently difficult to distinguish between a neutron star and a black hole.
Additional observations and modeling with more complex system geometries, such as binary sources are needed to resolve the puzzling nature of this object. 
For the remaining four candidates, the lens masses are $<2 M_\odot$ and they are unlikely to be black holes; two of the four are likely white dwarfs or neutron stars.
We compare the full sample of five candidates to theoretical expectations on the number of black holes in the Milky Way ($\sim 10^8$) and find reasonable agreement given the small sample size.

\end{abstract}


\section{Introduction \label{topsec:Introduction}}

Stellar-mass black holes are produced when massive stars collapse under their own gravity.
Observations of black holes (BHs) are a key ingredient for understanding outstanding questions in massive stellar evolution, such as which stars explode, which stars produce neutron stars vs. BHs, and whether there is a gap between the heaviest neutron stars (NSs) and the lightest BHs.

Black holes are abundant. 
There are predicted to be $10^7 - 10^9$ stellar-mass BHs in the Milky Way alone \citep{Shapiro:1983,Samland:1998,Timmes:1996,Agol_BHlens:2002, Sartore:2010}. 
However, only about two dozen have been definitively detected, all in binaries with dynamical mass measurements \citep{Corral-Santana:2016, Thompson:2019}. 
Beyond the Milky Way, over 80 binary BH mergers have been detected via gravitational waves, with component masses spanning the lower mass gap $\sim3 M_\odot$  to the lower intermediate-mass BH range $\sim 100 M_\odot$ \citep{Abbott_GW190814:2020}.

These BHs are not a representative sample of the population, as they are all in binary systems.
While most massive stars exist in binary or multiple systems \citep{Sana:2017}, the majority of the BH population is expected to be isolated due to the disruption of the progenitor systems  \citep{Belczynski:2004, Fender:2013, Wiktorowicz:2019}.

Isolated BHs in the Milky Way can be found and weighed using the technique of gravitational microlensing. 
When a foreground lens (e.g.~BH) passes in front of a background source star, the source light is temporarily bent and split into two unresolved images by the lens mass, producing a transient photometric and astrometric signal \citep{Paczynski:1986,Hog:1995,Miyamoto:1995,Walker:1995}. 
The characteristic cross-section of a microlensing event is set by the angular Einstein radius, $\theta_E = \sqrt{\kappa M_L (\pi_L - \pi_S)}$, and depends on the lens mass ($M_L$) and the parallax of the lens ($\pi_L$) and source ($\pi_S$), where $\kappa = 4G/(1 AU \cdot c^2) = 8.14$ mas $/$ $M_\odot $ is a constant.
A $\sim$10 $M_\odot$ BH in the Milky Way disk lensing a background bulge star typically has a $\theta_E = 1-3$ milliarcseconds.

Photometric light curves can measure the duration of the event, $t_E = \theta_E / \mu_{rel}$, where $\mu_{rel}$ is the relative source-lens proper motion, and the microlensing parallax, $\pi_E = (\pi_L - \pi_S) / \theta_E$. 
Precise, multi-epoch astrometry can measure $\theta_E$ directly and combined with photometry to measure the lens mass, $M_L = \theta_E / \kappa \pi_E$.

Over the past 25 years, numerous photometric microlensing surveys have been conducted to search for a wide variety of lenses, including massive astrophysical compact halo objects (MACHOs) that might make up dark matter, stars of all masses, and, most recently, exoplanets \citep{Paczynski:1986,Paczynski:1991,Griest:1991,Mao:1991}.
Current ground-based microlensing surveys  such as the Optical Gravitational Lensing Experiment \citep[OGLE,][]{Udalski:1994}, Microlensing Observations in Astrophysics  \citep[MOA,][]{Bond:2001}, and Korea Microlensing Telescope Network \citep[KMTNet,][]{Kim:2016} monitor hundreds of millions of stars toward the Galactic bulge, identifying thousands of photometric microlensing events each year.
Photometry-only searches for stellar-mass BHs have been attempted (e.g.~\cite{Bennett:2002,Mao:2002,Wyrzykowski:2016}),
but have only been able to identify BH candidates or place loose statistical constraints on the BH mass function.

In contrast to the now-routine measurements of photometric microlensing, detections of astrometric microlensing are still at the forefront of our technical capabilities \citep{Lu:2016,Kains:2017,Rybicki:2018}.
Only a handful of astrometric measurements of the gravitational deflection of light have ever been made, all for nearby ($<10$ pc) lenses that were astrometrically anticipated \citep{Eddington:1919,Sahu:2017,Zurlo:2018} and none of which were BHs. 

If there are $10^8$ BHs in the Milky Way, they should contribute only about 0.1\% to the stellar mass of the Milky Way.
In contrast, they would make up around 1\% of the Milky Way's microlensing events due to their larger lensing cross section. 
Thus, of the thousands of microlensing events detected each year, a few tens should be due to BHs \citep{Gould:2000, Lam:2020}.
However, a 1\% detection rate is akin to looking for BH needles in a Galactic haystack.
By limiting to long duration microlensing events with $t_E>120$ days, the probability of a microlensing event being a BH rises to $\sim 40\%$ \citep{Lam:2020}. 

We present a joint photometric and astrometric analysis of five candidate BH microlensing events.
We briefly summarize the observations in \S\ref{topsec:obs} and methods in \S\ref{topsec:methods}. 
A large non-linear astrometric microlensing signal was detected in one of the five candidates named OGLE-2011-BLG-0462/MOA-2011-BLG-191 (OB110462 for short) consistent with a compact object lens. 
The resulting mass-gap BH or NS nature of OB110462 is presented in \S\ref{topsec:Results} and discussed in a broader context in  \S\ref{topsec:Discussion}.
Final conclusions are presented in \S \ref{topsec:Conclusion}.
A complete description of the observations, methods, and detailed results and discussion for all five targets is presented in the Supplemental Materials within \citet{Lam:2022supp}.

\section{Observations and Analysis \label{topsec:obs}}

Five black hole candidate microlensing events were first identified photometrically in the OGLE-IV \citep{Udalski:2015} and MOA \citep{Hearnshaw:2006,Sumi:2008} Galactic Bulge surveys.
The target discussed in detail here is OB110462, which is located at (17:51:40.19,~-29:53:26.3) towards the Galactic bulge.
The remaining targets properties are described in Supplemental \S 3.1
of \citet{Lam:2022supp}.

The photometric light curves for each candidate span 7 to 11 years, with approximately daily cadence except for seasonal gaps from November to February, and photometric precision of $\sim$2\% for each measurement.
Astrometric observations were obtained with the Hubble Space Telescope (HST) in the F606W ($V$-band) and F814W ($I$-band) filters. 
Astrometric monitoring began around each event's photometric peak with a typical cadence of 1-2 times a year and an astrometric precision of $\sim0.3$~mas per epoch.
A more detailed description of the observations, data analysis, and multi-epoch astrometric alignment is presented in Supplemental 
\S 3
and 
\S 4
of \citet{Lam:2022supp}.

\section{Modeling Methods \label{topsec:methods}}

To measure the physical properties of the lens and source for each event, we simultaneously fit the ground-based photometry and HST photometry and astrometry with a point-source, point-lens (PSPL) microlensing model including source and lens parallax.
All microlensing quantities defined in this work are in the heliocentric reference frame. 
The model free parameters describing the lensing geometry include $t_E$ and $\pi_E$ as described in \S\ref{topsec:Introduction} as well as the time ($t_0$) and distance ($u_0$) of closest projected approach between the source and lens in the heliocentric frame, and the direction of the microlensing parallax vector $\boldsymbol{\pi}_E$. 
The direction of $\boldsymbol{\pi}_E$ is defined to be the same as the direction of the source-lens relative proper motion vector\footnote{Note that we define $\boldsymbol{\mu}_{rel}$ to be the source-lens relative proper motion (Supplementary \S 5.1
following the convention of astrometric microlensing papers \citep[e.g.][]{Hog:1995, Miyamoto:1995, Lu:2016}, while the exoplanet microlensing community typically defines $\boldsymbol{\mu}_{rel}$ to be lens-source relative proper motion.}, i.e. $\boldsymbol{\pi}_E \parallel \boldsymbol{\mu}_{rel}$.
For each photometric filter, the baseline brightness ($m_{base}$) and source flux fraction (or blend parameter, $b_{SFF}$) are fit.
We also fit the astrometric model parameters $\theta_E$ and the source's parallax ($\pi_S$), position at $t_0$ ($\boldsymbol{x}_{S0}$) and proper motion on the sky ($\boldsymbol{\mu}_{S}$).
A complete description of the microlens model is presented in Supplemental \S 5.1
of \citet{Lam:2022supp}.

The best-fit model parameters and uncertainties are estimated using Bayesian inference and the \texttt{MultiNest} nested sampling routine \citep[][see Supplemental \S 5
of \citet{Lam:2022supp} for details]{Feroz:2009, Skilling:2004}.
There is some tension between the preferred model parameters by the photometry vs. astrometry data for OB110462.
When using the default weight (DW) likelihood where each data point and corresponding measurement uncertainty contributes equally to the likelihood, the photometry dominates as it has $>100\times$ more data points than the astrometry dataset.
We also fit models with a likelihood that give equal weight (EW) to each independent dataset (see Supplemental \S 5.3
of \citet{Lam:2022supp} for details). 
With the EW likelihood, the astrometry has much more constraining power.  
Results from both models are described in \S\ref{topsec:Results}.

\begin{figure*}[t!]
    \centering
    \includegraphics[width=0.95\linewidth]{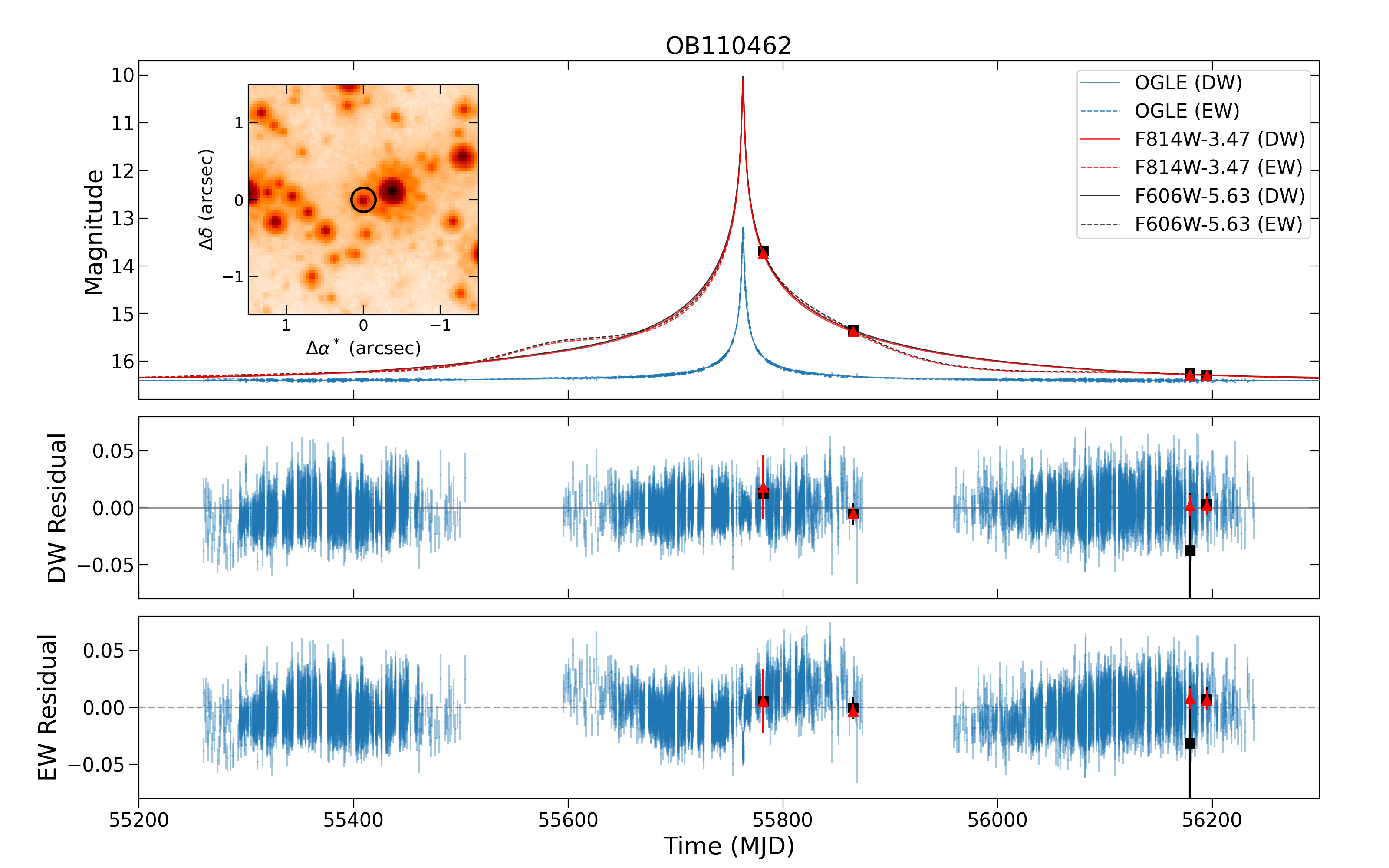} 
    \caption{HST image (\emph{inset}) and photometric lightcurve (\emph{top}) for OB110462.
    In the image, OB110462 is circled in black, and is shown in its unmagnified state.
    Observations are shown as points in blue (OGLE), red, and black (HST).
    The default weighted (DW; {\em solid}) and equal weighted (EW; {\em dashed}) maximum likelihood models are are shown along with their corresponding residuals (DW: {\em middle}, EW: {\em bottom}). 
    See Supplemental \S 5.3
    of \citet{Lam:2022supp} for more details on the two different models.
    \label{fig:OB110462_lightcurve}}
\end{figure*}

\begin{figure*}
    \centering
    \includegraphics[width=0.95\linewidth]{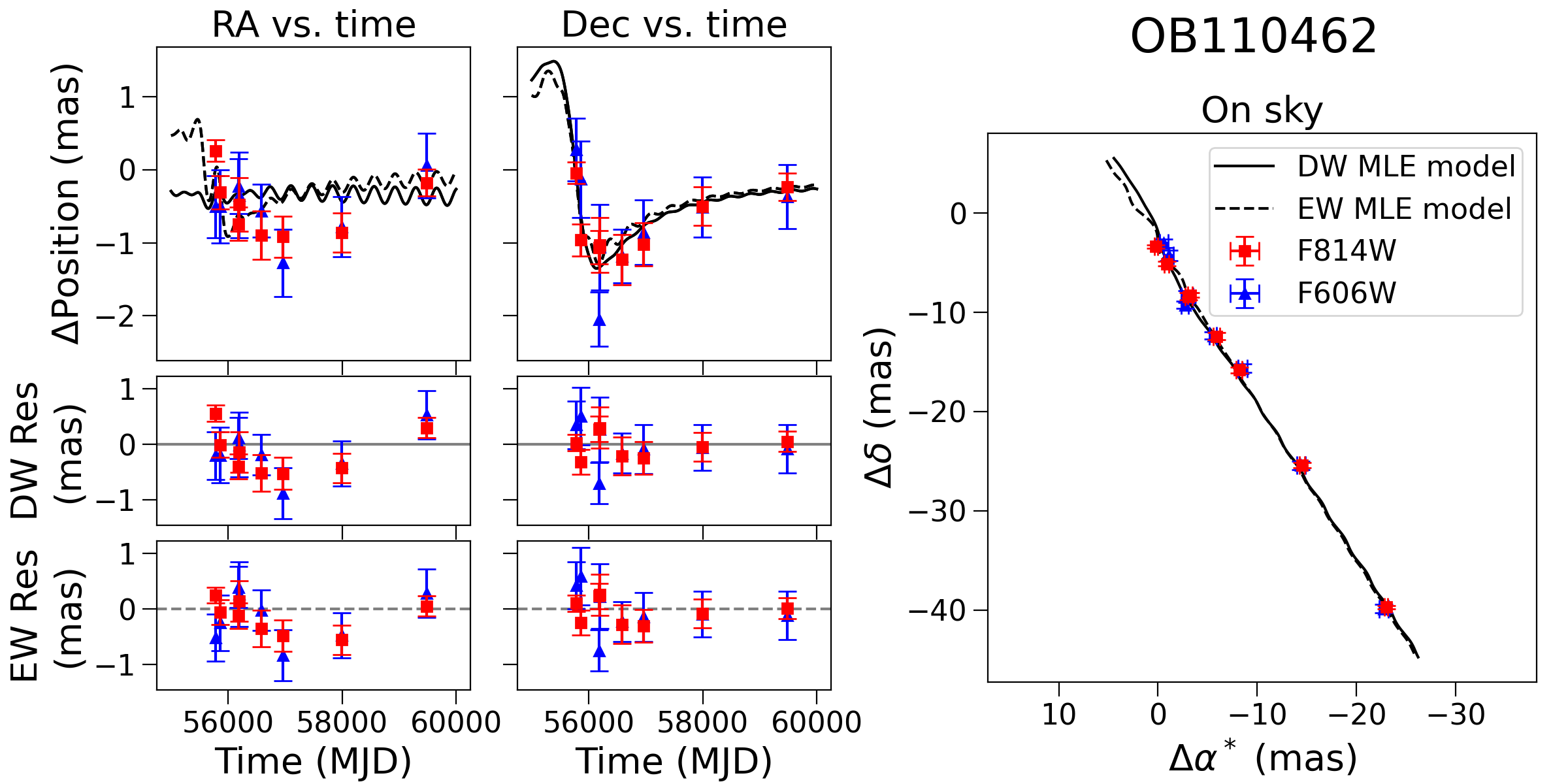}
    \caption{OB110462 astrometry, using the default weight (DW: {\em solid}) and equal weight (EW: {\em dashed}) likelihoods.
    \emph{Left column, top to bottom}: RA ($\Delta \alpha^*$) vs. time with source's unlensed motion subtracted; residuals to the maximum likelihood (MLE) model for $\Delta \alpha^*$ vs. time fit.
    HST F814W astrometry data is shown in red; HST F606W astrometry data is shown in blue.
    The MLE model is shown in black.
    \emph{Middle column, top to bottom}: Same as left column, except Dec ($\Delta \delta$) instead of $\Delta \alpha^*$.
    \emph{Right panel}: astrometry as seen on-sky, in the barycentric frame.
    OB110462 shows a strong $>1$~mas, non-linear astrometric microlensing signal.
    \label{fig:OB110462_astrom}}
\end{figure*}

\section{Results \label{topsec:Results}}

A large ($>1$ mas) astrometric microlensing signal is detected in OB110462 (\S \ref{topsec:OB110462}).
The four other candidates are presented in Supplemental \S 7
of \citet{Lam:2022supp} and either show no significant astrometric microlensing signal or have a low lens mass inconsistent with a black hole.

\subsection{OB110462 \label{topsec:OB110462}}

\begin{figure*}
\centering
    \includegraphics[width=0.45\linewidth]{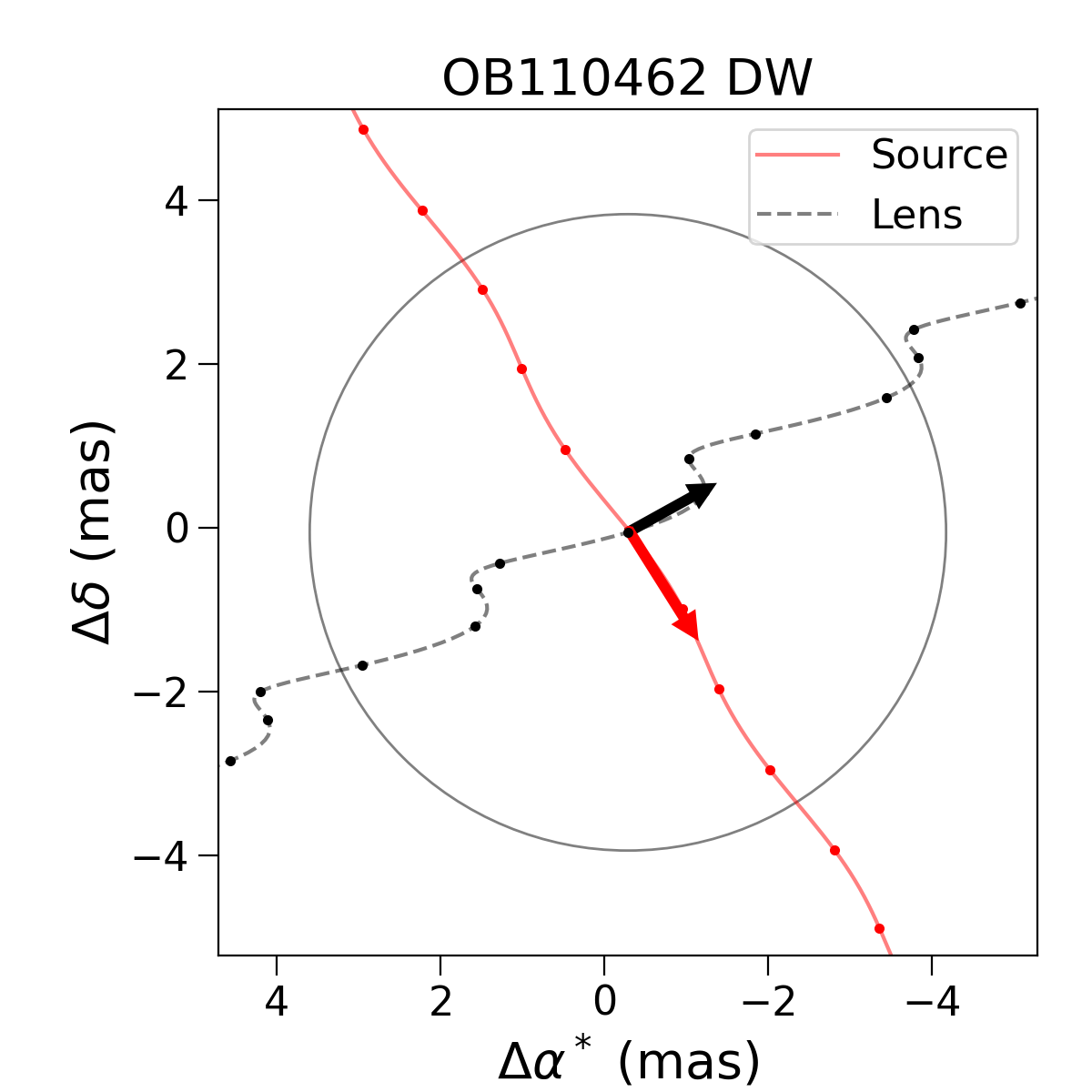}  
    \includegraphics[width=0.45\linewidth]{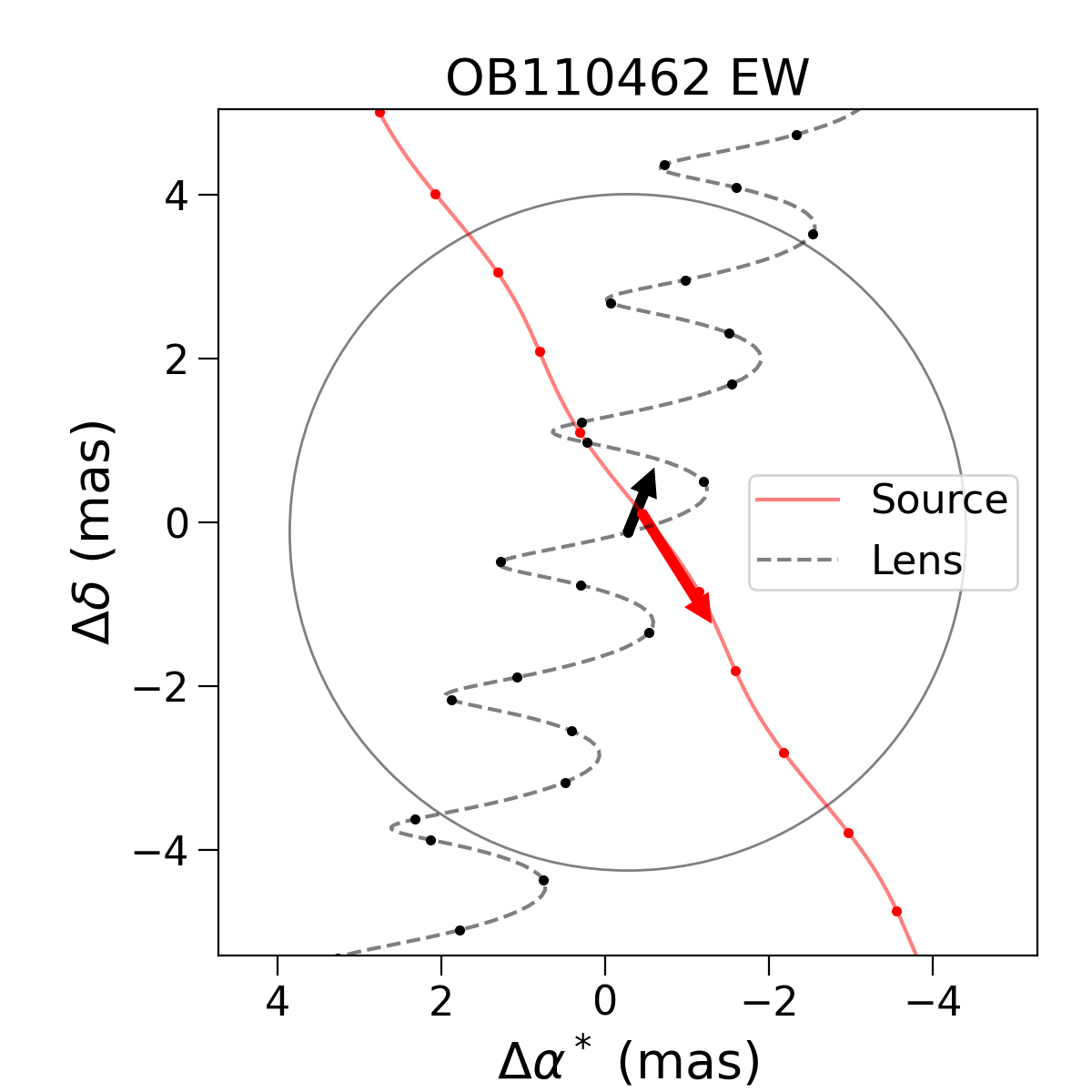}  
\caption{On-sky lensing geometry of OB110462 showing the resolved motion of the lens and source, as inferred from the default weight (DW) model (\emph{left}) and the equal weight (EW) model (\emph{right}); see Supplemental \S 5.3
of \citet{Lam:2022supp} for details about the different models.
The Einstein ring is shown as a gray circle of radius $\theta_E$.
The solid red line shows the trajectory of the source, while the dashed black line shows the trajectory of the lens.
Note that the red line shows the \emph{unlensed} position or the source, and not the centroid of the source's lensed images.
The dots on top of the trajectories are spaced at intervals of 100 days.
The red and black arrows indicate the proper motion of the source and lens, respectively.
The tail of the arrow is at the location of the source and lens at time $t_0$; the length of the arrows are proportional to the magnitude of the source and lens proper motions.
\label{fig:geometry}}
\end{figure*}

\begin{figure}
    \centering
    \includegraphics[width=1.0\linewidth]{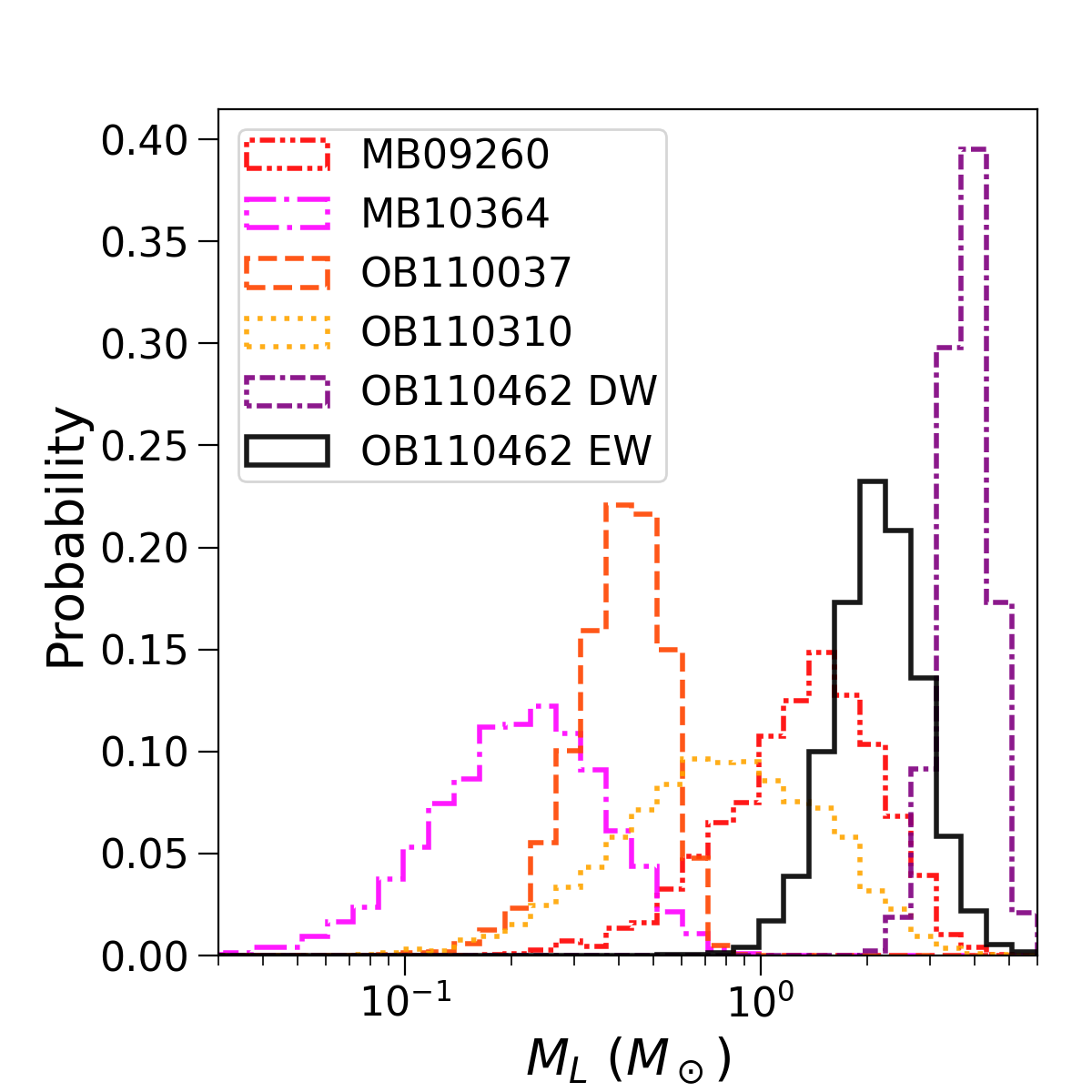} 
    \caption{Lens mass posterior probabilities for the five microlensing BH candidates.
    Two mass posteriors are shown for OB110462, one for each model (default weight (DW) and equal weight (EW)).
    See Supplemental \S 5.3
    of \citet{Lam:2022supp} for details on the two models.
    MB09260, MB10364, OB110037, and OB110310 are discussed in detail in the Supplemental paper.
    \label{fig:masses}}
\end{figure}

From the microlensing fit, we infer that OB110462 is a NS or a mass gap BH, depending on the likelihood function adopted (see Supplemental \S 5.3
of \citet{Lam:2022supp}).
The data and model for OB110462 with the default weight (DW) likelihood are shown in Figures
\ref{fig:OB110462_lightcurve} (photometry) and \ref{fig:OB110462_astrom}
(astrometry), and the fit posteriors are summarized in Table \ref{tab:ob110462_dw_fits}.
The mass posteriors of the lens are shown in Figure \ref{fig:masses}.
The inferred Einstein crossing time $t_E$ is $280.87^{+6.54}_{-5.96}$ days, the microlensing parallax $\pi_E$ is $0.12^{+0.01}_{-0.01}$, the Einstein radius $\theta_E$ is $3.89^{+1.12}_{-1.16}$, and the lens mass $M_L$ is $3.79^{+0.62}_{-0.57} M_\odot$.
The data and model for OB110462 with the equal weight (EW) likelihood are shown in Figures \ref{fig:OB110462_lightcurve} (photometry) and \ref{fig:OB110462_astrom} (astrometry), and the fit posteriors are summarized in Table \ref{tab:ob110462_ew_fits}.
The inferred Einstein crossing time $t_E$ is $278.56^{+12.52}_{-9.16}$ days, the microlensing parallax $\pi_E$ is $0.24^{+0.05}_{-0.05}$, the Einstein radius $\theta_E$ is $4.13^{+0.96}_{-0.91}$, and the lens mass $M_L$ is $2.15^{+0.67}_{-0.54} M_\odot$. 
Further, we find that the object is located relatively nearby at 0.70-1.92~kpc in the direction of the Galactic bulge and has a small transverse velocity of $<30$~km/s.
Figure \ref{fig:geometry} shows the on-sky lensing geometry of OB110462 inferred from the DW and EW likelihood models, showing the relative motions of the source and lens with respect to each other.

The probability that OB110462 is a dark lens is 100\%, ruling out the possibility of a stellar lens and making OB110462 the first detection of a compact object with astrometric microlensing.
Assuming there is a transition from white dwarfs to neutron stars at $1.2 M_\odot$ and neutron stars to BHs at $2.2 M_\odot$, the relative probabilities of WD:NS:BH are 0:0:100 for the default weighted (DW) fit and 6:50:44 for the equally weighted (EW) fit.

The microlensing fit also yields information about the distance and transverse velocity of the lens. 
The lens is relatively nearby at a distance of 1.47~-~1.92~kpc or 0.70~-~1.30~kpc for the DW and EW solutions, respectively. 
The inferred lens velocity is $<30$~km/s for both solutions with a slower velocity 2~-~12 km/s from the EW solution and a faster velocity of 21~-~27~km/s from the DW solution. 
In both cases, the velocities appear consistent with the compact object receiving little to no kick, although the line-of-sight velocity is not measurable from these observations.

Based on an analysis of the source position in the CMD, the OB110462 source star is around the main sequence turnoff on the redder and more luminous side of the main sequence, suggesting it is most likely a giant or sub-giant star.
However, a main sequence source could still be consistent.
The relative proper motion and parallax also favor a star in the near side of the Bulge (see Supplemental \S 7.6
of \citet{Lam:2022supp} for more details).

A point-source, point-lens (PSPL) model is not the end of the story for OB110462. 
There is no PSPL model which can simultaneously fit both the photometry and the astrometry.
Specifically, the direction of $\mu_{rel}$ preferred by the photometry and astrometry are different. 
The best fit PSPL model for the default weight (DW) likelihood fits the photometry very well, but leaves a significant $\sim 0.5$ mas coherent astrometric residual in RA (Figure \ref{fig:OB110462_astrom}).
The best fit PSPL model for the equal weight (EW) likelihood leaves a significant and coherent $\sim 0.03$ mag residual in the photometry, but fits the astrometry in RA better than the DW likelihood model, although some unexplained astrometric residuals still remain.
More complex microlensing geometry models, such as those involving a binary source or lens, should be explored.
As mentioned in Supplemental \S 4.2.5
of \cite{Lam:2022supp}, we apply a constant positional offset to the F606W data in order to make it match up with the F814W data.
However, this filter dependent positional difference may actually be astrophysical and consistent with a small contribution from a faint companion to the source.
Either way, both solutions indicate a NS or BH detection.

The alternative explanation to the tension between the photometry and astrometry of OB110462 is some type of systematic error in one or both sets observations.
This possibility is discussed in Supplemental \S 8.3
of \cite{Lam:2022supp}.

\newpage
\begin{deluxetable}{lccc}
\tablecaption{OB110462 DW Fit Values\label{tab:ob110462_dw_fits}}
\tablehead{
    \colhead{Parameter} &
    \colhead{Med$^{+1\sigma}_{-1\sigma}$} & 
    \colhead{MAP} &
    \colhead{MLE}}
\startdata
\input{OB110462_fit_table_joint.txt}
\enddata
\tablecomments{The columns list the median $\pm1\sigma$ (68\%) credible intervals, maximum a posteriori (MAP) solution, and and maximum likelihood estimator (MLE) solution.}
\end{deluxetable}

\begin{deluxetable}{lccc}
\tablecaption{OB110462 EW Fit Values\label{tab:ob110462_ew_fits}}
\tablehead{
    \colhead{Parameter} &
    \colhead{Med$^{+1\sigma}_{-1\sigma}$} & 
    \colhead{MAP} &
    \colhead{MLE}}
\startdata
\input{OB110462_ew_fit_table_joint.txt}
\enddata
\end{deluxetable}

\begin{figure*}
\centering
\includegraphics[width=0.95\linewidth]{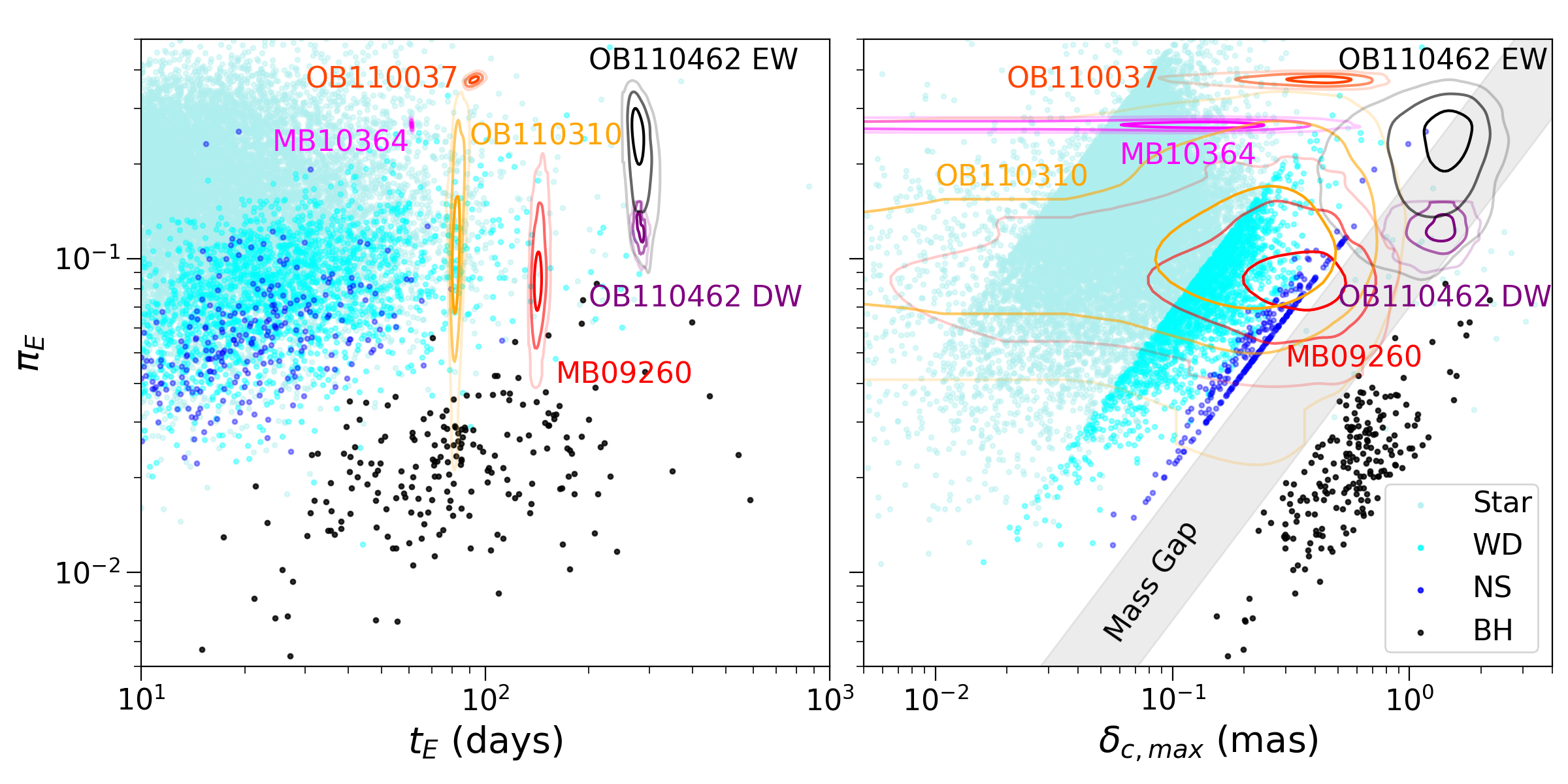}  
\caption{Microlensing parallax $\pi_E$ vs. Einstein crossing time $t_E$ \emph{(left)} and maximum astrometric shift $\delta_{c,max}$ \emph{(right)}.
Points are from the \texttt{PopSyCLE} simulation.
Contours are $1-2-3\sigma$ (39.3-86.5-98.9\%) credible regions from the microlensing model fits to the five BH candidates.
There are two fits for OB110462 (default weight (DW) and equal weight (EW); see Supplemental \S 5.3
of \citet{Lam:2022supp} for details).
The OB110462 DW solution has a smaller $\pi_E$ than the OB110462 EW solution, and has a correspondingly more massive lens mass.
Both solutions fall solidly in the NS-BH mass gap, making OB110462 the best BH-candidate.
MB09260 and OB110310 are most likely white dwarfs or neutron stars, although due to uncertainty in $\pi_E$ and $\delta_{c,max}$ higher and lower mass lenses cannot be definitively ruled out.
OB110037 and MB10364 are not BHs as they have very large $\pi_E$, as well as relatively short $t_E$ and small $\delta_{c,max}$.
MB09260, MB10364, OB110037, and OB110310 are discussed in detail in the Supplemental paper.
\label{fig:piE_tE_deltac}}
\end{figure*}

\subsection{Is OB110462 a BH or a NS?}

One means of further understanding the BH or NS nature of OB110462 would be to detect electromagnetic radiation from the lens. We searched existing X-ray and pulsar catalogs at the position of OB110462 and did not find any counterpart (see Supplemental \S 8.2 
of \citet{Lam:2022supp} for details). 
Unfortunately, OB110462 is not in the Gaia EDR3 catalog as it is too faint. 

Future observations of OB110462 will be useful to determine its true nature. 
Continued astrometric monitoring, including the remaining data from HST Cycle 29 program GO-16760 \citep{Lam:2021hst} will continue to improve the lens mass estimate as described in Supplemental \S 8.2
of \citet{Lam:2022supp}.
High contrast imaging observations in the next 5-10 years would also be worthwhile.
If the lens is indeed a solitary NS or BH, lack of an optical/infrared lens detection would bolster support of the dark isolated lens interpretation, but be unable to distinguish between NS or BH as the relative proper motion differentiating the DW and EW solutions cannot be measured in the case of a dark lens.
In that case, only a deep, targeted X-ray observation could help in differentiating between the NS vs. BH scenarios.
On the other hand, optical/infrared detection of a lens separate from the source would point to a binary lens scenario.
Binary source or binary lens models can be further explored through precision radial velocity searches.

\subsection{Number of detected BHs}

\begin{figure}
    \centering
    \includegraphics[width=1.0\linewidth]{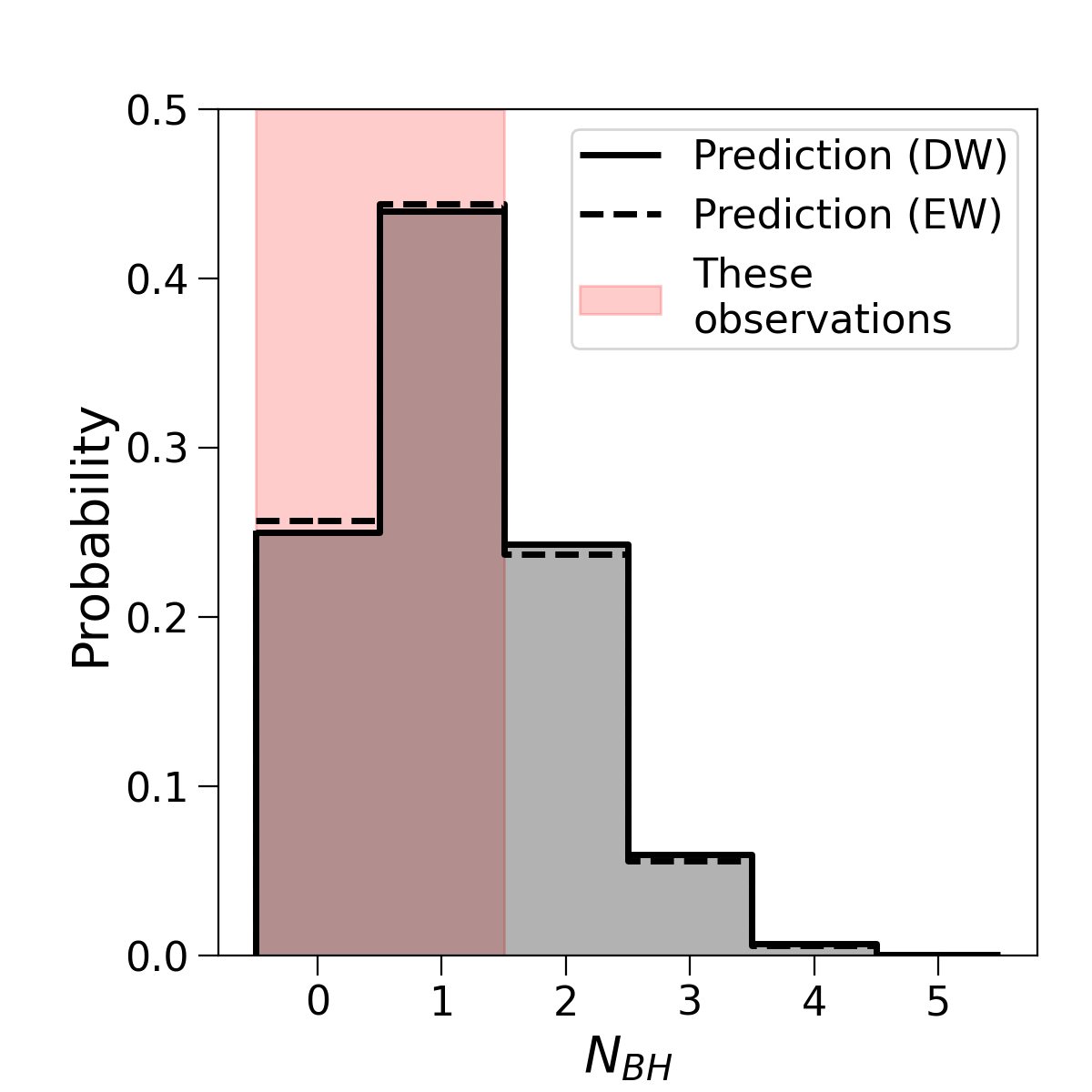} 
    \caption{Probability of detecting $N$ BHs as calculated from the \texttt{PopSyCLE} simulation.
    Two predictions are shown, depending on the likelihood used for OB110462 (default weight (DW) or equal weight (EW)).
    See Supplemental \S 5.3
    of \citet{Lam:2022supp} for details on the two models.
    Our observation of 0 or 1 BHs is consistent with either OB110462 model prediction.
    \label{fig:ndetect_BH_prob}}
\end{figure}

Next, we compare our observed BH yield to the theoretical expectation calculated using the \texttt{PopSyCLE} simulations.
\texttt{PopSyCLE}'s Galactic model contains $2 \times 10^8$ BHs ranging from $5 - 16 M_\odot$ \citep{Lam:2020}.
For a sample of simulated events that would be observable by OGLE (see Table 4 of \cite{Lam:2020}), we calculate the fraction of those events due to BHs as a function of the Einstein crossing time $t_E$
as described in Supplemental \S 7.9
in \citet{Lam:2022supp}.
Figure \ref{fig:ndetect_BH_prob} shows the expectation of detecting $N_{BH} = 0, 1,..., 5$ BHs within our sample of 5 targets.
The probability of detecting 0 or 1 BHs in our simulation is $\sim$25\% and $\sim$45\%, respectively.
This estimate is consistent with our single detection of a NS-BH object.
Note that in \texttt{PopSyCLE}, there are no $2-5 M_\odot$ mass gap NSs or BHs in the simulation, and hence no exact OB110462 analogue.

\section{Discussion \label{topsec:Discussion}}

OB110462 is the first definitive detection of a compact object discovered with astrometric microlensing.
Depending on the likelihoood function used to evaluate the fit (see Supplemental \S 5.3
of \citet{Lam:2022supp}), it is either a neutron star (50\% probability for the EW likelihood), a black hole (44\% probability for the EW likelihood, 100\% probability for the DW likelihood), or a white dwarf (6\% probability for the EW likelihood).
The other four candidates discussed in detail in Supplemental \S 7
of \citet{Lam:2022supp} are mostly likely stars, white dwarfs or neutron stars, although a NS-BH mass gap object cannot be ruled out in two cases.

Here we discuss the observed BH yield as compared to theoretical expectations (\S \ref{topsec:Comparison to simulations}) and and implications for the BH mass function (\S \ref{topsec:OB110462 in comparison to the BH population}).

\subsection{Comparison to simulations
\label{topsec:Comparison to simulations}}

\subsubsection{$\pi_E$-$t_E$-$\delta_{c,max}$}

As described in \citet{Lam:2020}, BH candidates can be identified photometrically by their long $t_E$ and small $\pi_E$, and confirmed astrometrically by measuring the maximum astrometric shift $\delta_{c,max} = \theta_E/\sqrt{8}$ (see Supplemental \S 5.1
).
Figure \ref{fig:piE_tE_deltac} shows the $1-2-3\sigma$ posterior contours of $\pi_E$ vs. $t_E$ \emph{(left)} and $\pi_E$ vs. $\delta_{c,max}$ \emph{(right)} of the microlensing models for the 5 targets, compared against simulated microlensing events generated by the \texttt{PopSyCLE} software \citep{Lam:2020}.
By comparing the $\pi_E-t_E$ and $\pi_E - \delta_{c,max}$ posteriors against simulation, we can gain a more intuitive understanding of the inferred lens types for the targets (Supplemental Table 9
of \cite{Lam:2022supp}).

Both EW and DW models for OB110462 fall solidly within the $2-5M_\odot$ mass gap shown in the $\pi_E - \delta_{c,max}$ parameter space.
Because the EW solution leads to a larger and more uncertain value of $\pi_E$ than the DW solution, a neutron star or even white dwarf lens is a possibility.
On the other hand, the DW solution prefers a smaller and more well constrained value of $\pi_E$ than the EW solution, leading to a much more definitive solution of a mass-gap BH.
However, both the EW and DW solution for OB110462 fall in a somewhat unusual part of the $\pi_E-t_E$ parameter space for BHs: typical BH $\pi_E$ are around 0.02, while for OB110462, $\pi_E$ is around 0.1.
This is because \texttt{PopSyCLE} simulations only contain BHs with masses from $\sim 5 - 16 M_\odot$; if OB110462 is truly a mass-gap BH, it would not correspond to any BHs in the simulation.

\subsection{OB110462 in comparison to the BH population
\label{topsec:OB110462 in comparison to the BH population}}

Several attempts have been made to determine the Milky Way BH mass function using dynamical mass measurements of BHs in binaries. The mass-gap between 3-5 $M_\odot$ was first observed in low-mass X-ray binaries \citep{Bailyn:1998,Ozel:2010,Farr:2011}. However, more recent detections of BHs in this mass range from both gravitational wave mergers and in non-interacting binaries suggests that the mass-gap may actually be filled with BHs \citep[e.g.][]{Abbott_GW170817:2017,Thompson:2019}. 
For a complete description of the Milky Way BHs found to date, see Supplemental \S 8.1
of \cite{Lam:2022supp}.
If OB110462 is a BH, it too shows that the mass function of BHs extends into the mass-gap regime. 

It is somewhat surprising that massive BHs have {\em not} been found in our microlensing search as well as in searches for BHs in wide binaries \citep{Rowan:2021,Thompson:2019,El-Badry:2022}.
Both microlensing and radial velocity searches should be biased toward finding 10 $M_\odot$ objects more easily than 3 $M_\odot$ objects as described in Supplemental \S 8.1
of \citet{Lam:2022supp}. 
It may be that the selection bias is offset by the steep mass function for massive stars and thus BHs. 
As the sample of BHs in the Milky Way grows, a more quantitative analysis of the sample selection will be essential to constrain the true BH mass function.

\section{Conclusion \label{topsec:Conclusion}}

We analyze five microlensing events with candidate BH lenses.
Combining HST astrometry and densely sampled ground-based photometry, we derive masses for these five lenses as well as their probability of being a BH.
Of the five targets, we make one definitive $ > 1$ mas detection of astrometric microlensing (OB110462).
The mass of the lens in OB110462 is in the range 1.6-4.4 $M_\odot$, making it the first detection of a compact object through astrometric microlensing.

We use our detection of a mass-gap BH or neutron star and the non-detections of BHs in the rest of the sample to observationally constrain the number of BHs in the Milky Way. 
Our observational BH yield currently agrees with simulations assuming $2 \times 10^8$ BHs in the Milky Way, albeit with very large uncertainties due to the small sample size.
The ability to place more stringent constraints on the number and mass distribution of Galactic BHs will require larger samples, such as those that may be delivered by the Roman Space Telescope's microlensing survey.

Astrometric microlensing holds the key to uncovering the hidden BH population.
Further pursuit and refinement of the event selection, observing, and modeling process will fulfill the full promise of this technique and its ability to reveal the properties of Galactic BHs.\\

Shortly prior to this work being submitted for review, we learned of an independent analysis of OB110462 carried out by \citet{Sahu:2022}.
Notably, they reach a different conclusion about the mass of the lens ($7.1 \pm 1.3 M_\odot$).
It is not clear whether the discrepancy is due to the use of different datasets (e.g.~we include an additional epoch of 2021 HST data), performing the analysis differently (e.g.~we explore solutions allowed by both photometry and astrometry using different likelihood weights), or a combination of both.
In addition, although both analyses make clear detections of an astrometric deflection, the direction of the deflections are in opposing directions in RA.
Preliminary work shows that different choice of reference stars across the two teams is not the source of the discrepancy.
However, significant further work is required to fully understand the differences between the two analyses.

\section*{Acknowledgements}

See the Supplement of \citet{Lam:2022supp} for a full list of acknowledgements.

\bibliography{sample63}{}
\bibliographystyle{aasjournal}

\end{document}


\title{Supplement: An isolated mass gap black hole or neutron star detected with astrometric microlensing}

\author[0000-0002-6406-1924]{Casey Y. Lam}
\correspondingauthor{Casey Y. Lam} 
\email{casey$\_$lam@berkeley.edu}
\affiliation{University of California, Berkeley, Department of Astronomy, Berkeley, CA 94720}

\author[0000-0001-9611-0009]{Jessica R. Lu}
\affiliation{University of California, Berkeley, Department of Astronomy, Berkeley, CA 94720}

\author[0000-0001-5207-5619]{Andrzej Udalski}
\altaffiliation{OGLE collaboration}
\affiliation{Astronomical Observatory, University of Warsaw, Al.~Ujazdowskie~4,00-478~Warszawa, Poland}

\author{Ian Bond}
\altaffiliation{MOA collaboration}
\affiliation{School of Mathematical and Computational Sciences, Massey University, Private Bag 102-904 North Shore Mail Centre, Auckland 0745, New Zealand}

\author[0000-0001-8043-8413]{David P. Bennett}
\altaffiliation{MOA collaboration}
\affiliation{Code 667, NASA Goddard Space Flight Center, Greenbelt, MD 20771, USA}
\affiliation{Department of Astronomy, University of Maryland, College Park, MD 20742, USA}

\author[0000-0002-2335-1730]{Jan Skowron}
\altaffiliation{OGLE collaboration}
\affiliation{Astronomical Observatory, University of Warsaw, Al.~Ujazdowskie~4,00-478~Warszawa, Poland}

\author[0000-0001-7016-1692]{Przemek Mr{\'o}z}
\altaffiliation{OGLE collaboration}
\affiliation{Astronomical Observatory, University of Warsaw, Al.~Ujazdowskie~4,00-478~Warszawa, Poland}

\author[0000-0002-9245-6368]{Radek Poleski}
\altaffiliation{OGLE collaboration}
\affiliation{Astronomical Observatory, University of Warsaw, Al.~Ujazdowskie~4,00-478~Warszawa, Poland}

\author{Takahiro Sumi}
\altaffiliation{MOA collaboration}
\affiliation{Department of Earth and Space Science,
Graduate School of Science,
Osaka University, Toyonaka,
Osaka 560-0043, Japan}

\author[0000-0002-0548-8995]{Micha\l \,K. Szyma{\'n}ski}
\altaffiliation{OGLE collaboration}
\affiliation{Astronomical Observatory, University of Warsaw, Al.~Ujazdowskie~4,00-478~Warszawa, Poland}

\author[0000-0003-4084-880X]{Szymon Koz{\l}owski}
\altaffiliation{OGLE collaboration}
\affiliation{Astronomical Observatory, University of Warsaw, Al.~Ujazdowskie~4,00-478~Warszawa, Poland}

\author[0000-0002-2339-5899]{Pawe{\l} Pietrukowicz}
\altaffiliation{OGLE collaboration}
\affiliation{Astronomical Observatory, University of Warsaw, Al.~Ujazdowskie~4,00-478~Warszawa, Poland}

\author[0000-0002-7777-0842]{Igor Soszy{\'n}ski}
\altaffiliation{OGLE collaboration}
\affiliation{Astronomical Observatory, University of Warsaw, Al.~Ujazdowskie~4,00-478~Warszawa, Poland}

\author[0000-0001-6364-408X]{Krzysztof Ulaczyk}
\altaffiliation{OGLE collaboration}
\affiliation{Astronomical Observatory, University of Warsaw, Al.~Ujazdowskie~4,00-478~Warszawa, Poland}
\affiliation{Department of Physics, University of Warwick, Gibbet Hill Road,Coventry, CV4~7AL,~UK}

\author[0000-0002-9658-6151]{{\L}ukasz Wyrzykowski}
\altaffiliation{OGLE collaboration}
\affiliation{Astronomical Observatory, University of Warsaw, Al.~Ujazdowskie~4,00-478~Warszawa, Poland}

\author{Shota Miyazaki}
\altaffiliation{MOA collaboration}
\affiliation{Department of Earth and Space Science,
Graduate School of Science,
Osaka University, Toyonaka,
Osaka 560-0043, Japan}

\author{Daisuke Suzuki}
\altaffiliation{MOA collaboration}
\affiliation{Department of Earth and Space Science,
Graduate School of Science,
Osaka University, Toyonaka,
Osaka 560-0043, Japan}

\author{Naoki Koshimoto}
\altaffiliation{MOA collaboration}
\affiliation{Code 667, NASA Goddard Space Flight Center, Greenbelt, MD 20771, USA}
\affiliation{Department of Astronomy, University of Maryland, College Park, MD 20742, USA}
\affiliation{Department of Astronomy, Graduate School of Science, The University of Tokyo, 7-3-1 Hongo, Bunkyo-ku, Tokyo 113-0033, Japan}

\author[0000-0001-5069-319X]{Nicholas J. Rattenbury}
\altaffiliation{MOA collaboration}
\affiliation{Department of Physics, University of Auckland, Private Bag 92019, Auckland, New Zealand}

\author[0000-0003-2874-1196]{Matthew W. Hosek, Jr.}
\affiliation{University of California, Los Angeles, Department of Astronomy, Los Angeles, CA 90095, USA}

\author{Fumio Abe}
\altaffiliation{MOA collaboration}
\affiliation{Institute for Space-Earth Environmental Research, Nagoya University, Nagoya 464-8601, Japan}

\author{Richard Barry}
\altaffiliation{MOA collaboration}
\affiliation{Code 667, NASA Goddard Space Flight Center, Greenbelt, MD 20771, USA}

\author{Aparna Bhattacharya}
\altaffiliation{MOA collaboration}
\affiliation{Code 667, NASA Goddard Space Flight Center, Greenbelt, MD 20771, USA}
\affiliation{Department of Astronomy, University of Maryland, College Park, MD 20742, USA}

\author{Akihiko Fukui}
\altaffiliation{MOA collaboration}
\affiliation{Department of Earth and Planetary Science, Graduate School of Science, The University of Tokyo, 7-3-1 Hongo, Bunkyo-ku, Tokyo 113-0033, Japan}
\affiliation{Instituto de Astrof\'isica de Canarias, V\'ia L\'actea s/n, E-38205 La Laguna, Tenerife, Spain}

\author{Hirosane Fujii}
\altaffiliation{MOA collaboration}
\affiliation{Institute for Space-Earth Environmental Research, Nagoya University, Nagoya 464-8601, Japan}

\author{Yuki Hirao}
\altaffiliation{MOA collaboration}
\affiliation{Department of Earth and Space Science, Graduate School of Science, Osaka University, Toyonaka, Osaka 560-0043, Japan}

\author{Yoshitaka Itow}
\altaffiliation{MOA collaboration}
\affiliation{Institute for Space-Earth Environmental Research, Nagoya University, Nagoya 464-8601, Japan}

\author{Rintaro Kirikawa}
\altaffiliation{MOA collaboration}
\affiliation{Department of Earth and Space Science, Graduate School of Science, Osaka University, Toyonaka, Osaka 560-0043, Japan}

\author{Iona Kondo}
\altaffiliation{MOA collaboration}
\affiliation{Department of Earth and Space Science, Graduate School of Science, Osaka University, Toyonaka, Osaka 560-0043, Japan}

\author{Yutaka Matsubara}
\altaffiliation{MOA collaboration}
\affiliation{Institute for Space-Earth Environmental Research, Nagoya University, Nagoya 464-8601, Japan}

\author{Sho Matsumoto}
\altaffiliation{MOA collaboration}
\affiliation{Department of Earth and Space Science, Graduate School of Science, Osaka University, Toyonaka, Osaka 560-0043, Japan}

\author{Yasushi Muraki}
\altaffiliation{MOA collaboration}
\affiliation{Institute for Space-Earth Environmental Research, Nagoya University, Nagoya 464-8601, Japan}

\author{Greg Olmschenk}
\altaffiliation{MOA collaboration}
\affiliation{Code 667, NASA Goddard Space Flight Center, Greenbelt, MD 20771, USA}

\author[0000-0003-2388-4534]{Cl\'ement Ranc}
\altaffiliation{MOA collaboration}
\affiliation{Zentrum f{\"u}r Astronomie der Universit{\"a}t Heidelberg, Astronomiches Rechen-Institut, M{\"o}nchhofstr.\ 12-14, 69120 Heidelberg, Germany}

\author{Arisa Okamura}
\affiliation{Department of Earth and Space Science, Graduate School of Science, Osaka University, Toyonaka, Osaka 560-0043, Japan}

\author{Yuki Satoh}
\affiliation{Department of Earth and Space Science, Graduate School of Science, Osaka University, Toyonaka, Osaka 560-0043, Japan}

\author{Stela Ishitani Silva}
\altaffiliation{MOA collaboration}
\affiliation{Department of Physics, The Catholic University of America, Washington, DC 20064, USA}
\affiliation{Code 667, NASA Goddard Space Flight Center, Greenbelt, MD 20771, USA}

\author{Taiga Toda}
\altaffiliation{MOA collaboration}
\affiliation{Department of Earth and Space Science, Graduate School of Science, Osaka University, Toyonaka, Osaka 560-0043, Japan}

\author{Paul J. Tristram}
\altaffiliation{MOA collaboration}
\affiliation{University of Canterbury Mt.\ John Observatory, P.O. Box 56, Lake Tekapo 8770, New Zealand}

\author{Aikaterini Vandorou}
\altaffiliation{MOA collaboration}
\affiliation{Code 667, NASA Goddard Space Flight Center, Greenbelt, MD 20771, USA}
\affiliation{Department of Astronomy, University of Maryland, College Park, MD 20742, USA}

\author{Hibiki Yama}
\altaffiliation{MOA collaboration}
\affiliation{Department of Earth and Space Science, Graduate School of Science, Osaka University, Toyonaka, Osaka 560-0043, Japan}

\author[0000-0002-0287-3783]{Natasha S. Abrams}
\affiliation{University of California, Berkeley, Department of Astronomy, Berkeley, CA 94720}

\author[0000-0002-2350-4610]{Shrihan Agarwal}
\affiliation{University of California, Berkeley, Department of Astronomy, Berkeley, CA 94720}

\author[0000-0003-4725-4481]{Sam Rose}
\affiliation{University of California, Berkeley, Department of Astronomy, Berkeley, CA 94720}

\author[0000-0002-5029-3257]{Sean K. Terry}
\affiliation{University of California, Berkeley, Department of Astronomy, Berkeley, CA 94720}

\begin{abstract}

This Supplement provides supporting material for \cite{Lam:2022lett}. We briefly summarize past gravitational microlensing searches for black holes and present details of the observations, analysis, and modeling of five black hole (BH) candidates 
observed with both ground-based photometric microlensing surveys and Hubble Space Telescope astrometry and photometry.
We present detailed results for four of the five candidates that show no or low probability for the lens to be a black hole. 
In these cases, the lens masses are $<2 M_\odot$ and two of the four are likely white dwarfs or neutron stars.
We also present detailed methods for comparing the full sample of five candidates to theoretical expectations of the number of BHs in the Milky Way ($\sim 10^8$).

\end{abstract}

\keywords{}

\section{Supplemental Introduction \label{sec:supp_introduction}}

This paper is a supplement to \citet{Lam:2022lett}, in which we present results from a search for stellar mass black holes in the Milky Way using gravitational microlensing. 
\citet{Lam:2022lett} and this supplement describe our analysis of five Milky Way black hole (BH) candidates, which constitute a sufficient sample to place preliminary constraints on the number of isolated black holes in the Milky Way.

This supplement is organized as follows.
In \S\ref{sec:Introduction}, we present a detailed review of past microlensing searches for BHs.
In \S \ref{sec:Observations}, the datasets and reduction processes for our BH search are described, and in \S \ref{sec:HST Data Analysis} the photometric and astrometric analysis of the HST data is explained.
In \S \ref{sec:Microlensing Modeling}, the microlensing modeling and fitting procedure are described in detail, while \S \ref{sec:Constraining the nature of the lens} describes how to combine the high-resolution imaging and results of the microlensing modeling to constrain the lens' luminosity.
\S\ref{sec:Results} presents the results of the modeling
for the five candidates.
Note that results for the single mass-gap black hole or neutron star candidate are presented in the main paper \citet{Lam:2022lett}.
Details on constraints on the Milky Way BH population and future BH microlensing searches are discussed in \S \ref{sec:Discussion}.

\section{Past Microlensing Searches for Black Holes \label{sec:Introduction}}

The advent of ground-based microlensing surveys  provided a new avenue to search for isolated black holes (BHs).
Notable BH candidates identified with photometric microlensing include MACHO-96-BLG-5, MACHO-98-BLG-6 \citep{Bennett:2002}, and MACHO-99-BLG-22/OGLE-1999-BUL-32 \citep{Mao:2002}. 
\cite{Poindexter:2005} found that MACHO-99-BLG-22 is likely a BH, MACHO-96-BLG-5 is possibly a BH, and MACHO-98-BLG-6 is most likely not a BH, with BH lens probabilities of 78\%, 37\%, and 2\%, respectively.
On the other hand, \cite{Abdurrahman:2021} found that MACHO-96-BLG-5 and MACHO-98-BLG-6 are still good BH candidates, ruling out non-BH lenses for source-lens relative proper motions larger than 2.5 mas/yr.
However, mass estimates for these lensing events cannot be made without invoking a Galactic model.
These candidates only had photometric microlensing observations, which alone cannot constrain the mass of the lens, unless rare higher-order effects such as finite-source effects are detected.

As mentioned briefly at the end of \cite{Paczynski:1986}, microlensing also has an astrometric signature, in which the centroid of the image is displaced from the source's true position \citep{Hog:1995, Miyamoto:1995, Walker:1995}.
In contrast to the now-routine measurements of photometric microlensing, detections of astrometric microlensing are still at the forefront of our technical capabilities.
Typical astrometric shifts toward the Bulge are $O(0.01 - 1)$ mas, and few existing facilities are currently capable of the astrometric precision to perform this measurement.
Only a handful of astrometric measurements of the gravitational deflection of light have ever been made, all for nearby ($<10$ pc) lenses that were astrometrically anticipated \citep{Eddington:1919, Sahu:2017, Zurlo:2018}. 
However, a combination of photometric and astrometric microlensing together can determine the mass of the lensing object, making detection of astrometric microlensing important for BH searches. 

To date, there have been two endeavors to measure lens masses by combining photometric and astrometric microlensing.
\cite{Lu:2016} attempted a measurement with Keck laser guide star adaptive optics (LGS AO), but no detections of astrometric microlensing were made. 
\cite{Kains:2017} reported a detection of astrometric microlensing made with the Hubble Space Telescope (HST), but the signal was very weak and no lens masses were well constrained. 
The currently operating Gaia mission is also anticipated to make measurements of astrometric microlensing, which can be searched for once per-epoch astrometry is released (\citet{McGill:2020} and references therein).
A handful of these deflections should be due to BHs \citep{Rybicki:2018}.

\section{Observations in Detail \label{sec:Observations}}

\subsection{Targets and Selection Criteria \label{sec:targets}}

\begin{deluxetable*}{lccll}
\tabletypesize{\footnotesize}
\tablecaption{Target Summary
\label{tab:target_summary}}
\tablehead{
    \colhead{Short Name} & 
    \colhead{OGLE Alert Name} & 
    \colhead{MOA Alert Name} & 
    \colhead{RA (J2000.0)} & 
    \colhead{Dec (J2000.0)}}
\startdata
\input{target_summary_table.txt}
\enddata
\end{deluxetable*}

Five BH candidate microlensing events\footnote{Three additional other targets were initially observed with HST, but dropped from the target list after a year \citep{Sahu:2012}.} were selected from the OGLE Early Warning System\footnote{\href{http://ogle.astrouw.edu.pl/ogle4/ews/ews.html}{http://ogle.astrouw.edu.pl/ogle4/ews/ews.html}} \citep{Udalski:2015} and MOA Alerts\footnote{\href{https://www.massey.ac.nz/~iabond/moa/alerts/}{https://www.massey.ac.nz/$\sim$iabond/moa/alerts/}} to be imaged with HST.
These targets were selected to have long ($t_E>200$ days) duration, no light contribution from the lens, and high magnifications to allow detection of parallax signals, making them good isolated BH candidates \citep{Sahu:2009}.
Preliminary results by \citet{Sahu:2017} reported that all five candidates
were low-mass ($<0.5 M_\odot$) stars, with no BH detections in the sample. 
We chose to re-analyze these targets in order to use both BH detections and non-detections in the sample to constrain the total number of BHs in the Milky Way.

Three of the targets, OGLE-2011-BLG-0037/MOA-2011-BLG-039, OGLE-2011-0310/MOA-2011-BLG-332, and OGLE-2011-BLG-0462/MOA-2011-BLG-191 (hereafter OB110037, OB110310, and OB110462), were alerted by both OGLE and MOA.
The other two targets, MOA-2009-BLG-260 and MOA-2010-BLG-364 (hereafter MB09260 and MB10364), were only alerted by MOA.
Table \ref{tab:target_summary} lists their coordinates.
Figures \ref{fig:MB09260_lightcurve} - \ref{fig:OB110462_EW_lightcurve} show the lightcurves of the targets.

\subsection{MOA \label{sec:MOA}}

The MOA-II survey is carried out with a 1.8-m telescope at Mt.~John University Observatory in New Zealand \citep{Hearnshaw:2006, Sumi:2008}.
The seeing at the site ranges from $\sim 1.8'' - 3.5''$, with the median seeing being $\sim 2.5''$.
The telescope has a 2.2 deg$^2$ FOV, with a 10-chip CCD camera with a plate scale $0.57''$/pixel.
The main observations are taken using the MOA-Red (630-1000 nm) filter \citep{Bond:2001}. 

MOA data for MB09260 and MB10364 were reduced as described in \cite{Bond:2017}.\footnote{OB110037, OB110310, and OB110462 also have MOA lightcurves.
For simplicity we only present the OGLE lightcurve fits for those events, since the seeing at OGLE is better.}
The MOA lightcurves are photometrically calibrated to the OGLE-III I-band.

\subsection{OGLE \label{sec:OGLE}}

The OGLE-IV survey is carried out at the 1.3-m Warsaw telescope at Las Campanas Observatory in Chile \citep{Udalski:2015}.
The seeing at the site ranges from $\sim 1.0'' - 2.0''$, with the median seeing being $\sim 1.3''.$
The telescope has a 1.4 deg$^2$ FOV, with a 32-chip CCD camera with a plate scale of $0.26''$/pixel.
The main observations are taken using the OGLE-I filter, which is similar to Cousins I-band.
The data was reduced using the Difference Image Analysis technique as implemented by \citet{Wozniak:2000}.
In addition, we rescaled the photometric uncertainties of OB110462 according to the method described in \citet{Skowron:2016}.

OGLE data is only available for OB110037, OB110310, and OB110462. 
The magnification of MB09260 was not observed by OGLE as it occurred during the OGLE-III to OGLE-IV upgrade.
MB10364 is located in a gap in the detectors of the OGLE camera.

\subsection{HST \label{sec:HST}}

\begin{deluxetable}{llllll}
\tabletypesize{\scriptsize}
\tablecaption{HST WFC3-UVIS Observations
\label{tab:hst_obs_table}}
\tablehead{
    \colhead{Target} &
    \colhead{Epoch (UT)} &
    \colhead{PA} &
    \colhead{Filter} &
    \colhead{$T_{exp}$} &
    \colhead{$N_{im}$} \\
    \colhead{} &
    \colhead{(yyyy-mm-dd)} &
    \colhead{(deg)} &
    \colhead{} &
    \colhead{(sec)} &
    \colhead{}}
\startdata
\input{hst_obs_table1.txt}
\enddata
\tablecomments{
Asterisk ($^*$) denotes observations excluded from analysis.}
\end{deluxetable}

\begin{deluxetable}{llllll}
\tabletypesize{\scriptsize}
\tablecaption{HST WFC3-UVIS Observations}
\tablehead{
    \colhead{Target} &
    \colhead{Epoch (UT)} &
    \colhead{PA} &
    \colhead{Filter} &
    \colhead{$T_{exp}$} &
    \colhead{$N_{im}$} \\
    \colhead{} &
    \colhead{(yyyy-mm-dd)} &
    \colhead{(deg)} &
    \colhead{} &
    \colhead{(sec)} &
    \colhead{}}
\startdata
\input{hst_obs_table2.txt}
\enddata
\tablecomments{
Asterisk ($^*$) denotes observations excluded from analysis.
}
\end{deluxetable}

HST observations come from a multi-year campaign following up these five targets (GO-11707, GO-12322, GO-12670, GO-12986, GO-13458, GO-14783; PI: K. C. Sahu).
Observations were taken with the UVIS channel on the Wide Field Camera 3 (WFC3) in two different wide-band filters, F606W ($V$-band) and F814W ($I$-band).
Table \ref{tab:hst_obs_table} summarizes the HST observations. 

The WFC3 UVIS channel is composed of two 2k$\times$4k CCDs and has a $162'' \times 162''$ field of view with a plate scale $0.04''$/pixel.
WFC3 UVIS supports sub-arraying, in which only a portion of the entire detector is read out, which can reduce data volume or exposure time and increase observational efficiency.
All observations prior to 2011-07-22 were taken with the UVIS1-2K4-SUB subarray mode.
Beginning HST Cycle 18, more subarray sizes were made available, and observations after 2011-07-22 were taken with the UVIS2-2K2C-SUB subarray mode, a 2k$\times$2k subarray.

Additional observations of OB110462 commenced in Cycle 29 (GO-16760; PI: C. Lam).
These were taken in as similar a configuration as possible to the later epochs of the archival program, using WFC3 UVIS in UVIS2-2K2C-SUB subarraying mode, with observations in F606W and F814W filters.
The first set of observations from this program was taken October 2021 and is presented here; an additional set of observations is anticipated to be taken Fall 2022 \citep{Lam:2021hst}.

\begin{figure*}
\centering
    \includegraphics[width=\linewidth]{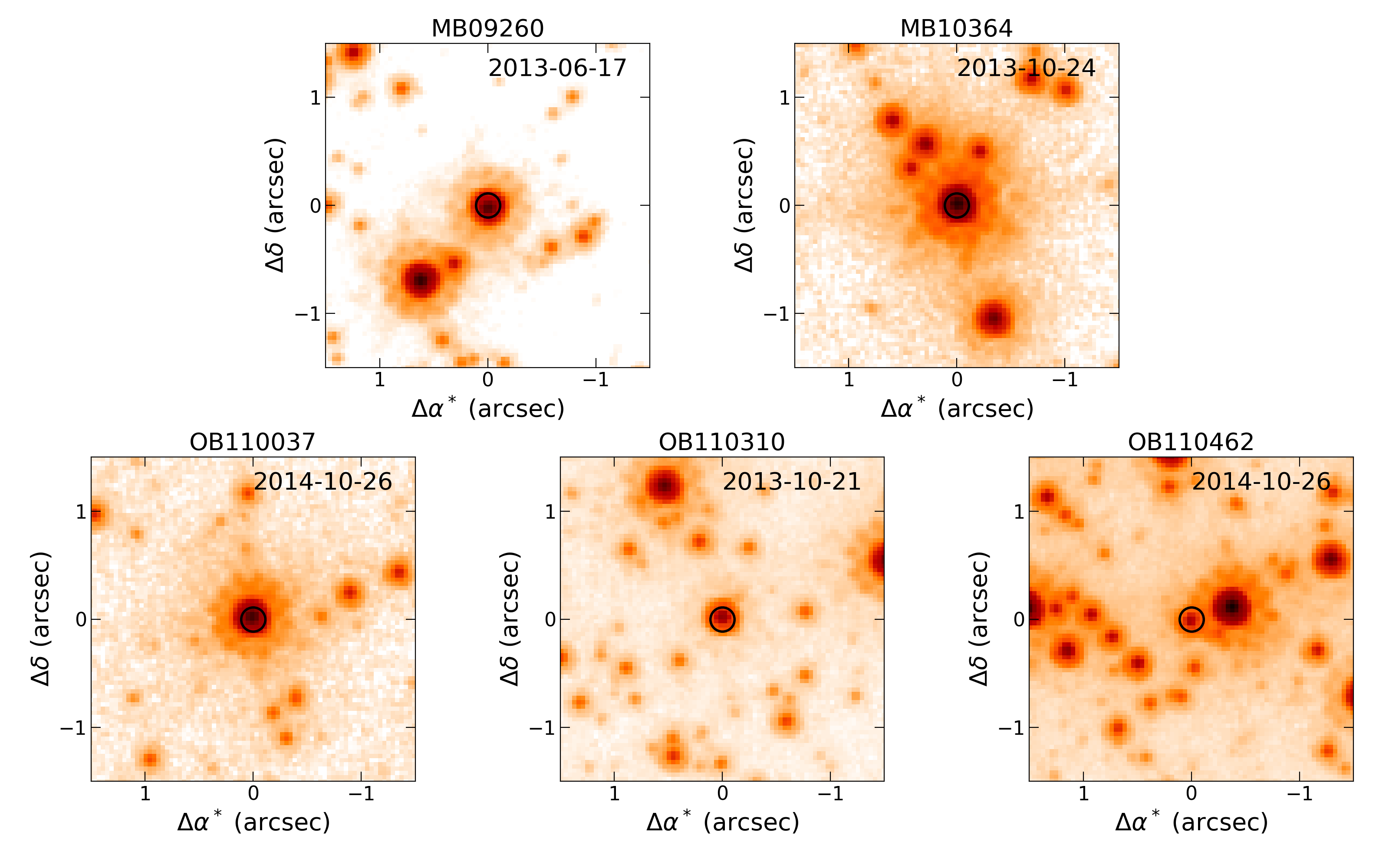}
\caption{\label{fig:images}
    Central $3'' \times 3''$ of HST WFC3-UVIS F814W combined images of the observed fields, centered on the target (circled).
    These images are of the target at or near baseline, i.e. unmagnified.
    The color stretch is logarithmic.
    Note that the color scale is not the same across panels.}
\end{figure*}

\begin{figure*}
\centering
    \includegraphics[width=\linewidth]{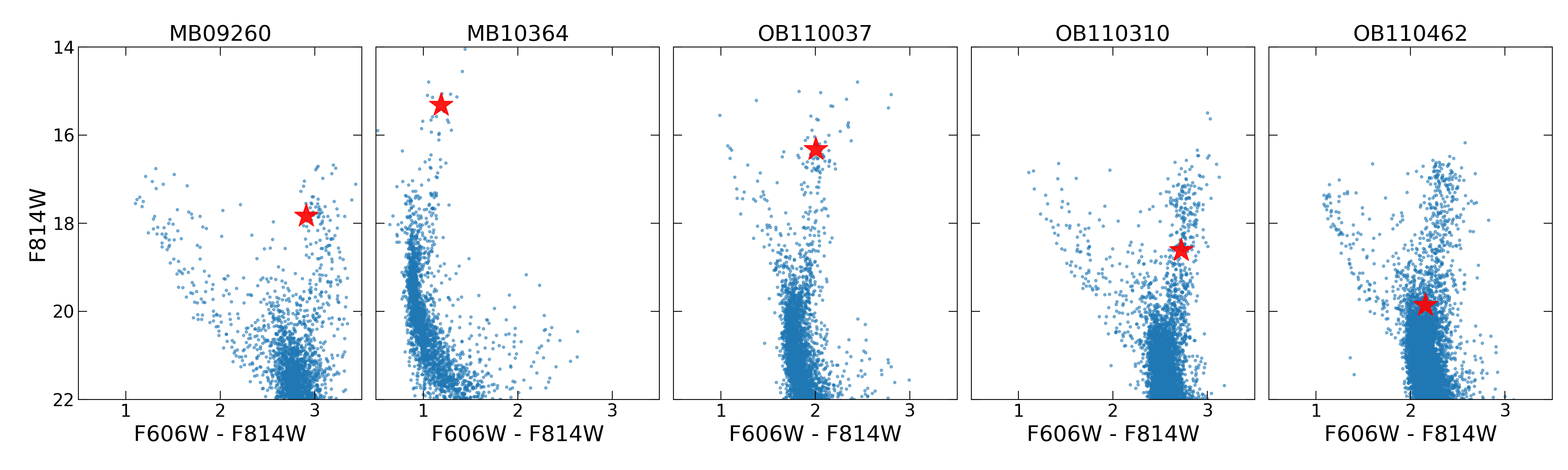}
\caption{\label{fig:CMDs}
    CMDs for each field.
     The target at baseline magnitude and color is marked as a red star.}
\end{figure*}

\subsection{Gaia \label{sec:Gaia}} 
Gaia is an all-sky scanning astrometric space mission \citep{Gaia:2016}.
All of the targets, with the exception of OB110462, are found in Gaia Early Data Release 3 (EDR3, Table \ref{tab:gaia}).
Gaia EDR3 covers the period from 25 July 2014 to 28 May 2017 \citep{Gaia:2021}.
MB10364, OB110037, and OB110310 have proper motions and parallaxes, while MB09260 does not.
OB110462 was not in Gaia as it is too faint.
We note that there is a Gaia source located $\sim 0.35''$ away from OB110462 (Gaia EDR3 Source ID 4056442477683080960), which coincides with the bright star directly west of the target seen in the HST images (Figure \ref{fig:images}).

\begin{deluxetable*}{lllll}
\tabletypesize{\footnotesize}
\tablecaption{Gaia EDR3 Values \label{tab:gaia}}
\tablehead{
    \colhead{Parameter} &
    \colhead{MB09260} &
    \colhead{MB10364} &
    \colhead{OB110037} &
    \colhead{OB110310}}
\startdata
\input{gaia_table.txt}
\enddata
\tablecomments{
Magnitude uncertainties are estimated from the Gaia reported flux errors.
Zero-point (ZP) correction comes from \citet{Lindegren:2021b}, $^*$ denotes values that are extrapolations.
For full descriptions we refer the reader to Gaia EDR3 documentation \citep{vanLeeuwen:2021}, Section 13.1.1 gaia\_source.
OB110462 is not in Gaia.
}
\end{deluxetable*}

\begin{figure*}[t!]
    \centering
    \includegraphics[width=0.95\linewidth]{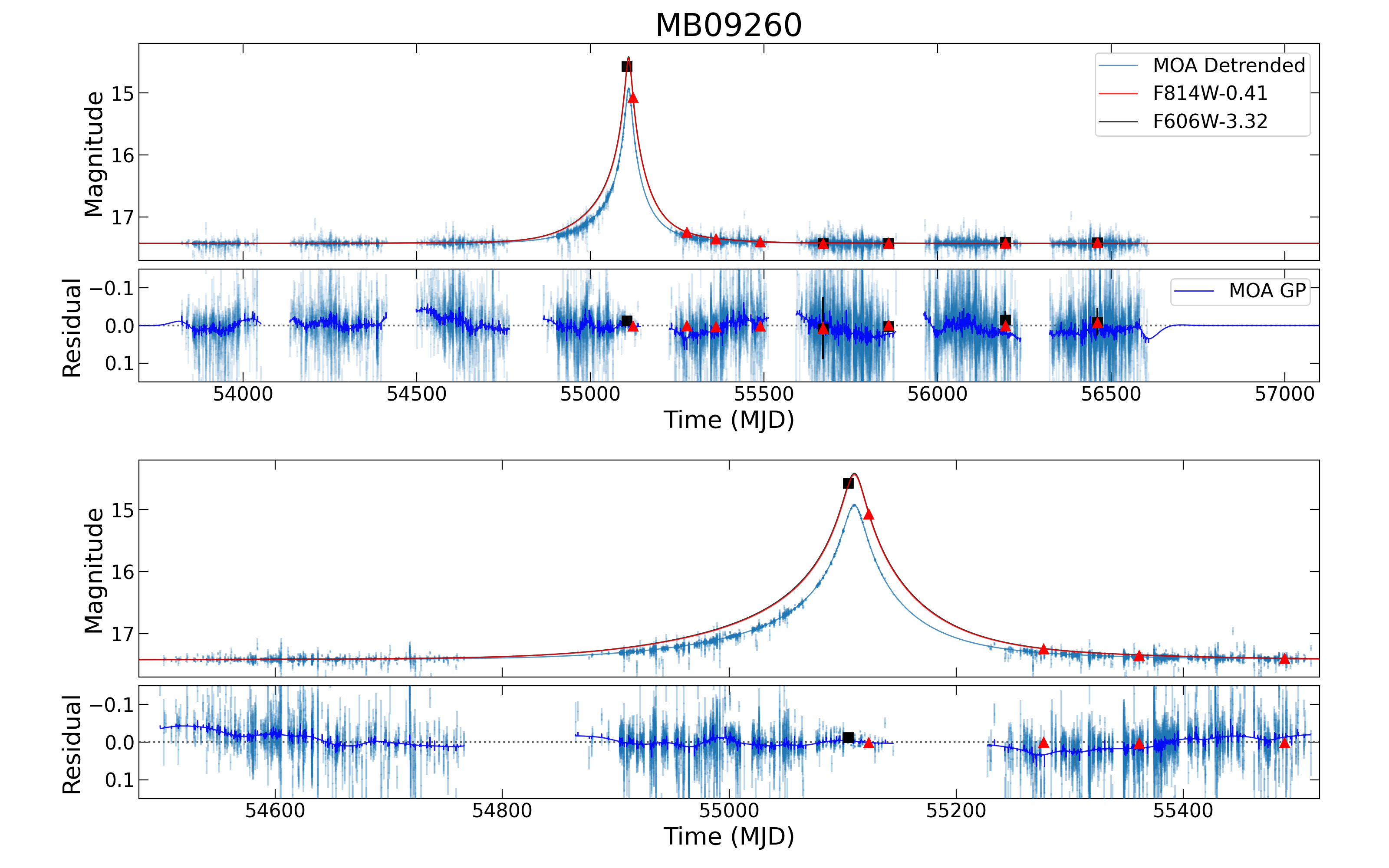}
    \caption{\emph{Top panel}: Detrended MB09260 lightcurve, as seen by MOA and HST F814W and F606W.
    The maximum likelihood model (MLE, described in \S \ref{sec:Results}) is plotted over the data.
    \emph{Second from top panel}: The residuals to the MLE model.
    The Gaussian Process (GP) model is plotted on top of the residual.
    We emphasize the residual is not independently fit by the GP, but is simultaneously fit with the model parameters; this is purely to visualize the data (also see \citet{Golovich:2022}).
    See \S \ref{sec:Microlensing Modeling} for more details about the fitting procedure.
    \emph{Second from bottom panel}: Same as top panel, but zoomed into the three most magnified years.
    \emph{Bottom panel}: Same as second from top panel, but zoomed into the three most magnified years.}
    \label{fig:MB09260_lightcurve}
\end{figure*}

\begin{figure*}[t!]
    \centering
    \includegraphics[width=0.95\linewidth]{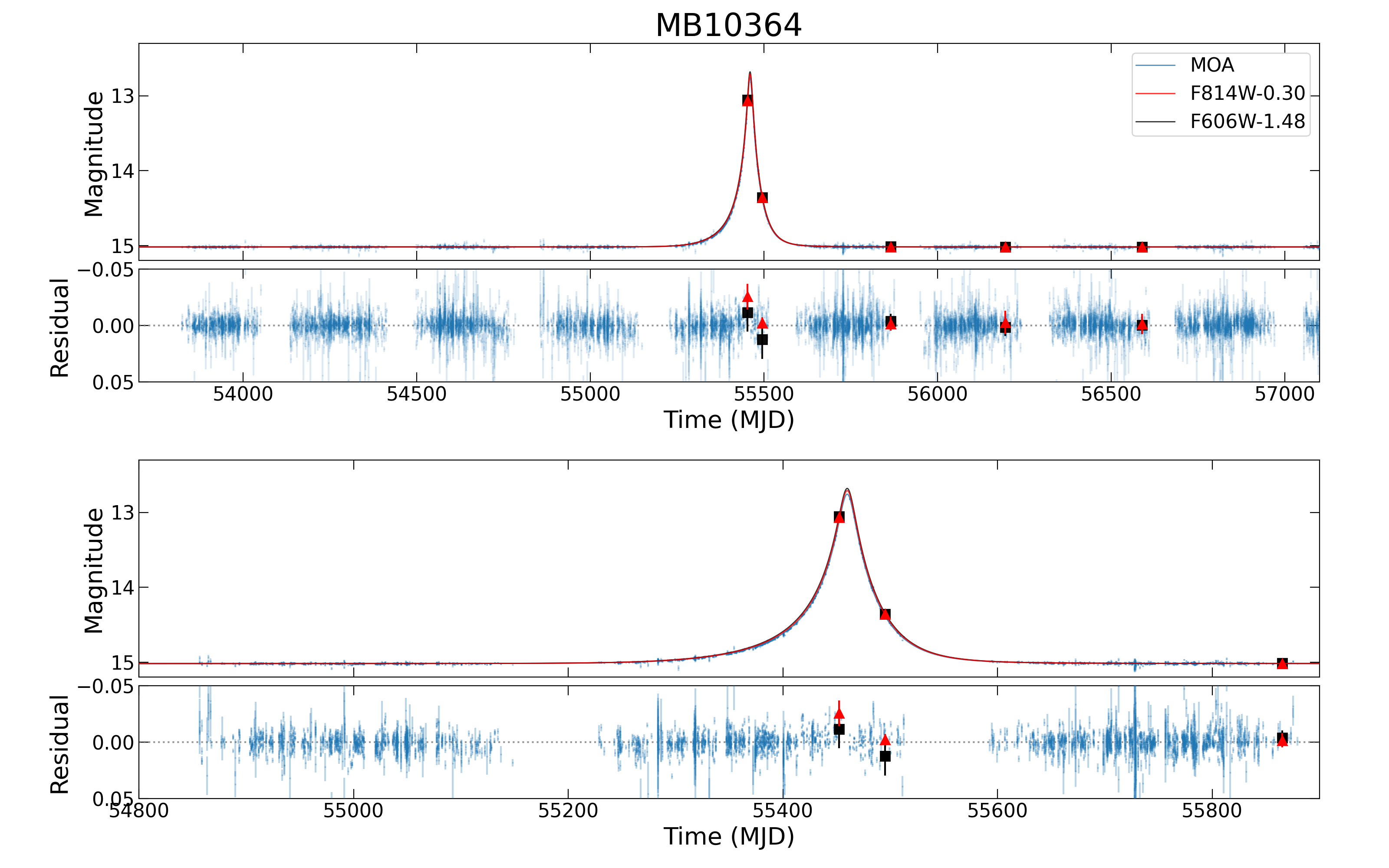} 
    \caption{Same as Figure \ref{fig:MB09260_lightcurve}, but for MB10364.
    Note we do not include GP in the MB10364 fit.}
    \label{fig:MB10364_lightcurve}
\end{figure*}

\begin{figure*}[t!]
    \centering
    \includegraphics[width=0.95\linewidth]{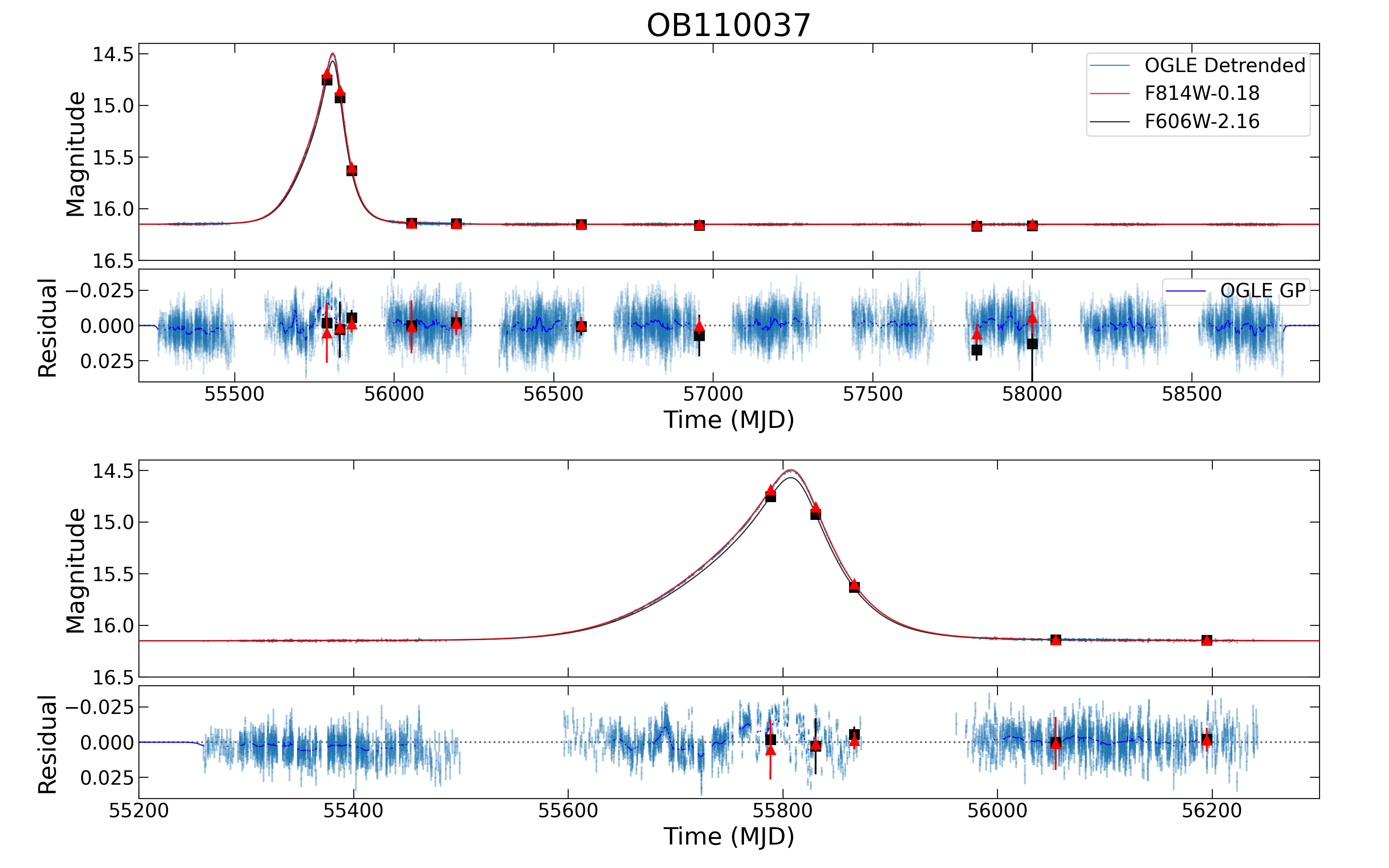} 
    \caption{Same as Figure \ref{fig:MB09260_lightcurve}, but for OB110037.
    Instead of MOA data, the blue data are OGLE data.}
    \label{fig:OB110037_lightcurve}
\end{figure*}

\begin{figure*}[t!]
    \centering
    \includegraphics[width=0.95\linewidth]{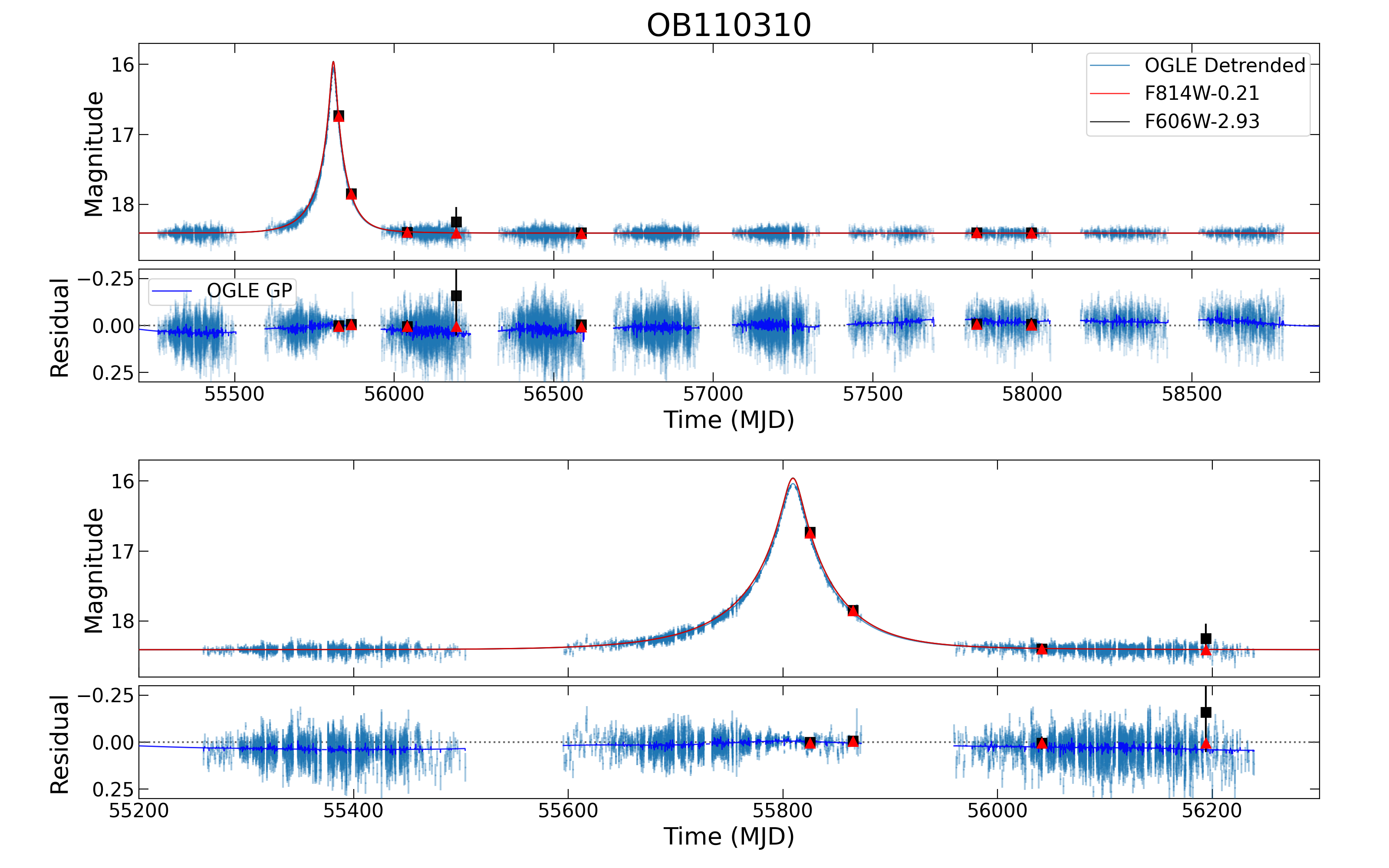} 
    \caption{Same as Figure \ref{fig:OB110037_lightcurve}, but for OB110310.}
    \label{fig:OB110310_lightcurve}
\end{figure*}

\begin{figure*}[t!]
    \centering
    \includegraphics[width=0.95\linewidth]{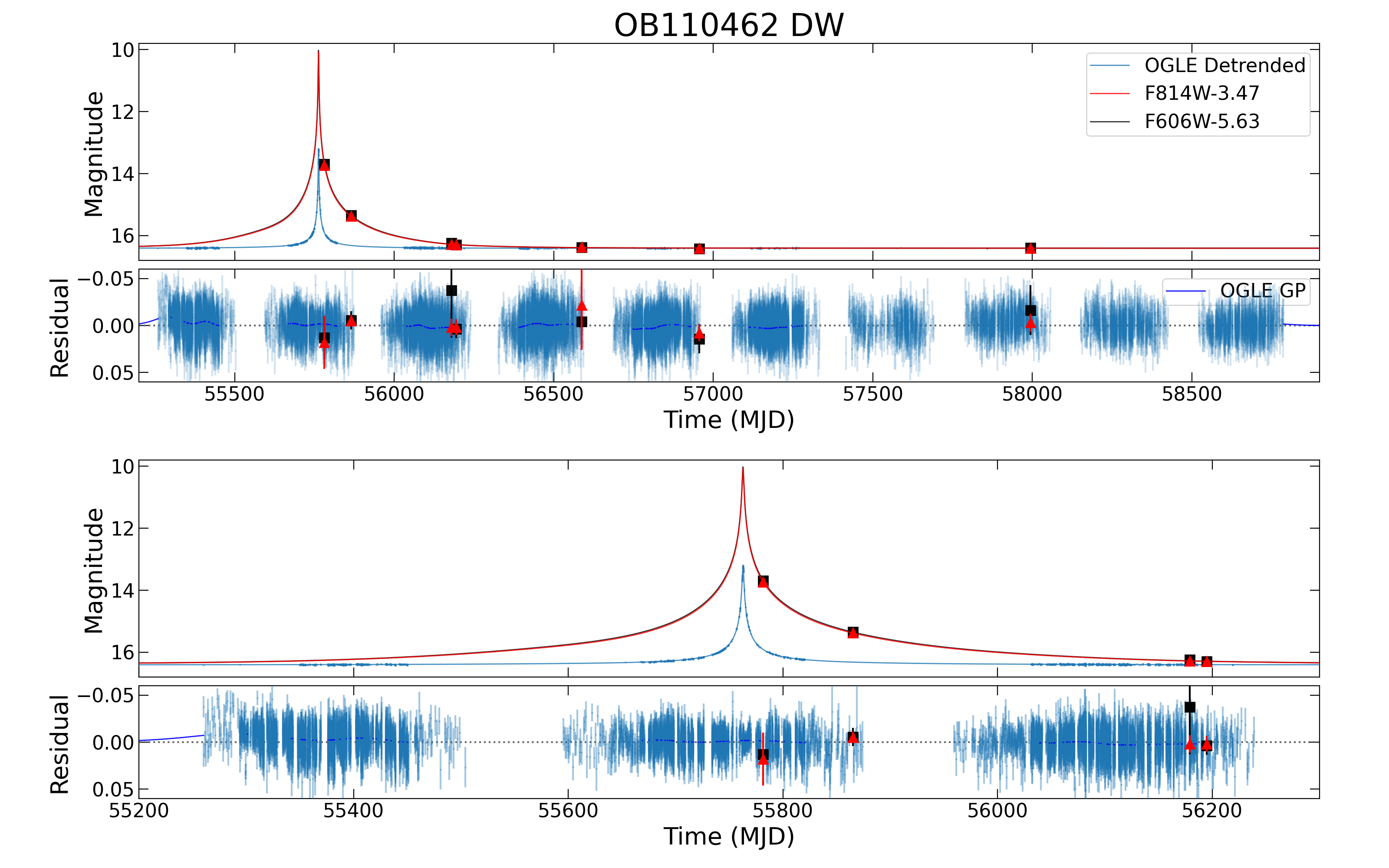} 
    \caption{Same as Figure \ref{fig:OB110037_lightcurve}, but for OB110462.
    This maximum likelihood model was calculated using the default weighted likelihood described in \S \ref{sec:Likelihood weighting}.}
    \label{fig:OB110462_DW_lightcurve}
\end{figure*}

\begin{figure*}[t!]
    \centering
    \includegraphics[width=0.95\linewidth]{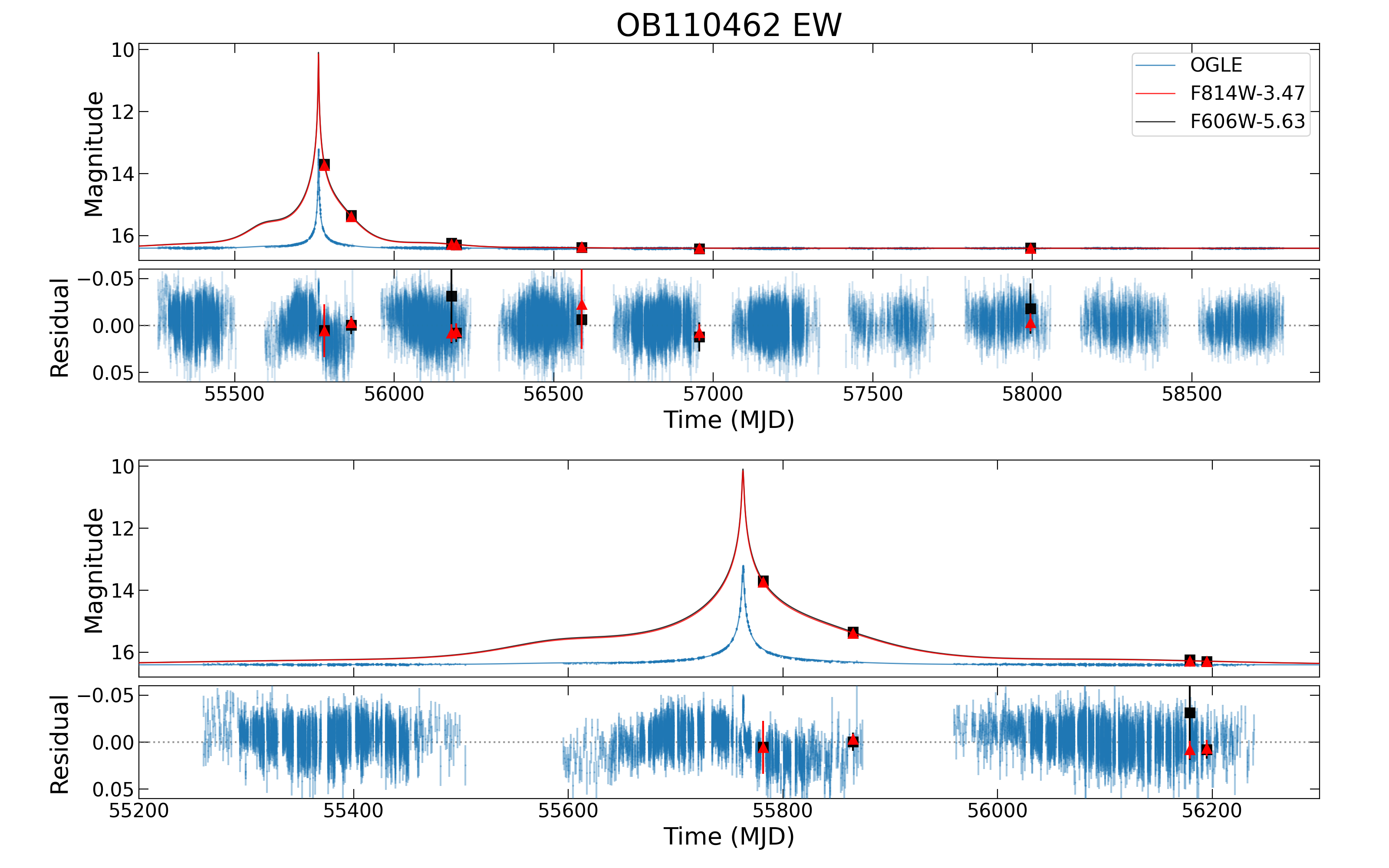} 
    \caption{Same as Figure \ref{fig:OB110037_lightcurve}, but for OB110462. This maximum likelihood model was calculated using the equal weight likelihood described in \S \ref{sec:Likelihood weighting}.}
    \label{fig:OB110462_EW_lightcurve}
\end{figure*}

\section{HST Data Analysis \label{sec:HST Data Analysis}}

\subsection{Reduction \label{sec:Reduction}}

The HST archival data was accessed from the Mikulski Archive for Space Telescopes\footnote{\url{https://archive.stsci.edu/hst/}} (MAST) in June 2021.
The Cycle 29 data was accessed October 2021.
For the following analysis, we employed the calibrated, flat-fielded, individual exposures corrected for charge transfer efficiency (CTE, HST files with suffix \texttt{\_flc}\footnote{See \cite{Gennaro:2018} for a full description of the different file name suffixes.}).
The archival data was processed with Version 3.6.0 (Dec-31-2020) of the \texttt{calwf3} pipeline, using Version 2.0 of the CTE correction algorithm.

CTE can alter astrometry at the milliarcsecond level, hence it is important to use \texttt{\_flc} files.
However, even the \texttt{\_flc} files do not necessarily fix all problems associated with CTE \citep{Kuhn:2021}.
Exploration of other methods of CTE correction will be explored in future work. 
At the present we mitigate CTE effects via other methods (\S \ref{sec:Reference star selection}) and validate our astrometry to ensure it is not distorted by CTE.

Images were converted into calibrated star lists via the following steps.
\begin{enumerate}
    \item \emph{Star list extraction from individual frames.} 
    Star lists were extracted from the individual \texttt{\_flc} exposures by modeling the PSFs of sources with \texttt{hst1pass}, an updated version of the software described in \cite{Anderson:2006}.
    Empirical filter-dependent PSF models as described in \cite{Anderson:2016} and geometric distortion solutions as described in \cite{Bellini:2011} were used when performing source extraction with \texttt{hst1pass}.

    \item \emph{Combined star list for one epoch.} 
    Within a single epoch and filter, multiple star lists were aligned to a common coordinate system in an iterative manner using \texttt{xym2mat} and \texttt{xym2bar} \citep{Anderson:2006}, which include the distortion solution for the WFC3 camera and filters \citep{Bellini:2011} to produce a single matched star list.
    
    \item \emph{Photometric calibration.} 
    Lastly, a zero-point is applied to the star lists to convert from instrumental to Vega magnitudes. 
    Star lists were calibrated against photometrically calibrated star lists on the Hubble Legacy Archive, Data Release 10 (HLA DR10).
    A magnitude offset is applied later during the astrometric alignment (\S \ref{sec:HST Photometric Analysis}) to obtain more precise relative photometry.
\end{enumerate}

Note that data taken in F606W and F814W filters are treated as independent measurements.
That is, observations taken on the same date are treated as distinct epochs, and are not combined into a single star list, as the importance of filter dependence in astrometry is not well established.
See \S \ref{sec:Astrometric color offset} and Appendix \ref{app:Astrometric color analysis} for further details.

Certain epochs were excluded from the analysis; these are marked with an asterisk in Table \ref{tab:hst_obs_table}.
The reason for their exclusion is detailed as follows.
\begin{itemize}
    \item \emph{Epochs with only a single frame.}
    Observations with only a single frame per filter cannot produce any useful photometric or astrometric constraints using \texttt{hst1pass}.
    This is the case for the MB09260 2009-10-01 F814W and 2009-10-19, 2010-03-22, and 2010-06-14 F606W epochs.
    In the MB09260 2010-10-20 F606W epoch, a cosmic ray in one of the exposures interfered with the extraction of the target, effectively leaving only a single usable frame.
    
    \item \emph{Multiple exposure times.}
    Although in principle mixing multiple exposure times in a single epoch is possible, in practice 
    most of the data was obtained with several long exposures and only a single frame with a shorter exposure.
    Rather than analyze the few short frames with different detection thresholds, PSF reference stars, and astrometric reference stars, which can lead to systematic errors, we choose to only use frames with the same exposure times within an epoch.
    For this reason, some frames from the MB09260 2009-10-01 F606W, MB10364 2010-09-13 F814W, OB110037 2011-08-15 F814W, OB110310 2013-10-21 F814W, and OB110462 2011-08-08 F606W and F814W epochs were not used.
    
    \item \emph{Saturation of target.}
    No useful astrometric limits can be placed when the target is saturated.
    The target is saturated in both filters in the MB10364 2011-04-13 and 2011-07-22 epochs.
    
    \item \emph{Telescope pointing issues.}
    The observations of OB110462 on 2017-08-11 suffered a telescope drifting issue, resulting in streaked images.
    
    \item \emph{Astrometric alignment systematics.}
    Although there are no standalone issues with the observations of OB110462 on 2013-05-13, astrometric systematics are apparent in the reference stars when this epoch is astrometrically aligned along with the other epochs using the methodology described in \S \ref{sec:HST Astrometric Analysis}. 
    This is due to the difference in position angle of the observations taken, as the 2013-05-13 epoch was taken at PA = 99.9 deg, while all the other epochs were taken with a PA different by $\sim180$ deg, with PA = 255.2 - 276.1 deg.
    Thus, the 2013-05-13 epoch is left out of the analysis.
    
    The other targets (MB09260, OB110037, OB110310) with $\sim180$ deg differences in PA across observations do not suffer this same problem, as there are multiple observations at each PA.
    This allows the systematics due to the $\sim180$ deg PA flip to be calibrated out during the astrometric alignment.
\end{itemize}

\subsection{HST Astrometric Analysis \label{sec:HST Astrometric Analysis}}

The positional measurements extracted from the different epochs of HST data (\S\ref{sec:Reduction}) must be transformed into a common reference frame in order to derive the motion of the target.
This is an iterative process, with multiple ``passes" at refining the reference frame, allowing for the best relative astrometry possible to be extracted.
We follow a similar procedure as described in \S 4.2 of \cite{Lu:2016}.
The final positions and magnitudes of the targets resulting from the analysis in this section is presented in Table \ref{tab:HST Calibrated Data}.
The photometry is shown in Figures \ref{fig:MB09260_lightcurve} - \ref{fig:OB110462_EW_lightcurve} and the astrometry is shown in Figures \ref{fig:OB110462_astrom_EW} - \ref{fig:OB110310_astrom}.

\subsubsection{Alignment procedure \label{sec:Alignment procedure}}

Following standard image processing techniques, a 2-D polynomial transformation of the form
\begin{align}
    x' &= a_0 + a_1 x + a_2 y + a_3 x^2 + a_4 xy + a_5 y^2 + ... \label{eq:xpoly} \\
    y' &= b_0 + b_1 x + b_2 y + b_3 x^2 + b_4 xy + b_5 y^2 + ... \label{eq:ypoly}
\end{align}
is applied to the images in order to match them to a reference image.
A first order 2-D polynomial transformation ($x' = a_0 + a_1 x + a_2 y$; $y' = b_0 + b_1 x + b_2 y$) is an affine transformation\footnote{An affine transformation maps points to points, lines to lines, planes to planes, and so on. 
Affine tranformations preserve collinearity and ratios of distances. 
Parallel lines also remain parallel after an affine transformation.}, which can be used to model translation, rotation, scaling, and shearing introduced by the camera.
A higher-order polynomial can correct for additional distortions, but going beyond second order generally does not improve results as the number of free parameters quickly increases and results in overfitting.

In the first pass, the HST images are aligned to the absolute reference frame of Gaia with a first order 2-D polynomial transformation to roughly establish the transformation.
The Gaia EDR3 catalog is matched to the HST catalog using the pattern matching algorithm of \cite{Groth:1986}.

In subsequent passes, the HST images are aligned to themselves, using a 2-D polynomial transformation going up to second order.
It is empirically determined that 3-4 passes gives optimal results. 
In each successive pass, the HST images are aligned to the reference frame derived in the previous pass, which continually refines the reference frame and derived proper motions.

To calculate the optimal transformation, a set of reference stars $r^R_{ref}(t_0)$ are selected from the stars in reference frame $R(t_0)$ observed at time $t_0$.
The reference stars $r^R_{ref}(t_0)$ are matched to corresponding stars $u^U_{ref}(t)$ in the untransformed $U(t)$ frame observed at time $t$.
The transformation $T:U(t) \rightarrow R(t)$ is found by least-squares minimization of the $x$ and $y$ position residuals from the alignment
\\

\begin{align}
    x_{res} &= \sum_i w_{x,i} (x^R_{r,i}(t) - T(x^U_{u,i}(t)))^2 \\
    y_{res} &= \sum_i w_{y,i} (y^R_{r,i}(t) - T(y^U_{u,i}(t)))^2
\end{align}

where $w_{(x,y),i}$ is the weight for the $i$-th reference star, $(x^R(t)_{r,i},y^R(t)_{r,i})$ and $(x^U_{u,i}(t),y^U_{u,i}(t))$ are the positions of the stars in the reference $R(t)$ and untransformed $U(t)$ frames at time $t$.
The positions of the reference stars $r^R_{ref}(t)$ in the reference frame $R$ at time $t$ are propagated from time $t_0$ using the proper motions
\begin{align}
    x^R_{r,i}(t) &= x^R_{r,i}(t_0) + v^R_{x,r,i}(t - t_0) \\
    y^R_{r,i}(t) &= y^R_{r,i}(t_0) + v^R_{y,r,i}(t - t_0) 
\end{align}
where $v^R_{(x,y),r,i}$ are the proper motions of the $i$-th reference stars in reference frame $R$.
After the transformation $T$ is derived, it is applied to all the stars $u^U$ in the $U$ frame to obtain a transformed star list $u^R(t) = T(u^U(t))$, where the stars $u^R(t)$ are now in the frame $R$.
This yields starlists for all $N$ epochs $u_1^R(t_1), \, ...  \, , u_N^R(t_N)$, where the positions of all the stars are now in the same reference frame $R$.
For each star $j$, a proper motion is derived by finding the best-fit straight line via non-linear least squares through the $n \leq N$ observations.\footnote{Some stars are not detected in all epochs, which is why it is possible to have $n < N$ observations.} 

\cite{Lu:2016} examined several different weighting schemes and showed that the resulting astrometry is not affected.
We choose to use weights $w_{(x,y),i} = 1/\sigma_{(x,y),i}$, where $\sigma_{(x,y),i}$ are the positional uncertainties of the stars in the untransformed frame.
For the positional uncertainties, instead of using the root mean square values  $\sigma_{RMS}$ directly returned by \texttt{hst1pass}, we follow \cite{Hosek:2015} and use the error on the mean $\sigma_{RMS}/\sqrt{N}$ where $N$ is the number of frames the source is detected in, with an additional empirical additive error.
The uncertainties, as well as the procedure used to determine them, are detailed in Appendix \ref{sec:Rescaling of uncertainties}.
The additive error term dominates over the $\sigma_{RMS}/\sqrt{N}$ term for bright stars, which makes the positional errors more uniform across epochs and magnitude as would be expected for systematic errors.

As we are interested in the astrometry of the target, the target itself is not used to establish the transformation into a common reference frame (\S \ref{sec:Reference star selection}) nor to judge the quality of the final transformation (Figure \ref{fig:chi2_xym}). 

\subsubsection{Reference star selection \label{sec:Reference star selection}}

Reference stars are stars assumed to have linear proper motions, which are used to derive the reference frame transformation and the motions of the other stars.
The selection of reference stars depends on multiple considerations, such as the stellar density, amount of geometric distortion, instrumental systematics, and number and brightness of targets of interest, to name a few.
The goal is to balance having enough stars to establish the reference frame, while excluding stars which would produce a non-stable reference frame.
The criteria for reference star selection for each target is summarized in Table \ref{tab:Reference star criteria}. 
We choose reference stars with brightness similar to the target, relatively large radial separations from our target of interest, and exclude likely foreground stars.
The target itself is also excluded from being a reference star.
We detail the reasoning for these choices below.

\emph{Brightness range}: Due to the nature of CTE, there are strong magnitude-dependent astrometric residuals, even when using the latest CTE-corrected \texttt{\_flc} images.
However, this is not unexpected \citep{Kuhn:2021}.
For this reason, stars are chosen to be in a brightness range similar to the one spanned by the target as narrow as possible.
For OB110310 and OB110462, which are relatively faint and where there are many stars of similar brightness, all stars falling within $\pm 0.1$ mag of the target's brightest and faintest in the HST data are used.
MB09260, on the other hand, is brighter, with less stars of similar brightness, so the range is larger, with all stars falling within $\pm 1$ mag of the target's brightest and faintest in the HST data are used.
MB10364 is so bright that many stars of comparable brightness are saturated in the longer exposures.
Because of this, only bright (F814W $< 18.0$, F606W $< 19.2$) and unsaturated stars were selected; the bright limit on the magnitude range differs between epochs because of the different exposure times.

\emph{Spatial separation}:
Only reference stars within $30''$, or $20''$ for the denser field around OB110462, are used as reference stars.
This minimizes the impact of geometric distortion residuals and spatially dependent PSF variations.

\emph{Foreground stars}: A key assumption in the astrometric alignment process is that reference stars have linear proper motions, and parallax effects are ignored.
For a typical bulge star 8 kpc from Earth, this is a reasonable assumption as the parallax will be 1/8000 arcsec = 0.125 mas, below our achievable astrometric precision. 
However, for nearby stars, ignoring parallax is an issue when trying to derive an accurate transformation.
As all the target fields are toward the highly extincted Galactic bulge, bright blue stars as identified on a CMD (Figure \ref{fig:CMDs}) are likely to be nearby and have a non-negligible parallax, and are excluded from the set of reference stars.
The color-magnitude exclusion criteria are listed in the last column of Table \ref{tab:Reference star criteria}. 
For MB10364, no bright blue stars were removed as reference stars, as all the observations came from within 6 weeks of the same time of year.
Hence, any type of yearly parallax signal would be negligible within this time span.

\emph{Number of detections}: We require reference stars to be detected in most, if not all, epochs.
If there are $N_{\textrm{ep}}$ total epochs, we require reference stars to be detected in $N_{\textrm{ep,detect}} = N_{\textrm{ep}} - 2$ epochs.

Lastly, as the motion of the target is the quantity we are interested in, we do not use it as a reference star.

\subsubsection{Derived stellar proper motions \label{sec:Derived stellar proper motions}}

\begin{figure*}
\centering
    \centering
    \includegraphics[width=1.0\linewidth]{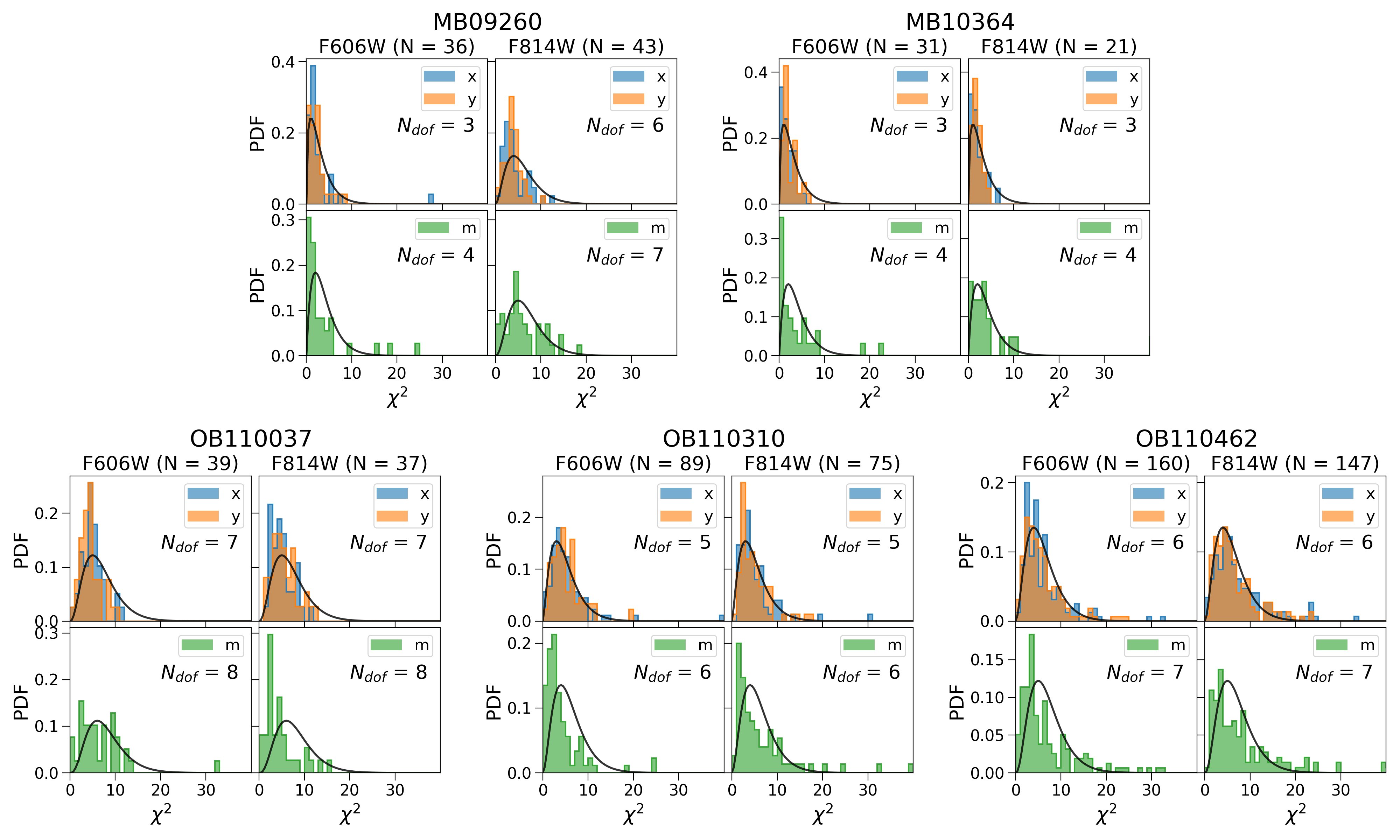} 
    \caption{Histogram of $\chi^2$ residual values to the linear fits with no parallax for the reference stars of each target (Table \ref{tab:Reference star criteria}).
    In each panel, the left column shows the distributions for reference stars in F606W, while the right column shows F814W.
    The number of reference stars is listed as $N$.
    In each panel, the top row shows the $\chi^2$ distribution of residuals of a linear fit to positions vs. time in $x \equiv$ RA and $y \equiv$ Dec.
    The bottom row shows the $\chi^2$ distribution of residuals of a constant fit to magnitude vs. time.
    The expected $\chi^2$ distributions are shown in black, with the number of degrees of freedom listed as $N_{dof}$.
    \label{fig:chi2_xym}}
\end{figure*}

To evaluate the goodness of the fits of the derived stellar proper motions, we consider the $\chi^2$ distributions of the position residuals
\begin{align}
    \chi^2_x &= \sum_t \Big( \frac{x_t - x_{t,fit}}{\sigma_{x_t}} \Big)^2 \\
    \chi^2_y &= \sum_t \Big( \frac{y_t - y_{t,fit}}{\sigma_{y_t}} \Big)^2
\end{align}
where $x \equiv$ RA, $y \equiv$ Dec, $(x,y)_t$ are the positions in the data, $(x,y)_{t,fit}$ are the positions as derived from the linear motion fit, and $\sigma_{(x,y),t}$ are the positional uncertainties at time $t$.
The distributions of $\chi^2_x$ and $\chi^2_y$ for the reference stars detected in all epochs $N_{ep}$ are shown in Figure \ref{fig:chi2_xym}, with the expected $\chi^2$ distribution overplotted on top.
The distributions for the positions in F814W and F606W are shown separately.
The expected residual distribution has $N_{ep,detect} - 2$ degrees of freedom, as there are two free parameters in the linear motion fit (initial position and proper motion).
Note that unlike Gaia, this linear model fit does not include parallax.
Parallax is only included when modeling the microlensing event.

\begin{deluxetable*}{lcccc}
\tablecaption{Reference star criteria \label{tab:Reference star criteria}}
\tablehead{
    \colhead{Target} &
    \colhead{Magnitude range} & 
    \colhead{Radius} & 
    \colhead{$N_{\textrm{ep,detect}}$} & 
    \colhead{EXCLUDED bright blue stars}}
\startdata
\input{ref_star_crit_table.txt}
\enddata
\tablecomments{
These are the criteria for the last pass.}
\end{deluxetable*}

\subsubsection{OB110462 bias correction \label{sec:OB110462 bias correction}}

There is a bright star (``the neighbor") $\sim$10 pixels ($\sim$0.4 arcsec) west of OB110462.
The neighbor is $\sim$3 magnitudes brighter than OB110462 at baseline (F814W = 16.7 mag, F606W = 19.0 mag).
Because of its proximity and high contrast, the neighbor's PSF might ``leak" onto OB110462 and alter its astrometry and photometry.
We perform injection and recovery tests to ascertain the reliability of faint source extraction near a bright source, in order to determine whether the astrometry and photometry of OB110462 as determined in \S\ref{sec:HST Astrometric Analysis} and \S\ref{sec:HST Photometric Analysis} is biased by the bright neighbor.

The methodology and results of the injection and recovery analysis are detailed in Appendix \ref{app:Injection and recovery tests}.
In summary, the positional bias is negligible when the target is highly magnified and of similar brightness to the neighbor.
However, in epochs where the target is no longer magnified, the bright star biases the position of the target.
In F814W, where the resolution is lower, the measured position of the target is biased toward the neighbor by $\sim 0.4$ mas along the target-neighbor separation vector.
In F606W, where the resolution is higher, the bias is less ($\sim 0.25$ mas) with the direction of bias more randomly oriented.
Similarly, the photometric bias is larger when the contrast is large, with the bright neighbor causing the extracted photometry of OB110462 to be brighter than the injected values.
The effect is again more severe in F814W than F606W because of the lower resolution. 

Using the results of the injection and recovery analysis, we calculate a bias correction to apply to OB110462 astrometry and photometry (Table \ref{tab:injection recovery bias}). 
The values in the table are added to astrometry and photometry derived in \S\ref{sec:HST Astrometric Analysis} and \S\ref{sec:HST Photometric Analysis}; the uncertainties are added in quadrature to the uncertainties in \S \ref{sec:HST Astrometric Analysis} and \S\ref{sec:HST Photometric Analysis}.

We only perform this analysis for OB110462, as it is the only faint target near a bright companion.
All the other targets are either bright with faint companions, isolated, or both bright and isolated.

\subsubsection{Astrometric color offset
\label{sec:Astrometric color offset}}

As mentioned in \S\ref{sec:Reduction} the data taken in F606W and F814W filters are treated as independent measurements.
For OB110037 and OB110462, the astrometric measurements in F606W and F814W do not agree within the uncertainties.

For OB110037, although the 2011 and 2012 epochs show good agreement, the 2013 to 2017 epochs become increasingly discrepant as time goes on.
We attribute this difference to binarity (\S\ref{sec:OB110037}).

In contrast, for OB110462 the astrometry in the F606W and F814W are discrepant in all datasets, but the difference appears to be a relatively constant offset with time.
This is true both before and after applying the bias correction in \S\ref{sec:OB110462 bias correction}.
Because the nature of the color difference appears to be a constant offset, we apply a constant shift to the OB110462 F606W astrometry
\begin{align}
    \label{eq:color_offset_x}
    \Delta_x &= \frac{\sum_t w_{x,t} (x_{{\rm F814W},t} - x_{{\rm F606W},t})}{\sum_t w_{x,t}} \\
    \label{eq:color_offset_y}
    \Delta_y &= \frac{\sum_t w_{y,t} (y_{{\rm F814W},t} - y_{{\rm F606W},t})}{\sum_t w_{y,t}} 
\end{align}
where 
\begin{align}
    w_{x,t} &= (\sigma_{x, {\rm F814W}, t}^2 + \sigma_{x, {\rm F606W}, t}^2)^{-1/2} \\
    w_{y,t} &= (\sigma_{y, {\rm F814W}, t}^2 + \sigma_{y, {\rm F606W}, t}^2)^{-1/2}.
\end{align}
and $t$ indexes the observation times.
Thus the modified astrometry for F606W is 
\begin{align}
    x_{{\rm F606W},t}' &= x_{{\rm F606W},t} + \Delta_x \\
    y_{{\rm F606W},t}' &= y_{{\rm F606W},t} + \Delta_y.
\end{align}
The values of the offset are $\Delta_x = -0.57$ mas and $\Delta_y = 0.39$ mas.
Note that these offsets are calculated using the bias-corrected astrometry.\footnote{Note that even before the bias correction of \S\ref{sec:OB110462 bias correction} is applied, this color offset is still present.
In fact, it is slightly larger, with $\Delta_x = -0.79$ mas and $\Delta_y = 0.52$ mas.
The bias correction is not the source of the color dependent astrometric offset; rather, it helps to slightly decrease the offset.}
See Appendix \ref{app:Astrometric color analysis} for further details and justification.

We also investigate stars nearby to determine whether any of them show similar behavior.
For the 70 stars within $3''$ of OB110462, we calculate the average positional offset between F814W and F606W in RA and Dec using Eqs.~\ref{eq:color_offset_x} and \ref{eq:color_offset_y}.
We then search for stars where the average positional uncertainty in F814W and F606W (whichever is larger) is smaller than the average positional offset to determine which differences are significant.
There are 4 stars where the average positional offset is greater than the average positional uncertainty in RA, and an additional 4 stars the average offset is greater than the average positional uncertainty in Dec.
Hence, a total of 8 out of the 70 stars near the target also show these significant offsets.
Thus, this effect is seen for roughly 10\% of the stars, and so is not very unusual.
Although we currently have no explanation for its significance, it appears random, and thus include the astrometric offset when analyzing the data, although it may be attributed to binarity (see \S \ref{sec:OB110037} and \S 4.1
in the main paper).
See Appendix \ref{app:Astrometric color analysis} for additional details.

\subsubsection{Comparison to Gaia proper motions and parallaxes
\label{sec:Comparison to Gaia proper motions and parallaxes}}

\begin{deluxetable*}{lccc}
\tablecaption{Gaia vs. HST Proper Motions}
\tablehead{
    \colhead{Target} &
    \colhead{$\mu_{\rm HST, L}$ (mas/yr)} &
    \colhead{$\mu_{\rm HST, S}$ (mas/yr)} & 
    \colhead{$\mu_{\rm Gaia}$ (mas/yr)}}
\startdata
\input{gaia_vs_hst_pm.txt}
\enddata
\tablecomments{The source and lens proper motions here have been transformed into the absolute Gaia proper motion frame, which is offset to the HST proper motion frame as described in Appendix \ref{sec:Absolute proper motion reference frame}. 
The uncertainties on $\mu_L$ and $\mu_S$ also reflect the uncertainty in the Gaia to HST proper motion transformation; the standard error on the mean of that transformation was added in quadrature to the uncertainties from the proper motion fits.
For this reason, the uncertainties for $\mu_L$ and $\mu_S$ in this table do not match those in Tables \ref{tab:mb10364_fits} - \ref{tab:ob110310_fits_modes}.
\label{tab:gaia_vs_hst_pm}}
\end{deluxetable*}

\begin{figure*}
\centering
    \includegraphics[width=\linewidth]{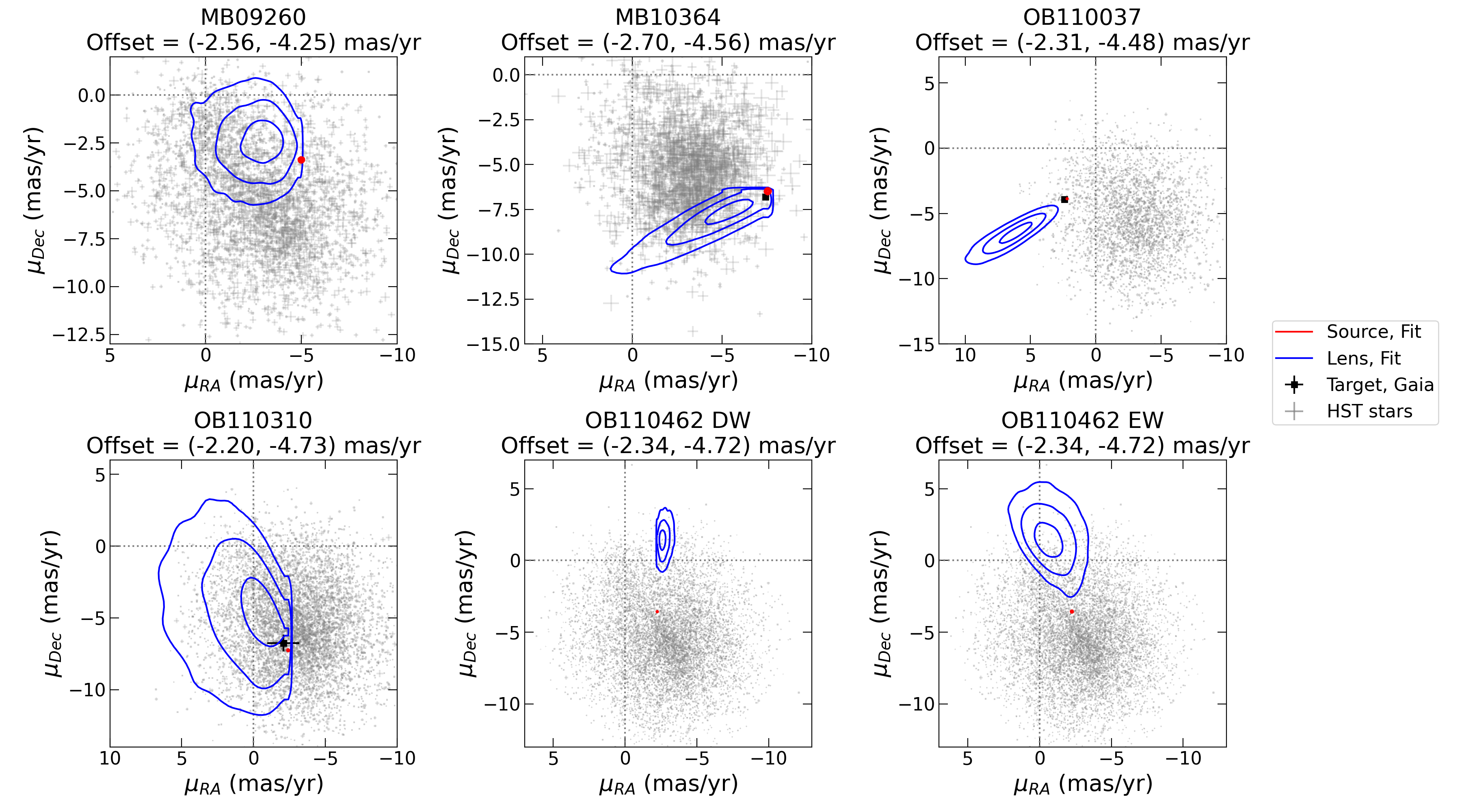}
\caption{
Proper motions for stars in the field of the target.
The proper motions derived from F814W HST observations for stars within $30''$ of the target are shown in gray with 1$\sigma$ uncertainties.
Only stars with F814W $< 23$ for MB09260, OB110310, OB110462 and F814W $< 22$ for MB10364, OB110037 are shown.
For OB110462 there are two models depending on the likelihood used (default weighted ``DW" or equal weighted ``EW",  see \S \ref{sec:Likelihood weighting} for details).
As the alignment procedure places the stars in a reference frame where the relative motion is zero, a constant offset must be added to obtain proper motions in the original Gaia absolute reference frame.
This offset is calculated by matching the stellar positions in HST to those in Gaia with \texttt{astrometric\_excess\_noise\_sig}$ < 2$, and then calculating the 3$\sigma$ clipped average weighted by the uncertainty in their difference.
The offset value is given in the title for each field.
For targets in Gaia where a single-star proper motion is estimated (MB10364, OB11037, OB110310), they are plotted as black squares.
The lens and source proper motion as determined from fitting the HST data with a microlensing model is shown in blue and red $1-2-3\sigma$ contours, respectively.
Note that the red contours are extremely small.
\label{fig:VPDs}}
\end{figure*}

\begin{figure}
\centering
    \includegraphics[width=\linewidth]{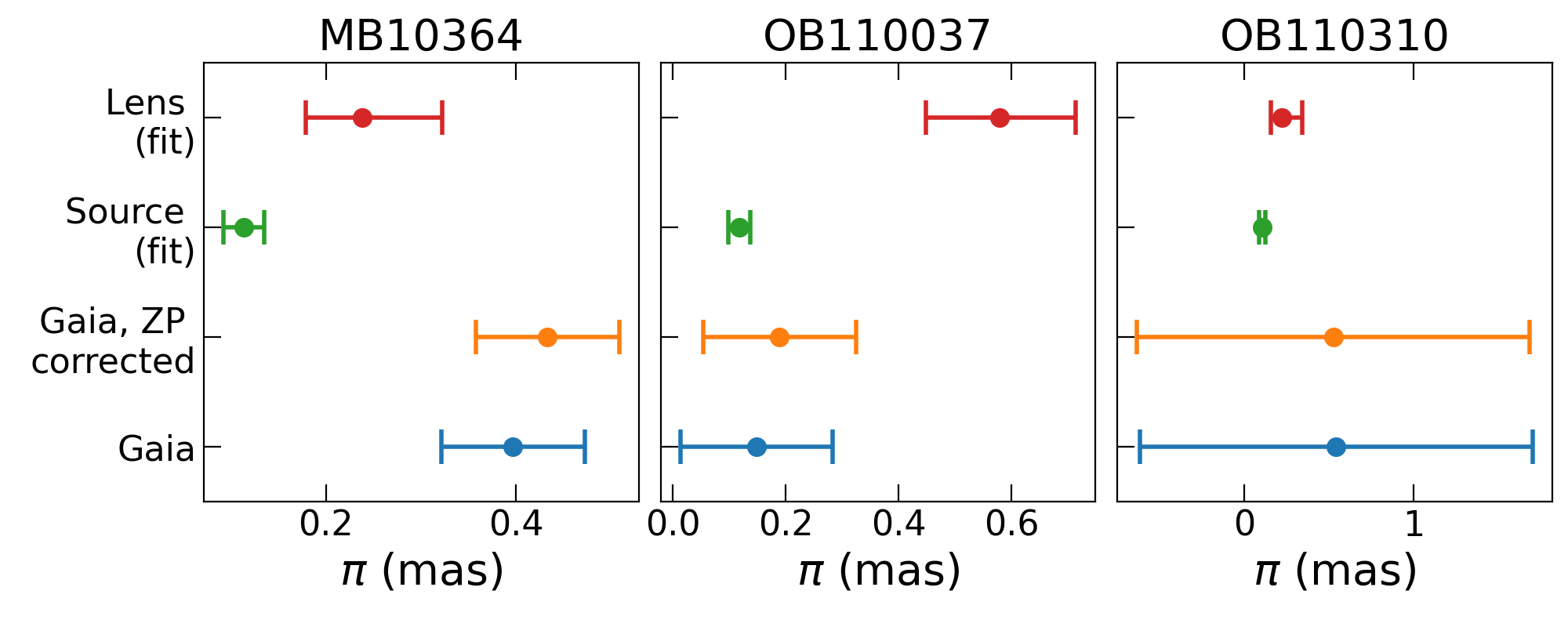}
\caption{Comparison of Gaia parallaxes with the fit lens and source parallax.
\label{fig:parallax}}
\end{figure}

\begin{figure*}
    \centering
    \includegraphics[width=1.0\linewidth]{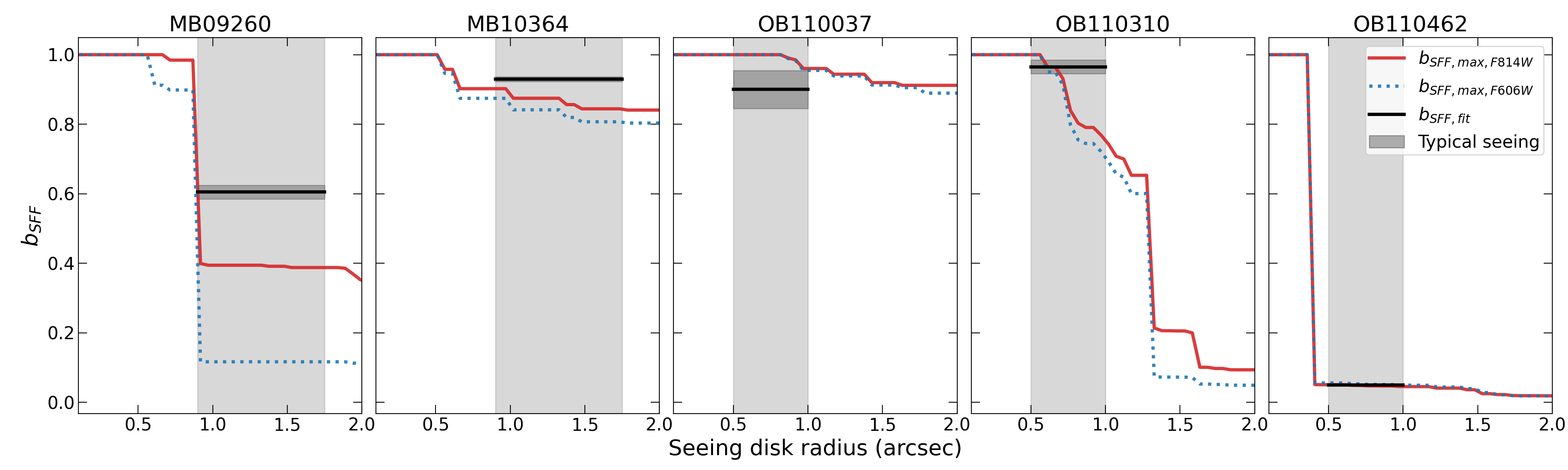} 
    \caption{Comparison of $b_{SFF}$ inferred from fitting microlensing events, as compared to upper limits inferred from the high-resolution HST images.
    The typical seeing disk \emph{radius} ranges are shown in light gray for MOA (MB09260, MB10364) and OGLE (OB110037, OB110310, OB110462).
    The inferred value of $b_{SFF}$ from fitting each microlensing event is shown as a black line, with uncertainties in dark gray.
    The upper limits on blending $b_{SFF,max}$ as a function of aperture radius, as estimated using the method described in \S \ref{sec:Verifying fit results with bSFF}, are shown as the solid red line (\emph{F814W}) and the dotted blue line (\emph{F606W}).
    For the three OGLE targets, at the relevant seeing disk radii, the inferred value of $b_{SFF}$ from the fit are at or below $b_{SFF,max}$ as inferred from the HST images.
    For the two MOA targets, the inferred values of $b_{SFF}$ are higher than the maximum value as inferred from the HST images.
    In the case of MB09260, this is likely due to the bright star around $1.0''$ from the target which causes an abrupt change in $b_{SFF}$ around that radius; proper convolution with a PSF would likely resolve the difference.
    In the case of MB10364, the difference is minimal, and likely just due to the imperfect nature of the comparison (for example, lack of proper PSF convolution, and F814W not exactly matching the red filter in MOA/OGLE).
    \label{fig:blend_vs_seeing_radius}}
\end{figure*}

For the three targets with astrometric solutions in Gaia EDR3 (MB10364, OB110037, and OB110310; Table \ref{tab:gaia}), the Gaia proper motions and parallaxes are compared to the fit proper motions and parallaxes presented in \S\ref{sec:Results} and Tables \ref{tab:mb10364_fits} - \ref{tab:ob110310_fits_modes}.
Direct comparisons are made in Table \ref{tab:gaia_vs_hst_pm} and Figures \ref{fig:VPDs} and \ref{fig:parallax}.
Note that we fit a proper motion and parallax to the source and lens along with an astrometric microlensing model.
On the other hand, Gaia fits a proper motion and parallax to the source and lens (the ``target") as they are unresolved, and assumes the target is a single star with parallax.
The effect of astrometric lensing in Gaia on the proper motions is negligible (Appendix \ref{app:Astrometric lensing in Gaia}).
Additional details about the various Gaia metrics discussed are in Appendix \ref{app:Gaia diagnostics} and \citet{Lindegren:2021b}.

To make proper motion comparisons between Gaia and HST, the proper motions from the HST frame in which the fitting was performed need to be transformed into the absolute Gaia frame, as the iterative astrometric alignment procedure described in \S \ref{sec:HST Astrometric Analysis} produces a reference frame that is at rest with the average proper motion of the aligned stars \citep{Lu:2016}.
See Appendix \ref{sec:Absolute proper motion reference frame} on how the proper motion offset between the HST and Gaia frames is calculated.
The source and lens proper motion in the Gaia frame, as well as the Gaia target proper motions, are listed in Table \ref{tab:gaia_vs_hst_pm} for MB10364, OB110037, and OB110310.
The vector point diagram for all stars from the HST observations transformed to the Gaia frame, along with the $1-2-3\sigma$ contours from the source and lens fit, are shown in Figure \ref{fig:VPDs}.
For MB10364, OB110037, and OB110310 the target proper motion from Gaia is also included.

\paragraph{MB10364}
MB10364's proper motion in Gaia is ($-7.43 \pm 0.08$, $-6.80 \pm 0.05$) mas/yr.
The fit from HST for the lens is ($-5.11_{-1.10}^{+1.62}$, $-7.78_{-0.89}^{+0.58}$) mas/yr, inconsistent with Gaia in RA and Dec at $\sim 2 \sigma$.
The fit from HST for the source is ($-7.56_{-0.12}^{+0.12}$, $-6.49_{-0.11}^{+0.11}$) mas/yr, inconsistent with Gaia in Dec at $\sim 2 \sigma$.
MB10364 has parallax $\pi = 0.43 \pm 0.08$ in Gaia.
The source and lens parallax from the MB10364 fits are $\pi_S = 0.11^{+0.02}_{-0.02}$, $\pi_L = 0.24^{+0.08}_{-0.06}$, neither of which are consistent with the Gaia value.
The Gaia fit for MB10364 has a large renormalized unit weight error (RUWE = 1.388), and a large astrometric excess noise ($\epsilon = 0.406$ mas) with high significance ($D = 12.020$), indicating the single star model is not providing a good fit. 
This mismatch is not due to astrometric microlensing (Appendix \ref{app:Astrometric lensing in Gaia}).
The most likely explanation for the discrepancies are crowding-- there are several stars close to MB10364 that would cause confusion (Figures \ref{fig:images} and \ref{fig:blend_vs_seeing_radius}).
As the source and lens are not resolvable, the fact that 15\% of the Image Parameters Determination (IPD) algorithm has identified a double peak is likely due to confusion. 
This means the Gaia measurement is not reliable.
See Appendix \ref{app:Gaia diagnostics} for further discussion.

\paragraph{OB110037}
OB110037's proper motion in Gaia is ($2.40 \pm 0.13$, $-3.91 \pm 0.09$) mas/yr.
The fit from HST for the lens is ($6.27_{-1.20}^{+1.27}$, $-6.56_{-0.81}^{+0.77}$) mas/yr, inconsistent with Gaia in both RA and Dec at $\sim 3 \sigma$. 
The fit from HST for the source is ($2.19_{-0.24}^{+0.24}$, $-3.87_{-0.20}^{+0.20}$) mas/yr, consistent with Gaia.
The caveat to this is that the fit to the HST F606W astrometry is poor (Figure \ref{fig:OB110037_astrom}).
However, the astrometric lensing model is mostly independent of the proper motion model, hence the proper motion value can still be believed.
OB110037 is very well measured and behaved in Gaia (it is the only target with a 5-parameter solution, see Table \ref{tab:gaia}).
Additionally from the image (Figures \ref{fig:images} and \ref{fig:blend_vs_seeing_radius}) OB110037 is relatively bright and isolated.
The fact that the fit source and Gaia proper motions are consistent would indicate that the lens is dim in comparison to the star in Gaia $G$ band.
This makes sense as the astrometric shift for OB110037 is undetectable by the time Gaia begins observing.

OB110037 is well modeled in Gaia by a single source with parallax.
Although the microlensing model is a poor fit to the astrometry due to a time-dependent color-offset, the proper motions from our model are in good agreement with Gaia.
OB110037 has parallax $\pi = 0.19 \pm 0.13$ in Gaia.
The source and lens parallax from the OB110037 fits are $\pi_S = 0.12^{+0.02}_{-0.02}$, $\pi_L = 0.58^{+0.14}_{-0.13}$.
The source parallax is consistent with the Gaia value.
We note that the source may appear well-behaved in Gaia because the astrometry is in a single filter. The multi-band HST astrometry may be useful in identifying binary companions.
See Appendix \ref{app:Gaia diagnostics} for further discussion.

\paragraph{OB110310}
OB110310's proper motion in Gaia is ($-2.08 \pm 1.12$, $-6.75 \pm 0.58$) mas/yr.
The fit from HST for the lens is ($-0.02_{-1.16}^{+1.93}$, $-4.68_{-2.13}^{+2.39}$) mas/yr, consistent with Gaia due to large uncertainties in both Gaia and the fit. 
The fit from HST for the source is ($-2.41_{-0.12}^{+0.12}$, $-7.26_{-0.08}^{+0.08}$) mas/yr, consistent with Gaia.
OB110310 has parallax $\pi = 0.53 \pm 1.16$ in Gaia (note the OB110310 zero point correction is an extrapolation), consistent with non-detection of parallax.
The source and lens parallax from the OB110310 fit are $\pi_S = 0.10^{+0.02}_{-0.02}$, $\pi_L = 0.22^{+0.12}_{-0.07}$, which are both consistent within the very wide uncertainties of Gaia.
The fact that the source and Gaia proper motions are consistent would indicate that the lens is dim in comparison to the star in Gaia $G$ band; blending in F814W and F606W also suggest a dark lens (Table \ref{tab:ob110310_fits_modes}).
The fact that in the IPD 55\% of transits have either truncation or multiple gates flagged in one or more windows indicates likely contamination.
OB110310 is not very bright and in a somewhat crowded region (Figures \ref{fig:images} and \ref{fig:blend_vs_seeing_radius}).
The astrometric noise is large ($\epsilon = 0.894$ mas) but the value is insignificant ($D = 0.332$).
Together these explain why the Gaia measurement does not produce very good constraints.
See Appendix \ref{app:Gaia diagnostics} for further discussion.

\subsection{HST Photometric Analysis \label{sec:HST Photometric Analysis}}

To obtain precise relative photometry, for each epoch a small constant magnitude offset is applied to the stars. 
The offset is calculated by assuming the reference stars have constant brightness in time, which we define as the $3\sigma$ clipped mean.
As with the positional uncertainties, for the magnitude uncertainties we used the error on the mean, with an additional additive error empirically determined during the astrometric alignment process; details are in Appendix \ref{sec:Rescaling of uncertainties}.

Analogous to the positional transformation, we evaluate the magnitude transformation by checking how well a constant magnitude describes the stars
\begin{equation}
    \chi^2_m = \sum_i \Big( \frac{m_i - m_{0}}{\sigma_{m_i}} \Big)^2
\end{equation}
where $m_i$ are the calibrated magnitudes, $m_0$ is the constant magnitude fit, and $\sigma_{m_i}$ is the uncertainty on the calibrated magnitude.  
The $\chi^2$ distributions for the magnitude residuals of the reference stars detected in all $N_{ep}$ epochs are shown in the bottom panels of Figure \ref{fig:chi2_xym}.
The expected residual distribution has $N_{ep,detect} - 1$ degrees of freedom, as there is one free parameter in the constant magnitude fit.

\begin{deluxetable*}{lccccc}
\tabletypesize{\tiny}
\tablecaption{HST Calibrated Data for Each Target \label{tab:HST Calibrated Data}}
\tablehead{
    \colhead{Target} &
    \colhead{Filter} &
    \colhead{Date} &
    \colhead{RA (mas)} &
    \colhead{Dec (mas)} &
    \colhead{Mag (Vega)}}
\startdata
\input{hst_data_table.txt}
\enddata
\tablecomments{
Relative positions and magnitude of the target by epoch and filter.
}
\end{deluxetable*}
 
\section{Microlensing Modeling \label{sec:Microlensing Modeling}}
The mass of the lens $M_L$ in a microlensing event is given by 
\begin{equation}
    M_L = \frac{\theta_E}{\kappa \pi_E},
    \label{eq:M_L}
\end{equation}
where $\theta_E$ is the angular Einstein radius (Equation \ref{eq:theta_E}), $\pi_E$ is the microlensing parallax (Equation \ref{eq:pi_E}), and $\kappa = 4G/(1 AU \cdot c^2) = 8.14 {\rm \, mas}/M_\odot$ is a constant.
Densely sampled photometric microlensing observations constrain $\pi_E$, while astrometric microlensing observations constrain $\theta_E$.\footnote{Theoretically, astrometric microlensing observations should also be able to constrain $\pi_E$, but due to the cadence of observations, this is currently unachievable.}

To measure these quantities of interest, we simultaneously fit the ground-based photometry and HST photometry and astrometry with a point-source point-lens (PSPL) microlensing model with parallax.
We do not consider models involving either binary lenses or sources, nor higher-order effects beyond parallax; these are beyond the scope of this work.
Discussion of the need for models more complex than PSPL with parallax are discussed in \S\ref{sec:Results}.

Throughout this section and the remainder of the paper, we report vector quantities decomposed into their RA and Dec components, subscripted by ``$\alpha *$" and ``$\delta$". 
Because we work solely in the Equatorial coordinate system, we also equivalently refer to RA as East and Dec as North, subscripted by ``E" or ``N", where RA increases to the East and Dec increases to the North.

\subsection{Microlensing Definitions \label{sec:Microlensing Definitions}}

All microlensing quantities defined in the following section are in the heliocentric reference frame.

By rearranging the terms in Equation \ref{eq:M_L}, the Einstein radius, which sets the angular scale of the microlensing event, can be written
\begin{equation}
    \label{eq:theta_E}
    \theta_E = \sqrt{\kappa \pi_{rel} M_L}
\end{equation}
where $\pi_{rel} = \pi_L - \pi_S$ is the relative parallax of the lens and source.

The Einstein crossing time $t_E$, the time it takes for the source to traverse the angular radius of the lens and sets the timescale of the events, is given by
\begin{equation}
    \label{eq:t_E}
    t_E = \frac{\theta_E}{\mu_{rel}}
\end{equation}
where $\mu_{rel}$ is the lens-source proper motion $|\boldsymbol{\mu}_S - \boldsymbol{\mu}_L|$. 

The source-lens separation on sky $\boldsymbol{\theta}_S - \boldsymbol{\theta}_L$ normalized by the Einstein radius is denoted $\boldsymbol{u}(t)$.
The minimum separation is denoted $\boldsymbol{u}(t_0) = \boldsymbol{u}_0$.
The impact parameter $u_0$ is the scalar quantity associated with $\boldsymbol{u}_0$.
We follow the convention of \citet{Gould:2004} where if $u_{0,E} > 0$, the source is to the East of the lens, and $u_0 > 0$; if $u_{0,E} < 0$, the source is to the West of the lens, and $u_0 < 0$.\footnote{Note that we define our coordinate system differently than \citet{Gould:2004}.
\citet{Gould:2004} works in a system where the position of the source relative to the lens is defined in a coordinate system that is right-handed in the relative proper motion and minimum separation vector.
However, we work in a system where the coordinate system is consistent on the sky; this means a coordinate system based on the relative proper motion and minimum separation vector does not always preserve handedness.}
If rectilinear motion is assumed, the lens-source separation is given by
\begin{equation}
    \boldsymbol{u}(t) = \boldsymbol{u}_0 + \frac{t - t_0}{t_E} \boldsymbol{\hat{\mu}}_{rel}.
\end{equation}
However, an Earthly observer's perspective of the lensing event is modulated by the Earth's motion around the Sun.
For events with long duration ($t_E \gtrsim 3$ months), the Earth's orbital motion violates this rectilinear assumption and must be taken into account. This modifies $\boldsymbol{u}(t)$ to
\begin{equation}
    \label{eq:vec_u(t)}
    \boldsymbol{u}(t) = \boldsymbol{u}_0 + \frac{t - t_0}{t_E} \boldsymbol{\hat{\mu}}_{rel} - \pi_E \boldsymbol{P}(t)
\end{equation}
where
\begin{equation}
    \label{eq:pi_E}
    \pi_E = \frac{\pi_{rel}}{\theta_E}
\end{equation}
is the microlensing parallax and $\boldsymbol{P}(t)$ is the parallax vector, defined to be the Sun-Earth separation vector normalized by 1 AU.
The microlensing parallax vector $\boldsymbol{\pi}_E$ (not to be confused with the parallax vector $\boldsymbol{P}(t)$) encodes the magnitude of the microlensing parallax and the direction of the relative source-lens proper motion: 
\begin{equation}
    \boldsymbol{\pi}_E = \pi_E \boldsymbol{\hat{\mu}}_{rel}.
\end{equation}

The photometric brightening of the source is given by
\begin{equation}
    \label{eq:amplification}
    A(u) = \frac{u^2 + 2}{u\sqrt{u^2 + 4}}.
\end{equation}
where the total flux $F(t)$ in the telescope aperture is
\begin{equation}
    \label{eq:F}
    F(t) = A(t) F_S + F_L + F_N
\end{equation}
where $F_S$, $F_L$, and $F_N$ are the fluxes of the source, lens, and neighboring un-lensed stars in the aperture of the telescope, respectively.
The source flux fraction is
\begin{equation}
    b_{SFF} = \frac{F_S}{F_S + F_L + F_N}
    \label{eq:bsff}
\end{equation}
and quantifies the fraction of light lensed in an observed microlensing event.
Note that $b_{SFF}$ depends on the observing wavelength and seeing/aperture.
The non-source flux $F_L + F_N$ is also called blend flux.
Blend flux decreases $b_{SFF}$ and dilutes the magnitude of both the photometric amplification and astrometric shift.

A PSPL photometric microlensing event is characterized by five geometric parameters: $t_0$, $u_0$, $t_E$, and $\boldsymbol{\pi}_E$.
For each telescope that observes this event, two additional parameters, $m_{base}$ and $b_{SFF}$, are needed to describe each lightcurve; these depend on the seeing and camera filter.
As all photometric microlensing observable quantities are normalized by $\theta_E$, the lens mass cannot be determined.

In contrast, astrometric microlensing provides a direct measurement of $\theta_E$.
The apparent position of the source, i.e. the centroid of the lensed source images $\boldsymbol{\theta}_{S,c}$, is given by
\begin{equation}
    \boldsymbol{\theta}_{S,c}(u, \theta_E) = \frac{(u^2 + 3)\boldsymbol{u} \theta_E}{u^2 + 2} .
\end{equation}
Assuming no blended light, the difference between the source's apparent and true positions, i.e. the astrometric shift, is given by 
\begin{equation}
    \label{eq:delta_c}
    \boldsymbol{\delta}_c(u, \theta_E) = \frac{\boldsymbol{u}\theta_E}{u^2 + 2}.
\end{equation}
The astrometric shift $\delta_c$ is maximized at $u = \sqrt{2}$.
This corresponds to the value of the maximum astrometric shift
\begin{equation}
    \label{eq:delta_c,max}
    \delta_{c,max} = \frac{\theta_E}{\sqrt{8}},
\end{equation} 
which is directly proportional to the Einstein radius.

Note that Equation \ref{eq:delta_c} is for an unblended event, i.e. $b_{SFF} = 1$.
If an event has $b_{SFF} < 1$, the non-source light would dilute the astrometric shift (see Equations 11 and 12 in \citet{Lam:2020}).
For the five candidates analyzed here, $b_{SFF} \sim 1$ in the HST filters, so assuming that the astrometry is unblended in HST is valid.

\subsection{Modeling framework}

We perform parameter estimation using a Bayesian framework.
Bayes' theorem
\begin{equation}
    \pi(\bTheta) \mathcal{L} (\by | \bTheta) = \mathcal{Z} (\by) \mathcal{P} (\bTheta | \by)
\end{equation}
relates the prior $\pi$ and likelihood $\mathcal{L}$ to the evidence $\mathcal{Z}$ and posterior $\mathcal{P}$.
The goal of parameter estimation is to calculate $\mathcal{P}$.
The likelihood $\mathcal{L} (\by | \bTheta)$ is presented in Appendix \ref{sec:Gaussian Process}, and the priors $\pi(\bTheta)$ are discussed in Appendix \ref{sec:Priors}. 

The data is fit using \texttt{MultiNest} \citep{Feroz:2009}, an implementation of the nested sampling algorithm \citep{Skilling:2004}.
Nested sampling produces an estimate of the evidence $\mathcal{Z} = \int \pi(\bTheta) \mathcal{L}(\bTheta) \, d \bTheta$, and as a by-product, the posterior $\mathcal{P}$.
In contrast to methods such as MCMC, nested sampling is designed to better explore multimodal likelihood spaces; however, care must still be taken to ensure that all local minima are explored.

\subsection{Likelihood weighting \label{sec:Likelihood weighting}}

There is a question of how best to combine the photometry and astrometry data sets, as they represent two different types of measurements.
In particular, the question is how much weight each dataset should receive given that there are several orders of magnitude more ground-based data points than HST data points.
Thus, the ground-based photometry has an outsize effect on the likelihood.
It can be argued that each data point should contribute equally to the likelihood.
We consider this to be ``default weight" (hereafter abbreviated as DW) likelihood, i.e.
\begin{equation}
    \log \mathcal{L}_{tot} = \log \mathcal{L}_{O,phot} + \log \mathcal{L}_{H,phot} + \log \mathcal{L}_{H,ast}
    \label{eq:default_weight_likelihood}
\end{equation}
where $\mathcal{L}_{O,phot}$, $\mathcal{L}_{H,phot}$, and $\mathcal{L}_{H,ast}$ are the likelihoods of the OGLE or MOA photometry, HST photometry, and HST astrometry respectively.

However, it could also be argued that each dataset is independent, and so should each contribute equally to the likelihood. We call this the ``equal weight" (hereafter abbreviated as EW) likelihood, i.e.
\begin{equation}
    \log \mathcal{L}_{tot} = \frac{\log \mathcal{L}_{O,phot}}{n_{O,phot}} + \frac{\log \mathcal{L}_{H,phot}}{n_{H,phot}}
    + \frac{\log \mathcal{L}_{H,ast}}{n_{H,ast}} 
    \label{eq:equal_weight_likelihood}
\end{equation}
where $n_{O,phot}$, $n_{H,phot}$, and $n_{H,ast}$ are the number of data points in in the OGLE photometry, HST photometry, and HST astrometry. 

These different likelihoods are essentially giving the different datasets different weights, in the case that they are inconsistent with each other. 
For all targets, we fit using the default weight likelihood; for OB110462 we additionally fit using the equal weight likelihood.

\section{Constraining the nature of the lens \label{sec:Constraining the nature of the lens}}

By modeling photometric and astrometric microlensing data as described in \S \ref{sec:Microlensing Modeling}, the lens' mass can be measured.
However, a mass measurement alone cannot distinguish the difference between a $5 M_\odot$ stellar lens and a $5 M_\odot$ BH.
Additional information about the lens' brightness is needed, which can be estimated using the source flux fractions in the high-resolution HST filters.
By calculating the brightest star allowed by the inferred source flux fractions, we can determine whether a luminous lens (i.e. stellar lens) or dark lens (i.e. compact object lens) scenario is more likely.

We follow a procedure similar to \cite{Wyrzykowski:2016} in order to calculate the probability of a dark lens.
The two main differences between the analysis of \cite{Wyrzykowski:2016} and ours is that they work with photometric data only and must invoke a Galactic model to obtain lens masses and distances, while in our case astrometric data allows us to fit lens mass and distances.
This greatly simplifies the analysis as we do not need to calculate the Jacobian to transform the PDFs between different variables.
In addition, we use realistic stellar and Galactic models to determine the lens luminosities, instead of relying on simple mass-luminosity scaling relations.

From fitting the microlensing data, we have posterior distributions for the lens mass $M_L$, distance $d_L$, baseline magnitude $m_{\rm base}$, and source flux fraction $b_{SFF}$.
In the following analysis we reassign any fit values with $b_{SFF} \geq 1$ to instead be equal to 0.999999.
This is because values of $b_{SFF} \geq 1$ would result in an infinite lens magnitude (Equation \ref{eq:lens_mag}).\footnote{One might worry this could bias the results.
However, a visual check comparing the posteriors (both individual and joint) for lens mass and distance don't show a change when excluding fits with $b_{SFF} > 1$.
In particular, the lens mass and distances are nearly independent of the lens magnitude.
In addition, if the true value of $b_{SFF}$ is 1, we would expect some scatter around that value, including unphysical $>1$ values.}

We draw a random sample of 1000 stars from our posterior.
For each star we calculate 1) the brightest lens allowed by $b_{SFF}$ and $m_{base}$, and 2) the brightest star allowed by $M_L$ and $d_L$.

To calculate 1), by assuming there are no contaminating neighbor stars, $b_{SFF} = F_S/(F_S + F_L)$, an upper limit can be placed on the brightness of the lens:
\begin{equation}
    \label{eq:lens_mag}
    m_L = m_{\rm base} - 2.5 \log_{10}(1 - b_{SFF}).
\end{equation}
We denote this $m_L(b_{SFF}, m_{base})$.

To calculate 2), we use the simple stellar population synthesis code \texttt{SPISEA} \citep{Hosek:2020} to generate a suite of simple stellar populations (SSPs) to simulate the possible lens population.
We use the \texttt{MISTv1.2} solar metallicity isochrones \citep{Choi:2016}, \texttt{get\_merged\_atmosphere} atmosphere model\footnote{This is a combination of the ATLAS9 \citep{Castelli:2004}, PHOENIX v16 \citep{Husser:2013}, BTSettl \citep{Baraffe:2015}, abd Koester10 \citep{Koester:2010} models. 
For further details see Appendix B of \citet{Hosek:2020}.}, \cite{Damineli:2016} reddening law, and \cite{Kroupa:2001} initial mass function (IMF) over the mass range $0.1 M_\odot < M < 120 M_\odot$.
Each cluster is $10^4 M_\odot$, in order to reduce stochastic effects in the sampling of the IMF.

SSPs are generated at the distances spanned by each target's lens distance posteriors, sampled every 0.25 kpc.
SSPs of ages 7.0 to 10.0 $\log_{10}$ years in increments of 0.5 $\log_{10}$ years are simulated at each sampled distance in order to cover the age range of disk and bulge stars.
The stellar age distribution as a function of distance is drawn from the \texttt{Galaxia} Milky Way stellar simulation \citep{Sharma:2011}, which implements a version of the Besan{\c{c}}on Galactic model of \cite{Robin:2003}.
As our target sight lines are toward the bulge, most of the stars are old, with $\gtrsim 85\%$ of stars being at least 9.0 $\log_{10}$ years old, although younger stars tend to be at closer distances as disk stars dominate.
To calculate the simulated stars' apparent magnitudes, we use the 3-D extinction map of \cite{Marshall:2006}, accessed via the \texttt{dustmaps} software package \citep{Green:2018}, to obtain an $A_{\rm K_s}$ value for each distance.

For each sample from the posterior, we sample one of the SSPs at the corresponding distance at a particular stellar age from the Galactic distribution. 
All stars in the simulated SSP with masses within 10\% of the sampled posterior mass are identified, and the brightest apparent magnitude of the star in this group is denoted $m_L(M_L, D_L, \bigstar)$.
If no stars are found within 10\% of the sampled posterior mass, this indicates that stars of that mass have all died and formed compact objects.

If $m_L(M_L, D_L, \bigstar) < m_L(b_{SFF}, m_{base})$, a star with the inferred stellar mass from stellar evolution models would be too bright to be hidden in the blended light allowed by the fit.
This means that the lens is not a star.
We take this to mean the lens is a compact object, and hence a dark lens.
Thus the lower limit on the probability of a dark lens in the observed filter is the fraction of samples where $m_L(M_L, D_L, \bigstar) < m_L(b_{SFF}, m_{base})$ or no mass match is found.

Objects with $m_L(M_L, D_L, \bigstar) > m_L(b_{SFF}, m_{base})$ are samples where a luminous stellar lens is consistent with the inferred amount of blending.
Note that $m_L(b_{SFF}, m_{base})$ is an upper limit on the brightness of the lens, as we assumed all the blended flux in Equation \ref{eq:lens_mag} is due to the lens; it could be due to unresolved unrelated neighbor stars.
Thus the upper limit on the probability of a stellar lens in the observed filter is thus the fraction of samples where $m_L(M_L, D_L, \bigstar) > m_L(b_{SFF}, m_{base})$.

For dark lenses (i.e. any objects where $m_L(M_L, D_L, \bigstar) < m_L(b_{SFF}, m_{base})$ or no star of the same mass was found in the simulated SSP), we categorize them as brown dwarfs (BD), white dwarfs (WD), neutron stars (NS), or BHs by simplistically sorting them by their masses:
\begin{itemize}
    \item Brown dwarfs: $M < 0.2 M_\odot$
    \item White dwarfs: $0.2 M_\odot < M < 1.2 M_\odot$
    \item Neutron stars: $1.2 M_\odot < M < 2.2 M_\odot$
    \item Black holes: $M > 2.2 M_\odot$.
\end{itemize}
In reality there is overlap between white dwarf and neutron star masses, and the overlap between neutron star and BH masses (if they overlap at all) is unknown.
In addition, the maximum brown dwarf mass set by stellar physics is $0.08M_\odot$, but we extend this up to $0.2M_\odot$ to have continuity between the lowest mass WDs of around $0.2 M_\odot$. 
Hence these values are only approximate.

The above analysis is performed for both the HST F606W and F814W filters.
The reported probabilities are the joint constraint.
A lens is dark if no masses consistent with a stellar lens are found in either filter.
A lens is also dark if the maximum inferred lens flux is insufficient to hide a star (i.e. $m_L(M_L, D_L, \bigstar) < m_L(b_{SFF}, m_{base})$) in either filter.
A lens is luminous if the maximum inferred lens flux is sufficient to hide a star (i.e. $m_L(M_L, D_L, \bigstar) > m_L(b_{SFF}, m_{base})$) in both filters.
We do not perform this analysis for the OGLE or MOA photometry parameters as the high resolution HST images show the seeing-limited apertures have unrelated neighbor stars in the blend, and hence the limits will all be weaker than using HST.

\begin{figure*}[t!]
    \centering
    \includegraphics[width=1.0\linewidth]{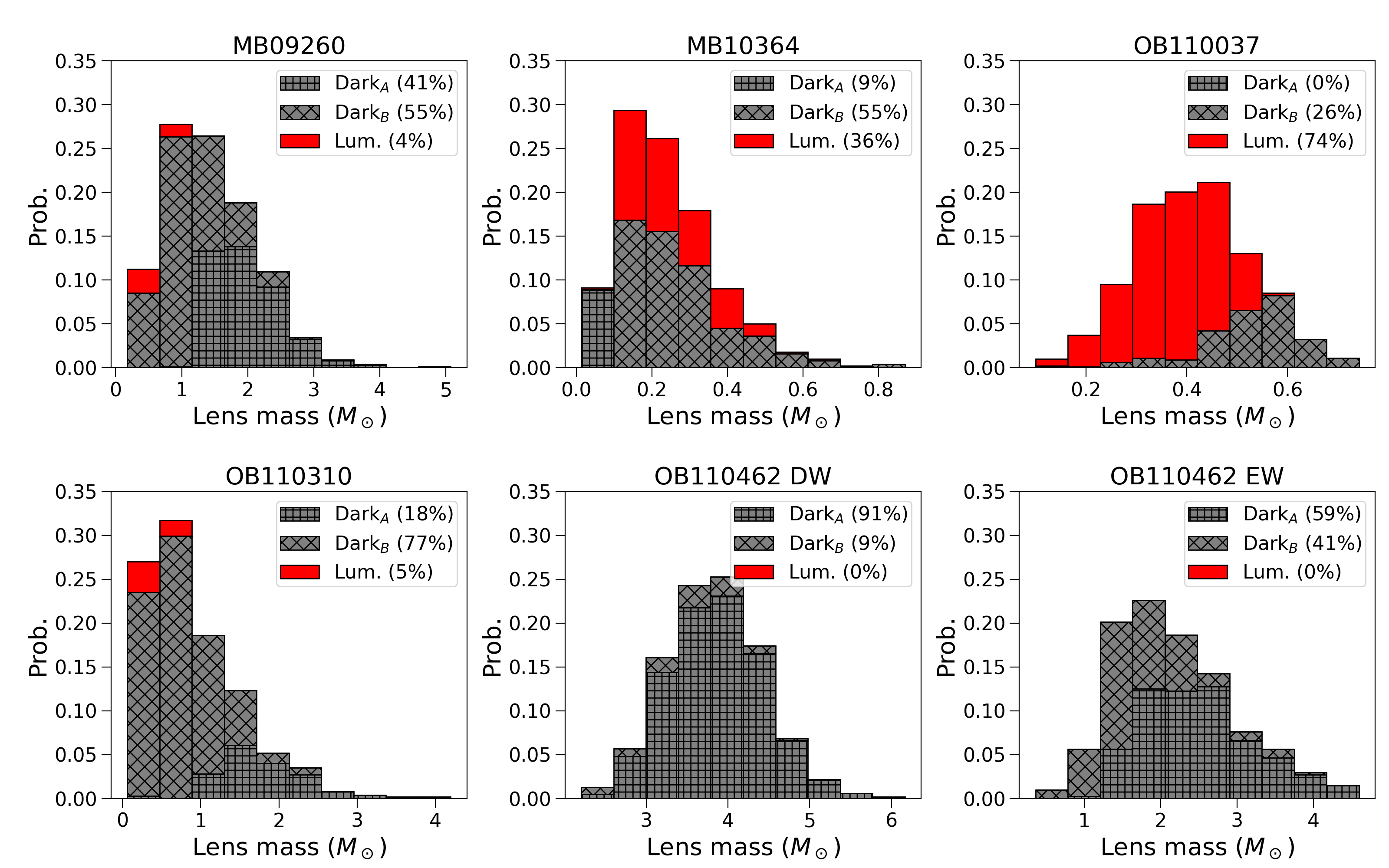}
    \caption{\label{fig:dark_lens_prob}
    Probabilities for dark (\emph{gray}) and luminous lenses (\emph{red}) 
    , as a function of lens mass for each target.
    Lenses that are dark because all stars of that mass have already evolved and died (implying they are compact objects) are subscripted with $A$ (\emph{square hatch}), while lenses that are dark because a star would be too bright to be hidden in the allowed lens flux are subscripted with $B$ (\emph{diamond hatch}).
    The probabilities for the luminous lenses are upper limits, while the probabilities for the dark lenses are lower limits, since the method described in \S\ref{sec:Constraining the nature of the lens} only places an upper limit on the brightness of the lens. Note that there are two fits for OB110462, one with equal weighting to the astrometry and photometry data (OB110462 EW) and one with the default weighting of the astrometry and photometry data (OB110462 DW). See \S\ref{sec:Likelihood weighting} for details.
    }
\end{figure*}

\begin{figure*}
\centering
\includegraphics[width=\linewidth]{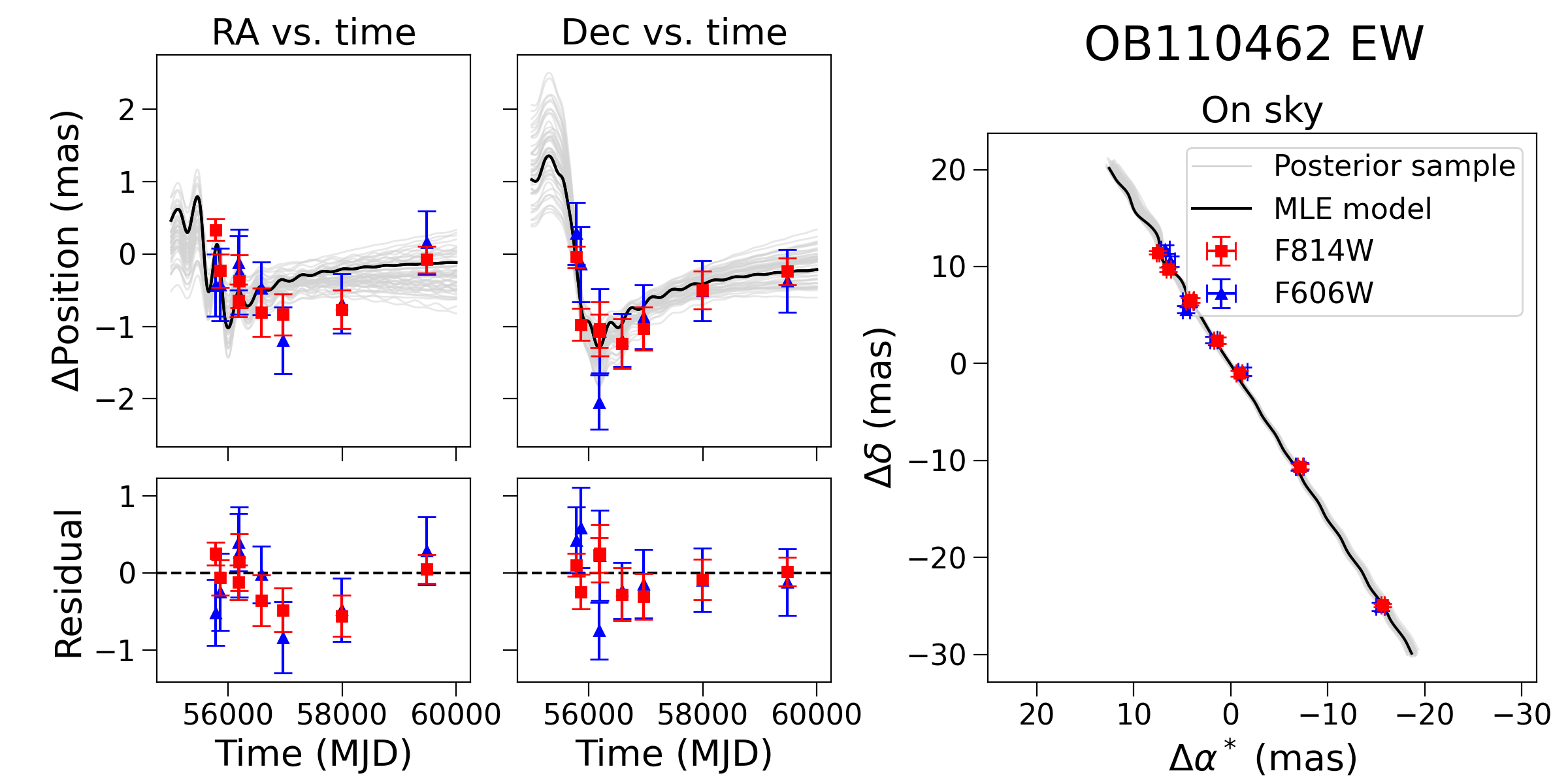}
\caption{OB110462 astrometry, using the equally weighted likelihood.
\emph{Left column, top to bottom}: RA vs. time with maximum likelihood (MLE) unlensed source motion model subtracted; residuals to the MLE model for RA vs. time fit.
HST F814W astrometry data is shown in red; HST F606W astrometry data is shown in blue.
The MLE model is shown in black.
Fifty random draws from the posterior distribution are shown in light gray.
\emph{Middle column, top to bottom}: Same as left column, except Dec instead of RA.
\emph{Right panel}: astrometry as seen on-sky, in the barycentric frame.
OB110462 shows a strong $>1$~mas astrometric microlensing signal.
\label{fig:OB110462_astrom_EW}}
\end{figure*}

\begin{figure*}
\centering
\includegraphics[width=\linewidth]{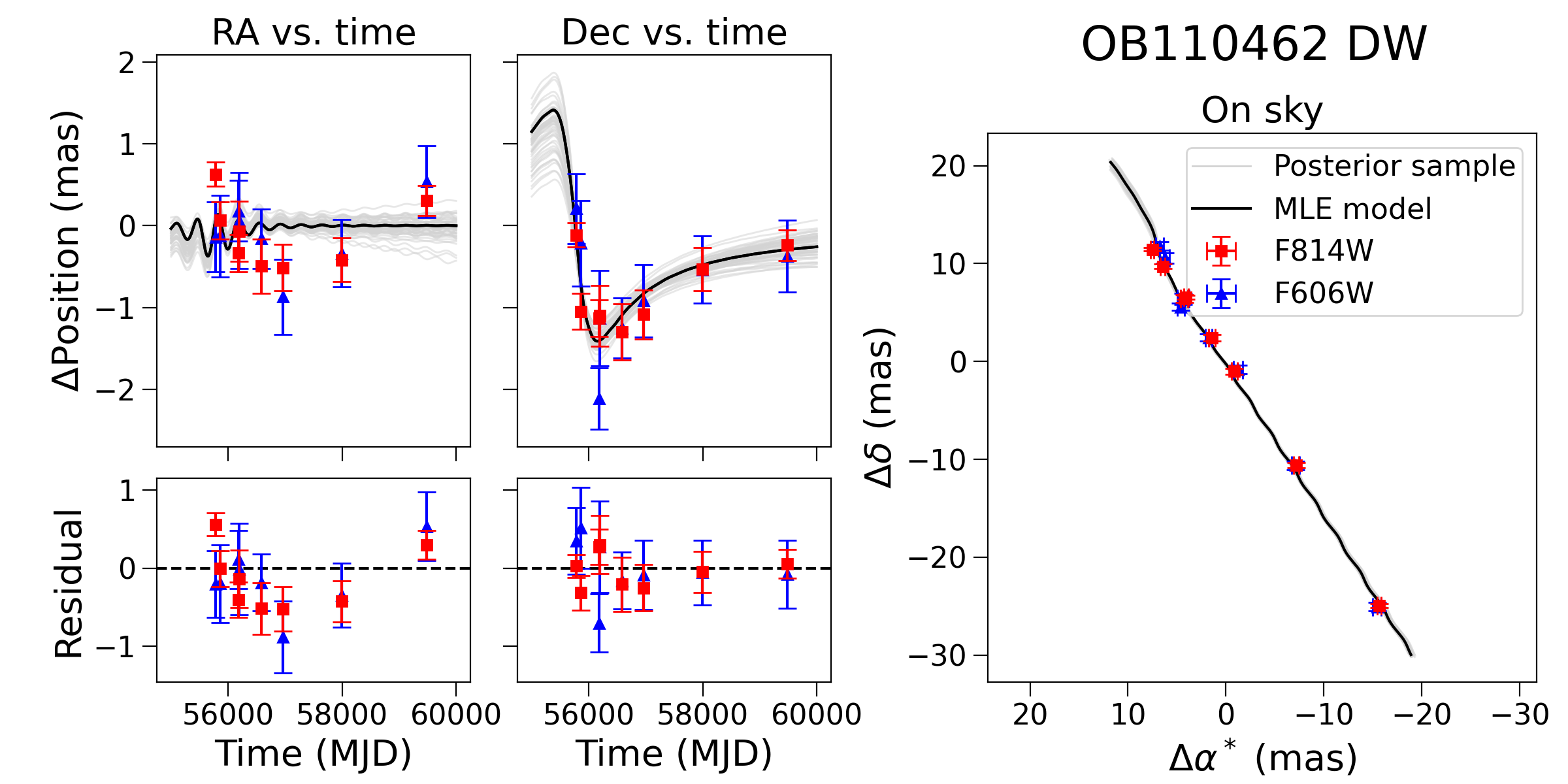}
\caption{Same as Figure \ref{fig:OB110462_astrom_EW}, but using the default weighted likelihood for OB110462. 
\label{fig:OB110462_astrom_DW}}
\end{figure*}

\begin{figure*}
\centering
    \includegraphics[width=\linewidth]{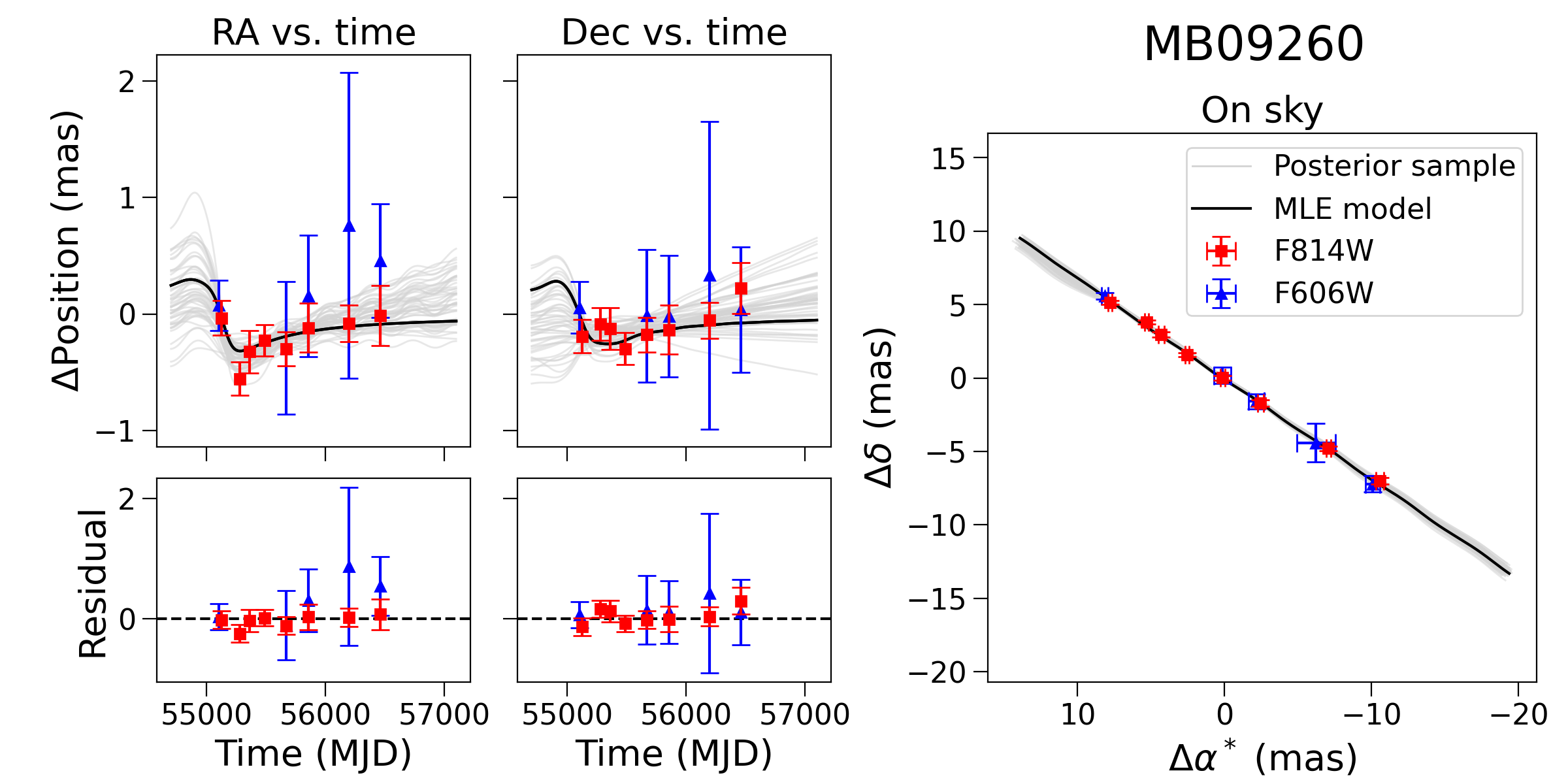}
\caption{MB09260 astrometry. 
Same as Figure \ref{fig:OB110462_astrom_EW}, but for MB09260.
The astrometric signal is small, and around the limit of the precision of the F814W measurements. 
\label{fig:MB09260_astrom}}
\end{figure*}

\begin{figure*}
\centering
    \includegraphics[width=\linewidth]{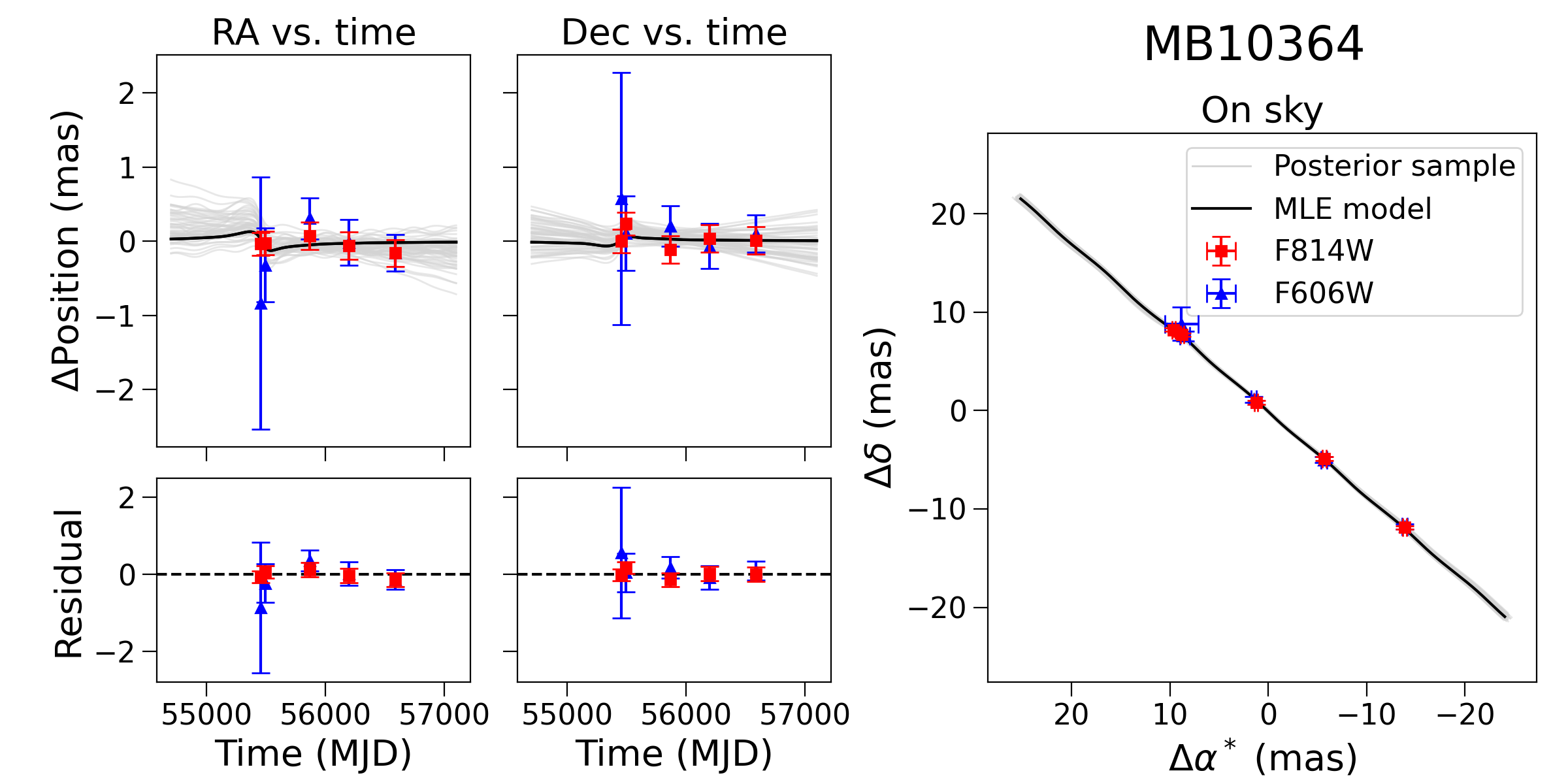}
\caption{Same as Figure \ref{fig:OB110462_astrom_EW}, but for
MB10364.
The astrometric signal is small, at or below the limit of the precision of the F814W measurements. 
\label{fig:MB10364_astrom}}
\end{figure*}

\begin{figure*}
\centering
    \includegraphics[width=\linewidth]{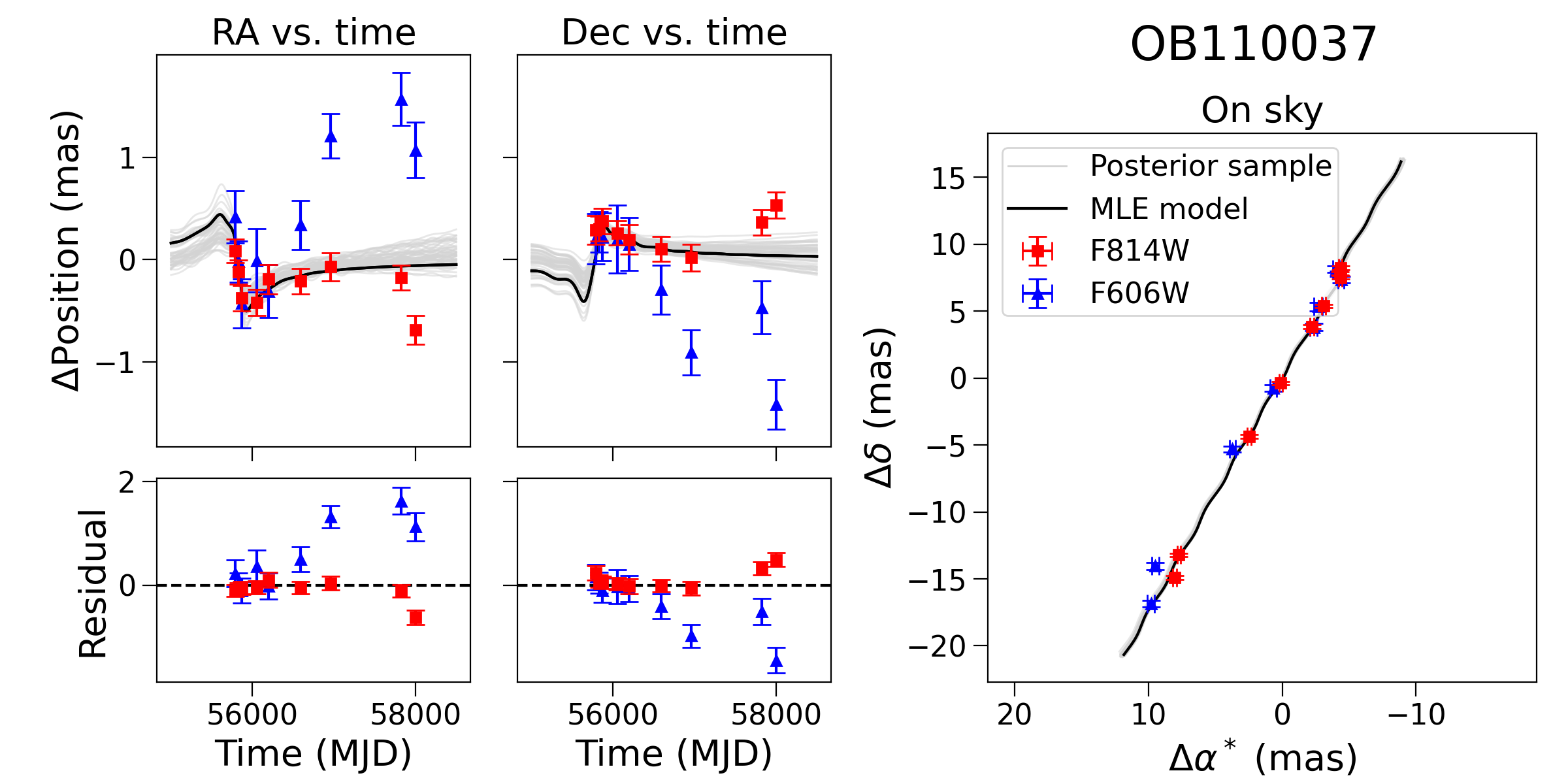}
\caption{Same as Figure \ref{fig:OB110462_astrom_EW}, but for
OB110037.
The photometry and astrometry seem to indicate this object is a binary.
\label{fig:OB110037_astrom}}
\end{figure*}

\begin{figure*}
    \includegraphics[width=\linewidth]{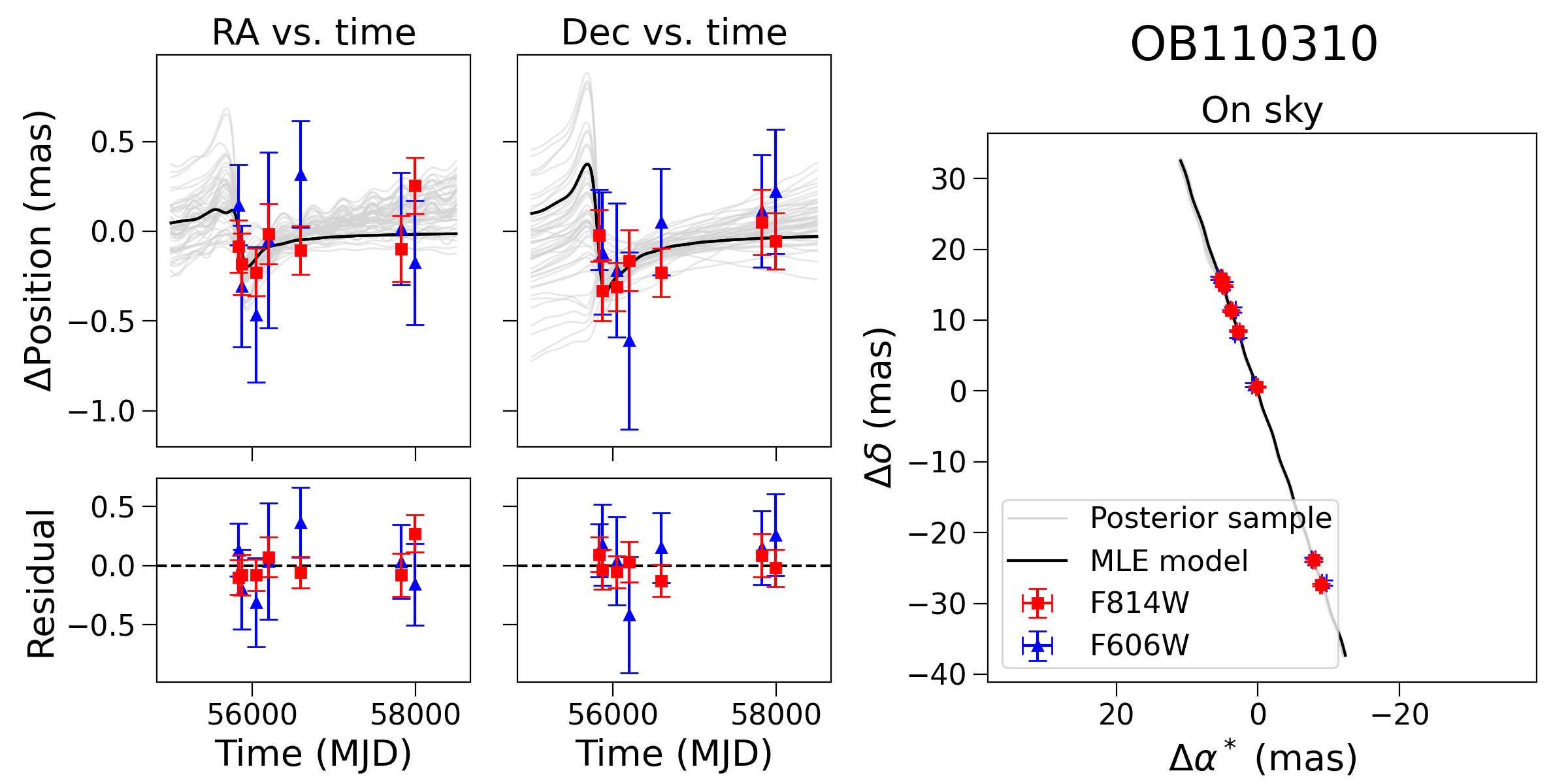}
\caption{Same as Figure \ref{fig:OB110462_astrom_EW}, but for
OB110310.
The astrometric signal is small, and around the limit of the precision of the F814W measurements. 
\label{fig:OB110310_astrom}}
\end{figure*}

\section{Results in Detail \label{sec:Results}}

\begin{deluxetable}{lcccc}
\tablecaption{99.73\% credible intervals/upper limits \label{tab:3sigma upper lims summary}}
\tablehead{
    \colhead{Parameter} &
    \colhead{$\theta_E$ (mas)} & 
    \colhead{$\delta_{c,max}$ (mas)} &
    \colhead{$M_L (M_\odot)$} &
    \colhead{$\pi_E$}}
\startdata
\input{upper_limits.txt}
\enddata
\tablecomments{There are two entries for OB110462, one based on a fit with ``default weighting" (OB110462 DW) and one based on a fit with ``equal weighting" (OB110462 EW). 
See \S\ref{sec:Likelihood weighting} for details.}
\end{deluxetable}

\begin{deluxetable}{l|ccccc}
\tablecaption{Lens type probabilities (\%)}
\tablehead{
    \colhead{Target} &
    \colhead{Star} & 
    \colhead{BD} & 
    \colhead{WD} & 
    \colhead{NS} & 
    \colhead{BH}}
\startdata
\input{lens_type_prob.txt}
\enddata
\tablecomments{\label{tab:lens type probabilities}
The Star probabilities are upper limits, while the brown dwarf (BD), white dwarf (WD), neutron star (NS), and BH probabilities are lower limits.
This is because the luminous lens probabilities are upper limits, and the dark lens probabilities are lower limits; see \S\ref{sec:Constraining the nature of the lens} for details.
Note that there are two entries for OB110462, one based on a fit with ``default weighting" (OB110462 DW) and one based on a fit with ``equal weighting" (OB110462 EW). See \S\ref{sec:Likelihood weighting} for details.}
\end{deluxetable}

A large ($>1$ mas) astrometric microlensing signal is detected in OB110462 (\S\ref{sec:OB110462}).
A filter-dependent astrometric signal is also detected in OB110037 (\S\ref{sec:OB110037}); averaging the astrometry across the F814W and F606W filters shows a small but significant 0.4 mas signal.
The remaining three targets MB09260 (\S\ref{sec:MB09260}), MB10364 (\S\ref{sec:MB10364}), and OB110310 (\S\ref{sec:OB110310}) have astrometric signals that are below HST's detection threshold.
Table \ref{tab:3sigma upper lims summary} reports the lens masses, Einstein radii, maximum astrometric shifts, and microlensing parallaxes either as measured values or upper limits. 

\S\ref{sec:OB110462} - \S\ref{sec:OB110310}, the posteriors of the joint photometry and astrometry microlensing fits for the individual targets are presented.
Tables \ref{tab:mb09260_fits_modes} - \ref{tab:ob110310_fits_modes} and Tables 1 - 2
in \citet{Lam:2022lett} 
list the median and $1\sigma$ (68\%) credible intervals for each parameter, in addition to the maximum a posteriori (MAP) and maximum likelihood (MLE)\footnote{For most of the targets, the MAP solution is equal to the MLE solution, i.e. values of the parameters for the mode of the posterior distribution are the same as those where the likelihood function is maximized.} solution.

Using the methodology described in \S \ref{sec:Constraining the nature of the lens}, constraints are placed on the lens types of each target.
We report the relative probabilities of Star:BD:WD:NS:BH in Table \ref{tab:lens type probabilities}.
Figure \ref{fig:dark_lens_prob} shows the distribution of dark vs. luminous lens probability as a function of mass for each target.
Table \ref{tab:lens type probabilities} lists the upper limit on the probability of a stellar lens and lower limits on the probabilities of different dark lenses for each target.
In \S\ref{sec:Number of detected BHs}, we compare the yield of BHs from our search to that expected from simulations assuming there are $\sim 10^8$ BHs in the Milky Way.

\subsection{OB110462 \label{sec:OB110462}}

The results for OB110462 are presented in the main paper, \S 4.1
of \cite{Lam:2022lett}.

\subsection{MB09260 \label{sec:MB09260}}

The data and model for MB09260 are shown in Figures \ref{fig:MB09260_lightcurve} (photometry) and \ref{fig:MB09260_astrom} (astrometry), and the fit posteriors are summarized in Table \ref{tab:mb09260_fits_modes}.
The inferred Einstein crossing time $t_E$ is $142.64_{-2.87}^{+3.49}$ days, the microlensing parallax $\pi_E$ is $0.09_{-0.01}^{+0.03}$, the Einstein radius $\theta_E$ is $1.04_{-0.39}^{+0.42}$, and the lens mass is $1.37_{-0.60}^{+0.74} M_\odot$.\footnote{These are the values for the posteriors before they are split into modes, which is why the values differ from what is reported in Table \ref{tab:mb09260_fits_modes}.
However, since these parameters are all globally unimodal, their distributions nearly identical across modes within the uncertainties, and neither mode is strongly favored, the values are not very different and we report the global median and uncertainties.}

The probability that MB09260 has a dark lens is at least 96\%, and the probability of a stellar lens is at most 4\%.
The relative probabilities of Star:BD:WD:NS:BH are 4:0:38:44:14.
Stellar lenses are only allowed below $1 M_\odot$.
A white dwarf or neutron star is the most probable type of lens, with black holes possible but less likely.

\begin{deluxetable*}{l|ccc|ccc}
\tablecaption{MB09260 Fit Values, Modes\label{tab:mb09260_fits_modes}}
\tablehead{
    \colhead{Parameter} &
    \multicolumn{3}{c}{Mode 1} & 
    \multicolumn{3}{c}{Mode 2} \\
    \colhead{} &
    \colhead{Med$^{+1\sigma}_{-1\sigma}$} & 
    \colhead{MAP} & 
    \colhead{MLE} &
    \colhead{Med$^{+1\sigma}_{-1\sigma}$} & 
    \colhead{MAP} & 
    \colhead{MLE}}
\startdata
\input{MB09260_split_fit_table_joint_modes.txt}
\enddata
\tablecomments{The columns list the median $\pm1\sigma$ (68\%) credible intervals, maximum a posteriori (MAP) solution, and and maximum likelihood estimator (MLE) solution for the microlensing parameters of MB09260.
The posterior is multimodal (primarily in $u_0$); it has been split and the parameters for each mode reported separately.
The fraction that each mode contributes to the whole posterior ($\Sigma w_i$) and log evidence ($\log \mathcal{Z}$) are listed for each mode at the end of the table.}
\end{deluxetable*}

\subsection{MB10364 \label{sec:MB10364}}

The data and model for MB10364 are shown in Figures \ref{fig:MB10364_lightcurve} (photometry) and \ref{fig:MB10364_astrom} (astrometry), and the fit posteriors are summarized in Table \ref{tab:mb10364_fits}.
The inferred Einstein crossing time $t_E$ is $61.11^{+0.24}_{-0.24}$ days, the microlensing parallax $\pi_E$ is $0.27^{+0.01}_{+0.01}$, the Einstein radius $\theta_E$ is $0.46^{+0.31}_{-0.21}$, and the lens mass is $0.21^{+0.14}_{-0.10} M_\odot$.

MB10364 is a low mass object, with the possibility of a neutron star or BH lens ruled out.
The relative probabilities of Star:BD:WD are 36:29:35.

\begin{deluxetable}{lccc}
\tablecaption{MB10364 Fit Values\label{tab:mb10364_fits}}
\tablehead{
    \colhead{Parameter} &
    \colhead{Med$^{+1\sigma}_{-1\sigma}$} & 
    \colhead{MAP} &
    \colhead{MLE}}
\startdata
\input{MB10364_fit_table_joint.txt}
\enddata
\tablecomments{Same as Table \ref{tab:mb09260_fits_modes}, but for MB10364. 
The solution is unimodal, hence there is only one column and $\Sigma w_i$ and $\log \mathcal{Z}$ are not reported.}
\end{deluxetable}

\subsection{OB110037 \label{sec:OB110037}}

\begin{deluxetable}{lccc}
\tablecaption{OB110037 Fit Values\label{tab:ob110037_fits}}
\tablehead{
    \colhead{Parameter} &
    \colhead{Med$^{+1\sigma}_{-1\sigma}$} & 
    \colhead{MAP} &
    \colhead{MLE}}
\startdata
\input{OB110037_fit_table_joint.txt}
\enddata
\tablecomments{Same as Table \ref{tab:mb10364_fits}, but for OB110037.}
\end{deluxetable}

The data and model for OB110037 are shown in Figures \ref{fig:OB110037_lightcurve} (photometry) and \ref{fig:OB110037_astrom} (astrometry), the fit posteriors are summarized in Table \ref{tab:ob110037_fits}.
The inferred Einstein crossing time $t_E$ is $92.78^{+2.63}_{-2.60}$ days, the microlensing parallax $\pi_E$ is $\pi_E$ is $0.37_{-0.01}^{+0.01}$, the Einstein radius $\theta_E$ is $1.24^{+0.36}_{-0.35}$, and the lens mass is $0.41^{+0.12}_{-0.12} M_\odot$.

The probability that OB110037 has a dark lens is at least 26\%, and the probability of a stellar lens is at most 74\%.
The relative probabilities of Star:BD:WD:NS:BH are 74:0:26:0:0.
Stellar lenses are only allowed below $0.6 M_\odot$, and white dwarfs are the only type of compact objects allowed.

The lightcurve of OB110037 appears to have some type of perturbation at MJD $\sim$ 55690.
This feature is also apparent in the MOA lightcurve, raising our confidence that the lightcurve feature is real. 
This perturbation may be attributed to a binary lens.

In addition, the astrometry fit, in particular for the F606W filter, is quite poor (Figure \ref{fig:OB110037_astrom}).
Although the first 5 observations from 2011-2012 seem to agree between the two filters, a drastic difference that increases as time goes on begins in 2013-2017.
This may be attributed to a binary source.

It is curious that the photometry may be better explained by a binary lens model, while the astrometry is likely better explained by a binary source model.
Re-analysis of this event with both types of binary models would be a worthwhile and interesting pursuit, but is beyond the current scope of this paper.

\subsection{OB110310 \label{sec:OB110310}}

The data and model for OB110310 are shown in Figures \ref{fig:OB110310_lightcurve} (photometry) and \ref{fig:OB110310_astrom} (astrometry), and the fit posteriors are summarized in Table \ref{tab:ob110310_fits_modes}.
The inferred Einstein crossing time $t_E$ is $82.64^{+2.18}_{-1.50}$ days, the microlensing parallax $\pi_E$ is $0.13_{-0.04}^{+0.08}$, and Einstein radius $\theta_E$ is $0.88^{+0.61}_{-0.42}$, and the lens mass is $0.78^{+0.71}_{-0.39} M_\odot$.\footnote{These are the values for the posteriors before they are split into modes, which is why the values differ from what is reported in Table \ref{tab:ob110310_fits_modes}.
However, since these parameters are all globally unimodal, their distributions nearly identical across modes within the uncertainties, and neither mode is strongly favored, the values are not very different and we report the global median and uncertainties.}

The probability that OB110310 has a dark lens is at least 95\%, and the probability of a stellar lens is at most 5\%.
The relative probabilities of Star:BD:WD:NS:BH are 5:3:65:22:5.
Stellar lenses are only allowed below $1 M_\odot$.
A white dwarf or neutron star is the most probable type of lens, although brown dwarfs and black holes are still allowed at the low and high mass ends, respectively.

\begin{deluxetable*}{l|ccc|ccc}
\tablecaption{OB110310 Fit Values, Modes\label{tab:ob110310_fits_modes}}
\tablehead{
    \colhead{Parameter} &
    \multicolumn{3}{c}{Mode 1} & 
    \multicolumn{3}{c}{Mode 2} \\
    \colhead{} &
    \colhead{Med$^{+1\sigma}_{-1\sigma}$} & 
    \colhead{MAP} & 
    \colhead{MLE} &
    \colhead{Med$^{+1\sigma}_{-1\sigma}$} & 
    \colhead{MAP} & 
    \colhead{MLE}}
\startdata
\input{OB110310_split_fit_table_joint_modes.txt}
\enddata
\tablecomments{Same as Table \ref{tab:mb09260_fits_modes}, but for OB110310.}
\end{deluxetable*}

\subsection{Source properties inferred from CMDs
\label{sec:Source properties inferred from CMDs}}

As there is very little blending for all the targets in the high resolution F814W and F606W filters ($b_{SFF} \sim 1$), the difference between the target and source on the CMD does not change much in color nor magnitude space ($\Delta$F814W $\lesssim 0.1$ mag and $\Delta$(F606W $-$ F814W) $\lesssim 0.1$ mag, Figure \ref{fig:CMDs}).
Based on a CMD analysis, the source stars in MB09260, MB10364, OB110037, and OB110310 are likely red giant stars in the bulge, as is typical for microlensing events in this part of the sky.

The source of OB110462 in the CMD is around the main sequence turnoff (MSTO) on the redder and more luminous side of the main sequence, suggesting it is most likely a giant or sub-giant star.
However, a main sequence source could still be consistent.

\begin{figure}
    \centering
    \includegraphics[width=1.0\linewidth]{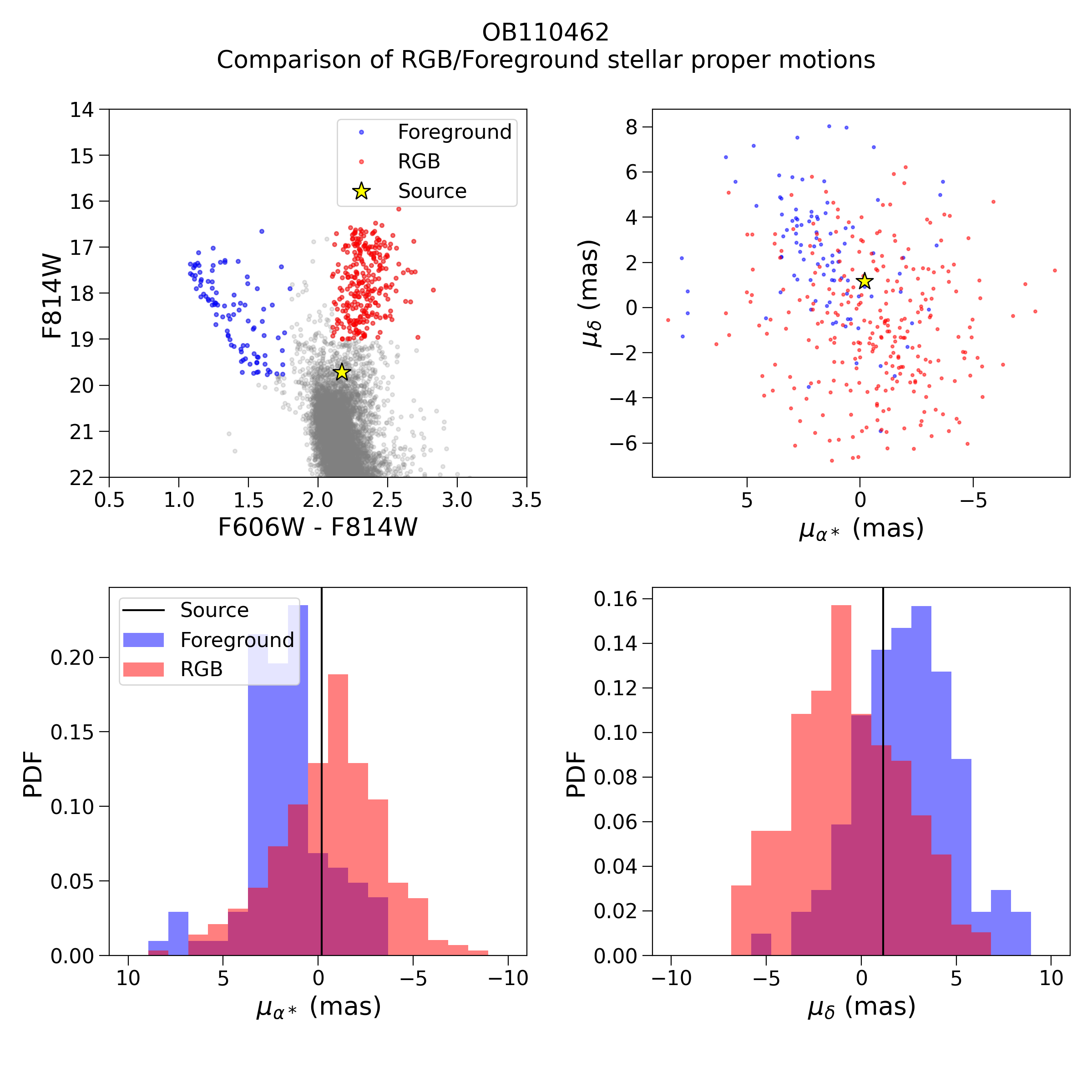} 
    \caption{\emph{Top left:} CMD with the blue foreground and red giant branch (RGB) stars marked.
    The source color and magnitude as inferred from the fit are shown as the yellow star.
    \emph{Top right:} VPD of the foreground and RGB stars.
    The source proper motion as inferred from the fit is shown as the yellow star.
    \emph{Bottom left:} Histogram of the foreground and RGB RA proper motions.
    The proper motion of the source is shown in the black line.
    \emph{Bottom right:} Same as bottom left, but for Dec.
    The source proper motion is consistent with either the foreground or RGB bulge population, although it falls closer to the bulk of the RGB population, hence we consider it is most likely a bulge star.
    \label{fig:ob110462_rgb_bfg_pm}}
\end{figure}

The region of the CMD around the MSTO contains both foreground stars from the disk as well as bulge stars.
We compare the proper motions of OB110462's source to stars in the bright blue foreground as well as in the bulge red giant branch to determine which population it most likely belongs to (Figure \ref{fig:ob110462_rgb_bfg_pm}).
The source is consistent with either population, although it falls within the bulk of the bulge population and more on the edge of the disk population, hence we consider it is most likely a bulge star. 
This is also consistent with the source parallax $\pi_S = 0.11 \pm 0.02$ inferred from the fit which also indicates the source is likely in the bulge.

MB09260 and OB110310 are in the the most highly reddened field, OB110462 and OB110037 are in intermediately reddened fields, and MB10364 is in the least reddened field.
This serves as a reminder that within the bulge the amount of extinction is highly variable, even over relatively small fields.

\subsection{Verifying fit results with $b_{SFF}$ \label{sec:Verifying fit results with bSFF}}

The fitting results are validated by comparing the best-fit OGLE or MOA I-band source flux fraction ($b_{SFF,O}$, $b_{SFF,M}$ in Tables 1 - 2
in the main paper, and Tables \ref{tab:mb09260_fits_modes} - \ref{tab:ob110310_fits_modes}) to the high-resolution F814W HST images (Figure \ref{fig:blend_vs_seeing_radius}).
We only compare the F814W images, as it is has a similar effective wavelength to OGLE I-band.
Assuming a seeing disk radius of $\sim 0.65''$ for OGLE and $\sim 1.25''$ for MOA, we add up the flux from all stars detected by \texttt{hst1pass} within this radius around the target.
Next, by assuming all the flux from the microlensing target at baseline as seen by HST is due to the source (i.e. no flux from the lens or blended neighbor stars within HST's diffraction limited aperture $\theta_{HST} \sim 0.09''$), we can estimate an upper limit on the source flux fraction for a given seeing disk radius $\theta_{see} > \theta_{HST}$ using
\begin{equation}
    b_{SFF,HST-derived} \leq \frac{F_{event}}{\sum_i F_i}
\end{equation}
where $F_i$ are the fluxes of the $i$ stars within $\theta_{see}$ and $F_{event}$ is the flux of the event at baseline in HST.
This estimate on the upper limit of the source flux fraction derived using HST, $b_{SFF,O,HST-derived}$ or $b_{SFF,M,HST-derived}$, is compared to the values inferred from the lightcurve fit, $b_{SFF,O}$ or $b_{SFF,M}$.
This approach is sufficient to cross check our results, considering other differences prevent a perfect comparison (e.g. HST F814W is not a perfect match to the OGLE I filter).

From the photometry fits, MB10364 has $b_{SFF,M} \sim 0.93$, OB110037 has $b_{SFF,O} \sim 0.90$, OB110310 has $b_{SFF,O} \sim 0.97$, and OB110462 has $b_{SFF,O} \sim 0.05$, which are all below or no more than $\sim 0.05$ higher than the upper limits inferred from the HST-derived values shown in Figure \ref{fig:blend_vs_seeing_radius}.

MB09260 has $b_{SFF,M} \sim 0.60$, which is higher than the upper limit of $b_{SFF,HST-derived} \sim 0.4$ in F814W at $1.25''$ estimated from the high resolution HST images.
However, there is an abrupt transition from $b_{SFF,HST-derived} \sim 1$ to $b_{SFF,HST-derived} \sim 0.4$ (indicating a very bright star) at a radius of 1'' from the source.
The best-fit $b_{SFF,M} \sim 0.60$ falls within these two values, indicating that properly treating this situation by convolving the HST image with a MOA PSF would result in a better match.
Hence, MB09260 seems consistent with the high resolution image.

The consistency between the source flux fraction inferred from the fits and the high resolution imaging provides an extra degree of confidence in the inferred fit values.

\subsection{Prospects for resolving lens and source \label{sec:Prospects for resolving lens and source}}
By definition, microlensing means the source and lens are not resolvable.
However, after many years, the lens and source can separate far enough to be resolved (e.g. \cite{Batista:2015, Bennett:2015}).
In the case of dark lenses, such as black holes, non-detections of the lens many years after the event can be used to place constraints on its properties (e.g. \cite{Abdurrahman:2021}).
For the five targets in this paper, we provide estimates to determine if and when taking late-time follow-up data would enable such analyses.

From the results of the fit, we can estimate the time necessary to resolve the source and lens $t_{\rm res}$ via
\begin{equation}
    t_{\rm res} = \theta_{\rm res}/\mu_{\rm rel}
\end{equation}
where the relative proper motion $\mu_{\rm rel}$ comes from fitting the data, and the minimum angular separation $\theta_{\rm res}$ can be estimated using the Rayleigh criterion.
For HST with a mirror diameter of 2.4m, $\theta_{\rm res}$ corresponds to 63.53 mas in F606W and 85.35 mas in F814W.

Assuming there are no contaminating stars, $b_{SFF} = F_S/(F_S + F_L)$.
Then the ratio of lens to source flux, or contrast, is $F_L/F_S = (1 - b_{SFF})/b_{SFF}$.
Note that $b_{SFF} > 1$ are allowed by some of the fits which would lead to an unphysical negative contrast.
$b_{SFF} > 1$ is often referred to as ``negative blending" because it means there is negative non-source flux (see Equation \ref{eq:bsff}).
Negative blending can occur in ground-based images if the background subtraction is imperfect \citep{Park:2004}.
For HST, where the photometry is not derived from difference imaging but rather PSF fitting, this is unlikely to be the case.
However, when fitting a microlensing model to data, negative blending can occur if $b_{SFF} \approx 1$, simply due to normal photometric uncertainties.
This is the most likely explanation for HST; all fits with negative blending values have posterior probabilities that encompass $b_{SFF} = 1$.
Thus, if a contrast value is negative, we cap it at 0.\footnote{This is technically not correct, analogous to truncating negative parallaxes in Gaia; a proper treatment would involve a full Bayesian analysis as explained in \citet{Luri:2018}. 
Such a treatment is beyond the scope of this paper.
We proceed with capping the contrast at 0, as if $b_{SFF} \sim 1$ then the target is likely very dim or dark, and the contrast unlikely to be detectable.}
In addition, the fit values for $\mu_{\rm rel}$ and $b_{SFF}$ are nearly independent, and so these results for $t_{\rm res}$ and $F_L/F_S$ can be thought of as independent.

The resolving time after photometric peak and maximum contrast in the F814W and F606W HST filters for each target are listed in Table \ref{tab:resolve_1sigma}.
Currently, only OB110037 could potentially have its source-lens pair resolved in F606W.
All the other targets have lens-source resolving times at least 15 years post-photometric peak or very low contrast.
In addition, since they all have very high source flux fractions $b_{SFF}$, their lens-source contrasts are very low, which means even after enough time has passed for the lens and source to separate, a luminous lens would be difficult to detect.
Thus, high resolution imaging of OB110037 in the near future could confirm the results of the fit if the separating source and lens could be detected.
For the other targets, the absence of any lens detection would imply consistency with the results presented here, but could not confirm them; however, any detection of a lens would imply the fit results here are incorrect.

\begin{deluxetable}{lcccc}
\tablecaption{Lens/source resolvability
\label{tab:resolve_1sigma}}
\tablehead{
    \colhead{} &
    \multicolumn{2}{c}{Resolving time $t_{\rm res}$ (yr)} & 
    \multicolumn{2}{c}{Contrast ($F_L$/$F_S$)} \\
    \colhead{Parameter} &
    \colhead{F814W} & 
    \colhead{F606W} & 
    \colhead{F814W} & 
    \colhead{F606W}}
\startdata
\input{resolve_1sigma.txt}
\enddata
\tablecomments{There are two entries for OB110462, one based on a fit with ``default weighting" (OB110462 DW) and one based on a fit with ``equal weighting" (OB110462 EW). 
See \S\ref{sec:Likelihood weighting} for details.}
\end{deluxetable}

\subsection{Number of detected BHs}
\label{sec:Number of detected BHs}

\begin{deluxetable}{lcc}
\tablecaption{Fraction of expected BH detections vs. $t_E$ from \texttt{PopSyCLE} simulation
\label{tab:tE_BH_prob}}
\tablehead{
    \colhead{Target} & 
    \colhead{$t_E$ range} & 
    \colhead{\% BH}}
\startdata
\input{tE_bh_prob.txt}
\enddata
\tablecomments{For each target, the $t_E$ range is the median $\pm 3 \sigma$. For OB110462 there are two entries, one with equal weighting to the astrometry and photometry data (OB110462 EW) and one with the default weighting of astrometry and photometry data (OB110462 DW).
See \S \ref{sec:Likelihood weighting} for details.}
\end{deluxetable}

Next, we compare our observed BH yield to the theoretical expectation calculated using the \texttt{PopSyCLE} simulations.
For a sample of simulated events that would be observable by OGLE (see Table 4 of \cite{Lam:2020}), we calculate the fraction of those events due to BHs as a function of the Einstein crossing time $t_E$.
Assuming the OGLE observability criterion for the MOA sample is not strictly correct; however, OGLE and MOA often observe an overlapping set of events, so this approximation suffices.

We wish to calculate the probability of detecting $k$ BHs in our sample given $n$ events, where the probability of detecting a BH in the $i$-th event is $p_i$.
This is described by a Poisson binomial distribution, which characterizes a ``success/no success" experiment with $n$ independent trials, where the $i$-th trial has probability $p_i$ of success \citep{Wang:1993}.
The probability of $k$ successes is given by
\begin{equation}
    \label{eq:Poisson binomial}
    P(K=k) = \sum_{A \in F_k} \prod_{i \in A} p_i \prod_{j \in A^c} (1 - p_j)
\end{equation}
where $F_{k}$ is the set of all subsets of $k$ integers that can be selected from the set $\{1,...,n\}$, i.e. 
\begin{equation*}
    F_k = \{A : A \subseteq \{1, ... , n\}, |A| = 0, 1, ... , n\}.
\end{equation*}
$|A|$ is the number of elements in $A$, and $A^c$ is the complement of $A$.
In the limit where all $p_i$ are equal, the Poisson binomial distribution is equivalent to the ordinary binomial distribution.

In our case, there are $n = 5$ independent trials (i.e. microlensing targets).
We calculate the probability of success $p_i$ (i.e. BH detection) using \texttt{PopSyCLE}.
We define the success probability for the $i$-th target as the fraction of BH lensing events in \texttt{PopSyCLE} over the range of $t_E$ inferred from the fit
\begin{equation*}
    {\rm med}(t_E) - 3\sigma < t_E < {\rm med}(t_E) + 3\sigma.
\end{equation*}
The probabilities of BH detection for each target are listed in Table \ref{tab:tE_BH_prob}.\footnote{The fact that BH probability does not increase monotonically as a function of $t_E$ appears to contradict the claims made in the main paper that events with longer $t_E$ are more likely to have a BH lens.
The issue is that the uncertainty in BH probability rapidly increases at longer $t_E$, as the number of events steeply drops off for $t_E \gtrsim 50$ days (Figure 4
in the main paper).
For example, for both OB110462 solutions, the estimates are based on $<10$ events.
Because of the small number of events, the estimate is highly sensitive to the particular $t_E$ range.}

The results of evaluating Eq. \ref{eq:Poisson binomial} for $k = 0, ... , 5$ BH detections assuming success probabilities $p_i$ is presented in Figure 6
in the main paper.
\texttt{PopSyCLE}'s Galactic model contains $2 \times 10^8$ BHs ranging from $5 - 16 M_\odot$ \citep{Lam:2020}.
There are no $2-5 M_\odot$ mass gap NSs or BHs in the simulation, and hence no exact OB110462 analogue.
Thus, we consider a mass gap detection as falling between 0 to 1 BH detections.
The probability of detecting 0 or 1 BHs are $\sim$25\% and $\sim$40\%, respectively.
This estimate is consistent with our single detection of a NS-BH mass gap object.

\section{Discussion in Detail \label{sec:Discussion}}

OB110462 is the first definitive detection of a compact object discovered with astrometric microlensing. 
Depending on the fit, it is either a neutron star (50\% probability for the EW fit), a black hole (44\% probability for the EW fit, 100\% probability for the DW fit), or a white dwarf (6\% probability for the EW fit).
MB09260 and OB110310 are mostly likely white dwarfs or neutron stars, although a NS-BH mass gap object cannot be ruled out.
MB10364 and OB110037 are definitively low mass objects; OB110037 is most likely a star or white dwarf, while MB10364 is either a star, brown dwarf, or white dwarf.

Discussion of the BH yield as compared to theoretical expectations is found in \S 5
of the main paper \citep{Lam:2022lett}. 
Here, we discuss in more detail our sample of BH candidates and the questions it raises about the Galactic BH population (\S\ref{sec:OB110462 in comparison to the BH population}), additional observations of OB110462 (\S\ref{sec:Auxiliary OB110462 observations}), potential sources of systematics in OB110462 observations (\S \ref{sec:Potential systematics in OB110462 observations}), and the future of BH microlensing searches (\S\ref{sec:Improving experimental strategy and design} and \S\ref{sec:BH searches with the Roman Space Telescope}).

\subsection{OB110462 in comparison to the BH population
\label{sec:OB110462 in comparison to the BH population}}

\subsubsection{Low-mass X-ray binaries}

Several attempts have been made to determine the Milky Way BH mass function using dynamical mass measurements of BHs in low-mass X-ray binaries (LMXBs).
Using a sample of 7 LMXBs, \cite{Bailyn:1998} found the BH mass function to be tightly centered around $7 M_\odot$ with a dearth of systems between $3-5 M_\odot$; they argued this dearth was not due to observational selection effects.
Later work by \cite{Ozel:2010} and \cite{Farr:2011} using samples of 15-16 LMXBs found similar results, providing further support to the idea of the NS-BH mass gap.
However, \cite{Kreidberg:2012} cautioned that systematic errors in the analysis of LMXB systems could push their inferred masses high, artificially creating the mass gap, and \cite{Jonker:2021} identified potential observational biases that prevent measurement of high-mass LXMB systems.
Additionally, LXMBs occupy a very small and specific part of BH evolutionary parameter space, and the BHs found in those systems are likely not representative of the Galactic BH population as a whole.

\subsubsection{Filling the NS-BH mass gap}

Measurements of BH masses in non-LMXB systems do not show evidence of a mass gap.
Gravitational wave searches have found mass gap objects both as the merger remnant ($\sim 3 M_\odot$ in GW170917 \citep{Abbott_GW170817:2017} and $\sim 3.4 M_\odot$ in GW190425 \citep{Abbott_GW190425:2020}), and in the merger components ($\sim 2.6 M_\odot$ in GW190814 \citep{Abbott_GW190814:2020}).
Non-interacting mass gap BHs of $\sim 3 M_\odot$ with red giant companions have also been detected in the Milky Way \citep{Thompson:2019}.
With a mass of 1.6 - 4.4 $M_\odot$, the lens of OB110462 is the first measured isolated Galactic NS or mass gap BH.
These detections of mass-gap objects will improve our theoretical understanding of compact object formation channels.

\subsubsection{A lack of higher-mass BH systems?}

In order to gain a full understanding of the Galactic BH population, BHs must be uncovered outside of closely interacting X-ray binary systems.

We first consider searches for isolated BHs using microlensing, as discussed in this paper.
From our sample of 5 events, we have a single detection of a mass gap object; all other lenses are lower mass non-BH detections.
In addition, we have a single detection of a $>1$~mas astrometric shift; most of the remaining detections are at the $\sim 0.5$~mas level, near the limit of HST's precision.
As discussed in \S 5.1
of \citet{Lam:2022lett}, the low yield of BHs in this sample is consistent with predictions by Galactic models.
However, this presents tentative evidence that Galactic BHs may be less massive than the $\sim 10 M_\odot$ expectation.
If the BH mass function truly peaks at 8$M_\odot$, then selecting candidates via long duration microlensing events should doubly bias us toward finding these high-mass lenses.
First, the lensing cross section $\sigma = \pi \theta_E^2$ is proportional to the mass of the lens $M_L$, since $\theta_E \propto \sqrt{M_L}$.
Thus, more massive objects are more likely to be microlenses.
Second, the Einstein crossing time is proportional to the square root of the lens mass, $t_E \propto \sqrt{M_L}$.  
Thus, long duration events are also more likely to be due to massive microlenses.

Next, we consider searches for BHs in detached/non-interacting binary systems.
The mass function $f$ of a single lined spectroscopic binary is given by
\begin{equation}
    f = \frac{P_{orb} K^3 (1 - e^2)^{3/2}}{2 \pi G} = \frac{M_2^3 \sin^3 i}{(M_1 + M_2)^2}
\end{equation}
where $P_{orb}$ is the orbital period, $K$ is the radial velocity (RV) semi-amplitude, $e$ is the orbital eccentricity, $i$ is the orbital inclination, $M_1$ is the mass of the visible component, and $M_2$ is the mass of the unseen component.
If measurements of $P_{orb}$, $K$, $e$, and $M_1$ can be obtained, then a minimum mass on $M_2$ can be derived.
If $M_2 > 5 M_\odot$ without evidence of luminosity, the unseen system is inferred to be a BH.
Since systems with larger RV semi-amplitudes have larger mass functions, they are most likely to host unseen BH companions.
To date, searches for large RV semi-amplitudes in spectroscopic catalogs have detected an object which fall within the mass gap, but no $\sim 10 M_\odot$ BHs, which suggest a paucity of higher-mass systems (\citet{Thompson:2019, El-Badry:2022}).
Complementary searches using ellipsoidal variables \citep{Rowan:2021} also suggest that higher-mass systems are rare.

Although both microlensing and RV searches should be biased toward finding $10 M_\odot$ objects more easily than $3 M_\odot$ objects, only the latter are being detected.
It may be that the selection bias for massive objects is cancelled out by the fact that the mass function of stars, WDs, NS, and BHs sharply decreases from low to high mass.
Additional work to quantify and compare these two competing effects, combined with larger sample sizes, will be needed to understand the Galactic BH mass function.

Finally, astrometric searches for detached binaries are also eagerly anticipated with Gaia (e.g. \cite{Yamaguchi:2018,  Yalinewich:2018, Wiktorowicz:2020}).
It will be very fruitful to compare the results of those searches to the X-ray transient, microlensing, and RV searches.

\subsection{Auxiliary OB110462 observations
\label{sec:Auxiliary OB110462 observations}}

Additional observations of OB110462 may assist in ascertaining its NS or BH nature.
Here we describe planned observations of OB110462, as well as data found in searches of archival catalogs.

The 2021 astrometric observations of OB110462 are crucial to the modeling as they extend the temporal baseline of the original archival observations by 50\%, from 6 years to 10 years.
The remaining data from HST Cycle 29 program GO-16760 to be taken in Fall 2022 \citep{Lam:2021hst} will further extend the baseline by another year and improve the characterization of the astrometric signal.
The astrometric microlensing shift is a deflection with respect to the unlensed position of source (Equation \ref{eq:delta_c}), and we can only have confidence in our measurement of that shift if we also have confidence in our measurement of the source's unlensed proper motion.
The astrometric shift due to lensing when $u \gg 1$ can be approximated as $\delta_c \approx \theta_E t_E / (t - t_0)$
(see Appendix \ref{app:Astrometric lensing in Gaia}).
For OB110462 ($t_E = 280$ days and $\theta_E = 4$ mas), the microlensing astrometric shift did not dip below HST's astrometric precision of $\sim 0.3$ mas until 2021, 10 years after source-lens closest approach.
This calculation illustrates the importance of having a long temporal baseline for OB110462 in order to properly measure the source's unlensed proper motion and characterize the astrometric microlensing signal.

Additional follow-up observations in the X-ray can place limits on accretion from the ISM \citep{Agol_Xray:2002}. 
For example, \cite{Maeda:2005} and \cite{Nucita:2006} looked for X-rays at the location of BH microlensing candidate MACHO-96-BLG-5 reported in \cite{Bennett:2002}, using ACIS on Chandra and EPIC on XMM-Newton, respectively.
Neither detected any X-rays.

We searched several X-ray catalogs that have observed at OB110462's coordinates to determine whether there are any coincident sources.
OB110462 was not detected as an X-ray source in any of the following catalogs:
\begin{itemize}
    \item Chandra Source Catalog 2.0\footnote{\href{https://cxc.harvard.edu/csc/}{https://cxc.harvard.edu/csc/}} \citep{Evans_AAS:2019, Evans_HEAD:2019}.
    The limiting sensitivity provides an upper limit of $1.91 \times 10^{-14}$ erg/s/cm$^2$ at 0.5-7.0 keV.
    
    \item XMM-Newton Science Archive\footnote{\href{http://nxsa.esac.esa.int/nxsa-web/\#search}{http://nxsa.esac.esa.int/nxsa-web/\#search}} \citep{Sarmiento:2019}.
    This provides an upper limit of $1.52 \times 10^{-14}$ erg/s/cm$^2$ and $< 1.38 \times 10^{-3}$ counts/s at 0.2 - 12.0 keV.
    
    \item Swift XRT Point Source Catalogue\footnote{\href{https://www.swift.ac.uk/2SXPS/}{https://www.swift.ac.uk/2SXPS/}} \citep{Evans:2020}.
    This provides an upper limit of $2.4 \times 10^{-3}$ counts/s at 0.3 - 10 keV.
\end{itemize}
OB110462's coordinates are not in the eROSITA-DE Early Data Release catalog\footnote{\href{https://erosita.mpe.mpg.de/edr/}{https://erosita.mpe.mpg.de/edr/}}.

We also searched the Australia Telescope National Facility Pulsar Catalogue\footnote{\href{https://www.atnf.csiro.au/research/pulsar/psrcat/}{https://www.atnf.csiro.au/research/pulsar/psrcat/}} \citep{Manchester:2005} Version 1.65 for any pulsars coincident with the target.
There are no coincident pulsars; the nearest pulsar is 0.55 deg away.

As mentioned in \S\ref{sec:Gaia}, OB110462 is not in the Gaia EDR3 catalog.
OB114062's baseline magnitude is F606W~$\approx 22$, while Gaia's nominal magnitude limit is $G \approx 20.7$, and will be even brighter in crowded regions \citep{Fabricius:2021}.

\subsection{Potential systematics in OB110462 observations
\label{sec:Potential systematics in OB110462 observations}}

We briefly discuss whether the tension between the photometry and astrometry could be due to systematics in the data.

First we consider the OGLE photometry.
Possible sources of systematic error include differential refraction, proper motion, or low-level stellar variabliliy are possibilities for such a long baseline.
However, no trends due to proper motion or stellar variability are seen in the lightcurve.
In addition, the F814W $-$ F606W color of the bright star within $\sim 0.3$~arcsec of OB110462 that are blended together in the OGLE images are within 0.1 of each other, and hence effects due to differential refraction would be undetectable. 
We also inspect the lightcurves of several stars near OB110462 that have similar magnitude to the baseline magnitude of OB110462.
No trends can be seen that resemble the residuals in the EW fit.
In addition, we explored re-scaling the photometric uncertainties of the OGLE data; however, even inflating all uncertainties by a factor of $\sim 2$ did not significantly change the structure of the residuals, nor the inferred microlensing parameters, including lens mass.

Potential sources of systematics in the HST astrometry are discussed in the next section.

\subsection{Improving experimental strategy and design
\label{sec:Improving experimental strategy and design}}

\subsubsection{Multi-filter astrometry}
In contrast to photometric observations, multi-filter astrometry is not routinely obtained, as astrometric observations are expensive and facilities with the requisite precision are rare. 
This is one of the first projects to explore the impact of different filters on relative astrometry. 
The nature of the difference between F606W and F814W filter astrometric observations of OB110037 and OB110462 is an open question, whether it be astrophysical (e.g. binaries sources with different color), systematic (e.g. uncorrected CTE), or statistical.
However, it demonstrates that multi-filter observations are worth continuing to pursue in future astrometric microlensing studies.
For example, it could help break degeneracies between certain types of binaries lenses or sources.

\subsubsection{CTE correction}

As mentioned in \S \ref{sec:Reduction}, the CTE correction in the \texttt{\_flc} files is not perfect.
Future pursuits will explore other methods of correction, such as a re-analysis of OB110462 that uses the newer and more accurate tabular correction for CTE \citep{Anderson_ctetab:2021}.
In addition, trying to fix CTE via a magnitude-dependent astrometric alignment is another avenue that is being explored. 

\subsubsection{Observational strategy}

For these precise astrometric measurements, taking good observations is critical.
The lens mass constraints for several of the targets are only upper limits, as the astrometric shifts were so small as to be undetectable at the precision of the measurements.
A dominant source of astrometric uncertainty with HST WFC3-UVIS observations is the undersampling of the PSF. 
It has been shown that there is a floor in the astrometric precision that can be achieved, even at high SNR, when only a few exposures are used \citep{Hosek:2015}. 
As the majority of the observations in each filter had 4 or less exposures, this limited the achievable astrometric precision for several of the targets, in particular OB110462.
Increasing the number of exposures and implementing small uncorrelated dithers to sample different pixel phases can reduce this floor as $\sqrt{N_{dithers}}$. 

In addition, the effects of CTE are worsening with time.
Actively mitigating CTE through careful planning of observations rather than trying to correct it afterwards is even more important than before (e.g. as described in \S7 of \citet{Anderson:2021}).

\subsubsection{Event selection}

Although all events presented in this work were selected to have $t_E > 200$ days \citep{Sahu:2009}, the inferred $t_E$ values for 4 of the 5 events did not satisfy this criteria.
As a result, the true $t_E$ range probed extended down to $t_E = 60$ days, and did not sample the $t_E$ range that maximizes the expected yield of BHs ($t_E \gtrsim 100$ days).
Only MB09260 and OB110462 had $t_E > 100$ days, weakening the constraints on the BH fraction.
We are currently attempting to determine whether prediction of $t_E$ before the photometric peak of the event could be improved.
If possible, this would enable improved selection of BH candidates for astrometric follow-up.

A secondary concern is the target field itself.
A sufficient number of reference stars is needed, hence the field in the immediate vicinity of the target must be sufficiently crowded.
However, the magnitude range of those nearby stars must also be similar to the target.
Because of the steepness of the luminosity function, bright targets or targets with high magnification are more difficult to analyze as they lack sufficient reference stars to perform relative astrometric alignment.
This is in tension with the need to have high photometric precision in order to precisely measure the microlensing parallax.
Although bright and highly magnified stars should still be followed up if they are long duration events, special care must be taken when designing observations to ensure good astrometric alignment.

\subsection{BH searches with the Roman Space Telescope
\label{sec:BH searches with the Roman Space Telescope}}

Although the initial idea and subsequent design requirements for the Roman Space Telescope (hereafter \emph{Roman}) microlensing survey are driven by exoplanet searches \citep{Penny:2019}, it also hails the next generation of astrometric microlensing campaigns searching for BHs.
Presently, each event must be followed up individually, with only two facilities (HST and Keck) capable of the precision in the near-infrared required to make such a measurement.
Such measurements are expensive (requiring an $\sim$orbit of HST or $\sim$hour of Keck time per measurement), prohibiting dense astrometric temporal sampling or a large sample of targets. 

\emph{Roman} will change this with its ability to simultaneously obtain precise photometry and astrometry over a wide field of view $100\times$ the area of HST and astrometric precision almost an order of magnitude better than HST \citep{Spergel:2015, WFIRST_AWG:2019}.
This will also allow the masses of NS-BH mass gap objects to be precisely measured, and allow a sample of 100-1000 BH candidates to be built up over the duration of the survey \citep{Lam:2020}.
In addition, \emph{Roman} will probe a large sample of shorter $t_E$ events which will place constraints on BH kicks.

To make \emph{Roman} as effective as possible for finding BHs, there are several considerations that need to be addressed.
Due to the placement of the observatory's solar panels, which dictates the available pointings of the telescope, the Galactic Bulge can only be observed twice a year during a 72 day window centered on the vernal and autumnal equinoxes.
In addition, for \emph{Roman}'s planned 5 year mission, only 6 of the 10 available Bulge seasons are to be dedicated to the microlensing survey.
These large temporal gaps are generally not a concern for exoplanets searches, as the transient portion of the lightcurve is nearly covered within the 72 day window.
However, for long duration events where the transient portion of the lightcurve is much longer than the window and where a measurement of small microlensing parallax is crucial, incomplete lightcurve coverage will mean the difference between a confirmed BH mass measurement and only an upper limit.
Observations filling in these gap will be crucial.
Collaboration with a ground-based telescope to provide imaging during the gaps, or a smaller independent follow-up efforts would be very important.

\section*{Acknowledgements}

We thank Dan Foreman-Mackey, Tharindu Jayasinghe, Tom Loredo, Greg Martinez, and Jeff Andrews for helpful and interesting conversations.
In addition, we thank the referee for feedback that improved this paper.
We also thank Kailash Sahu, Howard Bond, Jay Anderson, Martin Dominik, Philip Yock, and Annalisa Calamida for proposing and taking the archival HST observations used in this work.

C.Y.L. and J.R.L. acknowledge support by the National Science Foundation under Grant No. 1909641 and the National Aeronautics and Space Administration (NASA) under contract No. NNG16PJ26C issued through the WFIRST (now Roman) Science Investigation Teams Program.
C.Y.L. also acknowledges support from NASA FINESST grant No. 80NSSC21K2043.
D.P.B. was supported by NASA grants NASA-80NSSC18K0274 and 80GSFC17M0002.
{\L}.W. acknowledges support from the Polish National Science Centre (NCN) grants Harmonia No. 2018/30/M/ST9/00311 and Daina No. 2017/27/L/ST9/03221 as well as the European Union’s Horizon 2020 research and innovation programme under grant agreement No 101004719 (OPTICON-RadioNet Pilot, ORP) and MNiSW grant DIR/WK/2018/12.

Based on observations made with the NASA/ESA Hubble Space Telescope, obtained from the data archive at the Space Telescope Science Institute (STSci) operated by the Association of Universities for Research in Astronomy, Inc. under NASA contract NAS 5-26555, and obtained from the the Hubble Legacy Archive, a collaboration between STScI/NASA, the Space Telescope European Coordinating Facility (ST-ECF/ESA) and the Canadian Astronomy Data Centre (CADC/NRC/CSA).

This paper makes use of data obtained by the MOA collaboration with the 1.8 metre MOA-II telescope at the University of Canterbury Mount John Observatory, Lake Tekapo, New Zealand. 
The MOA collaboration is supported by JSPS KAKENHI (Grant Number JSPS24253004, JSPS26247023, JSPS23340064, JSPS15H00781, JP16H06287, and JP17H02871) and the Royal Society of New Zealand Marsden Fund.
The MOA project has received funding from the Royal Society of New Zealand, grant  MAU1901 to IB.

This work presents results from the European Space Agency (ESA) space mission Gaia. 
Gaia data are being processed by the Gaia Data Processing and Analysis Consortium (DPAC). 
Funding for the DPAC is provided by national institutions, in particular the institutions participating in the Gaia MultiLateral Agreement (MLA). 

This research has made use of data obtained from the Chandra Source Catalog, provided by the Chandra X-ray Center (CXC) as part of the Chandra Data Archive.

\software{Galaxia \citep{Sharma:2011}, astropy \citep{Astropy:2013, Astropy:2018}, Matplotlib \citep{Hunter:2007}, NumPy \citep{vdWalt:2011}, SciPy \citep{Scipy:2019}, SPISEA \citep{Hosek:2020}, PopSyCLE \citep{Lam:2020}, dynesty \citep{Speagle:2020}, PyMultiNest \citep{Buchner:2014, Feroz:2009}, dustmaps \citep{Green:2018}, hst1pass, xym2mat, xym2bar \citep{Anderson:2006}, ks2 \citep{Anderson:2008, Bellini_omegacen:2018}, celerite \citep{Foreman-Mackey:2017}}

\appendix

\section{Rescaling of uncertainties \label{sec:Rescaling of uncertainties}}

For each epoch, \texttt{hst1pass} returns the RMS error of extracted source positions and magnitudes over multiple frames $\sigma_x$, $\sigma_y$ and $\sigma_m$, respectively.
Following \citet{Hosek:2015}, instead of using the RMS errors for our uncertainties, we use the error on the mean $\sigma/\sqrt{N}$ where $N$ is the number of frames the star is detected in, inflated with a empirical additive error.
The empirical constant additive error on the positions $\Delta_{xy}$ and magnitudes $\Delta_m$ is calculated for each epoch and added in quadrature to the error on the mean.
This produces the final rescaled uncertainties used in the analysis
\begin{align}
    \sigma_x' &= \sqrt{\sigma_x^2/N + \Delta_{xy}^2} \\
    \sigma_y' &= \sqrt{\sigma_y^2/N + \Delta_{xy}^2} \\
    \sigma_m' &= \sqrt{\sigma_m^2/N + \Delta_m^2}.
\end{align}

To determine $\Delta_{xy}$ and $\Delta_m$, a sample of bright, unsaturated stars are selected.
The exact magnitude range constituting ``bright" is roughly saturation to 3-5 magnitudes fainter, with the exact range determined empirically through the astrometric alignment process.
In this sample, the additive error is selected such that the $\chi^2$ distribution of the reference stars position and magnitude fits is roughly consistent with the expected distribution (Figure \ref{fig:chi2_xym}).
The resulting values are listed in Table \ref{tab:Additive errors}.
Note that for MB09260 F606W, the same additive magnitude error was used across all epochs.

As mentioned in \S\ref{sec:Alignment procedure}, for the five microlensing targets, the RMS and rescaled astrometric uncertainties are generally similar.
However, adopting an additive error makes the resulting astrometric uncertainties more uniform across the field, particularly for bright stars.
This is critical as these uncertainties are used as weights in the astrometric alignment (\S\ref{sec:Derived stellar proper motions} and Figure \ref{fig:chi2_xym}).
The reference stars' uncertainties are used to determine how good the reference frame transformation is, which is ultimately used to measure the astrometry of the target.

\begin{deluxetable}{lllll}
\tabletypesize{\scriptsize}
\tablecaption{Additive errors for HST data\label{tab:Additive errors}}
\tablehead{
    \colhead{Epoch} &
    \multicolumn{2}{c}{Pos. error (mas)} & 
    \multicolumn{2}{c}{Mag. error (mmag)} \\
    \colhead{(yyyy-mm-dd)} &
    \colhead{F814W} &
    \colhead{F606W} &
    \colhead{F814W} &
    \colhead{F606W}}
\startdata
MB09260 & & & & \\
\hline
\input{MB09260_add_err.txt}
\\\hline
MB10364 & & & & \\
\hline
\input{MB10364_add_err.txt}
\\\hline
OB110037 & & & & \\
\hline
\input{OB110037_add_err.txt}
\\\hline
OB110310 & & & & \\
\hline
\input{OB110310_add_err.txt}
\\\hline
OB110462 & & & & \\
\hline
\input{OB110462_add_err.txt}
\enddata
\end{deluxetable}

\section{Injection and recovery tests}
\label{app:Injection and recovery tests}

We use the \texttt{ks2} software (\citet{Anderson:2008, Bellini_omegacen:2018}, see also \S 3 of \citet{Sabbi:2016} for a detailed description) to inject artificial stars into the OB110462 HST images to determine how well we can recover the magnitudes and positions of faint sources near bright sources.
Injection and recovery tests are performed for all OB110462 epochs in two different manners: 
\begin{enumerate}
    \item Sources are injected radially around the bright neighbor star at the radius of OB110462 at a variety of azimuths excluding the azimuth of OB110462 itself, as we cannot recover a source planted on top of a real star.
    \item Sources are injected near a star of similar brightness to the neighbor, in the same radial and azimuthal configuration relative to the star as OB110462 relative to the bright neighbor. 
\end{enumerate}
The first test directly probes the region around the neighbor itself but excludes the actual position of the OB110462, while the second test probes a region around a star similar to the neighbor, at the same separation and angle of OB110462 relative to the neighbor.

\begin{figure}[h!]
\centering
    \includegraphics[width=\linewidth]{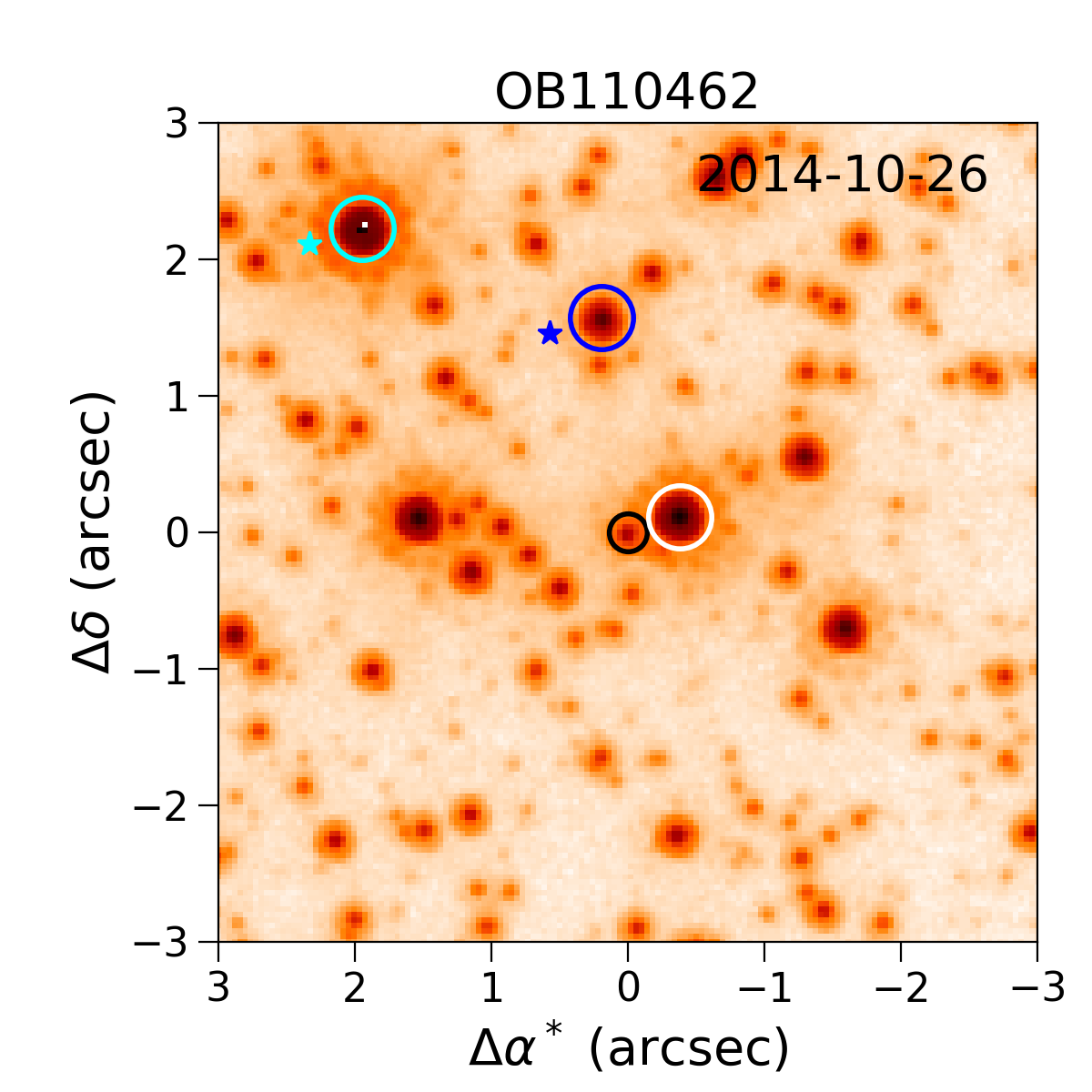} 
\caption{Injection and recovery for OB110462.
\emph{Black circle, center}: OB110462.
\emph{White circle}: the bright neighbor.
\emph{Blue circle}: neighbor-like star for F606W.
\emph{Cyan circle}: neighbor-like star for F814W.
Positions where stars are injected in reference to the neighbor-like star are shown as blue and cyan stars (F606W and F814W, respectively).
\label{fig:inject_recover_image}}
\end{figure}

The star we dub ``the neighbor" is $\sim$10 pixels ($\sim$0.4 arcsec) west of OB110462.
The star we dub ``the neighbor-like star" in F814W is $\sim$75 pixels ($\sim$3 arcsec) northeast of OB110462; for F606W the neighbor-like star is $\sim$40 pixels ($\sim$1.6 arcsec) north of OB110462 (Figure \ref{fig:inject_recover_image}).

In summary, both methods produce similar results (Figures \ref{fig:inject_recover_pos_bias} and \ref{fig:inject_recover_mag_bias}).
We use the single-azimuth results from injection around the neighbor-like star in order to capture any azimuthal dependence which would be lost by averaging over multiple azimuths (Table \ref{tab:injection recovery bias}).

In F814W, the bias is negligible in the first epoch, when the target is magnified and is of similar brightness to the neighbor.
However, in later epochs once the target is no longer magnified, the measured position of the target is biased by $\sim 0.4$ mas toward the bright neighbor star along the target-neighbor separation vector.
The magnitude and direction of the positional offset are comparable across the two approaches, except in 2014, where the injection around the neighbor-like star leads to a larger difference than when injecting around the neighbor. 
The uncertainties are larger in the approach of planting around the neighbor, as it averages over more azimuths and results in a larger uncertainty.

In F606W, the bias is smaller than F814W because the shorter wavelength results in higher resolution.
Like F814W, the bias is also negligible in the first epoch, and around $\sim 0.25$ mas in later epochs.
The uncertainties are also larger when averaging across many azimuths.
Unlike F814W, the bias is mixed between radial and azimuthal components when injecting around the neighbor-like star at a single azimuth.

\subsection{Injection around the neighbor \label{sec:Injection around the neighbor}}

For each epoch, we determine the separation $r$ of OB110462 and the bright neighbor. 
We inject three rings of stars surrounding the bright neighbor, of radii $r - 0.2$, $r$, and $r + 0.2$ pixels.
Each ring consists of 24 evenly spaced stars, resulting in one star every 15 degrees.
Because we cannot recover stars injected on top of the target itself, we do not attempt to recover injected stars that fall within 4 pixels of the target itself; this excludes three of the positions.
We thus inject a total of $(24 - 3) \times 3 = 63$ stars per epoch.

\begin{deluxetable}{lccc}
\tablecaption{Bias correction derived from injection and recovery
\label{tab:injection recovery bias}}
\tablehead{
    \colhead{Epoch} & 
    \colhead{$\Delta$RA (mas)} & 
    \colhead{$\Delta$Dec (mas)} &
    \colhead{$\Delta$Mag (mag)}}
\startdata
F606W & & & \\
\hline
\input{bright_f606w_flc_positional_bias.txt}
\\\hline
F814W & & & \\
\hline
\input{bright_f814w_flc_positional_bias.txt}
\enddata
\tablecomments{
Bias correction derived from injection/recovery around a star of comparable brightness at the same separation, azimuth, and magnitude difference as the target to its bright neighbor.}
\end{deluxetable}

\subsection{Injection around the neighbor-like star \label{sec:Injection around the neighbor-like star}}

The neighbor-like star we inject around is different for F814W and F606W, because the surrounding stars do not have the same colors as the neighbor and target.
The neighbor-like stars were chosen to have similar magnitude and saturation level to the neighbor.
In F814W, the neighbor tended to be saturated; the F814W neighbor-like star is also saturated.
On the other hand, in F606W the neighbor was not saturated, and the F606W neighbor-like star is also not saturated.
The F814W neighbor-like star is brighter than the neighbor in F814W by $\sim 0.6$ mag, and the F606W neighbor-like star is fainter than the neighbor in F606W by $\sim 0.5$ mag.

In each epoch, we inject three arcs of radii $r - 0.2$, $r$, and $r + 0.2$ pixels centered on the neighbor-like star, where $r$ is the target-neighbor separation ($\sim 10$ pixels).
Each arc consists of 15 stars at the azimuth of the target relative to the neighbor $\pm$ 0.2 pixels/(target - neighbor separation in pixels), which corresponds to a subtended angle of approximately 2.2 degrees.
This corresponds to a region of $\sim$0.04 pix$^2$ where $3 \times 3 \times 15 = 135$ stars are injected.

At each position we inject stars of magnitude $m_I - 0.1$, $m_I$, and $m_I + 0.1$, where $m_I$ is the magnitude that results in the same contrast with the neighbor-like star as OB110462 to the neighbor. 
That is, if OB110462 has magnitude $m_T$ and the neighbor $m_N$, and the injected star is $m_I$ around the neighbor-like star $m_C$, then $m_T - m_N = m_I - m_C$.

\subsection{Recovery of injected sources}

After planting fake stars into the image, we determine how well we can recover the positions and magnitudes.
To match the properties of our original dataset, we consider stars to be recovered if they are detected in at least $N$ frames, where $N$ is the number of frames that were used to calculate the position of the source.
Iterative 3-sigma clipping is performed to exclude outliers due to confusion, e.g. from the diffraction spike mask.
We then use the transformation parameters derived for this epoch (as described in \S \ref{sec:Alignment procedure}) to convert from (x,y) pixel positions to (RA, Dec) coordinates. 

\begin{figure}[h!]
\centering
    \includegraphics[width=\linewidth]{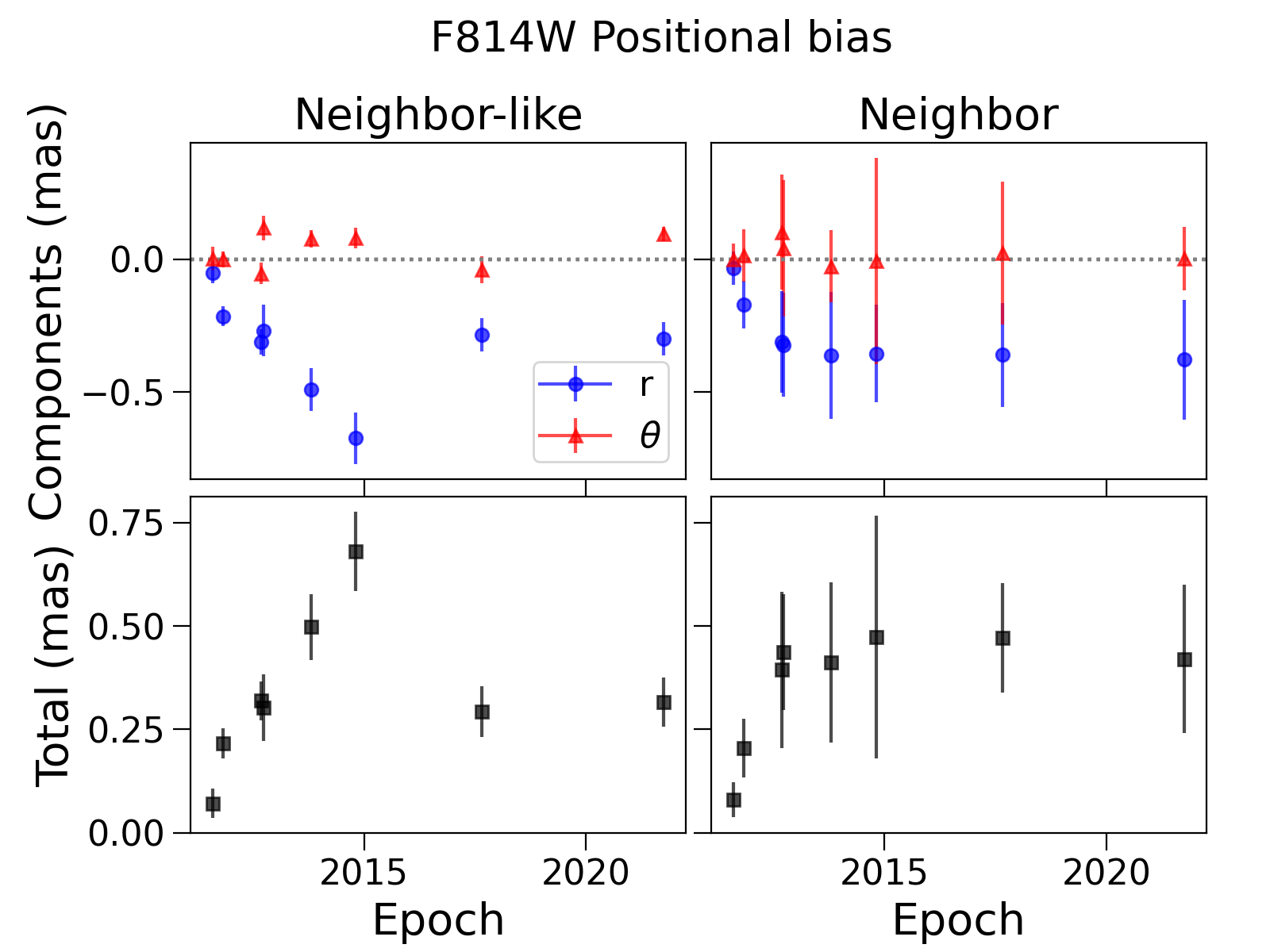} \\
    \includegraphics[width=\linewidth]{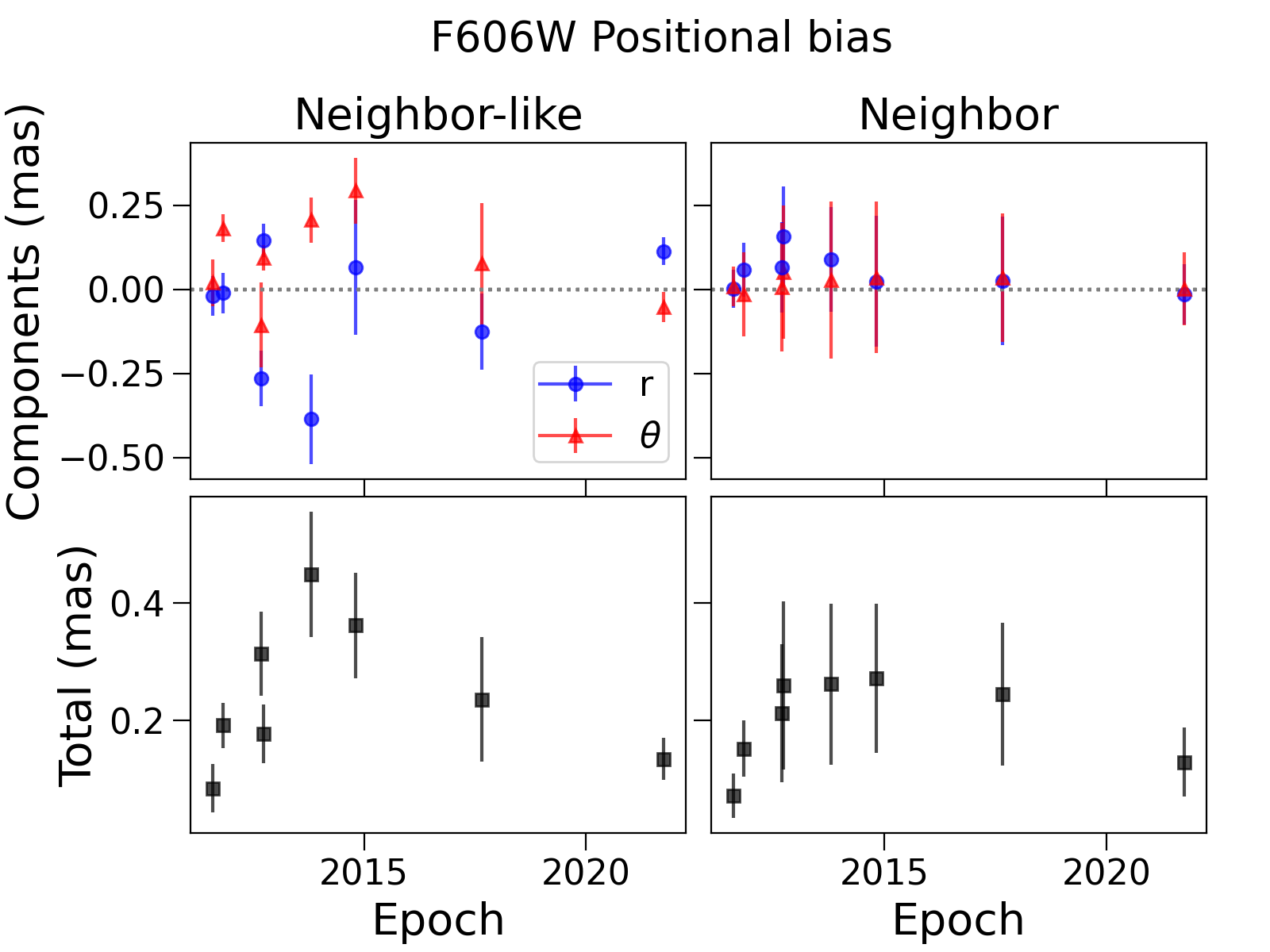}
\caption{Positional bias in F814W (\emph{top panel}) and F606W (\emph{bottom panel}) as calculated from injection and recovery tests, injecting around a neighbor-like star (\emph{left column}) or around the neighbor (\emph{right column}). 
The positional bias (position recovered minus the true position input) is shown as a function of radial $r$ and azimuthal $\theta$ components (\emph{top row}), as well as total positional bias (\emph{bottom row}).
\label{fig:inject_recover_pos_bias}}
\end{figure}

\begin{figure}[h!]
\centering
    \includegraphics[width=\linewidth]{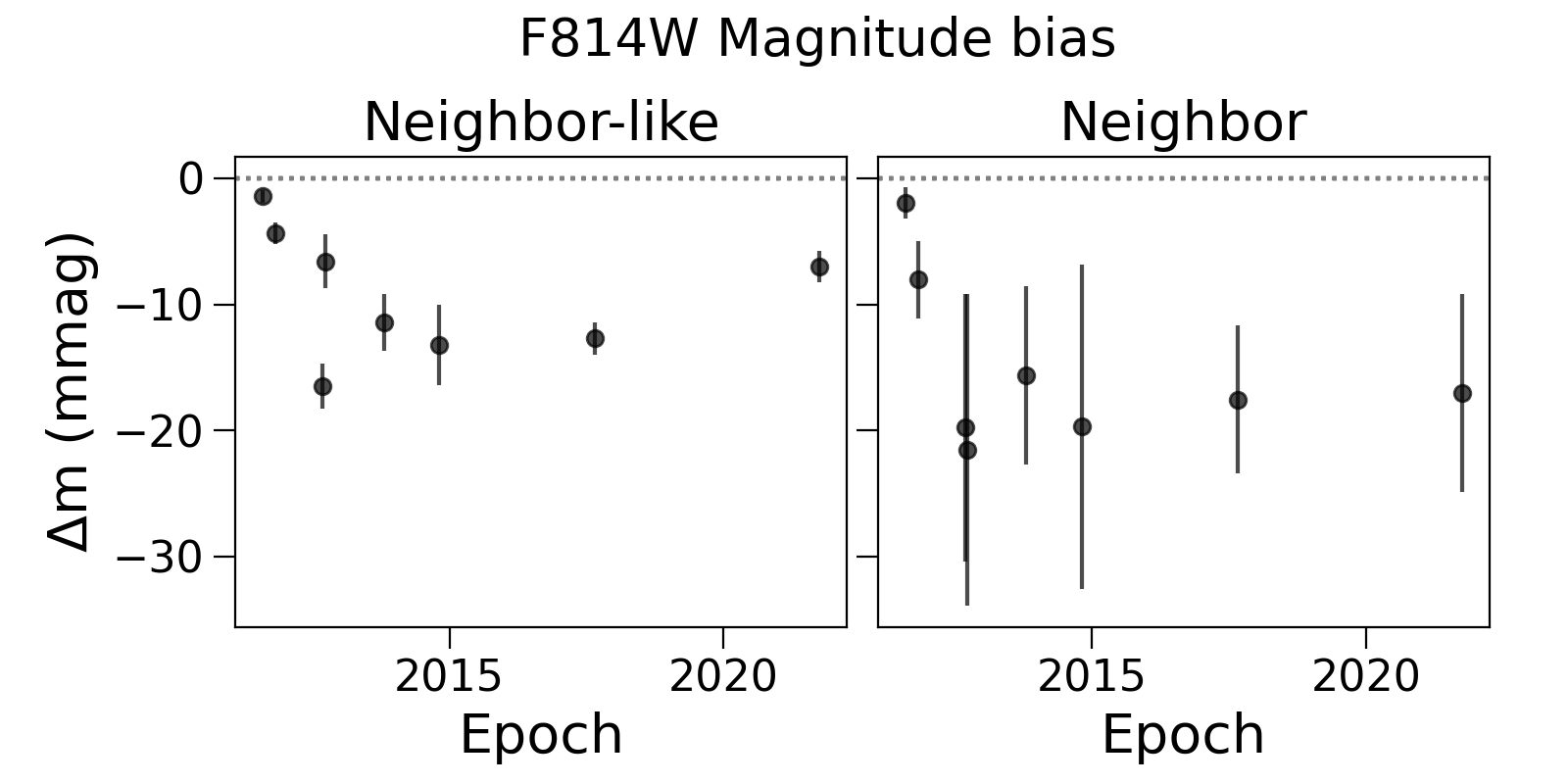} \\
    \includegraphics[width=\linewidth]{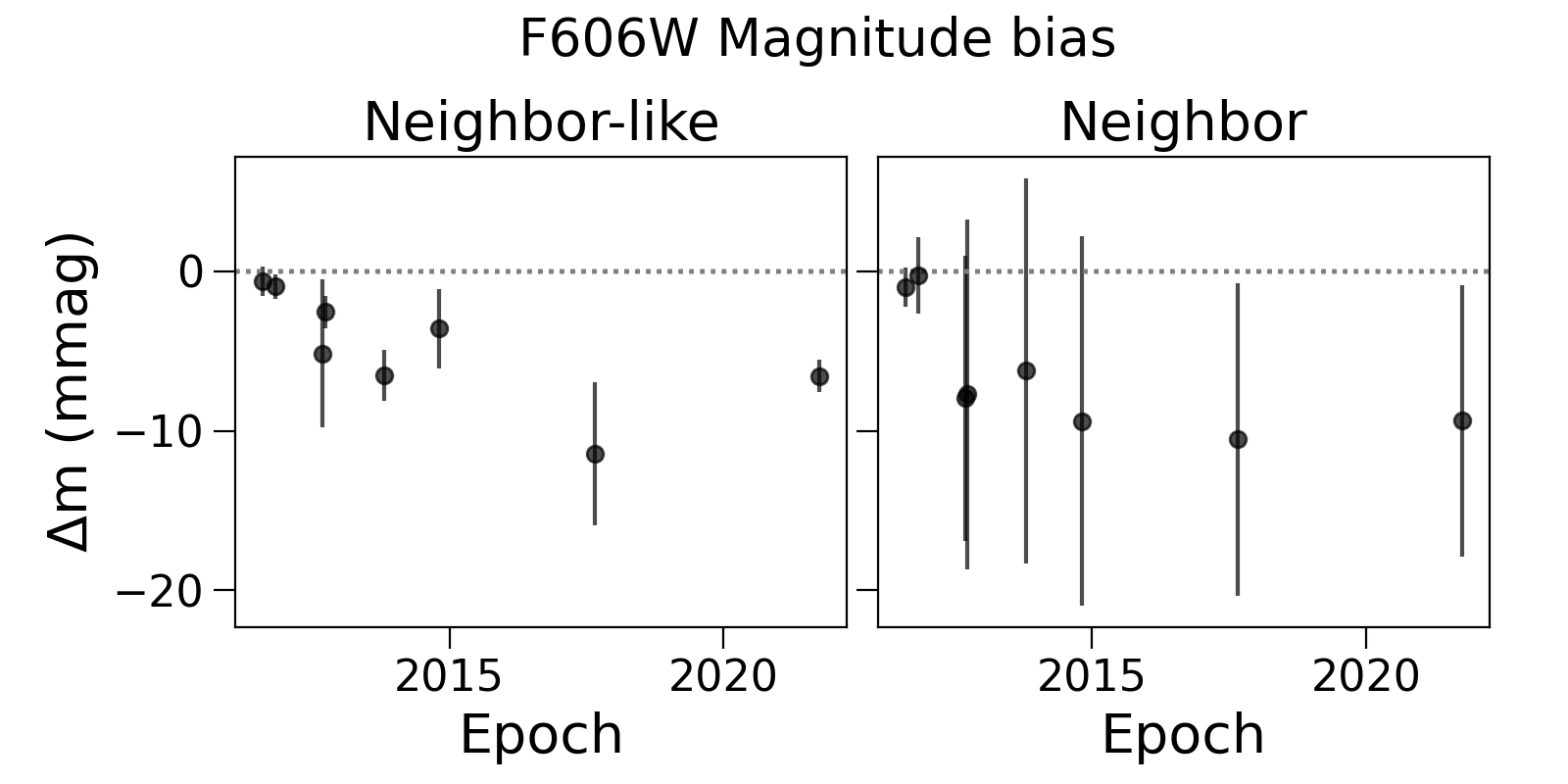}
\caption{Magnitude bias in F814W (\emph{top panel}) and F606W (\emph{bottom panel}) as calculated from injection and recovery tests, injecting around a neighbor-like star (\emph{left column}) or around the neighbor (\emph{right column}).
The magnitude bias is defined as the magnitude recovered minus the true magnitude input.
\label{fig:inject_recover_mag_bias}}
\end{figure}

We define a polar coordinate system with the origin located at the neighbor star for the analysis in \S \ref{sec:Injection around the neighbor} and located at the neighbor-like star for the analysis in \S \ref{sec:Injection around the neighbor-like star}.
The azimuthal direction is measured counterclockwise from the origin-OB110462 separation vector.
The average offset in the radial and azimuthal directions are $r$ and $\theta$ (Figures \ref{fig:inject_recover_pos_bias} and \ref{fig:inject_recover_mag_bias}).

The color of the neighbor and OB110462 are very similar (F606W - F814W = 2.25 and 2.15, respectively).
At baseline, the neighbor is about 3.1 magnitudes brighter than OB110462. 
During magnification in the first epoch (2011-08-08), the neighbor is only about 0.4 magnitudes brighter.
Since the resolution is higher at shorter wavelengths, it is not unexpected that the positional bias is less in F606W than F814W, since the separation is the same in both filters.
It is also not surprising that the bias is smallest in the first epoch when the magnitude difference between OB110462 and the neighbor is small, and larger in the remaining epochs when the magnitude difference is large. 

\section{Absolute proper motion reference frame}
\label{sec:Absolute proper motion reference frame}

The HST astrometry in \S\ref{sec:HST Astrometric Analysis} is derived in a procedure where the average motion of the stars is at rest with respect to the reference frame. 
To interpret the lens' proper motion or transverse velocity, we must place the relative astrometry into an absolute reference frame.
To do this, we calculate the proper motion offset between stars in the relative HST frame and the absolute Gaia frame.
We match all stars in the Gaia EDR3 catalog within 1 arcmin of the target that have \texttt{astrometric\_excess\_noise\_sig} $ < 2$ and a proper motion measurement (i.e. 5-parameter and 6-parameter solutions) to the bright stars in our HST proper motion catalog (F814W $<$ 22 for MB10364 and OB110037; F814W $<$ 23 for MB09260, OB110310, and OB110462).
The 1-iteration $3\sigma$-clipped uncertainty-weighted average difference in the proper motion between the cross-matched stars is calculated, then applied to the relative HST astrometry to place it into the absolute Gaia proper motion frame.
The values to convert between the HST and Gaia frames for each target's field are listed in Table \ref{tab:hst_gaia_offset}.

\begin{figure*}[h!]
\centering
    \includegraphics[width=0.4\linewidth]{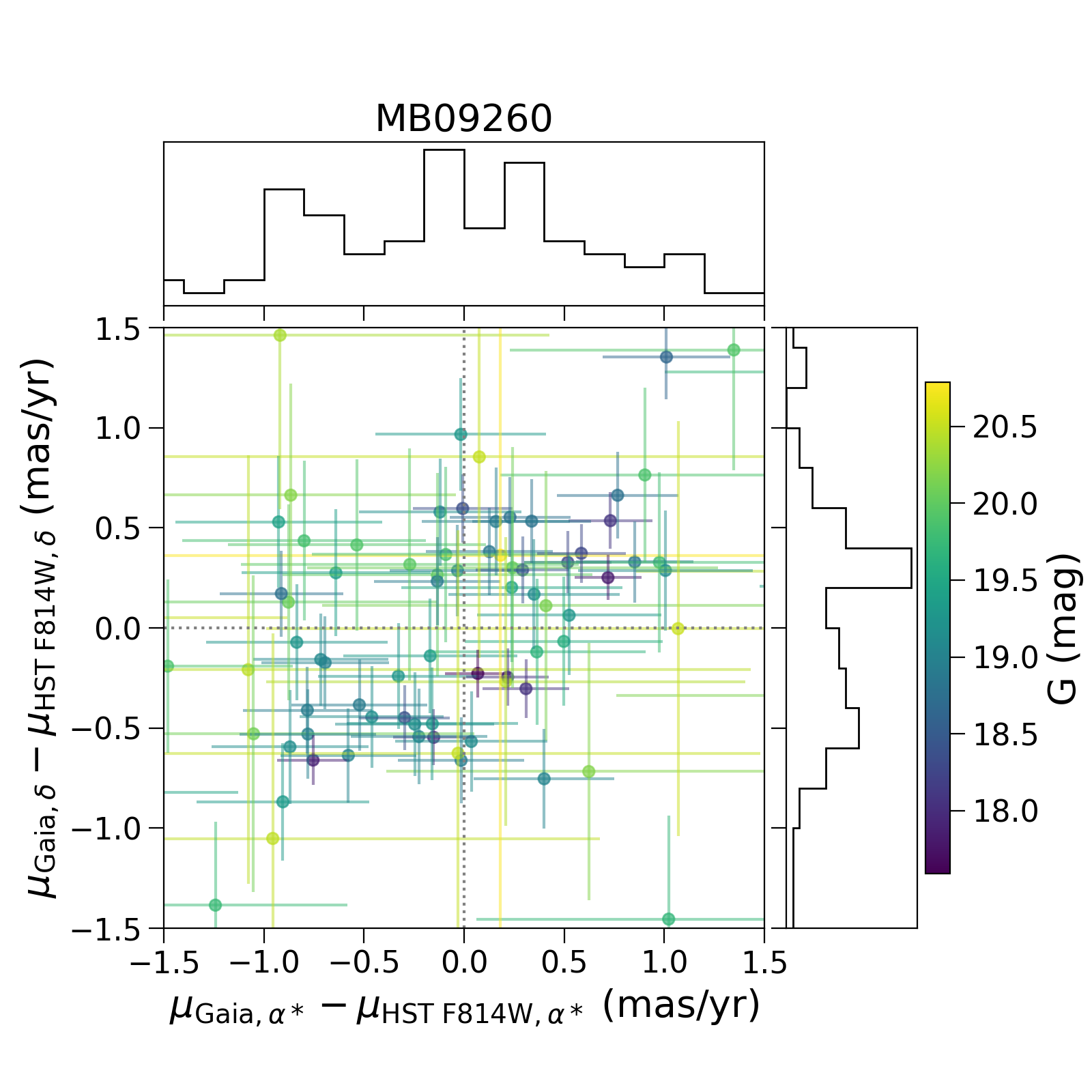} \\
    \includegraphics[width=0.4\linewidth]{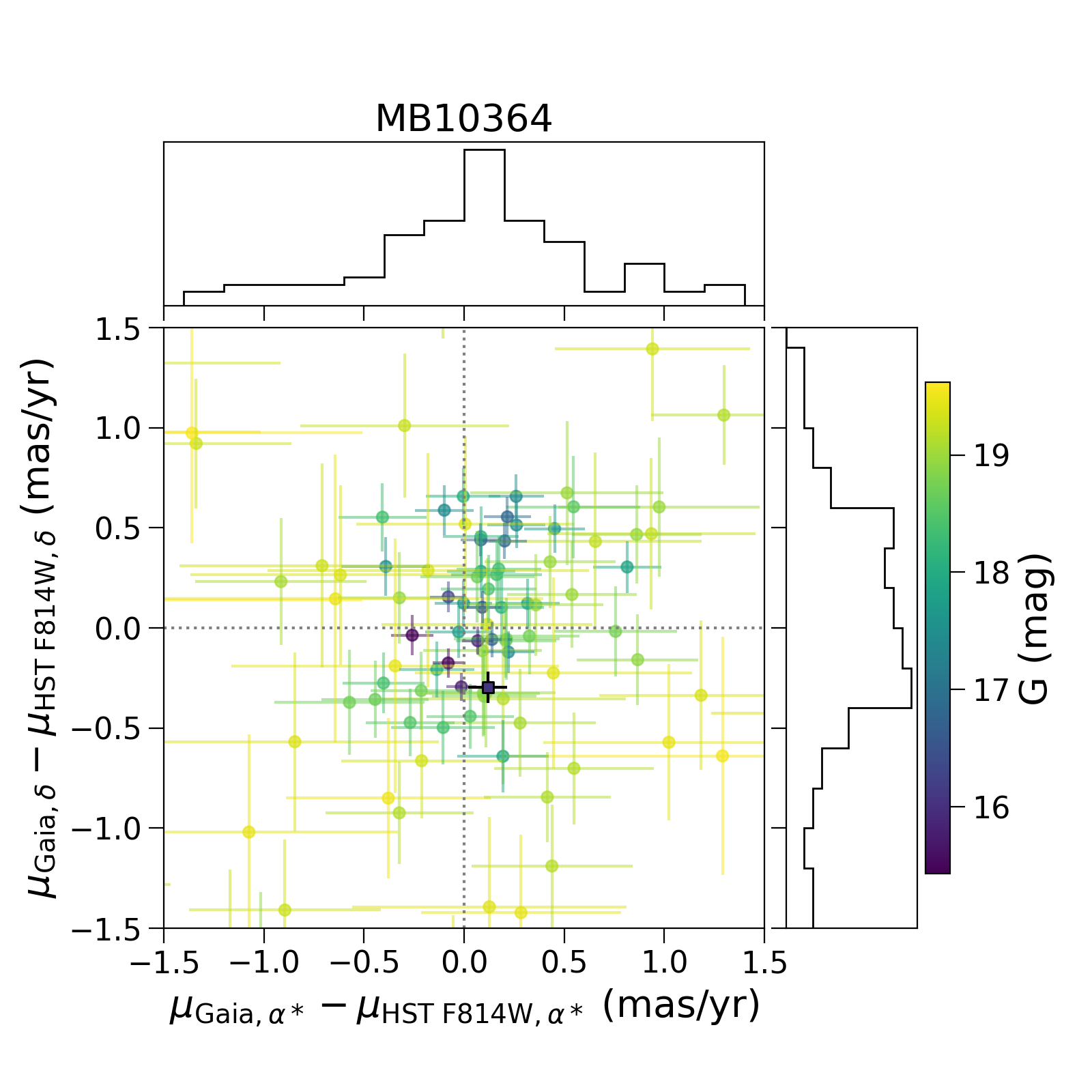}
    \includegraphics[width=0.4\linewidth]{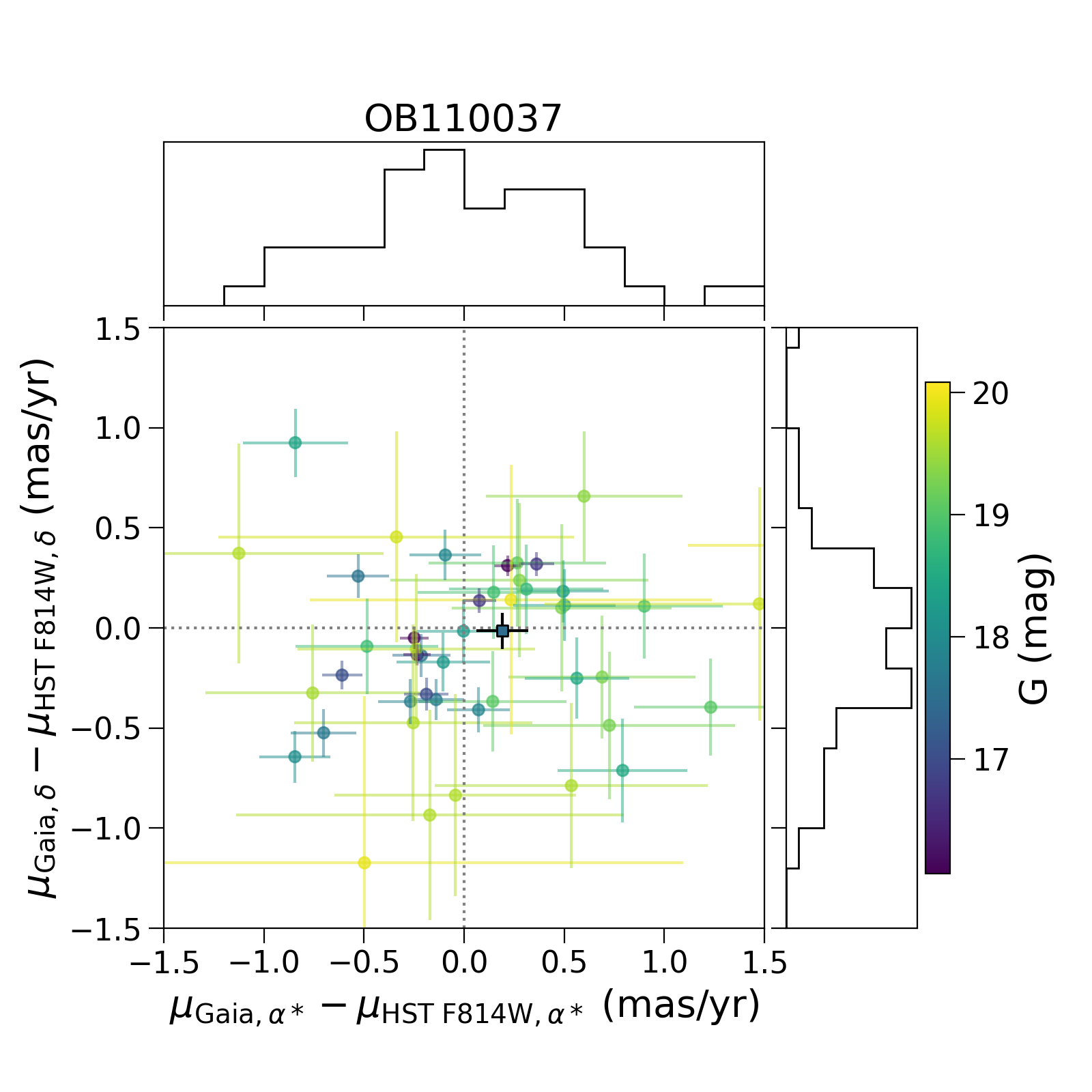} \\
    \includegraphics[width=0.4\linewidth]{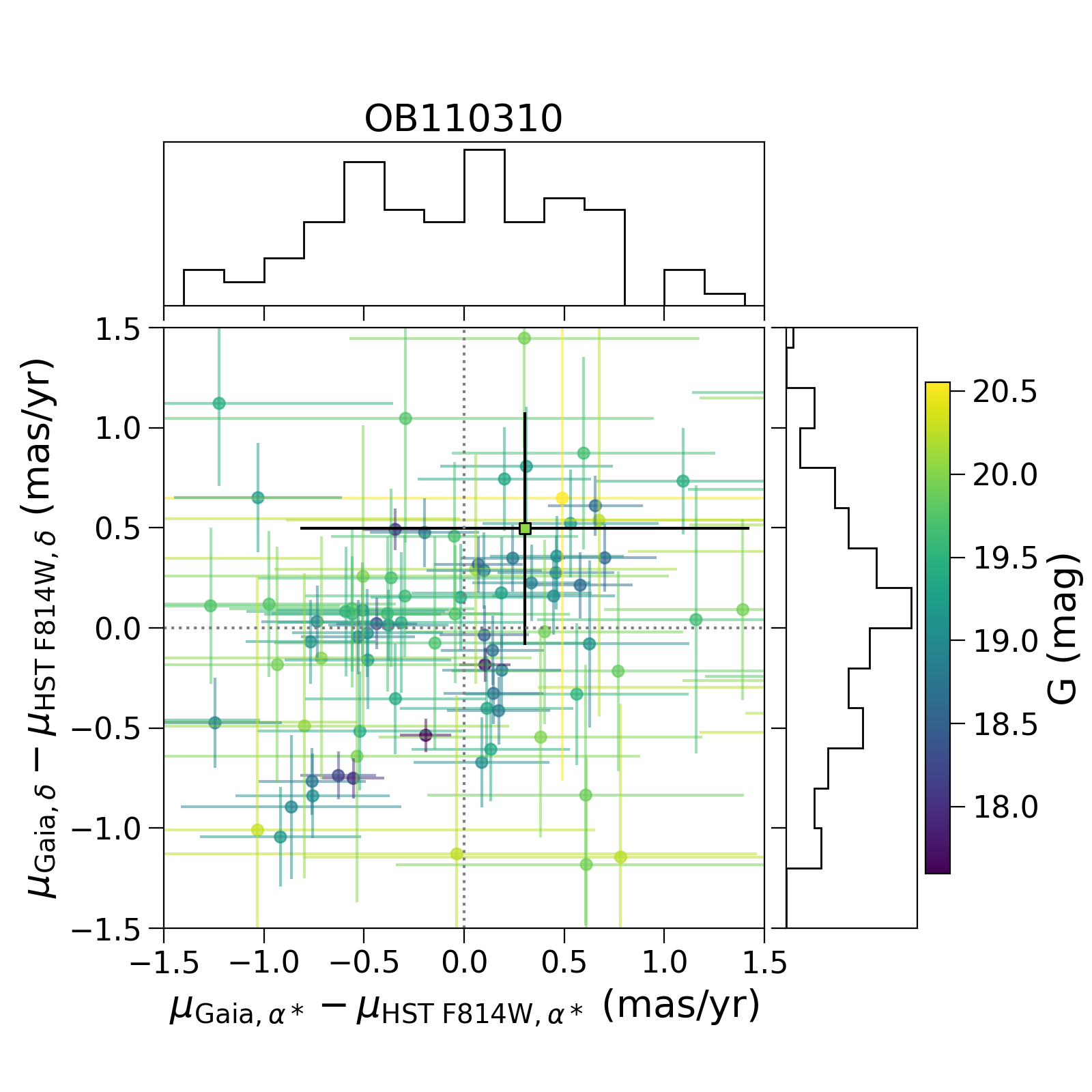}
    \includegraphics[width=0.4\linewidth]{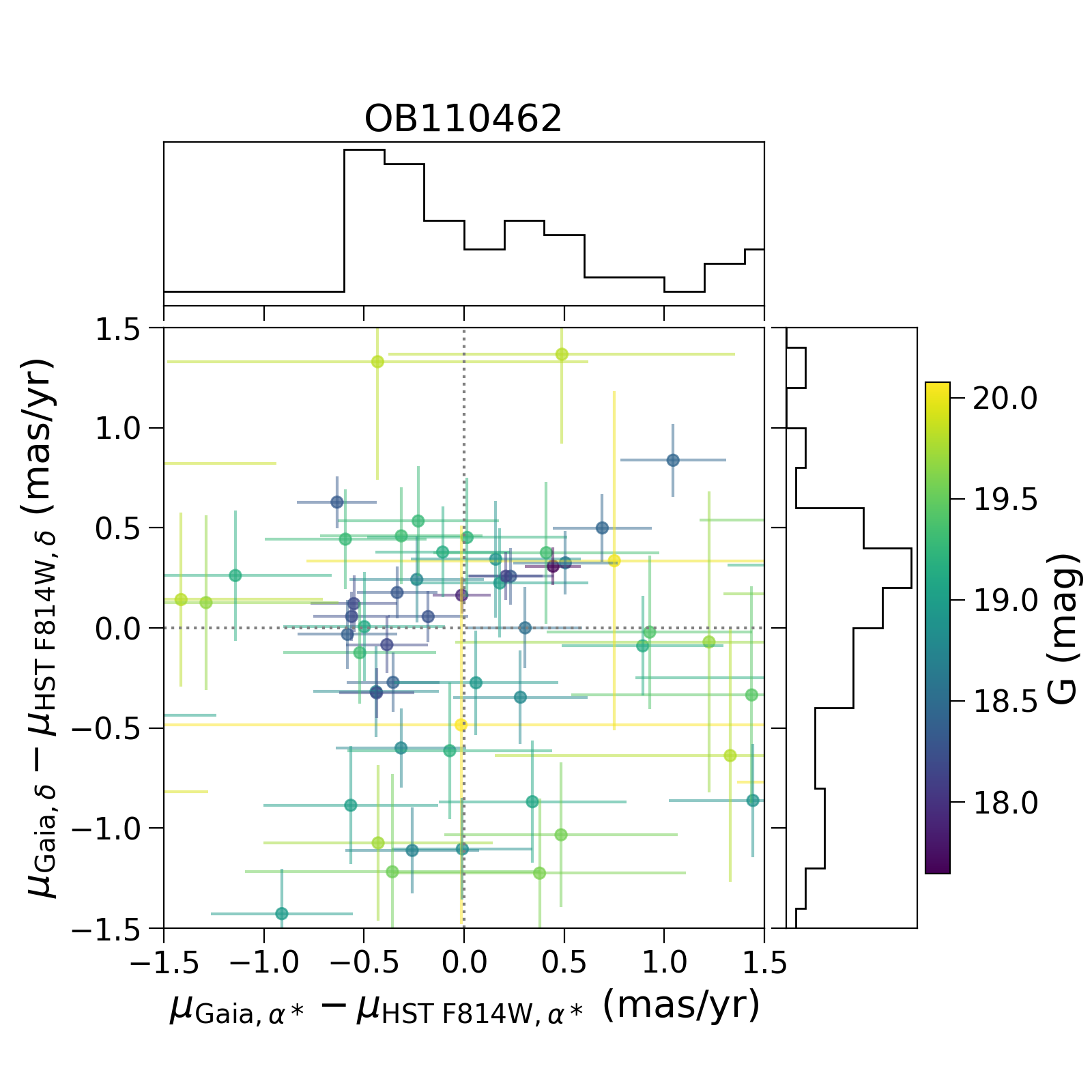}
\caption{Vector point diagram of proper motion differences between cross-matched sources in the Gaia and HST F814W catalogs within $60''$ of the targets, after applying the constant proper motion offset derived in Appendix \ref{sec:Absolute proper motion reference frame} to place the HST catalog into the Gaia reference frame.
The color of the point denotes the cross-matched source's magnitude in Gaia G-band.
MB10364, OB110037, and OB110310 are sources in Gaia; they are shown as the black-outlined squares and error bars.
\label{fig:HST_Gaia_PMs}}
\end{figure*}

The vector point diagram of proper motion differences between cross-matched sources in the Gaia and HST F814W catalogs, after applying the proper motion offset to place the HST catalog into the Gaia reference frame, is shown in Figure \ref{fig:HST_Gaia_PMs}.
In general, the proper motions of bright stars in Gaia are inconsistent with those derived using HST.
For fainter stars, the uncertainties are much larger, so there is more consistency between Gaia and HST; however, there is substantial scatter between the measurements of the two catalogs. 
The discrepancies between the HST and Gaia proper motions could indicate that the uncertainties are underestimated in one or both catalogs, or that there are higher-order distortions between the two reference frames that cannot be captured by a constant offset.
However, the most likely explanation is that the Gaia proper motions are not accurate, as it is clear that the Gaia EDR3 astrometry is unreliable in crowded regions like the Galactic Bulge (see Appendix \ref{app:Gaia diagnostics} and \citet{Rybizki:2022}).

\begin{deluxetable}{lcc}
\tablecaption{HST to Gaia Proper Motion Reference Frame Offset
\label{tab:hst_gaia_offset}}
\tablehead{
    \colhead{Target} &
    \colhead{$\langle \Delta \mu_{\alpha *} \rangle$ (mas/yr)} &
    \colhead{$\langle \Delta \mu_{\delta} \rangle$ (mas/yr)}} 
\startdata
\input{hst_gaia_offset.txt}
\enddata
\tablecomments{
$\Delta \mu_{\alpha *} = \mu_{{\rm Gaia},\alpha *} - \mu_{{\rm HST \; F814W},\alpha *}$
and $\Delta \mu_{\delta} = \mu_{{\rm Gaia},\delta} - \mu_{{\rm HST \; F814W},\delta}$.}
\end{deluxetable}

\section{Astrometric lensing in Gaia \label{app:Astrometric lensing in Gaia}}

As discussed in \S\ref{sec:Comparison to Gaia proper motions and parallaxes}, there are several differences between the proper motion modeling of microlensing targets in this paper as compared to Gaia.
First, we simultaneously model the proper motions of the lens and source with parallax within a microlensing model.
In contrast, Gaia models the proper motion of a single star with parallax.
As the source and lens are not resolved by Gaia during the observations, if the lens is luminous, the proper motion of the source and lens are being treated as a single object.
Depending on the relative source-lens motion and colors, this can produce different inferred proper motions.
Second, the temporal baseline of the data used to calculate the proper motions differ.
The HST time baseline spans 2010-2013 for MB10364 and 2011-2017 for OB110037 and OB110310, in contrast to 2014-2017 in Gaia.
As the astrometric lensing signal is time dependent, temporal baseline differences can also lead to different inferred proper motions.

With regard to the second point, we estimate the amount by which astrometric microlensing could affect the proper motion measurement.
The lens-source separation $\boldsymbol{u}(t)$ (Equation \ref{eq:vec_u(t)}) sets the time evolution of the astrometric shift $\delta_c(t)$ (Equation \ref{eq:delta_c}).
Three terms contribute to $\boldsymbol{u}(t)$: the impact parameter $\boldsymbol{u}_0$, the source-lens proper motion $(t - t_0)\boldsymbol{\hat{\mu}}_{rel}/t_E$, and the microlensing parallax  $\pi_E \boldsymbol{P}(t)$.
As the Earth's orbit around the Sun in nearly circular, $|\boldsymbol{P}(t)| \approx 1$ to within 3\%.
For typical microlensing events like MB10364, OB110037, and OB110310, $|\boldsymbol{u}_0| < 1$ and $\pi_E < 1$.
Thus, long after source-lens closest approach, i.e. $(t - t_0)/t_E \gg 1$, the contribution by the impact parameter and microlensing parallax terms to $\boldsymbol{u}(t)$ are subdominant to that of the source-lens relative proper motion, and the source-lens separation can be approximated 
\begin{equation}
    u(t) \approx \frac{t - t_0}{t_E} \gg 1.
\end{equation}
The astrometric shift for large source-lens separations $u \gg 1$ can be approximated as
\begin{equation}
    \delta_c(t) \approx \frac{\theta_E}{u(t)}.
\end{equation}
Putting these two equations together, we can approximate how much a proper motion measurement based on observations made at times $t_2$ and $t_1$ will change due to astrometric microlensing as
\begin{align}
    \Delta_{PM} &\approx \frac{\delta_c(t_2) - \delta_c(t_1)}{t_2 - t_1} \\
    &\approx \frac{\theta_E}{t_2 - t_1} \Bigg( \frac{t_E}{t_2 - t_0} - \frac{t_E}{t_1 - t_0} \Bigg),
\end{align}
where $(t_2 - t_0)/t_E$, $(t_1 - t_0)/t_E \gg 1$.
For MB10364, OB110037, and OB110310, for $t_1$ and $t_2$ corresponding to the start and end of Gaia EDR3 observations, $\Delta_{PM} \approx 0.003$, 0.02, and 0.01 mas/yr, respectively.
In other words, the effect of lensing on proper motions should be negligible in Gaia.
Thus, if the lens is dark, the proper motions of the source as measured by the HST data and the proper motion of the target reported by Gaia should be the same within the uncertainties.

\section{Gaia diagnostics}
\label{app:Gaia diagnostics}

\begin{deluxetable}{lcccc}
\tabletypesize{\scriptsize}
\tablecaption{Distribution of Gaia parameters towards fields of interest \label{tab:Gaia details}}
\tablehead{
    \colhead{Parameter} &
    \colhead{MB09260} &
    \colhead{MB10364} &
    \colhead{OB110037} &
    \colhead{OB110310}}
\startdata
$N_{\rm stars}$ & 611 & 1157 & 1056 & 884 \\ 
$f_{D > 2}$ & 0.44 & 0.51 & 0.59 & 0.55 \\ 
$f_{\epsilon = 0}$ & 0.81 & 0.77 & 0.85 & 0.86 \\ 
$f_{{\rm multi} = 0}$ & 0.79 & 0.41 & 0.53 & 0.70 \\ 
med(multi$_{>0}$) & 4.50 & 14.00 & 9.00 & 5.00 \\ 
multi$_{>0 [95\%]}$) & 37.55 & 54.00 & 42.30 & 43.00 \\ 
$f_{{\rm odd} = 0}$ & 0.76 & 0.61 & 0.64 & 0.64 \\ 
med(odd$_{>0}$) & 25.50 & 30.00 & 26.00 & 27.50 \\ 
odd$_{>0 [95\%]}$) & 83.55 & 78.20 & 80.00 & 78.70 \\ 
$\langle \log_{10}(\rm{amp}) \rangle$ & -1.24 & -0.89 & -1.03 & -1.13 \\ 
$\sigma_{\log_{10}(\rm{amp})}$ & 0.45 & 0.39 & 0.58 & 0.44 \\ 
$\langle \log_{10}(\rm{ruwe}) \rangle$ & 0.04 & 0.12 & 0.08 & 0.06 \\ 
$\sigma_{\log_{10}(\rm{ruwe})}$ & 0.10 & 0.16 & 0.11 & 0.11 \\ 
$\langle \log_{10}(C) \rangle$ & 0.23 & 0.22 & 0.25 & 0.26 \\ 
$\sigma_{\log_{10}(C)}$ & 0.12 & 0.16 & 0.15 & 0.15
\enddata
\tablecomments{$N_{\rm stars}$: Number of Gaia sources within 1 deg$^2$ of the target. \\
$f_{D > 2}$: Fraction of sources with \texttt{astrometric\_excess\_noise\_significance}$ > 2$. \\
$f_{\epsilon = 0}$: Fraction of sources with \texttt{astrometric\_excess\_noise}$ > 2$. \\
$f_{{\rm multi} = 0}$: Fraction of stars with \texttt{ipd\_frac\_multi\_peak} = 0. \\
med(multi$_{>0}$): Median of nonzero \texttt{ipd\_frac\_multi\_peak} values. \\
multi$_{>0 [95\%]}$): 95th quantile of nonzero \texttt{ipd\_frac\_multi\_peak} values. \\
$f_{{\rm odd} = 0}$: Fraction of stars with \texttt{ipd\_frac\_odd\_win}= 0. \\
med(odd$_{>0}$): Median of nonzero \texttt{ipd\_frac\_odd\_win}values. \\
odd$_{>0 [95\%]}$): 95th quantile of nonzero \texttt{ipd\_frac\_odd\_win} values. \\
$\langle \log_{10} (\rm{amp}) \rangle$: Mean of $\log_{10}$(\texttt{ipd\_gof\_harmonic\_amplitude}). \\
$\sigma_{\log_{10}(\rm{amp})}$: Standard deviation of $\log_{10}$(\texttt{ipd\_gof\_harmonic\_amplitude}). \\
$\langle \log_{10} (\rm{ruwe}) \rangle$: Mean of $\log_{10}$(ruwe). \\
$\sigma_{\log_{10}(\rm{ruwe})}$: Standard deviation of $\log_{10}$(ruwe). \\
$\langle \log_{10}(C) \rangle$ : Mean of $\log_{10}(\texttt{phot\_bp\_rp\_excess\_factor})$. \\
$\sigma_{\log_{10}(C)}$: Standard deviation of $\log_{10}(\texttt{phot\_bp\_rp\_excess\_factor})$. \\
}
\end{deluxetable}

We would ideally like to incorporate information from Gaia EDR3 into our target analysis (e.g. using the reported parallaxes and proper motions to inform the prior on the source parallax in the fit, or to compare to the posteriors as a cross-check to validate the results).
However, it is known that there are many as-of-yet-unresolved systematics in the Gaia EDR3 astrometry, especially toward the Bulge, and extra care must be taken to verify if a proper motion or parallax for a particular Gaia source is reliable.
Gaia EDR3 is much better than Gaia DR2 in terms of photometry, but the astrometry still has issues that need to be worked out \citep{Fabricius:2021}.
We investigate several different metrics for the Gaia solutions for our targets to determine the reliability of the reported parallax and proper motions.
A brief summary of the meaning of relevant Gaia statistics is presented in Table \ref{tab:gaia}; we refer the reader to the Gaia Early Data Release 3 documentation \citep{vanLeeuwen:2021} for details.

\subsection{Solution type}
Sources in Gaia have varying amounts of information available, described by the \texttt{astrometric\_params\_solved} parameter.
A value of 3 signifies a 2-parameter solution (position), a value of 31 signifies a 5-parameter solution (position, parallax, proper motion), and a value of 95 signifies a 6-parameter solution (position, parallax, proper motion, astrometrically estimated effective wavenumber).
5-parameter solutions are generally the most accurate, followed by 6-parameter, then 2-parameter solutions.
6-parameter solutions are worse than 5-parameter solutions because it means an assumption had to be made about the source color.
This usually indicates an issue with determining the properties of the source, which reduces the accuracy of the solution.
This can happen in very crowded regions, like the Bulge.
MB09260 has a 2-parameter solution, MB10364 and OB110310 have 6-parameter solutions, and only OB110037 has a 5-parameter solution.
Based on the high-resolution HST images (Figure \ref{fig:images}) this is not surprising, as OB110037 is the brightest object in its vicinity, while the other targets are near comparably bright stars.

\subsection{Image parameter determination parameters}
The Gaia EDR3 solution assumes each source is a single star.
Gaia's image parameter determination parameters (IPD) can be used to determine whether this assumption is valid.
Different types of binaries can be identified through large of IPD values: \texttt{ipd\_gof\_harmonic\_amplitude} (indicates partially resolved binaries, asymmetric images),
\texttt{ipd\_frac\_multi\_peak} (indicates resolved, close binaries),
\texttt{ipd\_frac\_odd\_win} (indicates another bright source, or observation window conflicts for wide pairs) \citep{Lindegren:2021a}.
Given that the targets all have lightcurves well described by point lenses (although see \S \ref{sec:OB110037} regarding indications of binarity in OB110037), we do not expect any of these parameters to be unusual for our targets; unusual values likely indicates problems with the astrometric solution itself due to crowding or contamination.

The \texttt{ipd\_gof\_harmonic\_amplitude} values for MB09260, MB10364, OB110037, and OB11030 (0.089, 0.064, 0.036, 0.042, respectively) are somewhat large compared to the median values reported in Tables 4-6 of \cite{Lindegren:2021a}.
However, this is to be expected toward the Bulge.
Compared to other sources within 1 deg$^2$ (Table \ref{tab:Gaia details}) of these targets, the values are not unusual.\footnote{Note the values reported in Table \ref{tab:Gaia details} are the logarithms of \texttt{ipd\_gof\_harmonic\_amplitude}.
We choose to report the logarithms as they are approximately normally distributed.}

MB10364 has \texttt{ipd\_frac\_multi\_peak} = 15.
The high-resolution HST images do not indicate MB10364 is a resolved binary.
The last HST image is from 2014, and Gaia EDR3 observations span 2014-2017. 
A possibility is that the lens and source separation after 2014 was large enough to resolve with Gaia EDR3. 
However, this is ruled out by considering the lens-source separation criteria (Table \ref{tab:resolve_1sigma}) and that Gaia's resolution is worse than HST's.
The most likely explanation is confusion in Gaia due to stars near MB10364 since the crowding within 1 arcsec of MB10364 is high.
The other targets have \texttt{ipd\_frac\_multi\_peak} = 0.

MB09260 has \texttt{ipd\_frac\_odd\_win} = 18 and OB110310 has \texttt{ipd\_frac\_odd\_win} = 55.
Similar to the case of MB10364 in the previous paragraph, this is likely due to bright stars within 1 arcsec of these targets that led to confusion in Gaia.
The other targets have \texttt{ipd\_frac\_odd\_win} = 0.

\subsection{Astrometric goodness-of-fit statistics}
Gaia also presents several relevant goodness-of-fit (GOF) statistics from the astrometric fit:
the renormalized unit weight error (RUWE; ideally 1),
the extra noise required per observation to explain the residual in the astrometric fit of the source (\texttt{astrometric\_excess\_noise}; ideally 0),
and the significance of this source noise (\texttt{astrometric\_excess\_noise\_sig}; insignificant if $< 2$).

MB09260 does not have a RUWE because it is not calculated for 2-parameter solutions.
OB110037 and OB110310 both have RUWE $\sim 1$ (0.971 and 0.981, respectively).
MB10364 has a large RUWE (1.388), although not unusual compared to the sources nearby (Table \ref{tab:Gaia details}).\footnote{Note the values reported in Table \ref{tab:Gaia details} are the logarithms of RUWE.
We choose to report the logarithms as they are approximately normally distributed.}

MB09260 and MB10364 both have significant astrometric excess noise (1.241 mas and 0.406 mas, respectively).
OB110037 does not have astrometric excess noise and the noise for OB110310 is not significant.

\begin{figure*}[h!]
\centering
    \includegraphics[width=0.49\linewidth]{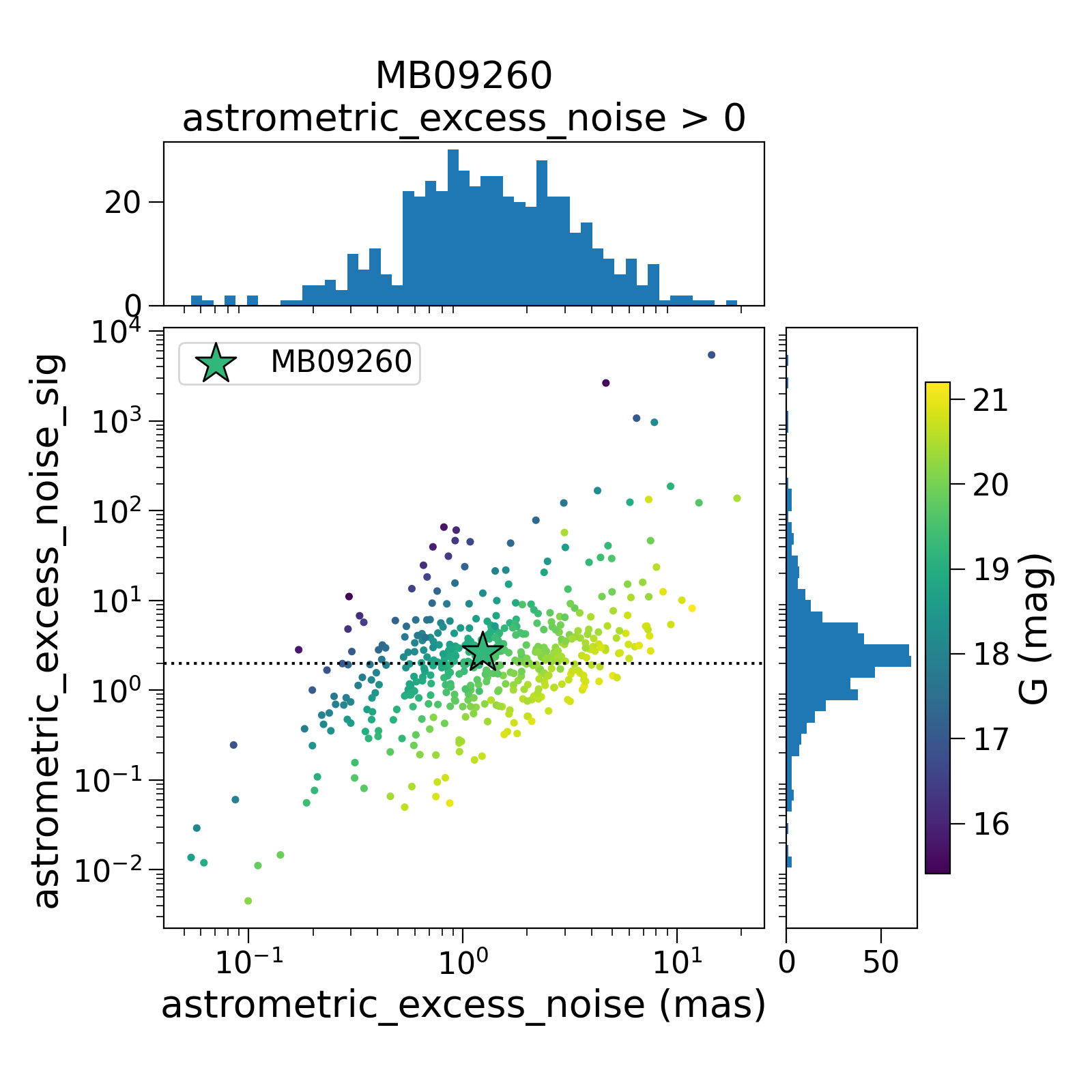}
    \includegraphics[width=0.49\linewidth]{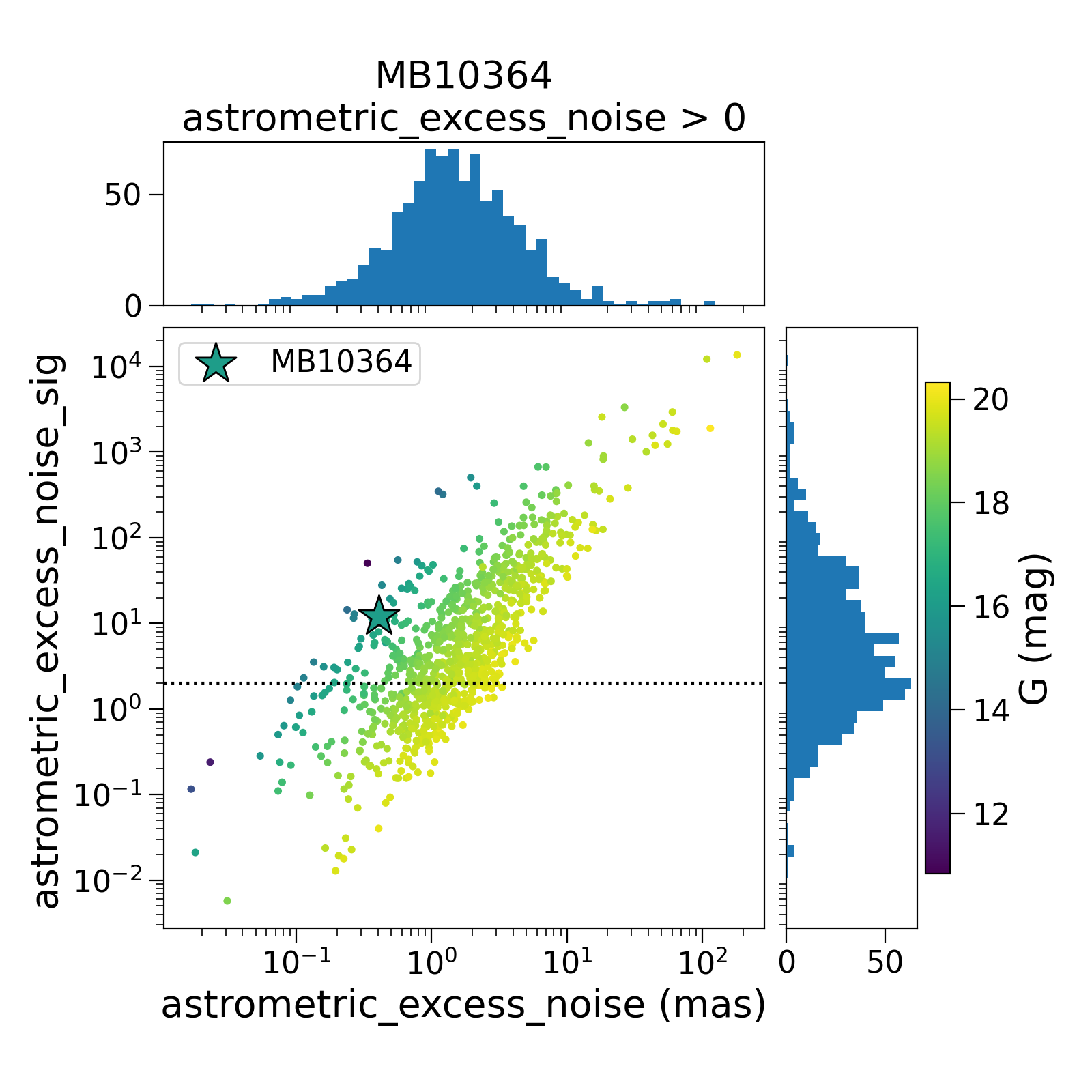}
    \includegraphics[width=0.49\linewidth]{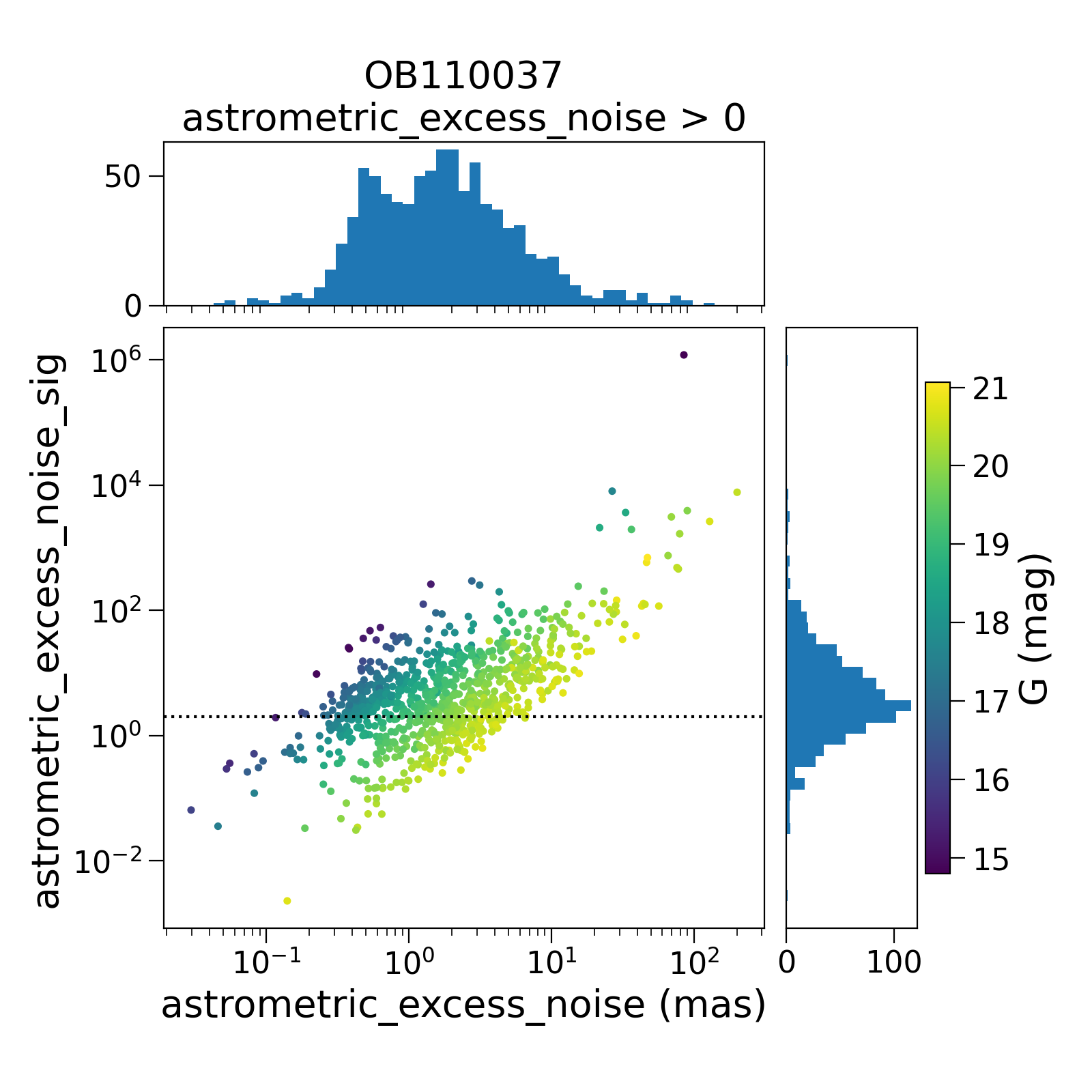}
    \includegraphics[width=0.49\linewidth]{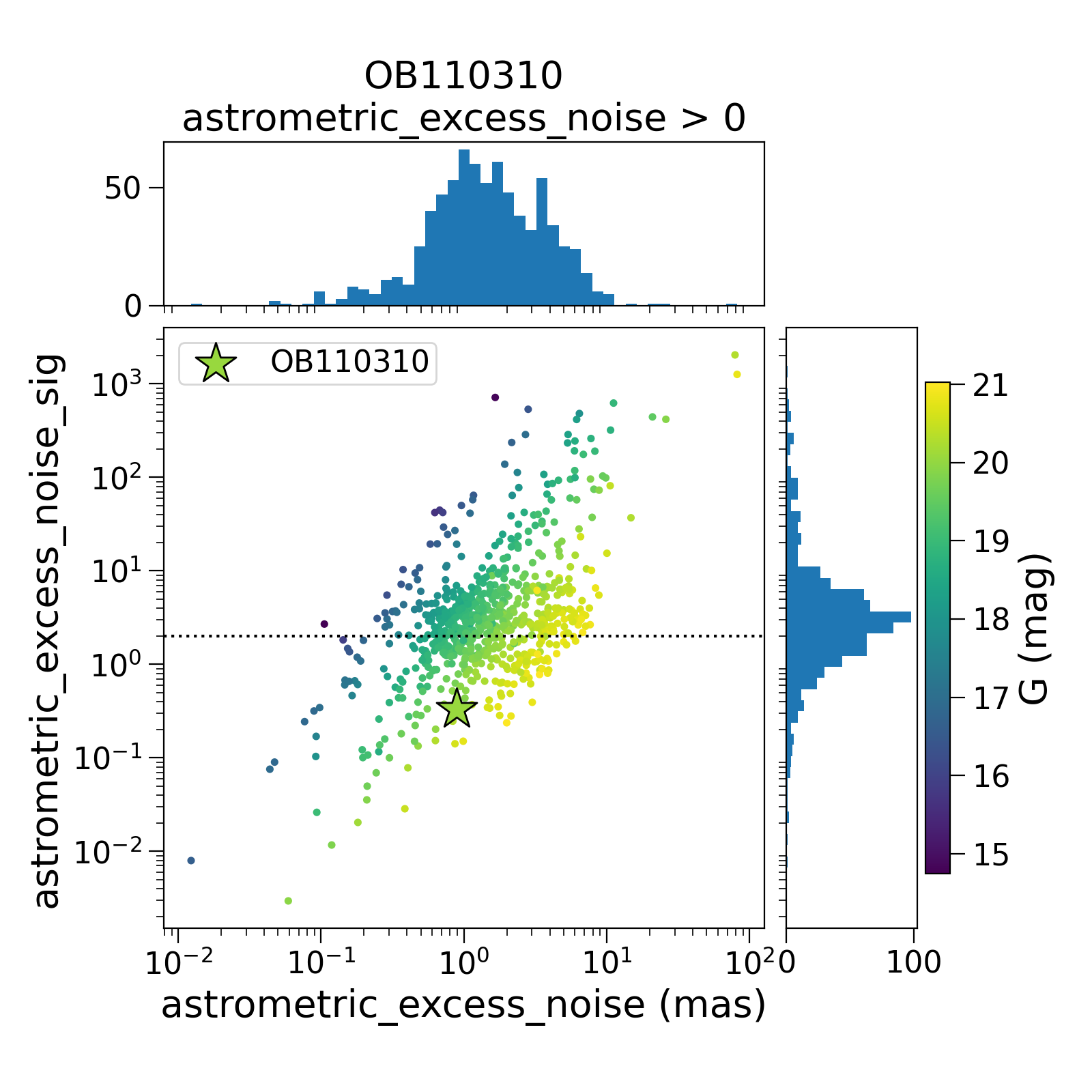}
\caption{Astrometric excess noise vs. astrometric excess noise significance as a function of $G$ magnitude, for sources with non-zero astrometric excess noise in the 1 deg$^2$ fields surrounding the microlensing targets found in Gaia.
The targets are marked as stars; OB110037 is not marked as it has astrometric excess noise 0.
The dotted horizontal line denotes \texttt{astrometric\_excess\_noise\_sig~$=2$}; values $<2$ are not significant.
\label{fig:Gaia diagnostics}}
\end{figure*}

Table \ref{tab:Gaia details} lists the distributions of $D=$~astrometric excess noise significance for the stars in Gaia within 1 deg$^2$ of our targets.
Gaia documentation notes that $D=0$ for roughly half the sources and $D>2$ for  a few percent of sources with well-behaved astrometric solutions \citep{vanLeeuwen:2021}.
Near MB09260, 84\% sources have $D > 0$ and 44\% have $D > 2$.
Near MB10364, 79\% sources have $D> 0$ and 51\% with have $D > 2$.
Near OB110037, 86\% sources have $D > 0$ and 59\% have $D > 2$.
Near OB110310, 87\% sources have $D > 0$ and 54\% have $D > 2$.
This indicates the astrometric solution is not well-behaved.
This is not surprising, as there still exist many systematics in the astrometry, especially toward the Bulge.
Figure \ref{fig:Gaia diagnostics} shows astrometric excess noise significance plotted against astrometric excess noise for systems where astrometric excess noise is nonzero.

\subsection{Color excess}

\cite{Fabricius:2021} use $C$ = \texttt{phot\_bp\_rp\_excess\_factor} as a proxy for contamination due to crowding.
Stars with large excess ($C > 5$) tend to have underestimated proper motion uncertainties by a factor of $\sim 1.7$.
All of our targets have $C < 5$; less than 1\% of stars within 1 deg$^2$ of the targets have $C > 5$.
This metric does not seem to capture crowding toward the Bulge.

\subsection{Summary}

Compared to the high resolution images from HST, the various metrics reported by Gaia make sense.
\begin{itemize}
    \item OB110037 is the brightest source in its vicinity, and hence has a good astrometric solution. 
    Its astrometric solution (parallax and proper motion) are likely to be reliable.
    The HST source proper motion agrees with the Gaia proper motion measurement.
    \item MB10364 is bright but is in a crowded field.
    OB110310 is somewhat isolated but is faint. 
    Their astrometric solutions are likely to be unreliable.
    \item MB09260 does not have enough visibility periods to generate an astrometric solution (which requires at least 9 visibility periods, while MB09260 only has 8).
\end{itemize} 

In conclusion, although Gaia is a dedicated astrometric mission, it is not optimized for the crowded and extincted Bulge, and the astrometric parameters are likely to be untrustworthy there \citep{Fabricius:2021, Rybizki:2022}.
Although Gaia EDR3 is also an improvement over DR2, those improvements are in the photometry and not the astrometry.
Placing too much weight on the Bulge astrometry in analyses (especially the uncertainties) should be done with caution.
To make use of Gaia data for these targets, we will need to wait for future data releases with improved astrometry in crowded fields as well as per-epoch astrometry.

\section{Gaussian Process}
\label{sec:Gaussian Process}

The photometric microlensing survey data contains uncharacterized systematics in the lightcurves, which may be due to unaccounted correlated noise from astrophysical processes or data acquisition and reduction.
Correlated noise can bias the results of parameter estimation.
\cite{Golovich:2022} fit a set of publicly released OGLE-III and OGLE-IV microlensing lightcurves using gaussian processes (GPs) to account for correlated noise; we follow that approach here.
We use the \texttt{celerite} package \citep{Foreman-Mackey:2017} to implement GPs in our microlensing model.

For a thorough reference about GPs and their application to machine learning and inference, the reader may consult sources such as \citet{Rasmussen:2006}.
In short, a GP is composed of two parts: the noise (the stochastic component) and the mean function (the deterministic component).
The properties of the stochastic component are governed by the covariance matrix, also known as the kernel, of the GP.
The notation
\begin{equation}
    y \sim \mathcal{GP}(\mu_{\btheta}(t), K_{\balpha}(t, t'))
\end{equation}
indicates a GP with mean function $\mu$ with parameters $\btheta$ and kernel $K$ with parameters $\balpha$.

When trying to infer some set of parameters $\btheta$ from data 
\begin{equation}
    \by = (y_1(t_1) \quad ... \quad y_N(t_N))^T
\end{equation}
the Gaussian log-likelihood is
\begin{equation}
    \label{eq:gaussian_likelihood_gp}
    \mathrm{log} \mathcal{L} (\by|\btheta, \balpha) = -\frac{1}{2} \br_{\btheta}^T K_{\balpha}^{-1} \br_{\btheta} - \frac{1}{2} \log |K_{\balpha}| - \frac{N}{2} \log (2\pi),
\end{equation}
where $N$ is the number of data points, and $\br_\theta$ is the residual vector
\begin{equation}
    \br_{\btheta} = (y_1 - \mu_{\btheta}(t_1) \quad ... \quad y_N - \mu_{\btheta}(t_N))^T.
\end{equation}
In the residual vector, $\mu_{\btheta}$ is the mean model, which in our case is the microlensing model.
The kernel or covariance matrix $K_{\balpha}$ describes the correlated errors.
If the errors are not correlated, $K_{\balpha}$ is diagonal and the likelihood reduces to the familiar form
\begin{equation}
\begin{aligned}
    \label{eq:gaussian_likelihood_indep}
    \mathrm{log} \mathcal{L} (\by|\btheta) &= -\frac{1}{2} \sum_i \Bigg( \mathrm{log}[2 \pi \sigma_i^2] + \frac{(y_i - \mu_{\btheta}(t_i))^2}{\sigma_i^2} \Bigg) \\
    &= -\frac{1}{2} \sum_i  \mathrm{log}[2 \pi \sigma_i^2] - \frac{1}{2} \chi^2 ,
\end{aligned}
\end{equation}
where $\sigma_i$ is the uncertainty on data point $y_i$ and $\chi^2$ is the ``goodness of fit".

Following \citet{Golovich:2022}, a damped driven simple harmonic oscillator (DDSHO) kernel $\kappa_{DDSHO}$ added to a Mat\'{e}rn-3/2 kernel $\kappa_{M3/2}$ is used to model the correlated noise in the photometric microlensing survey lightcurves.
Both these kernels are stationary, as they are a function of the differences of the times only:
\begin{equation}
    \tau_{ij} = |t_i - t_j|.
\end{equation}
The kernel is given by
\begin{equation}
    K_{\balpha}(\tau_{ij}) = \kappa_{DDSHO}(\tau_{ij}) +  \kappa_{M3/2}(\tau_{ij}) + \delta_{ij} \sigma_i^2.
\end{equation}
Qualitatively, the DDSHO kernel models smooth variations, while the Mat\'{e}rn-3/2 captures more irregular variations.

The DDSHO kernel is given by
\begin{equation}
    \kappa_{DDSHO}(\tau_{ij}) = S_0 \omega_0 e^{-\omega_0 \tau_{ij}/\sqrt{2}} \cos{\Bigg(\frac{\omega_0 \tau_{ij}}{\sqrt{2}} - \frac{\pi}{4}\Bigg)}
\end{equation}
where $S_0$ controls the amplitude of the deviation from the mean model and $\omega_0$ controls the variation frequency.
This kernel has been used in asteroseismic modeling (\citet{Li:2019} and references therein).

The Mat\'{e}rn-3/2 kernel is given by
\begin{equation}
    \kappa_{M3/2}(\tau_{ij}) = \sigma^2 e^{-\sqrt{3} \tau_{ij}/\rho} \Bigg(1 + \frac{\sqrt{3}\tau_{ij}}{\rho} \Bigg)
\end{equation}
where $\sigma$ determines the amount of deviation allowed from the mean model, and $\rho$ is the characteristic coherence scale.
The Mat\'{e}rn-3/2 kernel has been used to model correlated noise in the lightcurves of transiting and eclipsing exoplanets \citep{Gibson:2013, Evans:2015}, and in particular is appropriate for modeling non-smooth behaviors \citep{Gilbertson:2020}.
For numerical reasons \citep[see \S 4 of][]{Foreman-Mackey:2017}, the Mat\'{e}rn-3/2 kernel is approximated
\begin{equation}
    \kappa_{M3/2}(\tau_{ij}) = \sigma^2 \Bigg[ \Big(1 + \frac{1}{\epsilon} \Big) 
                                                e^{-\sqrt{3} (1 - \epsilon) \tau_{ij}/\rho} 
                                                \Big(1 - \frac{1}{\epsilon} \Big) 
                                                e^{-\sqrt{3} (1 + \epsilon) \tau_{ij}/\rho} \Bigg].
\end{equation}
In the limit $\epsilon \rightarrow 0$, this is exactly the Mat\'{e}rn-3/2 kernel.
We implement the approximation with $\epsilon = 0.01$.

The ground-based OGLE and MOA photometry are fit using the Gaussian likelihood with a full covariance matrix (Equation \ref{eq:gaussian_likelihood_gp}).
The HST photometry and astrometry are fit using the Gaussian likelihood assuming a diagonal covariance matrix (Equation \ref{eq:gaussian_likelihood_indep}).

Note for MB10364, instead of fitting the MOA lightcurve using a GP, we instead fit an additive error on the ground-based photometry. During nested sampling, the GP showed some numerical instability.

\section{Priors}
\label{sec:Priors}
The distributions for the priors $\pi$ are described in this sections.
$\mathcal{N}(\mu, \sigma)$ denotes a normal distribution with mean $\mu$ and standard deviation $\sigma$. 
$\mathcal{N}_T(\mu, \sigma, l_\sigma, u_\sigma)$ denotes a normal distribution with a low end truncation at $\mu + \sigma l_\sigma$ and a high end truncation at $\mu + \sigma u_\sigma$.
$\mathcal{U}(a,b)$ denotes a uniform distribution from $a$ to $b$. 
$\Gamma^{-1} (\alpha, \beta)$ is the inverse gamma distribution
\begin{equation}
    \Gamma^{-1}(x; \alpha, \beta) = \frac{\beta^\alpha}{\Gamma(\alpha)} x^{-\alpha-1} \exp[-\beta/x].
\end{equation}
The prior distributions for each target are summarized in Table \ref{tab:priors}.

\begin{deluxetable*}{lcccccc}
\tabletypesize{\scriptsize}
\tablecaption{Priors \label{tab:priors}}
\tablehead{
    \colhead{Parameter} &
    \colhead{MB09260} &
    \colhead{MB10364} &
    \colhead{OB110037} &
    \colhead{OB110310} &
    \colhead{OB110462 DW} &
    \colhead{OB110462 EW}}
\startdata
\input{priors.txt}
\enddata
\tablecomments{
For definitions of the different variables, see Appendix \ref{sec:Priors}. 
There are two fits for OB110462 depending on the likelihood used, ``equal weighting" (OB110462 EW) or ``default weighting" (OB110462 DW).
See \S\ref{sec:Likelihood weighting} for details.}
\end{deluxetable*}

\subsection{Photometry priors}

The five microlensing parameters in a PSPL with parallax fit are $t_0$, $u_0$, $t_E$, $\pi_{E,E}$, and $\pi_{E,N}$.

The prior on $t_0$ is a normal distribution centered on the time of peak magnification in the geocentric frame, with a spread of 75 days
\begin{equation}
    \pi (t_0) \sim \mathcal{N}(t_{peak, \oplus}, 75 \textrm{ days}).  
\end{equation}
Note that the time at peak magnification in the heliocentric frame $t_0$ is not necessarily the same as in the geocentric frame $t_{peak, \oplus}$, hence the large amount of spread in the prior. 

The prior on $u_0$ is a Gaussian with mean 0 and standard deviation 0.5
\begin{equation}
   \pi (u_0) \sim \mathcal{N}(0, 0.5)
\end{equation}
which takes into account that events with smaller $|u_0|$ are more likely to be detected, and that events with $|u_0| > 1.5$ are not robustly detectable with current ground-based surveys.\footnote{$|u_0| = 1.5$ corresponds to a brightening of no more than around 0.1 mag.
When selecting microlensing events, those with a brightening less than 0.1 mag are generally excluded in survey samples to prevent contamination from low-amplitude variables (e.g. \cite{Mroz:2019}).}

The prior on $t_E$ is a Gaussian centered at 200 days with a large spread of $\sigma = 100$ days. The distribution is truncated at $-1.8\sigma$ and $3\sigma$ (20 and 500 days, respectively):
\begin{equation}
   \pi (t_E) \sim \mathcal{N}_T(200, 50, -1.8, 10) \textrm{ days}.
\end{equation}

The priors on the microlensing parallax are estimated from the population of bulge microlensing events from the \texttt{PopSyCLE} simulation:
\begin{align}
   \pi (\pi_{E,E}) &\sim \mathcal{N}(-0.02, 0.12) \\
   \pi (\pi_{E,N}) &\sim \mathcal{N}(-0.03, 0.13).
\end{align}

For each dataset filter, $b_{SFF}$ and $m_{base}$ are also fit.
For the ground-based photometry, we use a prior
\begin{equation}
    b_{SFF,ground} \sim \mathcal{U}(0, 1.1)
\end{equation}
where the negative blend flux implied by $b_{SFF} > 1$ allows for some extra noise such as imperfect background subtraction.
Similarly for the HST astrometry, we use a uniform prior on $b_{SFF}$
\begin{equation}
    b_{SFF,HST} \sim \mathcal{U}(0, 1.05).
\end{equation}

$m_{base}$ is a normal distribution 
\begin{equation}
    m_{base} \sim \mathcal{N}(\overline{m}_{base}, \sigma_{\overline{m}_{base}})
\end{equation}
where $\overline{m}_{base}$ is the average magnitude during the un-magnified seasons, weighted by the measurement uncertainties, and $\sigma_{\overline{m}_{base}}$ is 0.1 for OGLE, 0.2 for MOA, and 0.05 for HST.

\subsection{Gaussian Process hyperparameter priors}

The ground-based photometry includes correlated noise we fit.
We follow a very similar parametrization to \citet{Golovich:2022} for the GP priors.
The main difference is that fit in magnitude space instead of flux space, and so our priors are also in magnitudes instead of fluxes.
 
For $\sigma$, we use the prior 
\begin{equation}
    \log (\sigma/ \textrm{mag}) \sim \mathcal{N}(0,5) 
\end{equation}
which allows a wide range of lightcurve amplitude variability.

For $\rho$, we use the prior
\begin{equation}
    \rho \sim \Gamma^{-1}(a,b)
\end{equation}
where 
$a$ and $b$ are the constants that satisfy the relation
\begin{align}
    0.01 &= \int_0^{\textrm{med}(\Delta t)} \Gamma^{-1}(x; a,b) \; dx \\
    0.01 &= 1 - \int_0^{\Delta T} \Gamma^{-1}(x; a,b) \; dx
\end{align}
where $\textrm{med}(\Delta t)$ is the median duration between observations and $\Delta T$ is the duration of full dataset.
This helps suppress values at extremely short or long timescales that might lead to ill-behaved models.\footnote{\scriptsize See \href{https://betanalpha.github.io/assets/case\_studies/gaussian\_processes.html} {betanalpha.github.io/assets/case\_studies/gaussian\_processes}}

For $S_0$ and $\omega_0$ we use the priors 
\begin{align}
    \log S_0 \omega_0^4 &\sim \mathcal{N}(\textrm{med}(\sigma_m^2), 5) \\
    \log \omega_0 &\sim \mathcal{N}(0,5).
\end{align}

\subsection{Astrometry priors}

The prior on the Einstein radius $\theta_E$ is a lognormal distribution estimated from \texttt{PopSyCLE} for events with $t_E > 50$ or $t_E > 120$ days as
\begin{align}
    \pi (\log_{10} (\theta_E/\textrm{mas})) &\sim \mathcal{N}(-0.2, 0.3) \\
    \pi (\log_{10} (\theta_E/\textrm{mas})) &\sim \mathcal{N}(0, 0.5),
\end{align}
respectively.
We use the prior from $t_E > 120$ days for OB110462 and $t_E > 50$ days for the other 4 targets.

The prior on the source parallax $\pi_S$ is estimated from the population of bulge microlensing events from the \texttt{PopSyCLE} simulation
\begin{equation}
    \pi (\pi_S) \sim \mathcal{N}_T(0.1126, 0.0213, -2.9390, 90.0) \textrm{ mas}
\end{equation}
which corresponds to source distances ranging from 0.5 to 20 kpc.

The prior on the source proper motion $\mu_{S,E}$ and $\mu_{S,N}$ are uniform distributions
\begin{align}
   \pi (\mu_{S,E}) &\sim \mathcal{U}(\overline{\mu}_{S,E} - f \sigma_{\overline{\mu}_{S,E}}, \overline{\mu}_{S,E} + f \sigma_{\overline{\mu}_{S,E}}) \\
   \pi (\mu_{S,N}) &\sim \mathcal{U}(\overline{\mu}_{S,N} - f \sigma_{\overline{\mu}_{S,N}}, \overline{\mu}_{S,N} + f \sigma_{\overline{\mu}_{S,N}})
\end{align}
where $\overline{\mu}_{S,E}$, $\overline{\mu}_{S,N}$ are the proper motions inferred from assuming straight-line motion (no parallax) from the F814W data,  $\sigma_{\overline{\mu}_{S,E}}$, $\sigma_{\overline{\mu}_{S,N}}$ are the uncertainties to that fit, and $f$ is an inflation factor.
To allow a wide range of proper motions we use $f = 100$.

The prior on the source position at $t_0$, $x_{0_S,E}$ and $x_{0_S,N}$, is
\begin{align}
   \pi (x_{0_S,E}) &\sim \mathcal{U}(\textrm{min}(x_E) - f \sigma_{x_E}, \textrm{max}(x_E) + f \sigma_{x_E}) \\
   \pi (x_{0_S,N}) &\sim \mathcal{U}(\textrm{min}(x_N) - f \sigma_{x_N}, \textrm{max}(x_N) + f \sigma_{x_N}) 
\end{align}
where $x_E$, $x_N$ are the positions in the F814W data of the target, $\sigma_{x_E}$, $\sigma_{x_N}$ is the standard deviation, and $f$ is an inflation factor.
We use $f=5$.

\section{Astrometric color analysis \label{app:Astrometric color analysis}}

For some stars, the astrometric measurements taken in the F814W and F606W filter are discrepant at the level of the reported uncertainties.
We explore this discrepancy specifically for OB110462, but this issue is also seen in OB110037 and reference stars for all targets.

First, we must quantify the discrepancy.
We consider several ways to measure the total offset between the astrometry of the two filters across all epochs.
$\overline{dx}$ is the average of the offsets in RA across all epochs for a particular star, and can be thought of measuring the amount of translation between F814W and F606W.
$\overline{|dx|}$ is the average of the magnitude of the offsets in RA across all epochs for a particular star, and can be thought of measuring the absolute amount of translation between F814W and F606W. 
$|\overline{dx}|$ is the absolute value of the average of the magnitude of the offsets in RA across all epochs for a particular star, and can be thought of as measuring the total amount of deviation between F814W and F606W. 
Note that $|\overline{dx}|$ is distinct from $\overline{|dx|}$.
The definitions for $\overline{dy}$, $\overline{|dy|}$, and $|\overline{dy}|$ are analogous, except they are the offsets in Dec.
We also consider these quantities in units of sigma, where the differences in each epoch $dx_i$ and $dy_i$ are normalized by the positional uncertainties $\sigma_{x,i}$ and $\sigma_{y,i}$.

\begin{figure*}[h!]
\centering
    \includegraphics[width=0.8\linewidth]{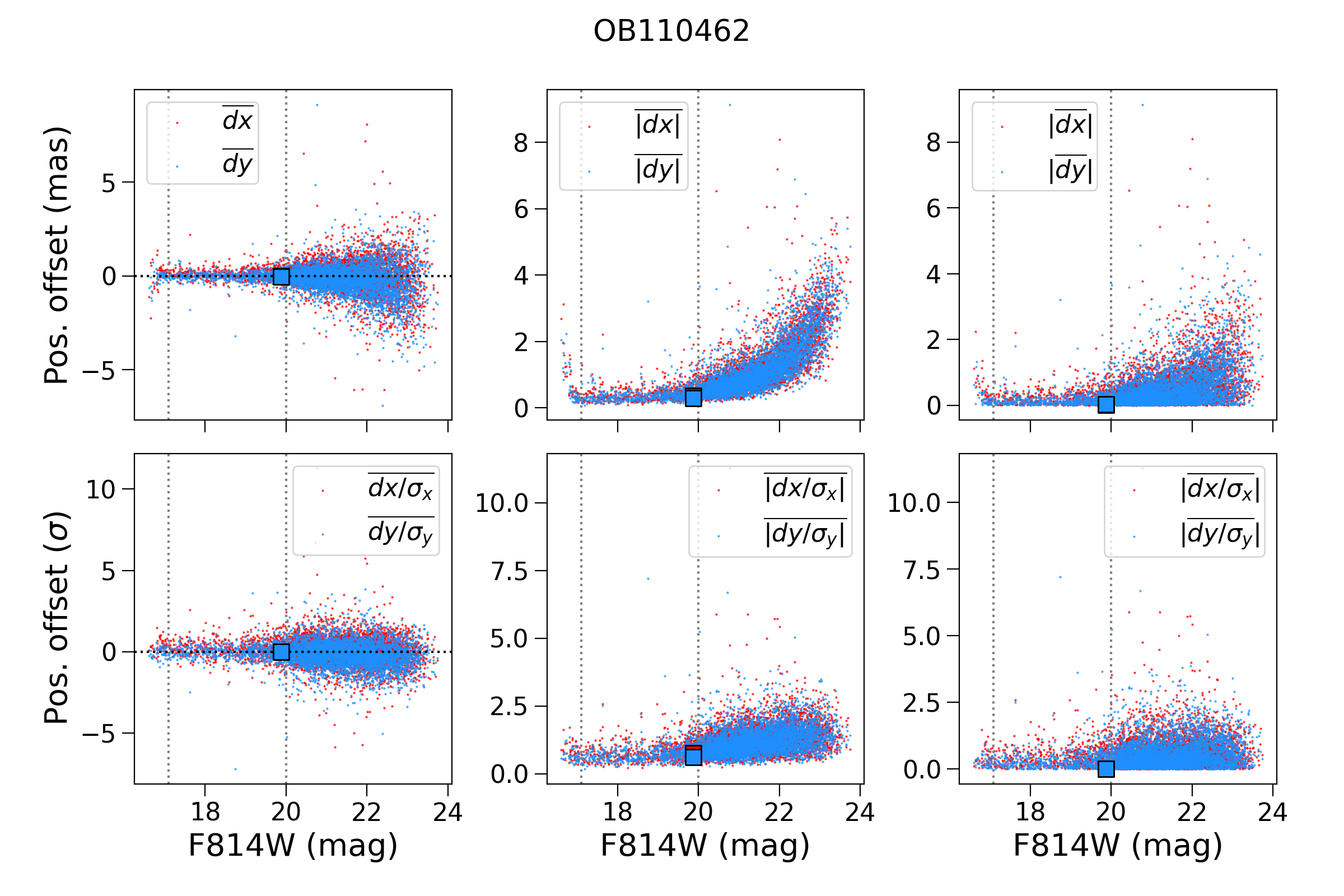}
\caption{\label{fig:color delta avg}
Difference in astrometry between observations taken in F814W  vs. F606W for stars in the OB110462 images, aggregated across all epochs.
Definitions of $\overline{dx}$, $\overline{|dx|}$, and $|\overline{dx}|$ are given in text of Appendix \ref{app:Astrometric color analysis}.
The squares mark OB110462.
The dashed lines indicate the magnitude range of the reference stars in the OB110462 images.}
\end{figure*}

In Figure \ref{fig:color delta avg}, we show the distributions of these quantities as a function of magnitude for stars within $30''$ of OB110462.
While not falling in the bulk of the distribution, OB110462 is not an extreme outlier.
Considering how large the variation in positional differences is, especially for fainter stars, OB110462 seems well within the other positions.
For this reason, we assume the positional differences are a systematic we can correct empirically.
We apply a constant positional offset to the F606W OB110462 observations to match the positions of the F814W observations as described in \S \ref{sec:Astrometric color offset}.
However, further investigation to determine whether the source of the filter dependent astrometry of OB110462 and other stars may actually be astrophysical is worth pursuing, as are more observational programs to study filter dependence on astrometry.

\section{Directly confronting the photometry and astrometry tension \label{app:Directly confronting the photometry and astrometry tension }}

\begin{figure*}[h!]
\centering
    \includegraphics[width=0.49\linewidth]{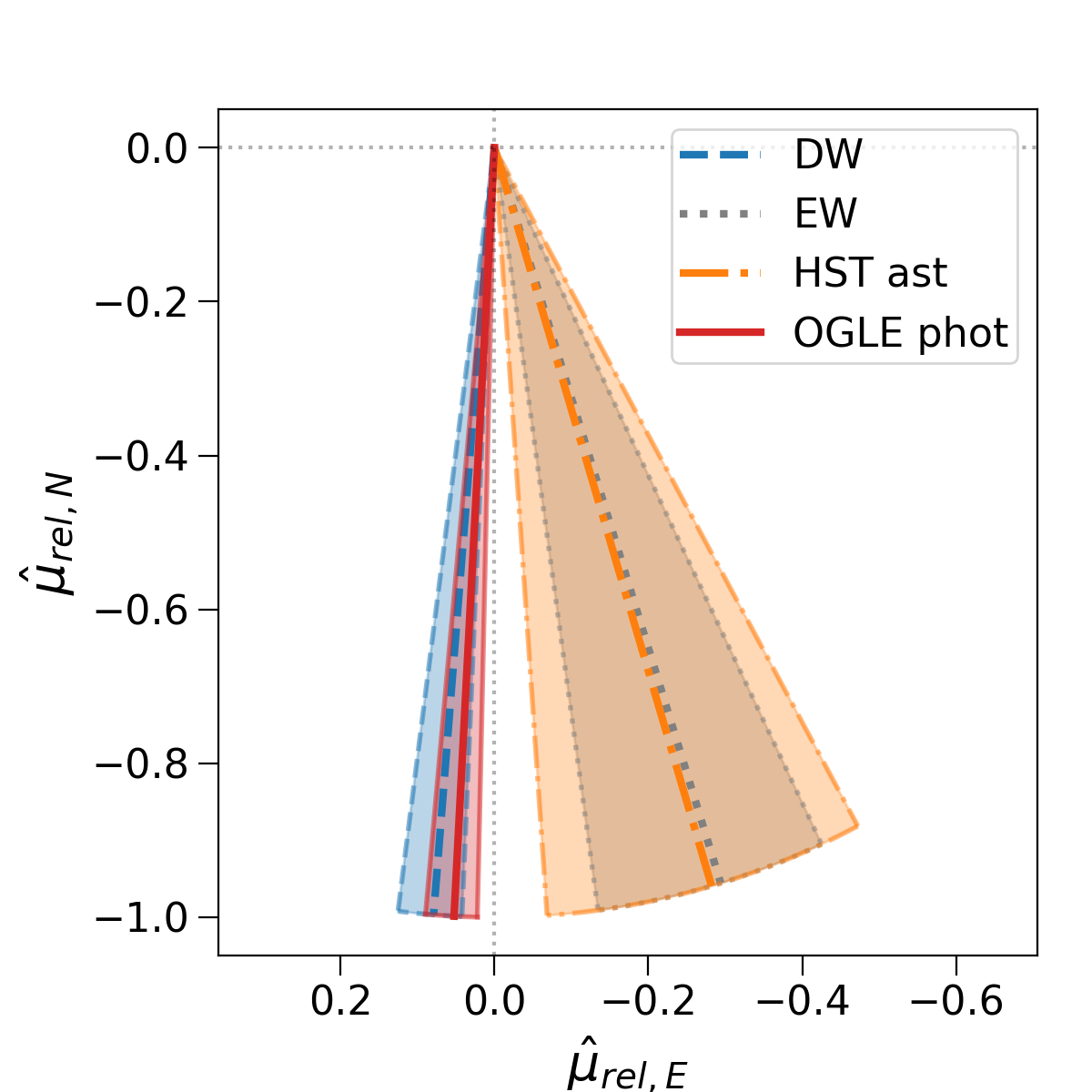}
    \includegraphics[width=0.49\linewidth]{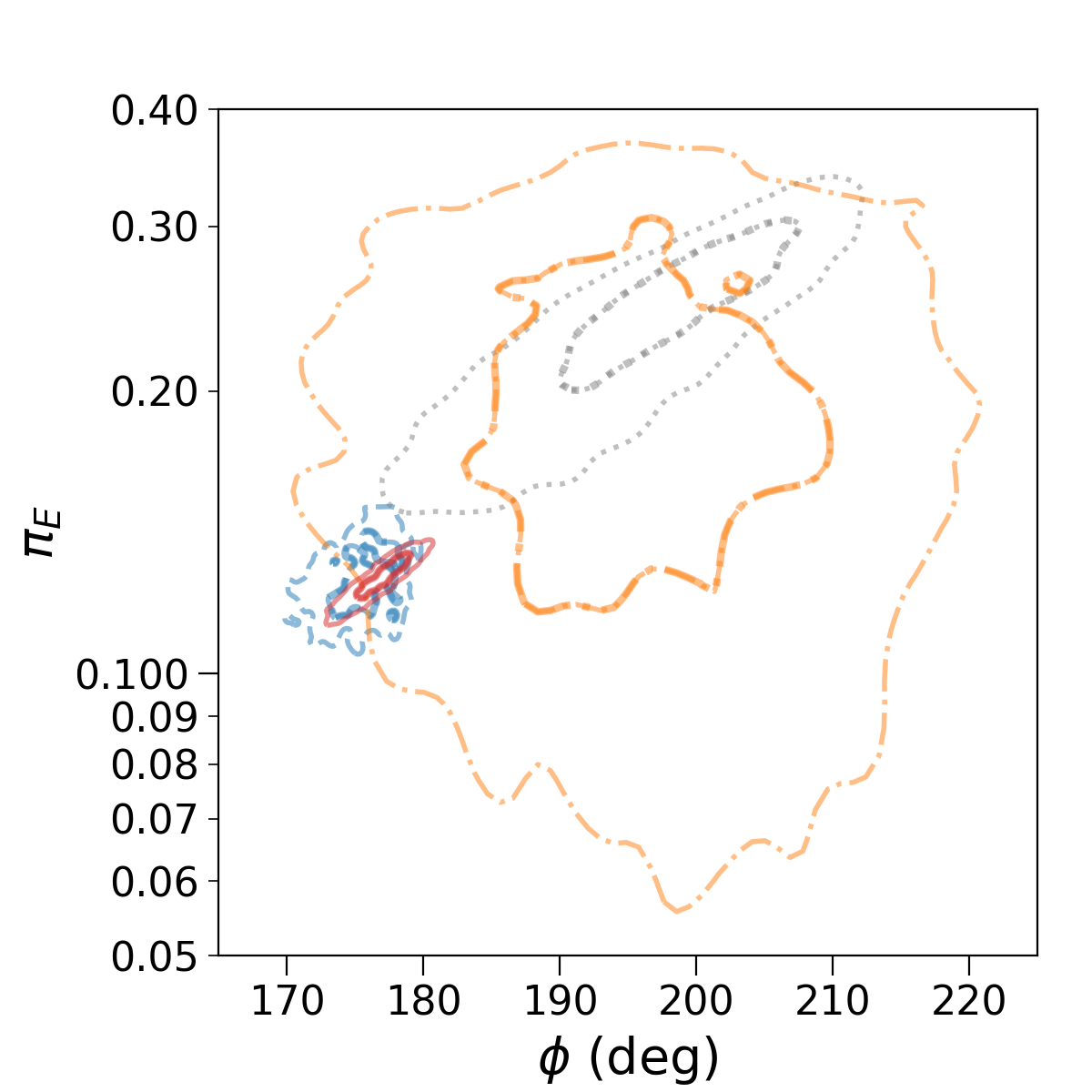}
\caption{\emph{Left:} Comparison of the source-lens relative proper motion inferred from the photometry and astrometry using the default weighted likelihood (\emph{DW}), the photometry and astrometry using the equal weighted likelihood (\emph{EW}), from the HST astrometry alone (\emph{HST ast}), and from the OGLE photometry alone (\emph{OGLE phot}). The median is shown as the thick line; the shaded region denotes the $1\sigma$ uncertainties.
\emph{Right}: 1 and $2\sigma$ contours of $\pi_E$ and direction of the source-lens relative proper motion $\phi$, for the 4 fits presented in the left panel.
the EW fit is consistent with the fit to the HST astrometry, although the astrometry alone has very little constraining power.
The DW and OGLE phot fits are consistent with each other, as expected.
\label{fig:tension}}
\end{figure*}

To try and elucidate the tension between the photometry and astrometry, we fit the OGLE photometry alone, and separately fit the HST astrometry alone.
The results are shown in Figure \ref{fig:tension}.
The HST astrometry alone does not have very much constraining power-- the uncertainties on parameters such as $t_0$, $t_E$, and $u_0$ are so wide that the fit itself is not useful.
However, the results are consistent with those of the EW fit.
The OGLE photometry has much more constraining power, but alone cannot constrain the lens mass.
The results are consistent with those of the DW fit.

\bibliography{sample63}{}
\bibliographystyle{aasjournal}

%% file: OB110462_fit_table_joint.txt
$t_0$ (MJD) & $55761.07_{{-0.96}}^{{+0.99}}$&55760.65&55759.15\\ 
$u_0$ & $-0.06_{{-0.009}}^{{+0.006}}$&-0.06&-0.07\\ 
$t_E$ (days) & $280.87_{{-5.96}}^{{+6.54}}$&284.94&277.47\\ 
$\log_{10}(\theta_E$/mas) & $0.59_{{-0.07}}^{{+0.05}}$&0.47&0.60\\ 
$\pi_S$ (mas) & $0.11_{{-0.02}}^{{+0.02}}$&0.12&0.12\\ 
$\pi_{E,E}$ & $0.010_{{-0.004}}^{{+0.005}}$&0.009&0.0007\\ 
$\pi_{E,N}$ & $-0.12_{{-0.01}}^{{+0.01}}$&-0.12&-0.14\\ 
$x_{S0,E}$ (mas) & $229.75_{{-0.08}}^{{+0.07}}$&229.82&229.80\\ 
$x_{S0,N}$ (mas) & $-214.28_{{-0.13}}^{{+0.11}}$&-214.58&-214.22\\ 
$\mu_{S,E}$ (mas/yr) & $-2.25_{{-0.02}}^{{+0.02}}$&-2.25&-2.25\\ 
$\mu_{S,N}$ (mas/yr) & $-3.57_{{-0.02}}^{{+0.02}}$&-3.55&-3.57\\ 
$b_{SFF,O}$ & $0.05_{{-0.0004}}^{{+0.0004}}$&0.05&0.05\\ 
$m_{base,O}$ (mag) & $16.41_{{-0.0001}}^{{+0.0001}}$&16.41&16.41\\ 
$b_{SFF,H8}$ & $0.90_{{-0.02}}^{{+0.02}}$&0.89&0.91\\ 
$m_{base,H8}$ (mag) & $19.86_{{-0.006}}^{{+0.006}}$&19.86&19.87\\ 
$b_{SFF,H6}$ & $0.94_{{-0.02}}^{{+0.02}}$&0.95&0.94\\ 
$m_{base,H6}$ (mag) & $22.04_{{-0.009}}^{{+0.009}}$&22.05&22.04\\ 
\hline 
$M_L$ ($M_\odot$) & $3.79_{{-0.57}}^{{+0.62}}$&3.01&3.58\\ 
$\pi_L$ (mas) & $0.60_{{-0.08}}^{{+0.08}}$&0.48&0.67\\ 
$\pi_{rel}$ (mas) & $0.48_{{-0.08}}^{{+0.08}}$&0.36&0.55\\ 
$\mu_{L,E}$ (mas/yr) & $-2.64_{{-0.24}}^{{+0.18}}$&-2.54&-2.28\\ 
$\mu_{L,N}$ (mas/yr) & $1.46_{{-0.71}}^{{+0.63}}$&0.26&1.69\\ 
$\mu_{rel,E}$ (mas/yr) & $0.40_{{-0.19}}^{{+0.23}}$&0.28&0.03\\ 
$\mu_{rel,N}$ (mas/yr) & $-5.02_{{-0.64}}^{{+0.71}}$&-3.81&-5.26\\ 
$\theta_E$ (mas) & $3.89_{{-1.16}}^{{+1.12}}$&2.98&4.00\\ 
$\pi_E$ & $0.12_{{-0.01}}^{{+0.01}}$&0.14&0.13\\ 
$\delta_{c,max}$ (mas) & $1.37_{{-0.41}}^{{+0.40}}$&1.05&1.41

%% file: OB110462_ew_fit_table_joint.txt
$t_0$ (MJD) & $55747.17_{{-7.55}}^{{+7.36}}$&55735.10&55735.10\\ 
$u_0$ & $-0.11_{{-0.01}}^{{+0.02}}$&-0.13&-0.13\\ 
$t_E$ (days) & $278.56_{{-9.16}}^{{+12.52}}$&267.37&267.37\\ 
$\log_{10}(\theta_E$/mas) & $0.62_{{-0.11}}^{{+0.09}}$&0.60&0.60\\ 
$\pi_S$ (mas) & $0.11_{{-0.02}}^{{+0.02}}$&0.10&0.10\\ 
$\pi_{E,E}$ & $-0.07_{{-0.05}}^{{+0.05}}$&-0.15&-0.15\\ 
$\pi_{E,N}$ & $-0.23_{{-0.03}}^{{+0.05}}$&-0.29&-0.29\\ 
$x_{S0,E}$ (mas) & $229.97_{{-0.17}}^{{+0.16}}$&230.07&230.07\\ 
$x_{S0,N}$ (mas) & $-214.31_{{-0.21}}^{{+0.21}}$&-214.37&-214.37\\ 
$\mu_{S,E}$ (mas/yr) & $-2.25_{{-0.03}}^{{+0.03}}$&-2.24&-2.24\\ 
$\mu_{S,N}$ (mas/yr) & $-3.56_{{-0.03}}^{{+0.03}}$&-3.57&-3.57\\ 
$b_{SFF,O}$ & $0.05_{{-0.004}}^{{+0.003}}$&0.06&0.06\\ 
$m_{base,O}$ (mag) & $16.41_{{-0.007}}^{{+0.007}}$&16.41&16.41\\ 
$b_{SFF,H8}$ & $0.95_{{-0.06}}^{{+0.05}}$&0.99&0.99\\ 
$m_{base,H8}$ (mag) & $19.88_{{-0.007}}^{{+0.008}}$&19.88&19.88\\ 
$b_{SFF,H6}$ & $0.99_{{-0.06}}^{{+0.04}}$&1.04&1.04\\ 
$m_{base,H6}$ (mag) & $22.04_{{-0.01}}^{{+0.01}}$&22.03&22.03\\ 
\hline 
$M_L$ ($M_\odot$) & $2.15_{{-0.54}}^{{+0.67}}$&1.51&1.51\\ 
$\pi_L$ (mas) & $1.09_{{-0.32}}^{{+0.34}}$&1.38&1.38\\ 
$\pi_{rel}$ (mas) & $0.98_{{-0.32}}^{{+0.34}}$&1.27&1.27\\ 
$\mu_{L,E}$ (mas/yr) & $-0.69_{{-0.94}}^{{+0.91}}$&0.25&0.25\\ 
$\mu_{L,N}$ (mas/yr) & $1.53_{{-1.12}}^{{+1.21}}$&1.22&1.22\\ 
$\mu_{rel,E}$ (mas/yr) & $-1.56_{{-0.91}}^{{+0.95}}$&-2.49&-2.49\\ 
$\mu_{rel,N}$ (mas/yr) & $-5.08_{{-1.22}}^{{+1.13}}$&-4.79&-4.79\\ 
$\theta_E$ (mas) & $4.13_{{-0.91}}^{{+0.96}}$&3.95&3.95\\ 
$\pi_E$ & $0.24_{{-0.05}}^{{+0.05}}$&0.32&0.32\\ 
$\delta_{c,max}$ (mas) & $1.46_{{-0.32}}^{{+0.34}}$&1.40&1.40

%% file: target_summary_table.txt
MB09260 & -- & MOA-2009-BLG-260 & 17:58:28.561 & -26:50:20.88 \\ 
MB10364 & -- & MOA-2010-BLG-364 & 17:57:05.401 & -34:27:05.01 \\ 
OB110037 & OGLE-2011-BLG-0037 & MOA-2011-BLG-039 & 17:55:55.83 & -30:33:39.7 \\ 
OB110310 & OGLE-2011-BLG-0310 & MOA-2011-BLG-332 & 17:51:25.39 & -30:24:35.0 \\ 
OB110462 & OGLE-2011-BLG-0462 & MOA-2011-BLG-191 & 17:51:40.19 & -29:53:26.3

%% file: hst_obs_table1.txt
MB09260 & 2009-10-01 & 275.0 & F606W & 10.0 & 4$^*$ \\ 
 & & & & 100.0 & 2 \\ 
 & & & F814W & 20.0 & 1$^*$ \\ 
 & 2009-10-19 & 275.0 & F606W & 310.0 & 1$^*$ \\ 
 & & & F814W & 72.0 & 6 \\ 
 & 2010-03-22 & 95.0 & F606W & 300.0 & 1$^*$ \\ 
 & & & F814W & 280.0 & 5 \\ 
 & 2010-06-14 & 95.0 & F606W & 200.0 & 1$^*$ \\ 
 & & & F814W & 275.0 & 5 \\ 
 & 2010-10-20 & 270.0 & F606W & 275.0 & 2$^*$ \\ 
 & & & F814W & 275.0 & 4 \\ 
 & 2011-04-19 & 90.0 & F606W & 275.0 & 2 \\ 
 & & & F814W & 275.0 & 4 \\ 
 & 2011-10-24 & 270.0 & F606W & 275.0 & 2 \\ 
 & & & F814W & 275.0 & 4 \\ 
 & 2012-09-25 & 270.0 & F606W & 275.0 & 2 \\ 
 & & & F814W & 275.0 & 4 \\ 
 & 2013-06-17 & 105.5 & F606W & 275.0 & 2 \\ 
 & & & F814W & 275.0 & 4 \\ 
\hline 
MB10364 & 2010-09-13 & 270.0 & F606W & 1.0 & 1 \\ 
 & & & & 2.0 & 1 \\ 
 & & & F814W & 3.0 & 5 \\ 
 & & & & 1.0 & 1$^*$ \\ 
 & 2010-10-26 & 277.4 & F606W & 12.0 & 2 \\ 
 & & & F814W & 12.0 & 6 \\ 
 & 2011-04-13 & 90.0 & F606W & 260.0 & 2$^*$ \\ 
 & & & F814W & 120.0 & 4$^*$ \\ 
 & 2011-07-22 & 260.5 & F606W & 160.0 & 4$^*$ \\ 
 & & & F814W & 90.0 & 4$^*$ \\ 
 & 2011-10-31 & 278.8 & F606W & 30.0 & 5 \\ 
 & & & F814W & 30.0 & 6 \\ 
 & 2012-09-25 & 270.9 & F606W & 30.0 & 5 \\ 
 & & & F814W & 30.0 & 6 \\ 
 & 2013-10-24 & 277.0 & F606W & 40.0 & 5 \\ 
 & & & F814W & 40.0 & 6 \\ 
\hline 
OB110037 & 2011-08-15 & 245.0 & F606W & 30.0 & 4 \\ 
 & & & F814W & 40.0 & 4 \\ 
 & & & & 20.0 & 3$^*$ \\ 
 & 2011-09-26 & 270.8 & F606W & 30.0 & 5 \\ 
 & & & F814W & 20.0 & 6 \\ 
 & 2011-11-01 & 276.1 & F606W & 50.0 & 5 \\ 
 & & & F814W & 30.0 & 5 \\ 
 & 2012-05-07 & 98.1 & F606W & 80.0 & 4 \\ 
 & & & F814W & 60.0 & 5 \\ 
 & 2012-09-25 & 270.8 & F606W & 80.0 & 4 \\ 
 & & & F814W & 60.0 & 5 \\ 
 & 2013-10-21 & 274.3 & F606W & 80.0 & 4 \\ 
 & & & F814W & 60.0 & 6 \\ 
 & 2014-10-26 & 275.1 & F606W & 60.0 & 4 \\ 
 & & & F814W & 55.0 & 6 \\ 
 & 2017-03-13 & 90.0 & F606W & 60.0 & 3 \\ 
 & & & F814W & 55.0 & 6 \\ 
 & 2017-09-04 & 256.9 & F606W & 60.0 & 3 \\ 
 & & & F814W & 55.0 & 6

%% file: hst_obs_table2.txt
OB110310 & 2011-09-21 & 270.0 & F606W & 75.0 & 4 \\ 
 & & & F814W & 75.0 & 5 \\ 
 & 2011-10-31 & 276.5 & F606W & 280.0 & 3 \\ 
 & & & F814W & 200.0 & 4 \\ 
 & 2012-04-24 & 96.0 & F606W & 280.0 & 3 \\ 
 & & & F814W & 230.0 & 4 \\ 
 & 2012-09-24 & 271.3 & F606W & 280.0 & 3 \\ 
 & & & F814W & 230.0 & 4 \\ 
 & 2013-10-21 & 274.8 & F606W & 280.0 & 3 \\ 
 & & & F814W & 68.0 & 1$^*$ \\ 
 & & & & 230.0 & 4 \\ 
 & 2017-03-14 & 90.4 & F606W & 270.0 & 3 \\ 
 & & & F814W & 230.0 & 4 \\ 
 & 2017-09-01 & 268.4 & F606W & 270.0 & 3 \\ 
 & & & F814W & 230.0 & 4 \\ 
\hline 
 OB110462 & 2011-08-08 & 270.0 & F606W & 60.0 & 1$^*$ \\ 
 & & & & 75.0 & 3 \\ 
 & & & F814W & 120.0 & 1$^*$ \\ 
 & & & & 60.0 & 1$^*$ \\ 
 & & & & 75.0 & 3 \\ 
 & 2011-10-31 & 276.1 & F606W & 280.0 & 3 \\ 
 & & & F814W & 200.0 & 4 \\ 
 & 2012-09-09 & 269.5 & F606W & 290.0 & 3 \\ 
 & & & F814W & 190.0 & 4 \\ 
 & 2012-09-25 & 271.3 & F606W & 280.0 & 3 \\ 
 & & & F814W & 200.0 & 4 \\ 
 & 2013-05-13 & 99.9 & F606W & 280.0 & 3$^*$ \\ 
 & & & F814W & 200.0 & 4$^*$ \\ 
 & 2013-10-22 & 274.6 & F606W & 285.0 & 3 \\ 
 & & & F814W & 285.0 & 4 \\ 
 & 2014-10-26 & 275.2 & F606W & 265.0 & 3 \\ 
 & & & F814W & 265.0 & 4 \\ 
 & 2017-08-11 & 255.2 & F606W & 250.0 & 3$^*$ \\ 
 & & & F814W & 250.0 & 4$^*$ \\ 
 & 2017-08-29 & 268.3 & F606W & 250.0 & 3 \\ 
 & & & F814W & 250.0 & 4 \\ 
 & 2021-10-01 & 272.0 & F606W & 407.0 & 5 \\ 
 & & & F814W & 307.0 & 6

%% file: gaia_table.txt
Source ID & 4064007633015639552 & 4042290560398692096 & 4056117808133831936 & 4056344036933003264 \\ 
RA (deg) & 269.619073405 $\pm$ 7.3e-07 & 269.272538687 $\pm$ 1.9e-08 & 268.982636391$\pm$ 3.0e-08 & 267.855757518 $\pm$ 2.3e-07  \\ 
Dec (deg) & -26.839323825 $\pm$ 5.9e-07 & -34.451415987 $\pm$ 1.5e-08 & -30.561059419$\pm$ 2.5e-08 & -30.409776355 $\pm$ 1.7e-07  \\ 
$\mu_{\alpha*}$ (mas/yr) & -- & -7.43 $\pm$ 0.08 & 2.40 $\pm$ 0.13 & -2.08 $\pm$ 1.12 \\ 
$\mu_{\delta}$ (mas/yr) & -- & -6.80 $\pm$ 0.05 & -3.91 $\pm$ 0.09 & -6.75 $\pm$ 0.58 \\ 
$\pi$ (mas) & -- & 0.40 $\pm$ 0.08 & 0.15 $\pm$ 0.13 & 0.54 $\pm$ 1.16 \\ 
ZP-corrected $\pi$ (mas) & -- & 0.43 & 0.19 & 0.53$^*$ \\ 
G (mag) & 19.216 $\pm$ 0.004 & 16.086 $\pm$ 0.002 & 17.477 $\pm$ 0.001 & 20.051 $\pm$ 0.010 \\ 
RP (mag) & -- & 14.929 $\pm$ 0.009 & 16.323 $\pm$ 0.010 & -- \\ 
BP (mag) & -- & 16.557 $\pm$ 0.011 & 19.049 $\pm$ 0.032 & -- \\ 
\hline 
ipd\_gof\_harmonic\_amplitude & 0.089 & 0.064 & 0.036 & 0.042 \\ 
ipd\_frac\_multi\_peak & 0 & 15 & 0 & 0 \\ 
ipd\_frac\_odd\_win & 18 & 0 & 0 & 55 \\ 
ruwe & -- & 1.388 & 0.971 & 0.981 \\ 
astrometric\_excess\_noise (mas) & 1.241 & 0.406 & 0.000 & 0.894 \\ 
astrometric\_excess\_noise\_sig & 2.657 & 12.020 & 0.000 & 0.332 \\ 
astrometric\_params\_solved & 3 & 95 & 31 & 95 \\ 
phot\_bp\_rp\_excess\_factor & -- & 1.69 & 1.39 & --

%% file: ref_star_crit_table.txt
MB09260  & Target range $\pm$ 1 mag:  & 30'' & 11 & F606W-F814W $<$ 2.6  \\ 
         & 14.5 $<$ F814W $<$ 18.8    &        &    & \& F814W $<$ 20.6   \\ 
         & 16.9 $<$ F606W $<$ 21.7    &        &    &                      \\ 
\hline 
MB10364  & 2010-09-13:                & 30'' & 8 & None excluded  \\ 
         & 12.5 $<$ F814W $<$ 18.0    &        &   &                \\ 
         & 11.5 $<$ F606W $<$ 19.2    &        &   &                \\ 
         & 2010-10-26:                &        &   &                \\ 
         & 15.1 $<$ F814W $<$ 18.0    &        &   &                \\ 
         & 13.4 $<$ F606W $<$ 19.2    &        &   &                \\ 
         & 2011-10-31:                &        &   &                \\ 
         & 15.8 $<$ F814W $<$ 18.0    &        &   &                \\ 
         & 14.4 $<$ F606W $<$ 19.2    &        &   &                \\ 
         & 2012-09-25:                &        &   &                \\ 
         & 16.0 $<$ F814W $<$ 18.0    &        &   &                \\ 
         & 14.5 $<$ F606W $<$ 19.2    &        &   &                \\ 
         & 2013-10-24:                &        &   &                \\ 
         & 16.1 $<$ F814W $<$ 18.0    &        &   &                \\ 
         & 14.7 $<$ F606W $<$ 19.2    &        &   &                \\ 
\hline 
OB110037 & Target range $\pm$ 0.5 mag:& 30'' & 12 & F606W-F814W $<$ 1.75  \\ 
         & 14.4 $<$ F814W $<$ 16.9    &        &    & \& F814W $<$ 19.6    \\ 
         & 16.4 $<$ F814W $<$ 18.8    &        &    &                       \\ 
\hline 
OB110310 & Target range $\pm$ 0.1 mag:& 30'' & 12 & F606W-F814W $<$ 2.4  \\ 
         & 16.9 $<$ F814W $<$ 18.7    &        &    & \& F814W $<$ 21.0   \\ 
         & 19.6 $<$ F606W $<$ 21.4    &        &    &                      \\ 
\hline 
OB110462 & Target range $\pm$ 0.1 mag:& 20'' & 14 & F606W-F814W $<$ 1.9  \\ 
         & 17.1 $<$ F814W $<$ 20.0    &        &    & \& F814W $<$ 20.6   \\ 
         & 19.2 $<$ F606W $<$ 22.1    &        &    &                      

%% file: gaia_vs_hst_pm.txt
MB10364 & ($-5.11_{-1.10}^{+1.62}$, $-7.78_{-0.89}^{+0.58}$) & ($-7.56_{-0.12}^{+0.12}$, $-6.49_{-0.11}^{+0.11}$) & ($-7.43 \pm 0.08$, $-6.80 \pm 0.05$) \\ 
OB110037 & ($6.27_{-1.20}^{+1.27}$, $-6.56_{-0.81}^{+0.77}$) & ($2.19_{-0.24}^{+0.24}$, $-3.87_{-0.20}^{+0.20}$) & ($2.40 \pm 0.13$, $-3.91 \pm 0.09$) \\ 
OB110310 & ($-0.02_{-1.16}^{+1.93}$, $-4.68_{-2.13}^{+2.39}$) & ($-2.41_{-0.12}^{+0.12}$, $-7.26_{-0.08}^{+0.08}$) & ($-2.08 \pm 1.12$, $-6.75 \pm 0.58$) 

%% file: hst_data_table.txt
MB09260 & F814W & 2009-10-19 & 7.80 $\pm$ 0.15 & 5.15 $\pm$ 0.14 & 15.484 $\pm$ 0.005 \\ 
 & & 2010-03-22 & 5.31 $\pm$ 0.14 & 3.80 $\pm$ 0.14 & 17.656 $\pm$ 0.005 \\ 
 & & 2010-06-14 & 4.31 $\pm$ 0.18 & 2.97 $\pm$ 0.18 & 17.760 $\pm$ 0.007 \\ 
 & & 2010-10-20 & 2.54 $\pm$ 0.14 & 1.60 $\pm$ 0.14 & 17.812 $\pm$ 0.014 \\ 
 & & 2011-04-19 & 0.12 $\pm$ 0.15 & 0.01 $\pm$ 0.15 & 17.838 $\pm$ 0.018 \\ 
 & & 2011-10-24 & -2.44 $\pm$ 0.21 & -1.71 $\pm$ 0.21 & 17.833 $\pm$ 0.005 \\ 
 & & 2012-09-25 & -7.07 $\pm$ 0.15 & -4.80 $\pm$ 0.15 & 17.836 $\pm$ 0.007 \\ 
 & & 2013-06-17 & -10.57 $\pm$ 0.26 & -7.02 $\pm$ 0.22 & 17.829 $\pm$ 0.011 \\ 
 & F606W & 2009-10-01 & 8.15 $\pm$ 0.22 & 5.57 $\pm$ 0.22 & 17.899 $\pm$ 0.013 \\ 
 & & 2011-04-19 & 0.13 $\pm$ 0.57 & 0.17 $\pm$ 0.57 & 20.750 $\pm$ 0.083 \\ 
 & & 2011-10-24 & -2.17 $\pm$ 0.52 & -1.59 $\pm$ 0.52 & 20.748 $\pm$ 0.007 \\ 
 & & 2012-09-25 & -6.23 $\pm$ 1.31 & -4.41 $\pm$ 1.32 & 20.733 $\pm$ 0.023 \\ 
 & & 2013-06-17 & -10.10 $\pm$ 0.49 & -7.21 $\pm$ 0.54 & 20.738 $\pm$ 0.036 \\ 
\hline 
MB10364 & F814W & 2010-09-13 & 9.61 $\pm$ 0.16 & 8.24 $\pm$ 0.16 & 13.366 $\pm$ 0.011 \\ 
 & & 2010-10-26 & 8.76 $\pm$ 0.16 & 7.72 $\pm$ 0.15 & 14.657 $\pm$ 0.005 \\ 
 & & 2011-10-31 & 1.23 $\pm$ 0.19 & 0.81 $\pm$ 0.18 & 15.315 $\pm$ 0.006 \\ 
 & & 2012-09-25 & -5.70 $\pm$ 0.19 & -4.89 $\pm$ 0.19 & 15.315 $\pm$ 0.010 \\ 
 & & 2013-10-24 & -13.90 $\pm$ 0.18 & -11.89 $\pm$ 0.18 & 15.316 $\pm$ 0.009 \\ 
 & F606W & 2010-09-13 & 8.81 $\pm$ 1.70 & 8.81 $\pm$ 1.70 & 14.538 $\pm$ 0.017 \\ 
 & & 2010-10-26 & 8.46 $\pm$ 0.50 & 7.59 $\pm$ 0.50 & 15.842 $\pm$ 0.017 \\ 
 & & 2011-10-31 & 1.47 $\pm$ 0.28 & 1.13 $\pm$ 0.28 & 16.498 $\pm$ 0.006 \\ 
 & & 2012-09-25 & -5.66 $\pm$ 0.31 & -4.99 $\pm$ 0.31 & 16.504 $\pm$ 0.008 \\ 
 & & 2013-10-24 & -13.89 $\pm$ 0.25 & -11.79 $\pm$ 0.25 & 16.502 $\pm$ 0.008 \\ 
\hline 
OB110037 & F814W & 2011-08-15 & -4.35 $\pm$ 0.11 & 8.24 $\pm$ 0.14 & 14.864 $\pm$ 0.021 \\ 
 & & 2011-09-26 & -4.33 $\pm$ 0.12 & 7.82 $\pm$ 0.12 & 15.029 $\pm$ 0.005 \\ 
 & & 2011-11-01 & -4.34 $\pm$ 0.12 & 7.52 $\pm$ 0.12 & 15.774 $\pm$ 0.006 \\ 
 & & 2012-05-07 & -3.10 $\pm$ 0.13 & 5.40 $\pm$ 0.12 & 16.315 $\pm$ 0.019 \\ 
 & & 2012-09-25 & -2.22 $\pm$ 0.14 & 3.86 $\pm$ 0.14 & 16.321 $\pm$ 0.009 \\ 
 & & 2013-10-21 & 0.10 $\pm$ 0.12 & -0.36 $\pm$ 0.12 & 16.327 $\pm$ 0.005 \\ 
 & & 2014-10-26 & 2.45 $\pm$ 0.14 & -4.36 $\pm$ 0.13 & 16.328 $\pm$ 0.006 \\ 
 & & 2017-03-13 & 7.74 $\pm$ 0.12 & -13.21 $\pm$ 0.12 & 16.334 $\pm$ 0.007 \\ 
 & & 2017-09-04 & 8.05 $\pm$ 0.14 & -14.90 $\pm$ 0.13 & 16.322 $\pm$ 0.011 \\ 
 & F606W & 2011-08-15 & -4.02 $\pm$ 0.25 & 8.15 $\pm$ 0.24 & 16.916 $\pm$ 0.011 \\ 
 & & 2011-09-26 & -4.23 $\pm$ 0.20 & 7.79 $\pm$ 0.20 & 17.086 $\pm$ 0.020 \\ 
 & & 2011-11-01 & -4.39 $\pm$ 0.24 & 7.36 $\pm$ 0.23 & 17.794 $\pm$ 0.005 \\ 
 & & 2012-05-07 & -2.69 $\pm$ 0.31 & 5.34 $\pm$ 0.33 & 18.301 $\pm$ 0.016 \\ 
 & & 2012-09-25 & -2.34 $\pm$ 0.26 & 3.81 $\pm$ 0.26 & 18.306 $\pm$ 0.006 \\ 
 & & 2013-10-21 & 0.65 $\pm$ 0.24 & -0.76 $\pm$ 0.24 & 18.314 $\pm$ 0.006 \\ 
 & & 2014-10-26 & 3.73 $\pm$ 0.22 & -5.29 $\pm$ 0.22 & 18.320 $\pm$ 0.015 \\ 
 & & 2017-03-13 & 9.48 $\pm$ 0.26 & -14.04 $\pm$ 0.26 & 18.331 $\pm$ 0.008 \\ 
 & & 2017-09-04 & 9.80 $\pm$ 0.27 & -16.84 $\pm$ 0.24 & 18.326 $\pm$ 0.027 \\ 
\hline 
OB110310 & F814W & 2011-09-21 & 5.20 $\pm$ 0.14 & 15.95 $\pm$ 0.14 & 16.945 $\pm$ 0.013 \\ 
 & & 2011-10-31 & 4.86 $\pm$ 0.17 & 14.85 $\pm$ 0.17 & 18.058 $\pm$ 0.004 \\ 
 & & 2012-04-24 & 3.79 $\pm$ 0.13 & 11.35 $\pm$ 0.13 & 18.602 $\pm$ 0.013 \\ 
 & & 2012-09-24 & 2.82 $\pm$ 0.17 & 8.47 $\pm$ 0.17 & 18.616 $\pm$ 0.011 \\ 
 & & 2013-10-21 & 0.14 $\pm$ 0.13 & 0.59 $\pm$ 0.14 & 18.621 $\pm$ 0.005 \\ 
 & & 2017-03-14 & -7.92 $\pm$ 0.18 & -23.84 $\pm$ 0.18 & 18.608 $\pm$ 0.005 \\ 
 & & 2017-09-01 & -8.88 $\pm$ 0.16 & -27.36 $\pm$ 0.16 & 18.613 $\pm$ 0.012 \\ 
 & F606W & 2011-09-21 & 5.43 $\pm$ 0.22 & 15.98 $\pm$ 0.22 & 19.663 $\pm$ 0.018 \\ 
 & & 2011-10-31 & 4.73 $\pm$ 0.34 & 15.06 $\pm$ 0.34 & 20.780 $\pm$ 0.007 \\ 
 & & 2012-04-24 & 3.55 $\pm$ 0.38 & 11.44 $\pm$ 0.37 & 21.329 $\pm$ 0.034 \\ 
 & & 2012-09-24 & 2.79 $\pm$ 0.49 & 8.02 $\pm$ 0.49 & 21.180 $\pm$ 0.212 \\ 
 & & 2013-10-21 & 0.56 $\pm$ 0.30 & 0.87 $\pm$ 0.30 & 21.339 $\pm$ 0.007 \\ 
 & & 2017-03-14 & -7.81 $\pm$ 0.31 & -23.78 $\pm$ 0.31 & 21.333 $\pm$ 0.022 \\ 
 & & 2017-09-01 & -9.31 $\pm$ 0.34 & -27.08 $\pm$ 0.35 & 21.335 $\pm$ 0.034 \\ 
\hline 
OB110462 & F814W & 2011-08-08 & 7.53 $\pm$ 0.15 & 11.45 $\pm$ 0.15 & 17.209 $\pm$ 0.028 \\ 
 & & 2011-10-31 & 6.44 $\pm$ 0.23 & 9.71 $\pm$ 0.22 & 18.849 $\pm$ 0.006 \\ 
 & & 2012-09-09 & 4.08 $\pm$ 0.23 & 6.55 $\pm$ 0.23 & 19.756 $\pm$ 0.009 \\ 
 & & 2012-09-25 & 4.25 $\pm$ 0.37 & 6.42 $\pm$ 0.37 & 19.767 $\pm$ 0.008 \\ 
 & & 2013-10-22 & 1.43 $\pm$ 0.33 & 2.40 $\pm$ 0.34 & 19.839 $\pm$ 0.048 \\ 
 & & 2014-10-26 & -0.87 $\pm$ 0.29 & -1.00 $\pm$ 0.30 & 19.881 $\pm$ 0.009 \\ 
 & & 2017-08-29 & -7.18 $\pm$ 0.27 & -10.61 $\pm$ 0.26 & 19.874 $\pm$ 0.009 \\ 
 & & 2021-10-01 & -15.67 $\pm$ 0.19 & -24.93 $\pm$ 0.19 & 19.865 $\pm$ 0.020 \\ 
 & F606W & 2011-08-08 & 6.76 $\pm$ 0.43 & 11.77 $\pm$ 0.43 & 19.313 $\pm$ 0.023 \\ 
 & & 2011-10-31 & 6.25 $\pm$ 0.50 & 10.54 $\pm$ 0.52 & 20.974 $\pm$ 0.010 \\ 
 & & 2012-09-09 & 4.60 $\pm$ 0.37 & 5.57 $\pm$ 0.37 & 21.867 $\pm$ 0.050 \\ 
 & & 2012-09-25 & 4.38 $\pm$ 0.59 & 6.40 $\pm$ 0.59 & 21.920 $\pm$ 0.010 \\ 
 & & 2013-10-22 & 1.76 $\pm$ 0.37 & 2.45 $\pm$ 0.36 & 22.011 $\pm$ 0.015 \\ 
 & & 2014-10-26 & -1.22 $\pm$ 0.46 & -0.83 $\pm$ 0.44 & 22.042 $\pm$ 0.015 \\ 
 & & 2017-08-29 & -7.10 $\pm$ 0.41 & -10.62 $\pm$ 0.41 & 22.017 $\pm$ 0.027 \\ 
 & & 2021-10-01 & -15.44 $\pm$ 0.44 & -25.06 $\pm$ 0.44 & 22.021 $\pm$ 0.017

%% file: upper_limits.txt
MB09260 & $<2.42$ & $<0.85$ & $1.37_{-1.16}^{+2.72}$ & $0.09_{-0.03}^{+0.13}$ \\ 
MB10364 & $<1.76$ & $<0.62$ & $0.21_{-0.18}^{+0.61}$ & $0.27_{-0.01}^{+0.02}$ \\ 
OB110037 & $1.24_{-0.90}^{+1.10}$ & $0.44_{-0.32}^{+0.39}$ & $0.41_{-0.30}^{+0.37}$ & $0.37_{-0.02}^{+0.02}$ \\ 
OB110310 & $<2.75$ & $<0.97$ & $0.78_{-0.68}^{+2.98}$ & $0.13_{-0.08}^{+0.20}$ \\ 
OB110462 (EW) & $4.13_{-3.02}^{+2.98}$ & $1.46_{-1.07}^{+1.05}$ & $2.15_{-1.43}^{+3.50}$ & $0.24_{-0.16}^{+0.11}$ \\ 
OB110462 (DW) & $3.89_{-1.61}^{+1.69}$ & $1.37_{-0.57}^{+0.60}$ & $3.79_{-1.64}^{+2.17}$ & $0.12_{-0.04}^{+0.03}$

%% file: lens_type_prob.txt
MB09260 & 4 & 0 & 38 & 44 & 14 \\ 
MB10364 & 36 & 29 & 35 & 0 & 0 \\ 
OB110037 & 74 & 0 & 26 & 0 & 0 \\ 
OB110310 & 5 & 3 & 65 & 22 & 5 \\ 
OB110462 DW & 0 & 0 & 0 & 0 & 100 \\ 
OB110462 EW & 0 & 0 & 6 & 50 & 44

%% file: MB09260_split_fit_table_joint_modes.txt
$t_0$ (MJD) & $55099.19_{{-1.37}}^{{+1.40}}$&55099.27&55099.27& $55099.33_{{-1.25}}^{{+1.26}}$&55099.99&55099.99\\ 
$u_0$ & $-0.09_{{-0.04}}^{{+0.02}}$&-0.07&-0.07& $0.02_{{-0.06}}^{{+0.03}}$&-0.0002&-0.0002\\ 
$t_E$ (days) & $143.16_{{-2.95}}^{{+3.35}}$&141.71&141.71& $142.37_{{-2.89}}^{{+3.43}}$&141.64&141.64\\ 
$\log_{10}(\theta_E$/mas) & $0.03_{{-0.20}}^{{+0.14}}$&0.24&0.24& $0.008_{{-0.21}}^{{+0.15}}$&0.06&0.06\\ 
$\pi_S$ (mas) & $0.10_{{-0.02}}^{{+0.02}}$&0.12&0.12& $0.10_{{-0.02}}^{{+0.02}}$&0.09&0.09\\ 
$\pi_{E,E}$ & $-0.08_{{-0.009}}^{{+0.010}}$&-0.08&-0.08& $-0.08_{{-0.009}}^{{+0.009}}$&-0.07&-0.07\\ 
$\pi_{E,N}$ & $-0.02_{{-0.04}}^{{+0.03}}$&-0.005&-0.005& $-0.04_{{-0.06}}^{{+0.04}}$&-0.06&-0.06\\ 
$x_{S0,E}$ (mas) & $236.25_{{-0.12}}^{{+0.13}}$&236.43&236.43& $236.20_{{-0.11}}^{{+0.13}}$&236.30&236.30\\ 
$x_{S0,N}$ (mas) & $-692.07_{{-0.10}}^{{+0.11}}$&-692.15&-692.15& $-692.01_{{-0.13}}^{{+0.12}}$&-691.91&-691.91\\ 
$\mu_{S,E}$ (mas/yr) & $-5.00_{{-0.05}}^{{+0.05}}$&-5.07&-5.07& $-4.99_{{-0.05}}^{{+0.05}}$&-5.05&-5.05\\ 
$\mu_{S,N}$ (mas/yr) & $-3.38_{{-0.05}}^{{+0.04}}$&-3.34&-3.34& $-3.39_{{-0.05}}^{{+0.05}}$&-3.44&-3.44\\ 
$b_{SFF,M}$ & $0.61_{{-0.02}}^{{+0.02}}$&0.60&0.60& $0.60_{{-0.02}}^{{+0.02}}$&0.61&0.61\\ 
$m_{base,M}$ (mag) & $17.43_{{-0.003}}^{{+0.003}}$&17.43&17.43& $17.43_{{-0.003}}^{{+0.003}}$&17.42&17.42\\ 
$b_{SFF,H8}$ & $1.00_{{-0.03}}^{{+0.02}}$&0.99&0.99& $0.99_{{-0.03}}^{{+0.02}}$&1.00&1.00\\ 
$m_{base,H8}$ (mag) & $17.84_{{-0.004}}^{{+0.003}}$&17.83&17.83& $17.84_{{-0.003}}^{{+0.003}}$&17.84&17.84\\ 
$b_{SFF,H6}$ & $1.03_{{-0.03}}^{{+0.02}}$&1.02&1.02& $1.02_{{-0.03}}^{{+0.02}}$&1.02&1.02\\ 
$m_{base,H6}$ (mag) & $20.75_{{-0.007}}^{{+0.007}}$&20.75&20.75& $20.75_{{-0.007}}^{{+0.006}}$&20.75&20.75\\ 
\hline 
$M_L$ ($M_\odot$) & $1.44_{{-0.59}}^{{+0.74}}$&2.70&2.70& $1.30_{{-0.58}}^{{+0.74}}$&1.44&1.44\\ 
$\pi_L$ (mas) & $0.20_{{-0.04}}^{{+0.05}}$&0.25&0.25& $0.20_{{-0.04}}^{{+0.06}}$&0.20&0.20\\ 
$\pi_{rel}$ (mas) & $0.09_{{-0.04}}^{{+0.04}}$&0.14&0.14& $0.10_{{-0.04}}^{{+0.05}}$&0.11&0.11\\ 
$\mu_{L,E}$ (mas/yr) & $-2.62_{{-0.93}}^{{+1.18}}$&-0.61&-0.61& $-2.88_{{-0.94}}^{{+1.15}}$&-2.82&-2.82\\ 
$\mu_{L,N}$ (mas/yr) & $-2.63_{{-0.91}}^{{+0.98}}$&-3.04&-3.04& $-2.40_{{-1.12}}^{{+1.09}}$&-1.49&-1.49\\ 
$\mu_{rel,E}$ (mas/yr) & $-2.40_{{-1.17}}^{{+0.95}}$&-4.46&-4.46& $-2.11_{{-1.16}}^{{+0.94}}$&-2.23&-2.23\\ 
$\mu_{rel,N}$ (mas/yr) & $-0.73_{{-1.01}}^{{+0.89}}$&-0.31&-0.31& $-1.01_{{-1.09}}^{{+1.15}}$&-1.95&-1.95\\ 
$\theta_E$ (mas) & $1.07_{{-0.40}}^{{+0.41}}$&1.73&1.73& $1.02_{{-0.39}}^{{+0.41}}$&1.15&1.15\\ 
$\pi_E$ & $0.09_{{-0.01}}^{{+0.02}}$&0.08&0.08& $0.09_{{-0.01}}^{{+0.04}}$&0.10&0.10\\ 
$\delta_{c,max}$ (mas) & $0.38_{{-0.14}}^{{+0.15}}$&0.61&0.61& $0.36_{{-0.14}}^{{+0.15}}$&0.41&0.41\\ 
\hline 
$\sum w_i$ & \multicolumn{3}{c}{0.42} & \multicolumn{3}{c}{0.58} \\ 
$log\mathcal{Z}$ & \multicolumn{3}{c}{31613.22} & \multicolumn{3}{c}{31613.55}

%% file: MB10364_fit_table_joint.txt
$t_0$ (MJD) & $55445.13_{{-0.12}}^{{+0.12}}$&55445.06&55445.06\\ 
$u_0$ & $-0.008_{{-0.01}}^{{+0.01}}$&-0.004&-0.004\\ 
$t_E$ (days) & $61.11_{{-0.24}}^{{+0.24}}$&61.06&61.06\\ 
$\log_{10}(\theta_E$/mas) & $-0.33_{{-0.25}}^{{+0.22}}$&-0.40&-0.40\\ 
$\pi_S$ (mas) & $0.11_{{-0.02}}^{{+0.02}}$&0.11&0.11\\ 
$\pi_{E,E}$ & $-0.24_{{-0.003}}^{{+0.003}}$&-0.24&-0.24\\ 
$\pi_{E,N}$ & $0.12_{{-0.01}}^{{+0.01}}$&0.12&0.12\\ 
$x_{S0,E}$ (mas) & $130.18_{{-0.10}}^{{+0.11}}$&130.13&130.13\\ 
$x_{S0,N}$ (mas) & $-78.98_{{-0.10}}^{{+0.11}}$&-79.02&-79.02\\ 
$\mu_{S,E}$ (mas/yr) & $-7.56_{{-0.06}}^{{+0.06}}$&-7.52&-7.52\\ 
$\mu_{S,N}$ (mas/yr) & $-6.49_{{-0.06}}^{{+0.06}}$&-6.47&-6.47\\ 
$b_{SFF,M}$ & $0.93_{{-0.007}}^{{+0.007}}$&0.93&0.93\\ 
$m_{base,M}$ (mag) & $15.02_{{-0.00006}}^{{+0.00006}}$&15.02&15.02\\ 
$b_{SFF,H8}$ & $0.99_{{-0.02}}^{{+0.02}}$&0.98&0.98\\ 
$m_{base,H8}$ (mag) & $15.32_{{-0.006}}^{{+0.006}}$&15.32&15.32\\ 
$b_{SFF,H6}$ & $1.00_{{-0.02}}^{{+0.02}}$&1.01&1.01\\ 
$m_{base,H6}$ (mag) & $16.50_{{-0.006}}^{{+0.006}}$&16.50&16.50\\ 
\hline 
$M_L$ ($M_\odot$) & $0.21_{{-0.10}}^{{+0.14}}$&0.18&0.18\\ 
$\pi_L$ (mas) & $0.24_{{-0.06}}^{{+0.08}}$&0.22&0.22\\ 
$\pi_{rel}$ (mas) & $0.12_{{-0.05}}^{{+0.08}}$&0.11&0.11\\ 
$\mu_{L,E}$ (mas/yr) & $-5.11_{{-1.09}}^{{+1.62}}$&-5.38&-5.38\\ 
$\mu_{L,N}$ (mas/yr) & $-7.78_{{-0.89}}^{{+0.57}}$&-7.56&-7.56\\ 
$\mu_{rel,E}$ (mas/yr) & $-2.46_{{-1.61}}^{{+1.09}}$&-2.13&-2.13\\ 
$\mu_{rel,N}$ (mas/yr) & $1.29_{{-0.57}}^{{+0.88}}$&1.09&1.09\\ 
$\theta_E$ (mas) & $0.46_{{-0.21}}^{{+0.31}}$&0.40&0.40\\ 
$\pi_E$ & $0.27_{{-0.005}}^{{+0.005}}$&0.27&0.27\\ 
$\delta_{c,max}$ (mas) & $0.16_{{-0.07}}^{{+0.11}}$&0.14&0.14

%% file: OB110037_fit_table_joint.txt
$t_0$ (MJD) & $55781.53_{{-0.30}}^{{+0.28}}$&55781.49&55781.49\\ 
$u_0$ & $-0.002_{{-0.02}}^{{+0.03}}$&-0.008&-0.008\\ 
$t_E$ (days) & $92.78_{{-2.60}}^{{+2.63}}$&93.31&93.31\\ 
$\log_{10}(\theta_E$/mas) & $0.09_{{-0.14}}^{{+0.11}}$&0.22&0.22\\ 
$\pi_S$ (mas) & $0.12_{{-0.02}}^{{+0.02}}$&0.11&0.11\\ 
$\pi_{E,E}$ & $-0.31_{{-0.005}}^{{+0.005}}$&-0.31&-0.31\\ 
$\pi_{E,N}$ & $0.21_{{-0.02}}^{{+0.01}}$&0.21&0.21\\ 
$x_{S0,E}$ (mas) & $15.21_{{-0.06}}^{{+0.06}}$&15.23&15.23\\ 
$x_{S0,N}$ (mas) & $-115.53_{{-0.07}}^{{+0.07}}$&-115.61&-115.61\\ 
$\mu_{S,E}$ (mas/yr) & $2.19_{{-0.02}}^{{+0.02}}$&2.18&2.18\\ 
$\mu_{S,N}$ (mas/yr) & $-3.87_{{-0.02}}^{{+0.02}}$&-3.86&-3.86\\ 
$b_{SFF,O}$ & $0.90_{{-0.05}}^{{+0.06}}$&0.89&0.89\\ 
$m_{base,O}$ (mag) & $16.15_{{-0.0003}}^{{+0.0003}}$&16.15&16.15\\ 
$b_{SFF,H8}$ & $0.91_{{-0.05}}^{{+0.06}}$&0.90&0.90\\ 
$m_{base,H8}$ (mag) & $16.33_{{-0.003}}^{{+0.003}}$&16.33&16.33\\ 
$b_{SFF,H6}$ & $0.84_{{-0.05}}^{{+0.06}}$&0.82&0.82\\ 
$m_{base,H6}$ (mag) & $18.32_{{-0.003}}^{{+0.004}}$&18.31&18.31\\ 
\hline 
$M_L$ ($M_\odot$) & $0.41_{{-0.12}}^{{+0.12}}$&0.55&0.55\\ 
$\pi_L$ (mas) & $0.58_{{-0.13}}^{{+0.14}}$&0.74&0.74\\ 
$\pi_{rel}$ (mas) & $0.46_{{-0.13}}^{{+0.14}}$&0.62&0.62\\ 
$\mu_{L,E}$ (mas/yr) & $6.27_{{-1.17}}^{{+1.25}}$&7.59&7.59\\ 
$\mu_{L,N}$ (mas/yr) & $-6.56_{{-0.78}}^{{+0.74}}$&-7.54&-7.54\\ 
$\mu_{rel,E}$ (mas/yr) & $-4.07_{{-1.25}}^{{+1.17}}$&-5.42&-5.42\\ 
$\mu_{rel,N}$ (mas/yr) & $2.69_{{-0.75}}^{{+0.78}}$&3.68&3.68\\ 
$\theta_E$ (mas) & $1.24_{{-0.35}}^{{+0.36}}$&1.67&1.67\\ 
$\pi_E$ & $0.37_{{-0.009}}^{{+0.008}}$&0.37&0.37\\ 
$\delta_{c,max}$ (mas) & $0.44_{{-0.12}}^{{+0.13}}$&0.59&0.59

%% file: OB110310_split_fit_table_joint_modes.txt
$t_0$ (MJD) & $55802.11_{{-1.57}}^{{+1.21}}$&55801.39&55801.39& $55802.66_{{-1.17}}^{{+1.15}}$&55801.95&55801.95\\ 
$u_0$ & $-0.18_{{-0.07}}^{{+0.05}}$&-0.24&-0.24& $-0.005_{{-0.08}}^{{+0.07}}$&-0.09&-0.09\\ 
$t_E$ (days) & $83.40_{{-1.83}}^{{+2.39}}$&83.23&83.23& $82.20_{{-1.33}}^{{+1.66}}$&81.47&81.47\\ 
$\log_{10}(\theta_E$/mas) & $-0.05_{{-0.28}}^{{+0.24}}$&0.04&0.04& $-0.06_{{-0.28}}^{{+0.21}}$&0.04&0.04\\ 
$\pi_S$ (mas) & $0.10_{{-0.02}}^{{+0.02}}$&0.10&0.10& $0.10_{{-0.02}}^{{+0.02}}$&0.09&0.09\\ 
$\pi_{E,E}$ & $-0.08_{{-0.01}}^{{+0.01}}$&-0.09&-0.09& $-0.09_{{-0.02}}^{{+0.01}}$&-0.10&-0.10\\ 
$\pi_{E,N}$ & $-0.08_{{-0.08}}^{{+0.06}}$&-0.14&-0.14& $-0.11_{{-0.09}}^{{+0.10}}$&-0.21&-0.21\\ 
$x_{S0,E}$ (mas) & $-104.56_{{-0.10}}^{{+0.12}}$&-104.62&-104.62& $-104.62_{{-0.09}}^{{+0.09}}$&-104.58&-104.58\\ 
$x_{S0,N}$ (mas) & $-183.61_{{-0.11}}^{{+0.13}}$&-183.53&-183.53& $-183.57_{{-0.14}}^{{+0.13}}$&-183.49&-183.49\\ 
$\mu_{S,E}$ (mas/yr) & $-2.41_{{-0.03}}^{{+0.02}}$&-2.39&-2.39& $-2.40_{{-0.02}}^{{+0.02}}$&-2.43&-2.43\\ 
$\mu_{S,N}$ (mas/yr) & $-7.26_{{-0.03}}^{{+0.03}}$&-7.26&-7.26& $-7.26_{{-0.03}}^{{+0.03}}$&-7.28&-7.28\\ 
$b_{SFF,O}$ & $0.97_{{-0.02}}^{{+0.02}}$&0.98&0.98& $0.96_{{-0.03}}^{{+0.02}}$&0.96&0.96\\ 
$m_{base,O}$ (mag) & $18.41_{{-0.005}}^{{+0.005}}$&18.41&18.41& $18.41_{{-0.005}}^{{+0.005}}$&18.41&18.41\\ 
$b_{SFF,H8}$ & $1.02_{{-0.03}}^{{+0.02}}$&1.04&1.04& $1.02_{{-0.03}}^{{+0.02}}$&1.04&1.04\\ 
$m_{base,H8}$ (mag) & $18.62_{{-0.003}}^{{+0.003}}$&18.61&18.61& $18.62_{{-0.003}}^{{+0.003}}$&18.62&18.62\\ 
$b_{SFF,H6}$ & $1.02_{{-0.03}}^{{+0.02}}$&1.04&1.04& $1.02_{{-0.03}}^{{+0.02}}$&1.05&1.05\\ 
$m_{base,H6}$ (mag) & $21.34_{{-0.006}}^{{+0.006}}$&21.34&21.34& $21.34_{{-0.006}}^{{+0.006}}$&21.34&21.34\\ 
\hline 
$M_L$ ($M_\odot$) & $0.90_{{-0.47}}^{{+0.77}}$&0.83&0.83& $0.71_{{-0.34}}^{{+0.62}}$&0.58&0.58\\ 
$\pi_L$ (mas) & $0.21_{{-0.06}}^{{+0.11}}$&0.28&0.28& $0.23_{{-0.07}}^{{+0.13}}$&0.35&0.35\\ 
$\pi_{rel}$ (mas) & $0.11_{{-0.06}}^{{+0.10}}$&0.18&0.18& $0.12_{{-0.07}}^{{+0.13}}$&0.25&0.25\\ 
$\mu_{L,E}$ (mas/yr) & $0.21_{{-1.30}}^{{+2.20}}$&0.22&0.22& $-0.16_{{-1.06}}^{{+1.67}}$&-0.30&-0.30\\ 
$\mu_{L,N}$ (mas/yr) & $-4.79_{{-1.96}}^{{+2.45}}$&-3.22&-3.22& $-4.57_{{-2.31}}^{{+2.32}}$&-2.84&-2.84\\ 
$\mu_{rel,E}$ (mas/yr) & $-2.61_{{-2.22}}^{{+1.30}}$&-2.60&-2.60& $-2.24_{{-1.69}}^{{+1.06}}$&-2.13&-2.13\\ 
$\mu_{rel,N}$ (mas/yr) & $-2.46_{{-2.48}}^{{+1.97}}$&-4.04&-4.04& $-2.69_{{-2.34}}^{{+2.33}}$&-4.44&-4.44\\ 
$\theta_E$ (mas) & $0.90_{{-0.43}}^{{+0.66}}$&1.10&1.10& $0.87_{{-0.41}}^{{+0.55}}$&1.10&1.10\\ 
$\pi_E$ & $0.12_{{-0.03}}^{{+0.06}}$&0.16&0.16& $0.14_{{-0.05}}^{{+0.08}}$&0.23&0.23\\ 
$\delta_{c,max}$ (mas) & $0.32_{{-0.15}}^{{+0.23}}$&0.39&0.39& $0.31_{{-0.14}}^{{+0.20}}$&0.39&0.39\\ 
\hline 
$\sum w_i$ & \multicolumn{3}{c}{0.43} & \multicolumn{3}{c}{0.57} \\ 
$log\mathcal{Z}$ & \multicolumn{3}{c}{24631.66} & \multicolumn{3}{c}{24631.92}

%% file: resolve_1sigma.txt
MB09260 & $32.08_{-9.20}^{+19.38}$ & $23.88_{-6.85}^{+14.42}$ & $0.01_{-0.00}^{+0.03}$ & $0.00_{-0.00}^{+0.01}$ \\ 
MB10364 & $30.72_{-12.19}^{+24.58}$ & $22.87_{-9.07}^{+18.29}$ & $0.01_{-0.00}^{+0.02}$ & $0.00_{-0.00}^{+0.02}$ \\ 
OB110037 & $17.45_{-4.01}^{+6.92}$ & $12.99_{-2.98}^{+5.15}$ & $0.10_{-0.07}^{+0.07}$ & $0.20_{-0.07}^{+0.08}$ \\ 
OB110310 & $22.03_{-8.95}^{+19.49}$ & $16.40_{-6.66}^{+14.51}$ & $0.00_{-0.00}^{+0.01}$ & $0.00_{-0.00}^{+0.01}$ \\ 
OB110462 DW & $16.91_{-1.89}^{+2.76}$ & $12.59_{-1.41}^{+2.05}$ & $0.07_{-0.04}^{+0.04}$ & $0.03_{-0.00}^{+0.04}$ \\ 
OB110462 EW & $15.81_{-3.06}^{+4.68}$ & $11.77_{-2.27}^{+3.49}$ & $0.06_{-0.05}^{+0.07}$ & $0.02_{-0.00}^{+0.07}$

%% file: tE_bh_prob.txt
MB09260 & $135 < t_E < 155$ days & 50 \\ 
MB10364 & $60 < t_E < 62$ days & 14 \\ 
OB110037 & $87 < t_E < 100$ days & 12 \\ 
OB110310 & $78 < t_E < 90$ days & 20 \\ 
OB110462 DW & $266 < t_E < 300$ days & 17 \\ 
OB110462 EW & $256 < t_E < 325$ days & 14

%% file: MB09260_add_err.txt
2009-10-01 & -- & 0.21 & -- & 4.0 \\ 
2009-10-19 & 0.12 & -- & 3.2 & -- \\ 
2010-03-22 & 0.14 & -- & 4.9 & -- \\ 
2010-06-14 & 0.14 & -- & 6.1 & -- \\ 
2010-10-20 & 0.13 & -- & 9.2 & -- \\ 
2011-04-19 & 0.14 & 0.36 & 12.8 & 4.0 \\ 
2011-10-24 & 0.15 & 0.23 & 4.0 & 4.0 \\ 
2012-09-25 & 0.15 & 0.27 & 4.3 & 4.0 \\ 
2013-06-17 & 0.16 & 0.28 & 9.8 & 4.0

%% file: MB10364_add_err.txt
2010-09-13 & 0.14 & 1.61 & 8.6 & 11.6 \\ 
2010-10-26 & 0.14 & 0.49 & 4.5 & 8.6 \\ 
2011-10-31 & 0.17 & 0.26 & 5.9 & 5.0 \\ 
2012-09-25 & 0.18 & 0.26 & 8.5 & 6.1 \\ 
2013-10-24 & 0.17 & 0.25 & 6.8 & 5.8

%% file: OB110037_add_err.txt
2011-08-15 & 0.11 & 0.22 & 16.0 & 8.5 \\ 
2011-09-26 & 0.11 & 0.18 & 4.1 & 14.0 \\ 
2011-11-01 & 0.10 & 0.21 & 4.0 & 3.8 \\ 
2012-05-07 & 0.10 & 0.26 & 12.9 & 12.2 \\ 
2012-09-25 & 0.10 & 0.22 & 6.7 & 4.3 \\ 
2013-10-21 & 0.10 & 0.23 & 3.9 & 4.9 \\ 
2014-10-26 & 0.10 & 0.21 & 4.2 & 9.7 \\ 
2017-03-13 & 0.10 & 0.21 & 4.9 & 5.1 \\ 
2017-09-04 & 0.11 & 0.21 & 8.5 & 19.6

%% file: OB110310_add_err.txt
2011-09-21 & 0.10 & 0.20 & 9.3 & 12.9 \\ 
2011-10-31 & 0.12 & 0.30 & 4.2 & 4.9 \\ 
2012-04-24 & 0.13 & 0.30 & 10.0 & 24.3 \\ 
2012-09-24 & 0.14 & 0.29 & 10.2 & 10.9 \\ 
2013-10-21 & 0.13 & 0.29 & 4.3 & 6.3 \\ 
2017-03-14 & 0.13 & 0.31 & 4.1 & 18.8 \\ 
2017-09-01 & 0.14 & 0.32 & 10.0 & 24.9

%% file: OB110462_add_err.txt
2011-08-08 & 0.13 & 0.41 & 18.1 & 17.0 \\ 
2011-10-31 & 0.19 & 0.37 & 3.9 & 5.4 \\ 
2012-09-09 & 0.19 & 0.36 & 4.7 & 7.2 \\ 
2012-09-25 & 0.19 & 0.36 & 6.8 & 7.2 \\ 
2013-10-22 & 0.19 & 0.36 & 3.6 & 4.8 \\ 
2014-10-26 & 0.19 & 0.39 & 3.8 & 10.7 \\ 
2017-08-29 & 0.20 & 0.40 & 5.4 & 14.7 \\ 
2021-10-01 & 0.17 & 0.36 & 7.5 & 12.9

%% file: bright_f606w_flc_positional_bias.txt
2011-08-08 & -0.025 $\pm$ 0.001 & 0.012 $\pm$ 0.001 & -0.001 $\pm$ 0.000 \\ 
2011-10-31 & -0.068 $\pm$ 0.002 & 0.168 $\pm$ 0.002 & -0.001 $\pm$ 0.000 \\ 
2012-09-09 & -0.216 $\pm$ 0.001 & -0.185 $\pm$ 0.001 & -0.005 $\pm$ 0.000 \\ 
2012-09-25 & 0.108 $\pm$ 0.002 & 0.136 $\pm$ 0.001 & -0.003 $\pm$ 0.000 \\ 
2013-10-22 & -0.431 $\pm$ 0.001 & 0.070 $\pm$ 0.001 & -0.007 $\pm$ 0.000 \\ 
2014-10-26 & -0.033 $\pm$ 0.001 & 0.298 $\pm$ 0.001 & -0.004 $\pm$ 0.000 \\ 
2017-08-29 & -0.144 $\pm$ 0.001 & 0.032 $\pm$ 0.001 & -0.011 $\pm$ 0.000 \\ 
2021-10-01 & 0.124 $\pm$ 0.002 & -0.014 $\pm$ 0.001 & -0.007 $\pm$ 0.000

%% file: bright_f814w_flc_positional_bias.txt
2011-08-08 & -0.049 $\pm$ 0.000 & -0.014 $\pm$ 0.000 & -0.001 $\pm$ 0.000 \\ 
2011-10-31 & -0.203 $\pm$ 0.001 & -0.070 $\pm$ 0.001 & -0.004 $\pm$ 0.000 \\ 
2012-09-09 & -0.278 $\pm$ 0.001 & -0.150 $\pm$ 0.001 & -0.016 $\pm$ 0.000 \\ 
2012-09-25 & -0.292 $\pm$ 0.001 & 0.025 $\pm$ 0.002 & -0.007 $\pm$ 0.000 \\ 
2013-10-22 & -0.490 $\pm$ 0.001 & -0.081 $\pm$ 0.001 & -0.011 $\pm$ 0.000 \\ 
2014-10-26 & -0.665 $\pm$ 0.001 & -0.138 $\pm$ 0.001 & -0.013 $\pm$ 0.000 \\ 
2017-08-29 & -0.258 $\pm$ 0.001 & -0.129 $\pm$ 0.001 & -0.013 $\pm$ 0.000 \\ 
2021-10-01 & -0.315 $\pm$ 0.000 & -0.001 $\pm$ 0.001 & -0.007 $\pm$ 0.000

%% file: hst_gaia_offset.txt
MB09260 & -2.56 $\pm$ 0.13 & -4.25 $\pm$ 0.10 \\ 
MB10364 & -2.70 $\pm$ 0.10 & -4.56 $\pm$ 0.10 \\ 
OB110037 & -2.31 $\pm$ 0.23 & -4.48 $\pm$ 0.20 \\ 
OB110310 & -2.20 $\pm$ 0.12 & -4.73 $\pm$ 0.08 \\ 
OB110462 & -2.34 $\pm$ 0.15 & -4.72 $\pm$ 0.11

%% file: priors.txt
$t_0$ (MJD) & $\mathcal{N}(55110, 75)$ & $\mathcal{N}(55460, 75)$ & $\mathcal{N}(55805, 75)$ & $\mathcal{N}(55810, 75)$ & $\mathcal{N}(55770, 75)$ & $\mathcal{N}(55770, 75)$\\
$u_0$ & $\mathcal{N}(0, 0.5)$ & " & " & " & " & " \\
$t_E$ (days) & $\mathcal{N}_T(200, 100, -1.8, 3)$  & " & " & " & " & "\\
$\pi_{E,E}$ & $\mathcal{N}(-0.02, 0.12)$ & " & " & " & " & " \\
$\pi_{E,N}$ & $\mathcal{N}(-0.03, 0.13)$ & " & " & " & " & " \\
$m_{base,O/M}$ (mag) & $\mathcal{N}(17.43, 0.2)$ & $\mathcal{N}(15.02, 0.2)$ & $\mathcal{N}(16.15, 0.1)$ & $\mathcal{N}(18.41,0.1)$ & $\mathcal{N}(16.41,0.1)$ & $\mathcal{N}(16.41,0.1)$ \\
$b_{SFF,O/M}$ & $\mathcal{U}(0,1.1)$ & " & "  & " & " & " \\
$m_{base,H8}$ (mag) & $\mathcal{N}(17.83, 0.05)$ & $\mathcal{N}(15.32, 0.05)$ & $\mathcal{N}(16.33, 0.05)$ & $\mathcal{N}(18.61, 0.05)$ & $\mathcal{N}(19.85, 0.05)$ & $\mathcal{N}(19.85, 0.05)$\\
$b_{SFF,H8}$ & $\mathcal{U}(0, 1.05)$ & " & " & " & " & "\\
$m_{base,H6}$ (mag) & $\mathcal{N}(20.74, 0.05)$ & $\mathcal{N}(16.50, 0.05)$ & $\mathcal{N}(18.33, 0.05)$ & $\mathcal{N}(21.34, 0.05)$ & $\mathcal{N}(22.03, 0.05)$ & $\mathcal{N}(22.03, 0.05)$ \\
$b_{SFF,H6}$ & $\mathcal{U}(0, 1.05)$ & " & " & " & " & " \\
\hline
$\log \sigma_{0,O/M}$ (mag) & $\mathcal{N}(0, 5)$ & " & " & " & --  & -- \\
$\rho_{O/M}$ (days) & $\Gamma^{-1}(0.448,0.063)$ & $\Gamma^{-1}(0.448,0.113)$ & $\Gamma^{-1}(0.473,0.162)$ & $\Gamma^{-1}(0.527,0.450)$ &  -- & -- \\
$\log \omega_{0,O/M}^4 S_{0,O/M}$ & $\mathcal{N}(3.53e-04, 5)$ & $\mathcal{N}(8.41e-06, 5)$ & $\mathcal{N}(3.60e-05, 5)$ & $\mathcal{N}(1.02e-03, 5)$ &  -- & --\\
(mag$^2$ days$^{{-3}}$) & & & & & & \\
$\log \omega_{0,O/M}$ (days$^{-1}$) & $\mathcal{N}(0, 5)$ & " & " & " & -- & --\\
\hline
$\log_{10}(\theta_E$) (mas) & $\mathcal{N}(-0.2, 0.3)$ & " & " & " & $\mathcal{N}(0.5, 0)$ & $\mathcal{N}(0.5, 0)$\\
$\pi_S$ (mas) & $\mathcal{N}_T(0.1126, 0.0213, -2.94, 90)$ & " & " & " & " & "\\
$x_{S0,E}$ (arcsec) & $\mathcal{U}(0.213,0.250)$ & $\mathcal{U}(0.086,0.158)$ & $\mathcal{U}(-0.034,0.091)$ & $\mathcal{U}(-0.108,-0.103)$ & $\mathcal{U}(0.227,0.233)$ & $\mathcal{U}(0.227,0.233)$\\
$x_{S0,N}$ (arcsec & $\mathcal{U}(-0.697,-0.683)$ & $\mathcal{U}(-0.096,-0.068)$ & $\mathcal{U}(-0.122,-0.104)$ & $\mathcal{U}(-0.228,-0.154)$ & $\mathcal{U}(-0.235,-0.183)$ & $\mathcal{U}(-0.235,-0.183)$\\
$\mu_{S,E}$ (mas/yr) & $\mathcal{U}(-5.96,1.12)$ & $\mathcal{U}(-7.93,-1.78)$ & $\mathcal{U}(0.87,8.05)$ & $\mathcal{U}(-1.30,0.95)$ & $\mathcal{U}(-4.82,4.99)$ & $\mathcal{U}(-4.82,4.99)$ \\
$\mu_{S,N}$ (mas/yr) & $\mathcal{U}(-2.37,4.17)$ & $\mathcal{U}(-7.28,3.41)$ & $\mathcal{U}(-1.67,2.94)$ & $\mathcal{U}(-4.55,-0.49)$ & $\mathcal{U}(-3.49,5.91)$ & $\mathcal{U}(-3.49,5.91)$